%% file: main.tex
\documentclass[letterpaper,12pt]{report}
\usepackage[dissertation]{USCthesis}

\usepackage{hyperref}
\usepackage{setspace}
\usepackage{tabularx}

\usepackage[margin=1in,footskip=.5in]{geometry}
\usepackage{graphicx}
\graphicspath{{./}{../}{figures/}{../figures/}}

\usepackage[
    style=nature,
    sorting=none,
    isbn=false,
    url=false,
    doi=true,
    eprint=false,
    date=year,
    maxnames=6,
    minnames=6
]{biblatex}
\AtEveryBibitem{\clearfield{eventtitle}} 
\AtEveryCitekey{\clearfield{eventtitle}}
\AtEveryBibitem{\clearfield{pagetotal}} 
\AtEveryCitekey{\clearfield{pagetotal}}
\usepackage[english]{babel}
\usepackage{blindtext}
\usepackage{amsmath,amssymb,amsthm,dsfont}
\usepackage{braket}
\usepackage{empheq}
\usepackage{subfigure}
\usepackage{cancel}
\newcommand{\omegac}{\omega\ped{c}}
\usepackage{siunitx}

\usepackage{cancel}
\usepackage{relsize}

\newcommand{\bes} {\begin{subequations}}
\newcommand{\ees} {\end{subequations}}
\newcommand{\beq}{\begin{equation}}
\newcommand{\eeq}{\end{equation}}
\newcommand{\ba}{\begin{eqnarray}}
\newcommand{\ea}{\end{eqnarray}}

\newcommand\norm[1]{\left\lVert#1\right\rVert}
\newcommand\eff{\mathrm{eff}}

\newcommand{\ketbra}[1]{|{#1}\rangle\langle#1|}
\newcommand{\vertiii}[1]{{\| #1\|}}
\newcommand{\ignore}[1]{}

\newtheorem*{theorem*}{Theorem}

\def\b{\beta}

\def\n{\nu}

\newcommand{\cL}{\mathcal{L}}

 \newcommand{\fs}{fluctuators }
\newcommand{\eps}{\varepsilon}
\newcommand{\ie}{i.\,e.}

\newcommand{\ped}[1]{_\text{#1}}
\newcommand{\api}[1]{^\text{#1}}

\newcommand{\rng}[2]{\ensuremath{[#1, #2]}}
\newcommand{\rngopen}[2]{\ensuremath{(#1, #2)}}
\newcommand{\rnglopen}[2]{\ensuremath{(#1, #2]}}

\newcommand{\nspin}{N}
\newcommand{\ham}{H}
\newcommand{\hamtf}{V\ped{TF}}

\newcommand{\hamtarget}{\ham_0}
\newcommand{\hamsysbath}{\ham_{SB}}
\newcommand{\tf}{\tau}
\newcommand{\sinv}{s\ped{inv}}
\newcommand{\tinv}{t\ped{inv}}
\newcommand{\scrit}{s\ped{c}}
\newcommand{\tcrit}{t\ped{c}}
\newcommand{\lpause}{t\ped{p}}
\newcommand{\mingap}{\Delta}
\newcommand{\sgap}{s_\mingap}
\newcommand{\tgap}{t_\mingap}
\newcommand{\pgs}{P_0}
\newcommand{\diss}{\mathcal{D}}
\newcommand{\iu}{i}

\newcommand*{\ev}[1]{\langle #1 \rangle}

\usepackage{algorithm}
\usepackage{algpseudocode}
\usepackage{setspace}
\usepackage{xcolor}

\begin{document}


\title{\textbf{\Large{Open-system modeling of quantum annealing: theory and applications}}}

\author{Ka Wa Yip}

\committee{A.~ & (Chair)\\*
           B.~\\*
           C.~\\*
           D.~\\*
           E.~Federico Spedalieri & (Outside Member)}

\majorfield{Physics}
\submitdate{May 2021}  



\begin{preface}


  \prefacesection{Acknowledgements}
  \input{acknowledgements}

  \tableofcontents
  \listoftables   
  \listoffigures

  \prefacesection{Abstract}
  \input{abstract}

\end{preface}


\input{Introduction}
\input{chapter1}
\input{chapter2}
\input{chapter3}

\input{chapter4}
\input{chapter5}

\input{chapter6}



\include{Conclusion}

\begin{singlespace}
\phantomsection
\addcontentsline{toc}{chapter}{References}%
\markboth{References}{References}%
\printbibliography[title=References]
\end{singlespace}

\phantomsection
\addcontentsline{toc}{chapter}{Appendices}%
\markboth{Appendices}{Appendices}%
\chapter*{Appendices}
\renewcommand\thesection{\Alph{section}}
\renewcommand*{\thesubsection}{\Alph{section}.\arabic{subsection}}
\begingroup
\numberwithin{equation}{section}
\input{appendix_proof}

\endgroup


\end{document}

%% file: acknowledgements.tex
I would like to thank my advisor Prof. Daniel Lidar for his supports, teachings and guidance throughout my Ph.D. This dissertation would be impossible without him, and his wisdom and leadership. Prof. Lidar is a great mentor and fatherly figure. He taught me a lot and I am honored to be his student. Next I would like to thank Dr. Tameem Albash for his guidance and discussions especially regarding all the details in simulations and physics, at the early stage of my doctoral studies. Finally, without naming anyone, I would like to thank all the friends, staff, and professors at USC. The staff at the department provided excellent administrative assistance for my doctoral studies.

%% file: abstract.tex
In this dissertation, we explore how quantum annealing (QA) and its applications behave in an open system setting. We give derivations and numerical recipes for  effective parallel simulation methods for time-dependent open dynamics of quantum annealing devices. We consider the weak-coupling limit to an environment and also the case of $1/f$ noise. The stochastic time-dependent quantum trajectories technique can be utilized in studying weak measurements and feedback error correction in quantum annealing. Then we focus on open-system descriptions of reverse annealing (RA), which is a promising variant and application of quantum annealing. We show that, with various simulation tools, reverse annealing can benefit from the interaction between the annealer and its environment.

%% file: Introduction.tex
\chapter{Introduction}
\label{cha:introduction}
Quantum annealing (QA)~\cite{kadowaki:qa, farhi2001quantum} and the broader class of adiabatic quantum computing (AQC)~\cite{albash:review-aqc} are quantum approaches to solving combinatorial optimization problems and sampling problems. Quantum annealing obtains the global solution(s) by using quantum fluctuations to escape from local minima~\cite{kadowaki:qa, Santoro, RevModPhys.80.1061, morita:mathematical-foundation-qa,albash:review-aqc, Hauke2020}. Quantum annealing is generally considered to be more noise resilient than gate-based quantum computing, and there are many ongoing developments (e.g.~\cite{novikov2018exploring}) on how to further improve the coherence of quantum annealers.

Accurate open quantum system theories of QA along with efficient simulation procedures serve at least two purposes: 1) to aid in the design of experimental quantum annealers and derivation of annealing protocols/applications such as advanced annealing schedules and reverse annealing, 2) to help the development of error correction techniques tailored to noisy quantum annealers.

\begin{figure}
\centering
\includegraphics[width=0.5\linewidth]{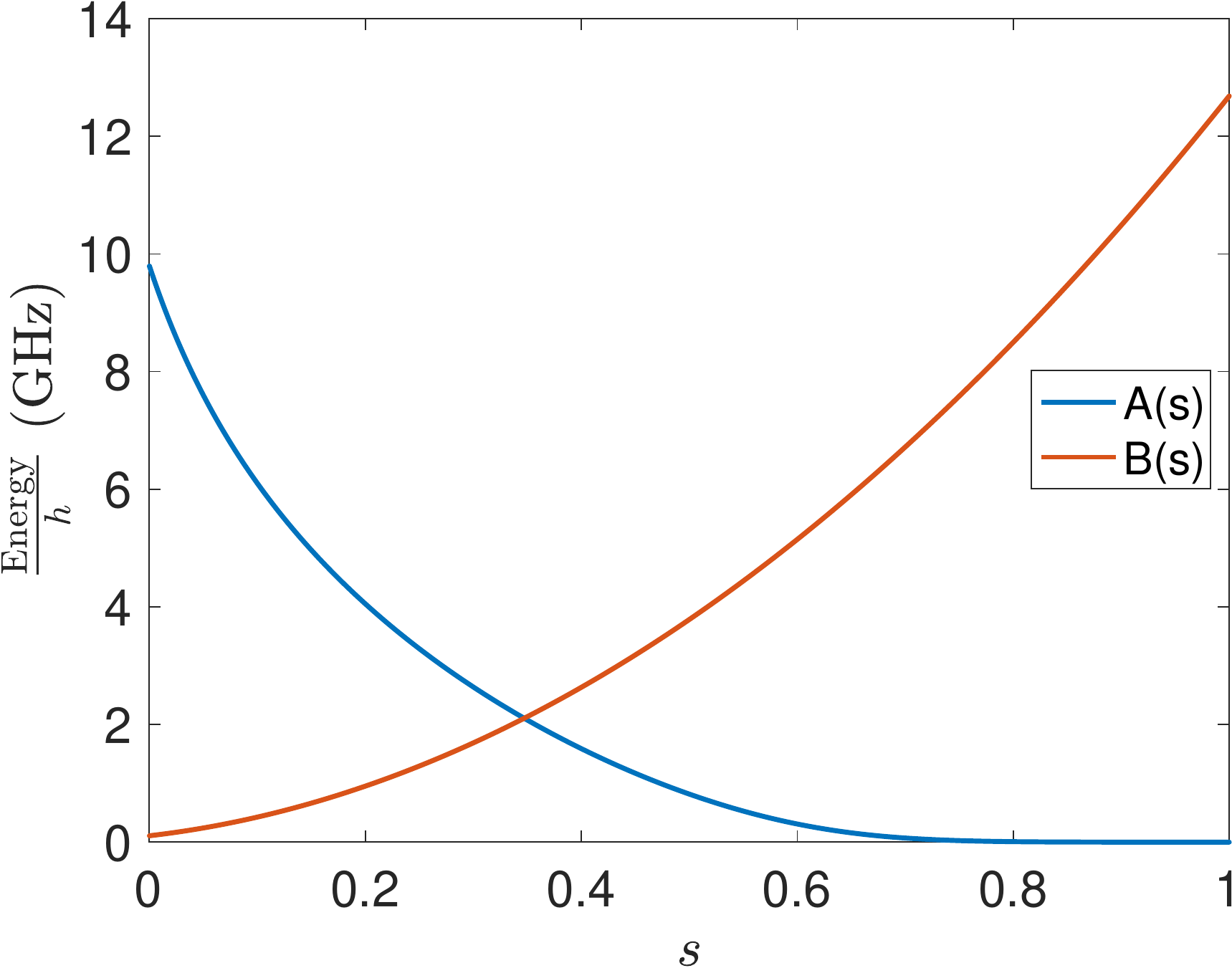}
\caption{D-Wave 2000Q NASA annealing schedules.}
\label{fig: DWavenasaschedule}
\end{figure}

The dynamics in quantum annealers is driven by a time-dependent system Hamiltonian, which in this thesis We will refer to specifically as the annealing Hamiltonian. In standard forward quantum annealing, it takes the following form:
\begin{equation}
H(s) = -\frac{A(s)}{2} V_{\text{TF}} + \frac{B(s)}{2}H_\text{p} \,.
\label{eq:Hising}
\end{equation}
$V_{\text{TF}} = \sum_{i}^{N} \sigma^{i}_x$ is the transverse field, $N$ is the number of qubits, and $H_\text{p}$ is the problem Hamiltonian. Here $s = t/t_f$ is the normalized time, with $t_f$ the total physical annealing time. The functions $A(s)$ and $B(s)$ are the annealing schedules (see for example Fig.~\ref{fig: DWavenasaschedule}.)

One metric for measuring the success of quantum annealing is the ground state population at the end of the anneal. This is equivalent to the success probability of reaching the correct solution, often simply called the success probability. One can model and estimate this population by a variety of open-system analysis tools. There are many previous studies about the open-system modeling of quantum annealing~\cite{amin_decoherence_2009, smirnov2018theory, davies1978open, Albash:2015nx}. Notably, a time-dependent adiabatic master equation was developed to describe decoherence of quantum annealers in the weak-coupling limit~\cite{ABLZ:12-SI, Albash:2015nx}. Under this master equation the decoherence occurs in the instantaneous eigenbasis of $H(s)$. However, it suffers from two problems: 1.~Solving the master equation for a density matrix requires storing and updating $N^2-1$ real numbers, which becomes infeasible for large $N$; 2.~The assumed weak-coupling limit does not apply to slow noise such as $1/f$ noise, which is present, e.g., in superconducting-qubit based quantum annealers. While there exists literature~\cite{amin2008macroscopic, smirnov2018theory} addressing the modeling of slow noise in current annealing device, knowledge about how in particular the $1/f$ power spectrum affects QA and efficient modeling of $1/f$-noise for QA is lacking. The first two chapters of this dissertation propose new and original solutions to these two problems. The first chapter proposes a time-dependent quantum trajectories method adapted to QA, which gives a factor of $N$ reduction in the simulation cost. The cost of running many trajectories can further be minimized with parallel computing. The quantum trajectories method for QA also provides insight into individual quantum jump trajectories and their jump statistics, thus shedding light on open system quantum adiabatic evolution beyond the master equation. The second chapter proposes stochastic trajectories methods for the slow noise of $1/f$ power spectrum, with the time-dependent system Hamiltonian of QA. We also analyze how different elements such as temperature and $1/f$ noise range play a role in the decoherence in QA.

Moreover, past research in weak measurement and feedback-based error correction focused on time-independent Hamiltonians and gated-based quantum computing. Based on the time-dependent trajectories techniques we develop, we further investigate how to perform weak measurement and feedback error correction on quantum annealers.

Standard forward annealing with the annealing Hamiltonian (Eq.~\ref{eq:Hising}) has the major drawback of suffering from thermal excitations, which take a certain portion of population out of the solution states at the end of the anneal. There care at least two remedies to this drawback.  One approach is to start in a random excited state of the problem Hamiltonian $H_p$ and simply make use of relaxation to get to the ground state of $H_p$, which has been studied in~\cite{campos2016relaxation}. However, one can do more. Unlike forward annealing, where $s$ runs from $0$ to $1$, one can start from an excited state of the problem Hamiltonian $H_p$ at $s=1$,\footnote{In Eq.~\eqref{eq:Hising}, we do not assume that $s$ goes from $0$ to $1$.} \textit{reverse anneal} to an inversion point $s=s^*$, and then forward anneal back to $s=1$. This approach combines both relaxation from excited states and annealing, and allows most of the relaxation to occur at the smaller gap at $s=s^*$. As for the second remedy, we observe that the annealing Hamiltonian (Eq.~\ref{eq:Hising}) is not the unique form of a Hamiltonian drive for quantum annealers. One can proceed with standard forward annealing, but with an extra term in Eq.~\eqref{eq:Hising}, sometimes called initialization Hamiltonian or catalyst Hamiltonian~\cite{perdomo:sombrero,Albash-Lidar:RMP,albash2021diagonal}. This additional term provides an easier annealing path to the solution and thus leads to a higher population in the solution state at the end of the anneal. The success of adding the initialization/catalyst Hamiltonian often depends on how close it is to the correct solution. The two remedies can be summarized within the general framework of reverse annealing. The first is ``reversed" in the sense that it starts at $s=1$; the second is ``reversed" in the sense that the catalyst term depends on our knowledge about the solution states at $s=1$.

Reverse annealing~\cite{perdomo:sombrero,chancellor:reverse,King:2018aa,Ottaviani2018,nishimori:reverse-pspin, nishimori:reverse-pspin-2,Venturelli2019,marshall} in general can therefore be separated into two formalisms: iterated reverse annealing (IRA) and  adiabatic reverse annealing (ARA). As mentioned before, one can start at $s=1$ with an initial state that is usually the excited state of $H_p$, anneal reversely to an inversion point $s=s^*$, and then anneal forwardly to $s=1$. The procedure can be performed more than once and may be iterated with the last output state as the new input state. It is called iterative reverse annealing (IRA) and its function has been included in the current generation of quantum annealers built by D-Wave~\cite{Crosson2020}. Adiabatic reverse annealing (ARA) is similar to standard forward annealing, but with an extra initialization/catalyst Hamiltonian.  The two formalisms have been studied previously in a closed system setting. In this dissertation we address the problem of how to model and predict their performance in an open-system settings, which has so far been lacking in the  literature. We show that both reverse annealing variants can benefit from the interaction between the annealer and the bath, in that the final ground state population can be higher than in a closed system setting. Therefore, a quantum annealer can take advantage of noise to improve its performance in the sense of increased success probabilities.

This dissertation is organized as follows. In chapter~\ref{chap: qt}, we derive the time-dependent quantum trajectories solution to adiabatic master equation, which yields a quadratic saving in time cost compared to the master equation approach. In chapter~\ref{chap: 1f}, we extend the stochastic trajectory formulation to the simulation of slow noise with a $1/f$ power spectrum.
In chapter~\ref{chap: feedback}, we describe and study the weak measurement and feedback error correction protocol tailored to QA. Chapters~\ref{chap:ira}-\ref{chap: ara} focus on reverse annealing applications. In Chapter~\ref{chap:ira}, we focus on the open-system simulation of iterative reverse annealing (IRA) and show how it benefits from environmental-induced relaxation. In chapter~\ref{chap: IRA_exp}, we study the iterative reverse annealing protocol in D-Wave and perform open-system simulations that agree with the experimental data. In chapter~\ref{chap: ara}, we focus on the open-system description of another variant of reverse annealing called adiabatic reverse annealing (ARA). In chapter~\ref{chap: conclusions} we provide a brief summary along with an outlook and perspective on the significance of the results obtained in this thesis.

%% file: chapter1.tex
\chapter{Quantum trajectories for time-dependent adiabatic master equations}
\label{chap: qt}
\section{Introduction}
In this chapter we describe a quantum trajectories technique for the unraveling of the quantum adiabatic master equation in Lindblad form.  The goal is that by evolving a complex state vector of dimension $N$ instead of a complex density matrix of dimension $N^2$, simulations of larger system sizes become feasible. The cost of running many trajectories, which is required to recover the master equation evolution, can be minimized by running the trajectories in parallel, making this method suitable for high performance computing clusters. In general, the trajectories method can provide up to a factor $N$ advantage over directly solving the master equation. In special cases where only the expectation values of certain observables are desired, an advantage of up to a factor $N^2$ is possible. We test the method by demonstrating agreement with direct solution of the quantum adiabatic master equation for $8$-qubit quantum annealing examples. We also apply the quantum trajectories method to a $16$-qubit example originally introduced to demonstrate the role of tunneling in quantum annealing, which is significantly more time consuming to solve directly using the master equation. The quantum trajectories method provides insight into individual quantum jump trajectories and their statistics, thus shedding light on open system quantum adiabatic evolution beyond the master equation. This chapter is based on~\cite{yip:mcwf}.

With the growing ability to control and measure ever-larger quantum systems, understanding how to model the interactions between open quantum systems and their environment has become exceedingly important \cite{Wiseman:book}. 
The open system dynamics is often described in terms of a master equation in Lindblad form, describing the effective dynamics of the quantum system after the environmental degrees of freedom have been traced out \cite{Breuer:2002}.  An equivalent approach is that of quantum trajectories (also known as the Monte Carlo wavefunction method)~\cite{carmichael2009statistical,dum1992monte,molmer1993monte}, which can be understood as an unraveling of the master equation in Lindblad form, and which generates a stochastic process whose average is fully equivalent to the master equation (for a review, see Ref.~\cite{daley2014quantum}).  Each trajectory in this approach can also be viewed as the result of continuous indirect measurements of the environment in a certain basis \cite{brun:719}. A quantum trajectories approach exists also for non-Markovian master equations \cite{Imamoglu:94,Breuer:2004wq}. 

While a vast literature exists on the topic of quantum trajectories for time-independent master equations, much less is known for the case of time-dependent master equations (see, e.g., Ref.~\cite{Caiaffa:2017aa}), which is our focus here. Specifically, we focus on the case of open systems evolving adiabatically according to a time-dependent Hamiltonian, weakly coupled to the environment \cite{springerlink:10.1007/BF01011696,ABLZ:12-SI}. This is particularly relevant in the context of quantum annealing and more generally adiabatic quantum computing, whereby the computation proceeds via a time-dependent Hamiltonian and the result of the computation is encoded in the ground state of the final Hamiltonian (for reviews see Refs.~\cite{RevModPhys.80.1061,Albash-Lidar:RMP}). A large body of literature exists on the use of quantum Monte Carlo methods in the context of quantum annealing (see, e.g., Refs.~\cite{Santoro,Heim:2014jf,Albash:2017aa}), but these methods focus on equilibrium properties, while here we are interested in dynamics. The interplay between the key quantities that determine adiabaticity and non-unitary dynamics has not been previously explored in the setting of Monte Carlo wavefunction methods, and here we resolve this question by finding an upper bound on the size of the Monte Carlo time-step. Nor has the question of reducing the computational cost of simulations of the adiabatic master equation via quantum trajectories been discussed so far, and we address this here.

Thus, here we develop the first treatment of a quantum trajectories unravelling of a time-dependent adiabatic master equation (AME).
We make a formal comparison between the quantum trajectory unraveling of the Lindblad master equation with time-independent and time-dependent operators, and discuss the validity of applying it to the unraveling of the AME.   While our analysis closely follows the standard time-independent approach, the time-dependent case results in new additional validity conditions that must be satisfied.

The individual trajectories of the AME shed new light on how the average case captured by the AME emerges.  When the quantum state is a pure energy eigenstate and the unitary evolution is adiabatic, the drift term in the quantum trajectories approach vanishes, and the quantum state follows the instantaneous eigenstates until a jump occurs.  We can associate these jumps with an excitation or relaxation process, depending on the direction of the jump.  This provides an intuitive (yet rigorous) picture for how the averaged dynamics of the AME arises.

An important advantage of the quantum trajectories approach is that for an $N$-dimensional system, one quantum trajectory requires storing and updating $2N-1$ real numbers, while solving the master equation for a density matrix requires storing and updating $N^2-1$ real numbers. This quadratic saving allows simulations of systems with sizes that are infeasible by directly solving the master equation. The tradeoff is that many trajectories must be run in order to accurately approximate the solution of the master equation, but this tradeoff can be reduced by using many parallel processes to represent each trajectory.

Our presentation is organized as follows. In Sec.~\ref{sec:II} we briefly review the AME. We unravel the AME in Sec.~\ref{sec:III} into quantum trajectories taking the form of quantum jumps, allowing for an arbitrary time-dependence of the Hamiltonian and Lindblad operators. In Sec.~\ref{sec:IV} we provide an algorithmic implementation for our adiabatic quantum trajectories and in Sec.~\ref{sec:V} we present three case studies. We perform a cost comparison between the direct simulation of the AME and the quantum trajectories method in Sec.~\ref{sec:timequantitative}. Additional technical details and proofs are provided in the Appendices. 

\section{Adiabatic master equation in Lindblad form}
\label{sec:II}
We focus on the AME in Lindblad form, which can be derived with suitable approximations (in the weak coupling limit after performing the Born-Markov, rotating wave, and adiabatic  approximation) from first principles starting
from the system Hamiltonian $H_{\textrm{S}}$, 
the environment Hamiltonian $H_B$, and the interaction Hamiltonian $H_I = g \sum_\alpha A_{\alpha} \otimes B_{\alpha}$, with system operators $A_\alpha$, environment operators $B_{\alpha}$, and system-bath coupling strength $g$~\cite{ABLZ:12-SI}.  
The adiabatic (Lindblad) master equation describes the evolution of the system density matrix $\rho(t)$ and has

\beq
\label{eqt:ME2-H}
\frac{d}{dt} \rho(t) = -i \left[H_{\textrm{S}}(t) + H_{\textrm{LS}}(t), \rho(t) \right] + \mathcal{L}_{\textrm{WCL}} [\rho(t)] \ , \\
\eeq
where $H_{\textrm{LS}}(t)$, which commutes with $H_{\textrm{S}}(t)$, is a Lamb shift Hamiltonian arising from the interaction with the environment.  The dissipative term  $\mathcal{L}_{\mathrm{WCL}}$ takes the form:

\begin{eqnarray}
\mathcal{L}_{\textrm{WCL}}  [\rho(t)]   &\equiv& \sum_{\alpha, \beta} \sum_{\omega} \gamma_{\alpha \beta}(\omega) \left(L_{\beta,\omega}(t) \rho(t) L_{\alpha, \omega}^{\dagger}(t)\phantom{\frac{1}{2}} \right. \notag \\
&& \hspace{-0.5cm} \left. \qquad\qquad - \frac{1}{2} \left\{ L_{\alpha,\omega}^{\dagger}(t) L_{\beta,\omega}(t) , \rho(t) \right\} \right) \ , \label{eqt:ME2-L}
\end{eqnarray}
where the sum over $\omega$ is over the Bohr frequencies (eigenenergy differences) of $H_{S}$, $\gamma_{\alpha \beta}(\omega)$ is an element of the positive matrix $\gamma$, and satisfies the Kubo-Martin-Schwinger (KMS) condition if the bath is in a thermal state with inverse temperature $\beta = 1/T$:
\beq
\gamma_{\alpha \beta}(- \omega) = e^{-\beta\omega} \gamma_{\beta\alpha }(\omega)\ .
\label{eq:KMS}
\eeq
The time-dependent Lindblad operators are given by:
\begin{align}
L_{\alpha, \omega}(t) &= \sum_{a,b} \delta_{\omega, \eps_b(t) - \eps_a(t)} \bra{\eps_a(t)} A_\alpha \ket{\eps_b(t)} | \eps_a(t) \rangle  \langle \eps_b(t)| \ ,
\label{eq:Lindblad2}
\end{align}
where $\ket{\varepsilon_a(t)}$ is the $a$-th instantaneous energy eigenstate of $H_{\textrm{S}}(t)$ with eigenvalue $\varepsilon_a(t)$. 
With this form for the Lindblad operators, decoherence can be understood as occurring in the instantaneous energy eigenbasis~\cite{Albash:2015nx}. 

For the purpose of unravelling the above master equation into quantum trajectories, it is convenient to diagonalize the matrix $\gamma$ 
by an appropriate unitary transformation $u(\omega)$:
\begin{equation}
\sum_{\alpha, \beta} u_{i, \alpha}(\omega) \gamma_{\alpha \beta}(\omega) u_{j, \beta}(\omega)^{\dagger} = \left(\begin{matrix} \gamma'_1(\omega) & 0 &\hdots \\0 & \gamma'_2(\omega) & \hdots \\ \vdots & \vdots & \ddots &
\end{matrix}\right)_{i, j} \ ,
\label{eq:unitarytransform}
\end{equation}
and to define new operators $A_{i,\omega}(t)$ given by
\begin{equation}
L_{\alpha, \omega}(t) = \sum_{i}u_{i,\alpha}(\omega) A_{i,\omega}(t) \ .
\label{eq:L}
\end{equation}
In this basis, we can write the dissipative part in diagonal form as:
\begin{eqnarray}
\label{eqt:dissdiag}
{\mathcal{L}}_{\textrm{WCL}}  [\rho(t)]  & = &\sum_{i}\sum_{\omega}\gamma'_i(\omega)\left(
A_{i,\omega}(t) \rho(t) A^\dagger_{i,\omega}(t)\phantom{\frac{1}{2}} \right. \notag \\
&& \hspace{-0.5cm} \left. \qquad\qquad - \frac{1}{2} \left\{ A_{i,\omega}^{\dagger}(t) A_{i,\omega}(t) , \rho(t) \right\} \right) \ .
\end{eqnarray}

\section{Stochastic Schr\"odinger Equation}
\label{sec:III}
With Eq.~\eqref{eqt:dissdiag}, the master equation Eq.~\eqref{eqt:ME2-H} is in diagonal form and can be unravelled into quantum trajectories. The trajectory is described by a stochastic differential equation (SDE) in the form of jumps or diffusion. Let us consider the case where the coefficients $\gamma'_i(\omega)$ in Eq.~\eqref{eqt:dissdiag} also depend on time. If all $\gamma'_i(\omega, t) \geq 0$, then the dynamics is completely positive (CP)-divisible~\cite{laine2010measure}, and the master equation can be unravelled by using the known unravelling of the the time-independent SDE case~\cite{Breuer:2002,gardiner2004quantum,brun:719}, simply by replacing the time-independent operators and coefficients by the time-dependent ones.

Such an unravelling is also possible, but with modifications, when the dynamics is positive (P)-divisible, i.e., where $\gamma'_i(\omega, t)$ need not be all positive.\footnote{The condition on $\gamma'_i(\omega, t)$ such that the map is P-divisible can be found in the proof given in~\cite{breuer2009stochastic} or Eq.~(25) in~\cite{Caiaffa:2017aa}.} This can be in the form of:
\begin{itemize}
\item {Jump} trajectories: the master equation is unravelled via the non-Markovian quantum jump method (NMQJ)~\cite{piilo2008non, breuer2009stochastic,piilo2009open,harkonen2010jump}, where terms with negative coefficients $\gamma'_i(\omega, t)$ describe the negative channel.
\item {Diffusive} trajectories: recent work on diffusive trajectories~\cite{Caiaffa:2017aa} replaces $\gamma'_i(\omega, t)$ and the operators by the eigenvalues and eigenvectors of a positive transition rate operator $W$ (Eq.~(11) in~\cite{Caiaffa:2017aa}). P-divisible dynamics can be unravelled into a SDE in terms of such eigenvalues and eigenvectors.
\end{itemize}

In the following, we focus on the case of CP maps [with all $\gamma'_i(\omega)\geq 0$] and unravel the master equation in the quantum jumps picture.

\subsection{Unravelling the master equation}
First we absorb the $\gamma'$ coefficients into the definition of $A_i$:
\beq 
\sqrt{\gamma'_i(\omega)}A_{i,\omega}(t)\rightarrow A_i(t) \ .
\label{eqt:Aired}
\eeq
In this redefinition, the index $i$ now includes the Bohr frequencies.
We write Eq.~\eqref{eqt:ME2-H} in terms of an effective non-Hermitian Hamiltonian $H_{\eff}$:
\begin{align}
\label{eqt:effdiagonalform}
\frac{d}{dt}{\rho}(t) &= - i\left(H_{\text{eff}}(t)\rho_S(t) - \rho_S(t)H^{\dagger}_{\text{eff}}(t)\right) \nonumber\\
&\phantom{{}=}+ \sum_{i}A_{i}(t) \rho_S(t) A^\dagger_{i}(t)   \,, 
\end{align}
where 
\begin{equation}
\label{eq:Heff}
H_{\text{eff}}(t) = H_{\textrm{S}}(t) + H_{\text{\textrm{LS}}}(t) - \frac{i}{2} \sum\limits_{i}{A}_{i}^{\dagger}(t) {A}_{i}(t)\,.
\end{equation}%
Equation~\eqref{eqt:effdiagonalform} can be unravelled into quantum trajectories in the quantum jumps picture, where each trajectory describes the stochastic evolution of a pure state (if the initial state $\rho$ is mixed, $\rho = \sum_i p_i \ketbra{\psi_i}$, then the evolution can be performed on each initial pure state).  The stochastic evolution of the pure state can be written in terms of a stochastic Schr\"{o}dinger equation (in It$\hat{\text{o}}$ form), the ensemble average of which is equivalent to the master equation:
\begin{align}
\label{eqt:sse}
d\ket{\psi(t)} &= \left(- i{H}_{\text{eff}}(t) + \frac{1}{2}\sum_{i}\braket{{A}_{i}^{\dagger}(t){A}_{i}(t)}\right) dt \ket{\psi(t)} \nonumber \\
&\phantom{{}=}+ \sum_{i} dN_{i}(t)\left(  \frac{{A}_{i}(t)}{\sqrt{\braket{{A}^{\dagger}_{i}(t){A}_{i}(t)}}} - \mathds{1}        \right)\ket{\psi(t)}
\end{align}
where 
\beq
\braket{{A}_{i}^{\dagger}(t){A}_{i}(t)} \equiv \braket{\psi(t)|{A}_{i}^{\dagger}(t){A}_{i}(t)|\psi(t)} = \lVert{A}_{i}(t)\ket{\psi(t)}\rVert^2\ .
\label{eq:12}
\eeq 
We give a derivation of Eq.~\eqref{eqt:sse} below.
The first term on the r.h.s. of Eq.~\eqref{eqt:sse} gives a deterministic evolution composed of a Hermitian component [$-i(H_{\text{S}}(t) + H_{\text{\textrm{LS}}}(t))$] and a ``drift'' component 
\beq
D(t) \equiv  \frac{1}{2}\sum_{i} A_{i}^{\dagger}(t)A_{i}(t) - \braket{A_{i}^{\dagger}(t)A_{i}(t)} \ ,
\eeq
and the second term describes the stochastic jump process.  The stochastic variable $dN_{i}(t)\equiv N_i(t+dt) - N_i(t)$ is the number of jumps of type $i$ in the interval $dt$, where we have denoted by $N_i(t)$ the number of jumps of type $i$ up to time $t$.  The expectation value of the stochastic variable is given by  \cite{Breuer:2002}:
\beq
E[dN_{i}(t)] = \braket{{A}_{i}^{\dagger}(t){A}_{i}(t)}dt \,.
\eeq
Since the probability of a jump occurring scales linearly with $dt$, the probability of having more than one jump vanishes faster than $dt$, so as $dt\to 0$ only one jump out of all possible types during $dt$ is permitted. Therefore we can write~\cite{gardiner2004quantum}:

\begin{equation}
\label{eqt:stochastic1}
dN_{i}(t) = \left\{
                \begin{array}{ll}
                  1 &\text{with prob.                } \braket{{A}_{i}^{\dagger}(t){A}_{i}(t)}dt\\
                  0 &\text{with prob.                }  1 - \braket{{A}_{i}^{\dagger}(t){A}_{i}(t)}dt
                \end{array}
\right.
\end{equation}
with the It$\hat{\text{o}}$ table:
\bes
\label{eqt:stochastic2}
\begin{align}
dN_{j}(t)dN_{k}(t) &= \delta_{jk}dN_{j}(t)  \\
dN_{j}(t)dt &= 0 \,.
\end{align}
\ees
From Eq.~\eqref{eqt:stochastic1}, the probability of any jump occurring, $\sum_{i}\braket{{A}_{i}^{\dagger}(t){A}_{i}(t)}dt$, is small compared to the probability of no jump occurring, so $\sum_{i}dN_{i}(t) = 0$ most of the time. During the infinitesimal time-step $dt$, if $\sum_{i}dN_{i}(t) = 0$, then only the deterministic evolution takes places; if however $\sum_{i}dN_{i}(t) = 1$, then a jump occurs. When a jump occurs, it dominates over the deterministic evolution, which is proportional to $d t$, and the deterministic part can be ignored.

\subsection{Deterministic evolution and jump process}
\label{ssec:A}
\begin{figure}[t!]
\centering
\includegraphics[width=4.7in]{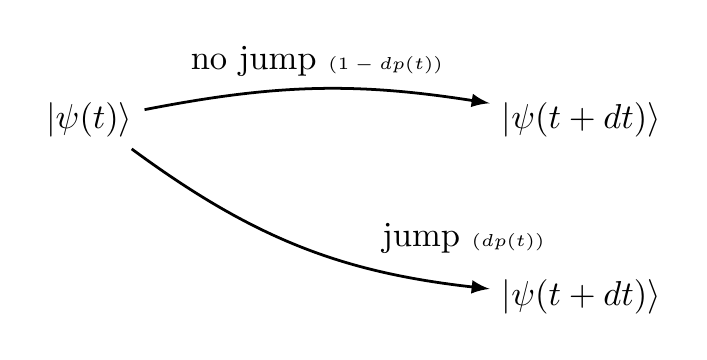}  
\caption{A depiction of the stochastic evolution of the state $\ket{\psi}$ by an infinitesimal time state at time $t$.}  
\label{fig:branch}
\end{figure}

We now derive Eq.~\eqref{eqt:sse} by explaining how each probability element appears.  Let us denote by $\ket{\psi(t)}$ and $\ket{\tilde{\psi}(t)}$ the normalized and unnormalized state vectors respectively, and assume they are equal at time $t$, i.e., $\ket{\tilde{\psi}(t)} = \ket{\psi(t)}$.  
For the infinitesimal time-step from $t$ to $t+dt$, the state vector evolution from $\ket{\psi(t)}$ to $\ket{\psi(t+dt)}$ involves two possibilities: either no jump occurring (with probability $1- dp$) or a jump occurring (with probability $dp$).  This is depicted in Fig.~\ref{fig:branch}.

When no jump occurs, the evolution is described by the Schr\"{o}dinger equation associated with $H_\eff$, and since the effective Hamiltonian is non-Hermitian the norm of the state vector is not preserved during the evolution:
\begin{equation}
\label{eqt:Schr}
\frac{d\ket{\tilde{\psi}(t)}}{dt} = -iH_{\text{eff}}(t) \ket{\tilde{\psi}(t)} \,.
\end{equation}
The resulting state after one infinitesimal time-step $dt$ is:
\bes
\label{eqt:detun}
\begin{align}
\ket{\tilde{\psi}(t+dt)} &= \exp\left[-iH_{\text{eff}}(t)dt\right]\ket{\psi(t)} \\   
&= \left[\mathds{I} - i dt H_{\text{eff}}(t)   + O(dt^2)\right] \ket{\psi(t)} \ .
\end{align}
\ees
The norm squared $\|\tilde{\psi}(t+dt)\|^2$ is the probability of the conditional evolution under $H_\eff$, 
so that (as we show explicitly in Appendix~\ref{app:A}) the jump probability is given by
\beq
1-\|\tilde{\psi}(t+dt)\|^2 = dt  \sum_{i}\braket{{A}_i^{\dagger}(t){A}_i(t)} + O(dt^2) 
\label{eqt:squareofnorm}
\eeq
[recall Eq.~\eqref{eq:12} and note that $H_{\text{\textrm{S}}}(t)+H_{\text{\textrm{LS}}}(t)$ cancels out to first order].
Therefore we can identify the infinitesimal jump probability $dp(t)$ with the r.h.s. of Eq.~\eqref{eqt:squareofnorm}, i.e., to first order in $dt$:
\bes
\label{eq:jumprate}
\begin{align}
\label{eqt:jumpprob}
dp(t) &= \sum_{i} dp_i (t) = dt \lambda(t) \\
\label{eqt:typeprob}
dp_i(t) &= dt \braket{{A}_i^{\dagger}(t){A}_i(t)} \ ,
\end{align}
\ees
where $\lambda(t) = \dot{p}(t)$ is the jump rate, and $dp_i(t)$ is the probability of the jump of type $i$.  Note that since our definition of the $A_i$ operators includes the rates $\gamma'$ [recall Eq.~\eqref{eqt:Aired}], the jump rate depends on the instantaneous Bohr frequencies and the KMS condition.

When the jump of type $i$ occurs the state is updated as:
\begin{equation}
\label{eqt:jumpun}
\ket{\tilde{\psi}(t+dt)} ={A}_i(t)\ket{\psi(t)} \,.
\end{equation}

We can unify the two possibilities in Eq.~\eqref{eqt:detun} and Eq.~\eqref{eqt:jumpun} as a stochastic Schr\"{o}dinger equation for the unnormalized state vector where only terms of order $dt$ are kept:
\bes
\label{eqt:sse2}
\begin{align}
d\ket{\tilde{\psi}(t)} &= \ket{\tilde{\psi}(t + dt)} - \ket{\tilde{\psi}(t)} \\
&= -i dt H_{\text{eff}}(t)\ket{\tilde{\psi}(t)}  + \nonumber\\
&\phantom{=}\sum_{i}dN_{i}(t)\left({A}_i(t)-\mathds{1}\right)\ket{\tilde{\psi}(t)} \,,
\end{align}
\ees
where we used $\ket{\tilde{\psi(t)}} = \ket{\psi(t)}$. The stochastic element $dN_{i}(t)$ has the properties given in Eqs.~\eqref{eqt:stochastic1} and~\eqref{eqt:stochastic2}; $\mathds{1}$ is subtracted since when the jump  occurs $\sum_{i}dN_{i}(t)\ket{\tilde{\psi}(t)} = \ket{\tilde{\psi}(t)}$ and the term involving $H_{\text{eff}}(t)$ is absent, so in this manner we ensure that $\ket{\tilde{\psi}(t)}$ is appropriately subtracted from the r.h.s.

We can write a similar expression for the normalized state vector $\ket{\psi(t)}$ by normalizing Eqs.~\eqref{eqt:detun} and~\eqref{eqt:jumpun}.  If a deterministic evolution occurs, we have
\bes
\label{eqt:deterministicequivalence}
\begin{align}
\label{eqt:deterministicequivalence-1}
\ket{\psi(t + dt)} &= \frac{\exp\left[-iH_{\text{eff}}(t)dt\right]\ket{\psi(t)}}{\norm{\exp\left[-iH_{\text{eff}}(t)dt\right]\ket{\psi(t)}}} \\
&\hspace{-1cm}= \frac{\left(\mathds{1} - i dt H_{\text{eff}}(t)+O(dt^2)\right)\ket{\psi(t)}}{\sqrt{1 - dt \sum_{i}\braket{{A}_i^{\dagger}(t){A}_i(t) } + O(dt^2)}}\\
&\hspace{-1cm} = \left(\mathds{1} - i dt H_{\text{eff}}(t) + \frac{1}{2}\sum_{i} \braket{{A}_i^{\dagger}(t){A}_i(t)}dt \right)\ket{\psi(t)}  \notag \\
& + {O}(dt^2)\ .
\end{align}
\ees
If a jump of type $i$ occurs, we have
\begin{equation}
\ket{\psi(t + dt)} = \frac{{A}_i(t)\ket{\psi(t)}}{\norm{{A}_i(t)\ket{\psi(t)}}} = \frac{{A}_i(t)\ket{\psi(t)}}{\sqrt{\braket{{A}^{\dagger}_{i}(t){A}_{i}(t)}}}\,.
\label{eqt:jump}
\end{equation}
Therefore, in analogy to Eq.~\eqref{eqt:sse2} we can write the stochastic Schr\"odinger equation for the normalized state as in Eq.~\eqref{eqt:sse}.

\section{Simulation procedure for adiabatic quantum trajectories}
\label{sec:IV}
In this section we formulate an algorithm for implementing adiabatic quantum trajectories. We start by noticing that the update in Eq.~\eqref{eqt:deterministicequivalence-1} corresponds to the evolution by the first part of the stochastic Schr\"{o}dinger equation. Therefore, the deterministic evolution in the first term of Eq.~\eqref{eqt:sse} is equivalent to propagating the state vector via the Schr\"{o}dinger equation with $H_\eff(t)$ and then renormalizing it. 

When a jump occurs, one of the operators $A_i(t)$ is applied.  The relative weight of each $A_i(t)$ is $dp_i(t)$, given in Eq.~\eqref{eqt:typeprob}. In this case, the state is evolved as in Eq.~\eqref{eqt:jumpun}, and the normalized state is given in Eq.~\eqref{eqt:jump}. 
The update in Eq.~\eqref{eqt:jump} corresponds to the evolution by the second term of Eq.~\eqref{eqt:sse}. 

This provides a direct way to algorithmically implement the quantum trajectories method.   Starting from a known normalized initial state, the state is evolved via a sequence of deterministic evolutions and jumps, as in Eqs.~\eqref{eqt:detun} and~\eqref{eqt:jump}, by drawing a random number at each finite but small time-step $\Delta t$ and determining which of the two choices to take.  Compared to the standard time-independent case, the size of the time-step must satisfy additional conditions in order for the approximations to hold:
\beq
\Delta t \ll \min_t \left\{ \frac{2\vertiii{H_{\text{eff}}(t)}}{\| \dot{H}_{\text{eff}}(t)\|},
\frac{1}{\vertiii{H_{\text{eff}}(t)}},
\left|\frac{ \lambda(t) }{ \lambda^2(t)-\dot{\lambda}(t)}\right| \right\} \ ,
\label{eq:newcond}
\eeq
where $\|\cdot\|$ is the operator norm (largest singular value).  We give a proof of this new bound in Appendix~\ref{app:B}.  While the second and third terms reduce to the known conditions for the time-independent case, the first term in Eq.~\eqref{eq:newcond} is unique to the time-dependent case and reflects the error associated with time evolution under a time-dependent effective Hamiltonian.  This term highlights the fact that the faster the effective Hamiltonian and its eigenstates vary in time, the smaller is the time-step required to properly follow the trajectory. Eq.~\eqref{eq:newcond} can also be viewed as the physical timestep upper bound in weak measurement.

However, drawing a random number at each time-step is computationally expensive, 
so it is more efficient to use the waiting time distribution~\cite{Breuer:2002}
to determine the first jump event.  As we mentioned before [Eq.~\eqref{eqt:squareofnorm}] the square norm of the unnormalized wavefunction at $t + dt$ gives the probability of no jump during the infinitesimal interval $[t,t+dt]$. We show in Appendix~\ref{sec:C} that starting from the normalized state $\ket{\psi(t)}$, the probability of no jump occurring in the finite (not necessarily small) time interval $[t,t+\tau]$ is given by
\begin{equation}
\|\tilde{\psi}(t+\tau)\|^2 =  \exp\left(-\int_t^{t+\tau}\lambda(s)ds \right) \,,
\end{equation}
where the jump rate $\lambda(t)$ is given in Eq.~\eqref{eqt:jumpprob}. 
With this, the simulation procedure for one single trajectory is as follows, starting from $t$:
\begin{itemize}
\label{item:algo}
\item Draw a random number $r$.
\item Propagate the unnormalized wavefunction by solving the Schr\"{o}dinger equation with $H_\eff$ [Eq.~\eqref{eqt:Schr}] until the jump condition is reached at $t+\tau$, i.e., for $\tau$ such that $\braket{\tilde{\psi}(t+\tau)|\tilde{\psi}(t+\tau)} \leq  r$. (Recall that the norm of the unnormalized wavefunction will keep decreasing in this process.)
\item Determine which jump occurs by drawing another random number and update the wavefunction by applying jump operators, and renormalize.
\item Repeat the above steps with the new normalized state.  
\item Repeat until the final simulation time is reached.
\end{itemize}

We prove that averaging over quantum trajectories recovers the master equation in Appendix~\ref{app:B}. Specifically, we show there that if we denote the state of the $k$-th trajectory at time $t$ by $\ket{\psi_k(t)}$, then we can approximate the master equation solution for the density matrix $\rho(t)$ as $\frac{1}{n}\sum_{k=1}^{n}\ketbra{\psi_k(t)}$ for large $n$. Choosing a basis $\{\ket{z_i}\}$ for the system Hilbert space, we can thus approximate the density matrix element $\bra{z_i} \rho(t) \ket{z_j}$ as $\frac{1}{n}\sum_{k=1}^{n}\langle{z_i} \ketbra{\psi_k(t)} {z_j}\rangle$ for large $n$.

\section{V: Case studies}
\label{sec:V}
We consider a system of $N$ qubits with a transverse-field Ising Hamiltonian given by
\begin{subequations}
\label{eqt:H_S}
\begin{eqnarray}
\label{eqt:H_Sa}
{H_{\textrm{S}}(t)} &=& A(t) H_{\textrm{S}}^X + B(t) H_{\textrm{S}}^Z {\ , } \\
H_{\textrm{S}}^X &\equiv& -\sum_{i=1}^N \sigma_i^x { \ ,}\\
H_{\textrm{S}}^Z &\equiv&  - \sum_{i=1}^N h_i
\sigma_i^z + \sum_{i>j=1}^N  J_{i j} \sigma_i^z \sigma_j^z \ .
\end{eqnarray}
\end{subequations}

\begin{figure}[h!]
\centering
\subfigure[\ ]{\includegraphics[width=0.69\textwidth]{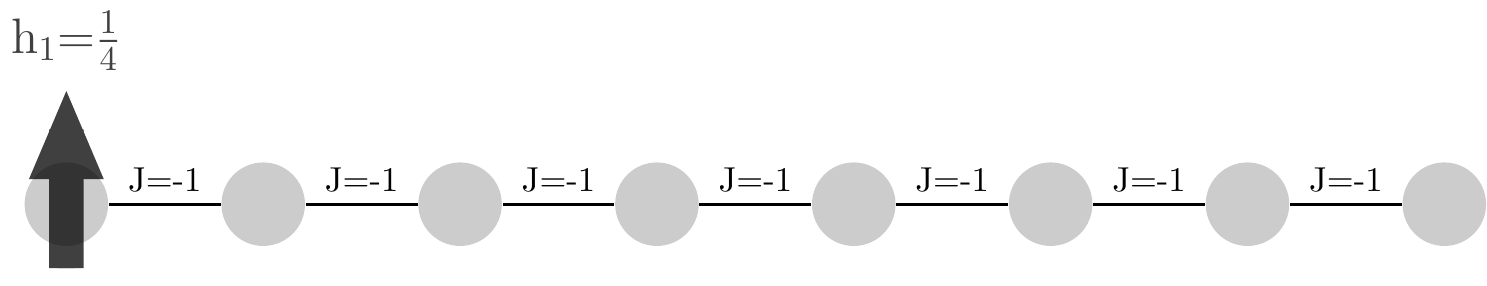}\label{fig:4a}} 
\subfigure[\ ]{\includegraphics[width=0.18\textwidth]{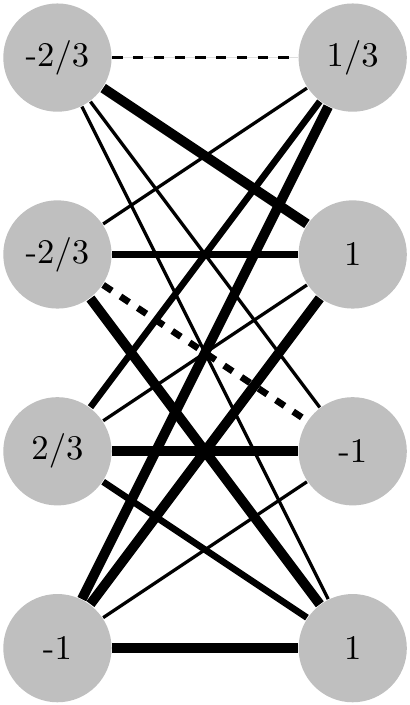}\label{fig:4b}} 
\subfigure[\ ]{\includegraphics[width=0.56\textwidth]{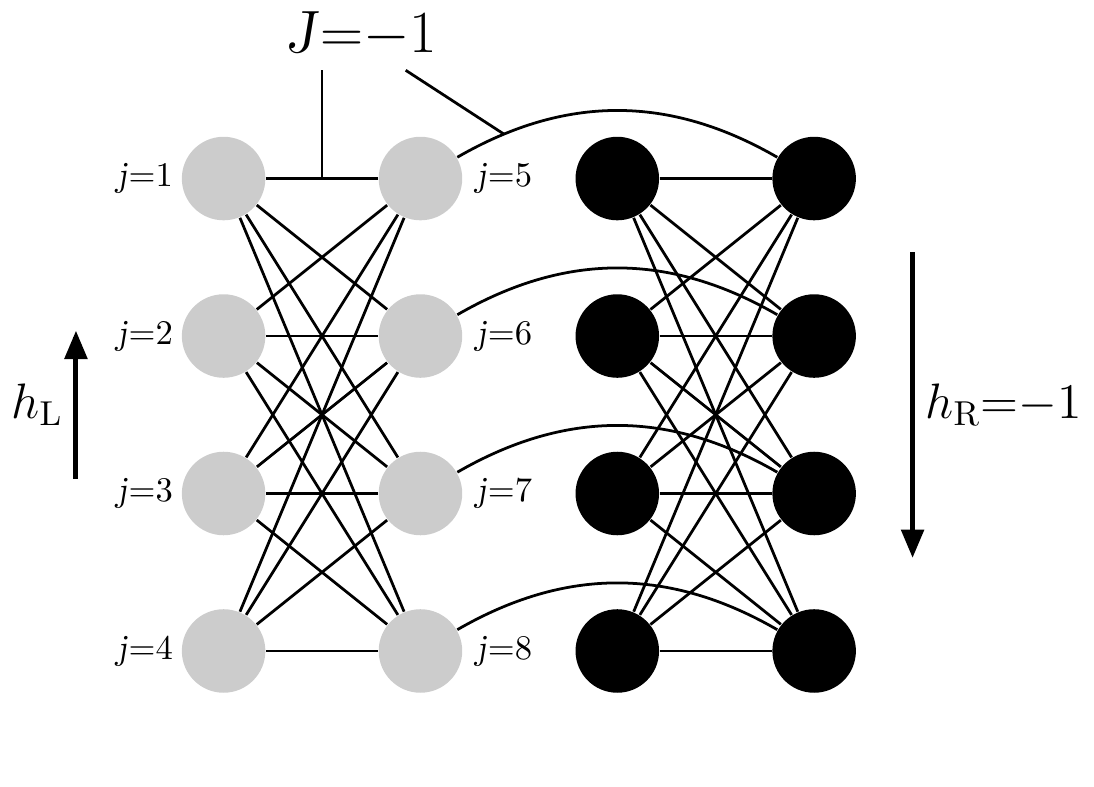}\label{fig:4c}} 
\caption{Graphs of (a) the $8$-qubit chain, (b) the $8$-qubit Hamiltonian exhibiting a small gap, and (c) the $16$-qubit ``tunneling-probe" Hamiltonian of Ref.~\cite{Boixo:2014yu}. (a) Only the first qubit is subjected to an applied field and each qubit is ferromagnetically coupled with $J = -1$. (b)  Solid lines corresponds to ferromagnetic coupling and dashed lines corresponds to  antiferromagnetic coupling.  The thickness denotes the strength of the coupling.  Local fields are shown inside the circles.  Full parameters are given in Eq.~\eqref{eqt:Case0989}.  (c) The left $8$-qubit cell and right $8$-qubit cell are each subjected to applied fields with opposite direction. Each qubit is ferromagnetically coupled to others as shown by the lines, with $J = -1$.}
\end{figure}

\begin{figure}[h!]
\begin{center}
{\includegraphics[width=0.8\textwidth]{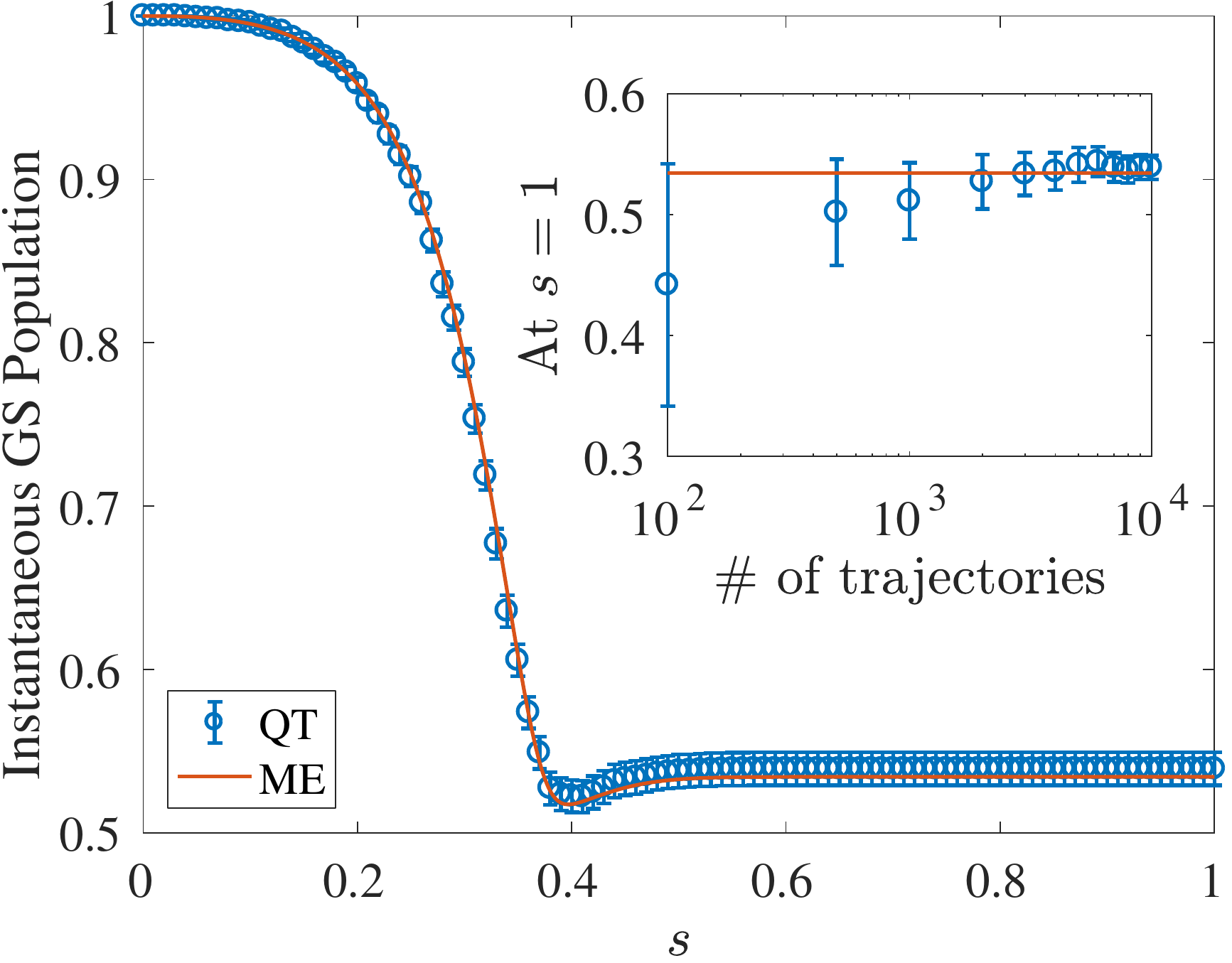}} 
\end{center}
\caption{The evolution of the population in the instantaneous ground state for the $8$-qubit problem in Eqs.~\eqref{eqt:H_S} and \eqref{eqt:Ising} for a total time of $t_f = 10 \mu s$ and temperature $2.6$GHz (in $\hbar\equiv 1$ units), as a function of the normalized time $s=t/t_f$.  The quantum trajectories results with $10^4$ trajectories (QT, blue circles) are in excellent agreement with the adiabatic master equation (ME, red solid line).   Inset: the convergence of the ground state population (averaged over quantum trajectories) towards the master equation result as a function of the number of trajectories,  at $s = 1$.  The error bars represent $2 \sigma$ confidence intervals, where $\sigma$ is the standard deviation of the mean generated by taking $10^3$ bootstraps over the number of trajectories. }
  \label{fig:3qubits8qubits}
\end{figure}

We assume that the qubit-system is coupled to independent, identical bosonic baths, with the bath and interaction Hamiltonian being 
\bes
\begin{align}
\label{eq:SBm}
H_B &= \sum_{i=1}^N \sum_{k=1}^\infty \omega_k b_{k,i}^\dagger b_{k,i} \ ,  \\
 H_{I} &= g \sum_{i=1}^N
\sigma_i^z \otimes \sum_k \left(b_{k,i}^\dagger + b_{k,i} \right) \ ,
\end{align}
\ees
where $b_{k,i}^\dagger$ and $b_{k,i}$ are, respectively, raising and lowering operators for the $k$-th
oscillator mode with natural frequency $\omega_k$. 
The bath correlation functions appearing in Eq.~\eqref{eqt:ME2-L} are given by
\begin{equation}
\gamma_{ij}(\omega) = 2\pi g^2\frac{\omega e^{-|\omega|/\omega_c}}{1-e^{-\b\omega}} \delta_{ij} \ ,
\end{equation}
arising from an Ohmic spectral density~\cite{ABLZ:12-SI}, and satisfy the KMS condition~\eqref{eq:KMS}.

\subsection{$8$-qubit chain} \label{sec:8qubitchain}
As a first illustrative example and as a consistency check, we reproduce the master equation evolution of the $8$-qubit ferromagnetic Ising spin chain in a transverse field studied in Ref.~\cite{ABLZ:12-SI}. For this problem, the Ising parameters are given by [also shown in Fig.~\ref{fig:4a}]
\begin{equation} \label{eqt:Ising}
 h_1 = \frac{1}{4} \ , \quad h_{i>1}=0  \ , \quad  J_{i, i+1} = -1 \ , \quad i = 1, \dots, 8 \ .
\end{equation}
The functional form of the functions $A(t)$ and $B(t)$ is given in Appendix \ref{app:Chain} (they are the annealing schedule of the D-Wave One ``Rainier" processor, described in detail, e.g., in Ref.~\cite{q108}, and were also used in the original AME study of the $8$-qubit chain~\cite{ABLZ:12-SI}). 
The effect of changing the schedule is small; for comparison we provide in Appendix~\ref{app:Chain} results for a linear annealing schedule with the same bath parameters. As shown in Fig.~\ref{fig:3qubits8qubits}, we recover the master equation solution within the error bars.  The initial state is the ground state of $H_{S}(0)$, which is the uniform superposition state.  

\begin{figure}[t] 
\centering
\includegraphics[width=0.8\textwidth]{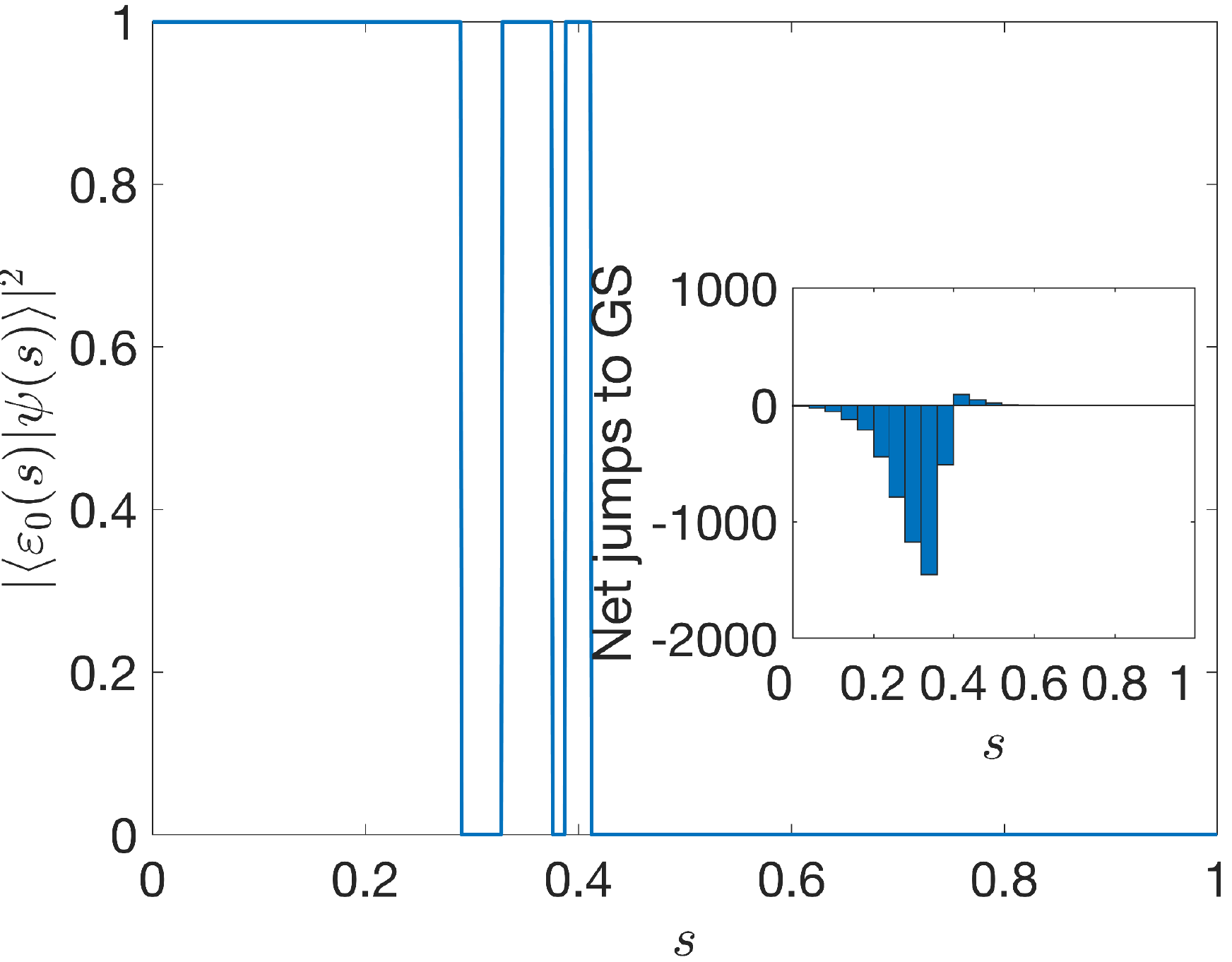}
\caption{The overlap squared of the (normalized) state with the instantaneous ground state of $H_{S}(t)$ for a typical single trajectory of the 8 qubit chain in Sec.~\ref{sec:8qubitchain}, with $t_f  = 10\mu s$ and temperature $2.62$GHz, as a function of the normalized time $s=t/t_f$. The sudden changes in the overlap are due to the action of the jump operators $\left\{A_i(t) \right\}$, taking the state from one eigenstate to another. This is to be contrasted with the smooth behavior of Fig.~\ref{fig:3qubits8qubits} when we average over different trajectories.  Inset: a histogram of the net number of jumps to the instantaneous ground state (GS).  A negative number indicates a jump out of the ground state, and a positive number indicates a jump towards the ground state. The change from negative to positive net jumps occurs at the minimum gap point.}
\label{fig:8qubitssingle}
\end{figure}

It is illustrative to see how a single trajectory differs from the averaged case, and we show this in Fig.~\ref{fig:8qubitssingle}.  Instead of the smooth change in the population as observed in the averaged case, the single trajectory behaves like a step-function. This is explained by the fact that the drift term vanishes if $\ket{\psi(t)}$ is a nondegenerate eigenstate, as shown in Appendix~\ref{app:eigenstate-proof}. Therefore changes in the state's overlap with the instantaneous ground state occur only due to the jump operators.  In this picture, the ground state population revival observed after the minimum gap is crossed is associated with jumps from the first excited state (or higher states for large $T$) to the ground state. After the minimum gap, there are more transitions back to the ground state than out of the ground state (see the inset of Fig.~\ref{fig:8qubitssingle}). Such a difference (divided by the number of trajectories) leads to the rise of the ground state population. 

Using Eq.~\eqref{eq:jumprate}, we can give an explicit expression for the jump rate from the first excited state back to ground state. As shown in  Appendix~\ref{app:lindbladfermi}, this is given by:
\begin{equation}
\label{eq:inctunnrate}
\lambda_{1\rightarrow0}(t) = 
\sum_{\alpha = 1}^{8} \gamma_\alpha(\omega_{10})|\langle\psi_0(t)|\sigma^{z}_{\alpha} |\psi_1(t)\rangle|^2\ .
\end{equation}

\begin{figure}[t]
\begin{center}
{\includegraphics[width=0.8\textwidth]{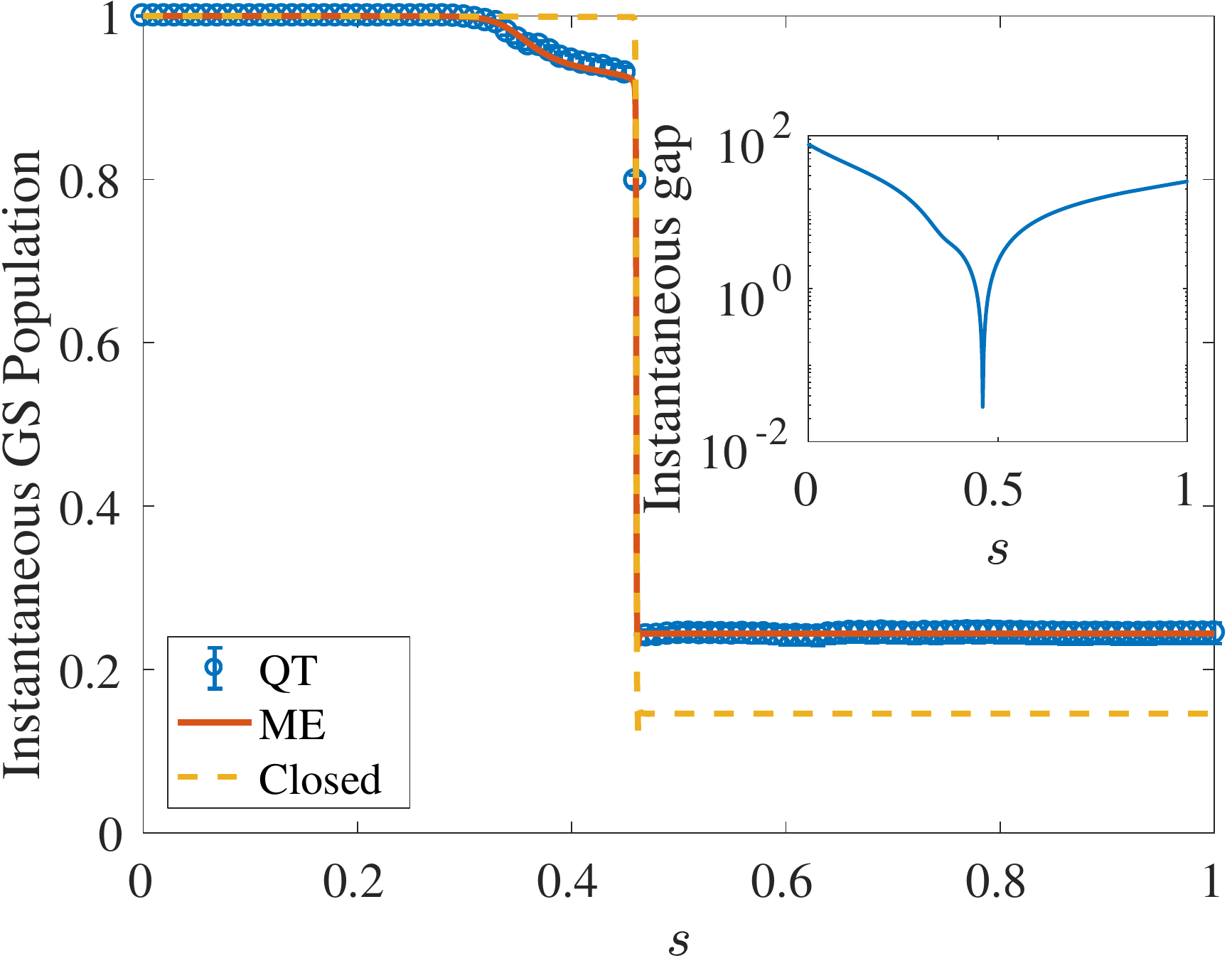}} 
\end{center}
\caption{The evolution of the population in the instantaneous ground state for the $8$-qubit problem in Eqs.~\eqref{eqt:H_S} and \eqref{eqt:Case0989} for a total time of $t_f = 10 \mu s$ and temperature $1.57$GHz, as a function of the normalized time $s=t/t_f$.  The quantum trajectories results with $5\times 10^3$ trajectories (blue circles) are in excellent agreement with the master equation (red solid line). We also show the closed-system evolution (yellow dashed line) to highlight that the evolution is not adiabatic.   Inset: the instantaneous energy gap between the first excited state and the ground state during the anneal. The minimum occurs at $s^*=0.46$, coinciding with the sharp discontinuity observed in the instantaneous ground state population.}
   \label{fig:case0989}
\end{figure}
\subsection{$8$-qubit non-adiabatic example} 
\label{sec:case0989}

We now consider an $8$-qubit problem with a sufficiently small minimum gap such that the closed-system evolution is not adiabatic even with $t_f = 10 \mu s$ and using the DW2X annealing schedule (described in detail, e.g., in Ref.~\cite{Albash:2017aa}).  While this strictly violates the assumptions under which the AME is derived,%
\footnote{Equation~(27) in Ref.~\cite{ABLZ:12-SI} is a necessary condition for the validity of the AME. It states that $\frac{h}{\Delta^2 t_f}\ll 1$, where $\Delta$ is the ground state gap and $h = \max_{s\in [0,1];a,b} |\bra{\epsilon_a(s)}|\partial_sH_S(s)\ket{\epsilon_b(s)}|$, with $s=t/t_f$ and $\ket{\epsilon_a(s)}$ being the instantaneous $a$-th eigenstate of the system Hamiltonian $H_S(s)$. We find that for $t_f = 10 \mu s$, the l.h.s.$\approx 5$.}
we can ask about the dynamics associated with the master equation irrespective of its origins. We are interested in this example since it illustrates some aspects of the quantum trajectories picture which are not visible in the adiabatic limit, as explained below.

The Ising Hamiltonian $H_S^Z$ is defined with parameters:
\bes \label{eqt:Case0989}
\begin{align}
3 \vec{h} &= (-2, -2, 2, -3, 1, 3, -3, 3) \ , \\
 3J_{0,4} &= \phantom{-}1, 3J_{0,5} = -3,  3J_{0,6} = -1,  3J_{0,7} = -1,  \\
 3J_{1,4} &= -1, 3J_{1,5} = -2,  3J_{1,6} = \phantom{-}2,  3J_{1,7} = -3,  \\
 3J_{2,4} &= -2, 3J_{2,5} = -1,  3J_{2,6} = -3,  3J_{2,7} = -2, \\
 3J_{3,4} &= -3, 3J_{3,5} = -3,  3J_{3,6} = -1,  3J_{3,7} = -3, 
\end{align}
\ees
Figure~\ref{fig:case0989} shows our simulation results, obtained by solving the AME directly and by using the trajectories approach. Reassuringly, the agreement between the two is excellent. Also plotted are the closed system results for this problem, which exhibit a sharp diabatic transition out of the ground state at the minimum gap point (the small gap is shown in the inset). The AME and trajectories results show that the ground state population loss starts before the diabatic transition, due to thermal excitations, but that the ground state population loss is partially mitigated by the presence of the thermal bath, with the open system ending up with a higher ground state population than the closed system. 

\begin{figure}[h] 
\centering
\includegraphics[width=0.8\textwidth]{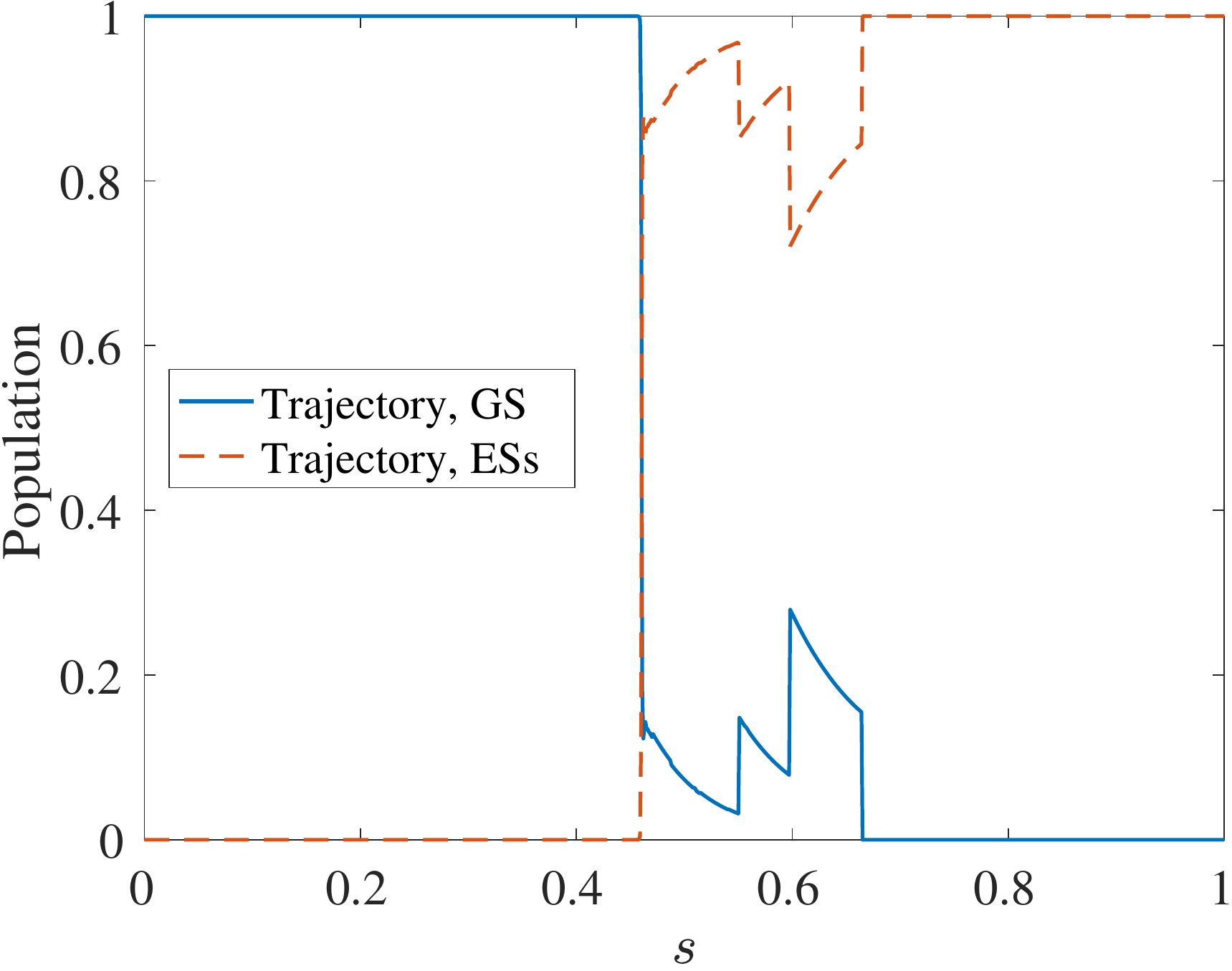}
\caption{The overlap squared of the (normalized) state with the instantaneous ground state of $H_{S}(t)$ (blue solid curve) and the sum of the first three instantaneous excited states (red dashed curve) for a typical single trajectory of the $8$-qubit problem in Sec.~\ref{sec:case0989}, with $t_f  = 10\mu s$ and temperature $1.57$GHz, as a function of the normalized time $s=t/t_f$. A diabatic transition occurs at $s^* = 0.46$. The continuous decay immediately afterward and between the `incomplete' jumps is to be contrasted with the step-function like single trajectories of the adiabatic case seen in Fig.~\ref{fig:8qubitssingle}. 
}
\label{fig:Case0989single}
\end{figure}

The diabatic transition results in different trajectories than those observed for the adiabatic case in Sec.~\ref{sec:8qubitchain}.  We show such a case in Fig.~\ref{fig:Case0989single}.  Instead of the pulse-like structure seen in Fig.~\ref{fig:8qubitssingle}, we observe a combination of both drifts and jumps.  Because the diabatic transition generates a non-eigenstate that is a coherent superposition of the ground state and first excited state, drifts caused by the environment show up in the subsequent evolution.  Furthermore, this superposition also means that the Lindblad operator associated with $\omega = 0$, if having different component weights in Eq.~\eqref{eq:Lindblad2}, can also induce jumps (e.g., the jumps around $s = 0.6$ in Fig.~\ref{fig:Case0989single}), an effect that is completely absent in the adiabatic case.  These jumps need not project the state completely onto an instantaneous energy eigenbasis state, but they can change the relative weights on the different occupied eigenstates, which manifest themselves as `incomplete' jumps in the trajectories.

We reemphasize that due to the violation of the conditions under which the AME is derived, the observations we have reported for this example are strictly valid only when the AME is taken at face value, and do not necessarily reflect actual physical dynamics.

\begin{figure}[h!] 
   \centering
   \includegraphics[width=0.8\textwidth]{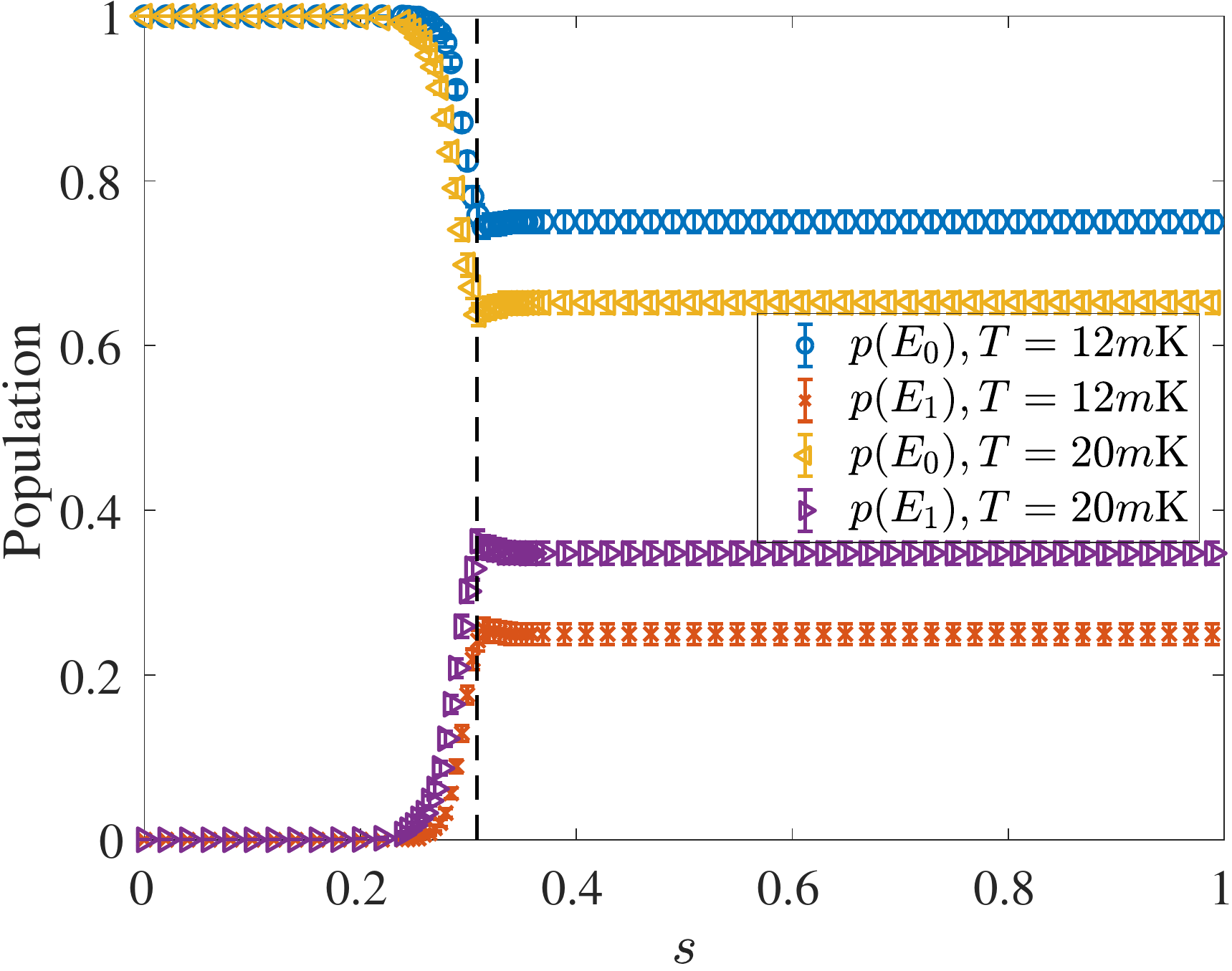} 
   \caption{Quantum trajectory results for a temperature of $12m$K and $20m$K using the DW2X annealing schedule. Results are averaged over $5000$ trajectories. A revival of instantaneous ground state population occurs after the minimum gap (shown by the dashed vertical line).}
   \label{fig:GoogleGadget}
\end{figure}
%

\subsection{$16$-qubit ``tunneling-probe" Hamiltonian}
\label{sec:16qubit}
In order to demonstrate the computational utility of the trajectories approach over the master equation approach, we now give results for a $16$-qubit system first studied in Ref.~\cite{Boixo:2014yu} for the purpose of probing tunneling in quantum annealing. 

For this problem, the parameters of Eq.~\eqref{eqt:H_S} are [also shown in Fig.~\ref{fig:4b}]:
\begin{equation} 
\label{eqt:H_S16}
 h_{\text{L}} = 0.44\ , \quad h_{\text{R}} = -1  \ , \quad  J_{i, i+1} = -1 \ , \quad i = 1, \dots, 16 \ .
\end{equation}
where the sets $L$ and $R$ range over $i = 1,\dots, 8$ and $i = 9,\dots, 16$, respectively. Ref.~\cite{Boixo:2014yu} chose the value of $h_{\text{L}}$ to ensure that the minimum ground state gap is lower than the temperature $T=15.5$mK, in order to study a non-trivial interplay between tunneling and thermal activation.\footnote{The problem features one global minimum and one local (false) minimum, which are separated by a tall energy barrier at around the minimum gap point (see Fig.~2 in Ref.~\cite{Boixo:2014yu}); to reach the global minimum from the false minimum, the system state has to transverse the barrier. Such transitions can be modeled as quantum jumps in the quantum jump picture.} This parameter choice means that incoherent effects play a relatively strong role in this problem, which are not well captured by the AME. Thus, similarly to the previous example, the AME is being used here outside of its strict validity domain. We are interested in testing whether it can nevertheless qualitatively capture the correct physical effects. Moreover, direct master equation simulations for such a large system take longer than 24 hours (which is a standard time-window on high-performance clusters), while each quantum trajectory takes less than 24 hours.  We can then exploit many CPU cores to perform many trajectories in parallel. To this end we used $320$ CPU cores and repeated the simulation $16$ times for a total of over $5000$ trajectories.
 
Our simulations (see Fig.~\ref{fig:GoogleGadget}) show how population is lost from the instantaneous ground state to the first excited state near the minimum gap point $s \approx 0.308$.
It also shows a small population revival after the minimum gap is crossed.  As in Sec.~\ref{sec:8qubitchain}, this revival is associated with jumps from the first excited state (or higher states for large $T$) back to the ground state. Encouragingly, despite the perturbative nature of the AME, this revival is qualitatively in agreement with the results of Ref.~\cite{Boixo:2014yu} (see their Fig.~4). The latter work found a stronger revival on the basis of the non-perturbative, non-interacting blip approximation (NIBA), which more accurately captures additional transitions that occur when the energy level broadening is larger than the energy gap between energy levels.

\section{Simulation cost comparison}
\label{sec:timequantitative}

We now provide a cost comparison between the simulations cost of directly solving the AME and the quantum trajectories method. The first two subsections in this section follow Ref.~\cite{Breuer:2002} closely (while adding some details), and we borrow the notation used in that reference. In subsection~\ref{sec:parallel} we provide a new analysis that reveals that the quantum trajectories method can exhibit a scaling advantage ranging between $O(N)$ and $O(N^2)$ over the direct solution of the AME.

 \subsection{Number of trajectories}
The number of trajectories needed can be found from the standard error of the sample mean. As an example, let us consider the standard error $\hat{\sigma}_{t}$ associated with the instantaneous ground state population  $\langle\psi_{0}(t)|\rho(t)|\psi_{0}(t)\rangle$:
\begin{align}
\hat{\sigma}_{t}^2 &= \frac{1}{R(R-1)}\sum_{r=1}^{R}\left(|\langle\psi_{0}(t)|\psi^{r}(t)\rangle|^2  - \hat{M}_t  \right)^2  \label{eq:stderror}\, ,
\end{align}
where $\ket{\psi^r(t)}$ denotes the state associated with trajectory $r$ at time $t$, $\hat{M}_t = \frac{1}{R}\sum_{r=1}^{R}|\langle\psi_{0}(t)|\psi^{r}(t)\rangle|^2$, and $R$ is the total number of trajectories.  By fixing the value of the standard error $\hat{\sigma}_{t}$, the number of necessary trajectories $R$ can then be determined.

\subsection{Cost comparison}
Since we expect $\hat{\sigma}_{t} \sim \frac{1}{\sqrt{R}}$, let us write
\begin{equation}
\label{eq:factor}
\hat{\sigma}_{t}^2 =  \frac{\lambda_{B}(N)}{R} \sim \frac{1}{R}\,,
\end{equation}
where $\lambda_{B}(N) = \frac{1}{R-1}\sum_{r=1}^{R}(\langle\psi^r(t)| B|\psi^r(t)\rangle  - \hat{M}_t)^2$ for an observable $B$ and mean value $\hat{M}_t = \frac{1}{R}\sum_{r=1}^{R}\langle\psi^{r}(t)|B|\psi^{r}(t)\rangle$. The factor $\lambda_{B}(N)$ is a non-increasing function of the system dimension $N$ \cite{Breuer:2002}: 
\begin{equation}
\label{eq:lambdaB}
\lambda_{B}(N) \sim N^{-x} \,,
\end{equation}
where the scaling $x$ depends on the observable: 
\begin{equation}
0 \left(\text{not self-averaging}\right) \leq x \leq 1 \left(\text{strongly self-averaging}\right) \,.
\end{equation}
Thus, to obtain the same standard error for increasing dimension, the number of trajectories need not be increased in general.  This is another advantage of the trajectories method for growing system dimension. Such a phenomenon has also been observed in time-dependent stochastic density functional theory~\cite{gao2015sublinear}.
From Ref.~\cite{Breuer:2002}, the total serial CPU time required for the simulation of the master equation, denoted $T_{\text{AME}}$, versus the stochastic method with $R$ trajectories, denoted $T_{\text{StS}}$, is:
\bes
\label{eq:timecost}
\begin{align}
T_{\text{AME}} &= k_{1} s_{1}(N) N^{\beta} \,, \label{eq:T-AME}\\
T_{\text{StS}} &= k_{2} R(N) s_{2}(N) N^{\alpha} \,,
\label{eq:T-StS}
\end{align}
\ees
where $k_1$ and $k_2$ are constants depending on the specific implementation of each method, $s_{1}(N)$ is the total number of evaluations of $\mathcal{L}_{\textrm{WCL}}[\rho(t)]$ [Eq.~\eqref{eqt:ME2-L}] using the master equation method, and $s_{2}(N)$ is the total number of evaluations of $H_{\text{eff}}(t)\ket{\psi(t)}$ [Eq.~\eqref{eq:Heff}] in a single trajectory.

$R(N)$ in Eq.~\eqref{eq:T-StS} is the minimum number of trajectories needed to obtain a standard error lower than a particular chosen value. To account for the constraint that $R(N)\geq 1$, we rewrite Eq.~\eqref{eq:lambdaB} as $\lambda_B(N) = \Lambda_{B} N^{-x}$, and Eq.~\eqref{eq:factor} as:
\begin{equation}
R(N) =
\begin{dcases} 
      \Bigl\lceil \frac{\Lambda_{B} N^{-x}}{\hat{\sigma}_{t}^2} \Bigr\rceil& \text{\space  for } N < N^{*} \\
      1 & \text{\space  for } N\geq N^{*} 
   \end{dcases}
\end{equation}
where $N^{*} =  \lceil\left({\Lambda_{B}}/{\hat{\sigma}_{t}^2}\right)^\frac{1}{x}\rceil$.   
For $x > 0$, the required number of trajectories decreases with $N$ until $N^{*}$, after which one trajectory gives the expectation value within the desired accuracy.

In general, the number of operations needed to evaluate $\mathcal{L}_{\textrm{WCL}}[\rho(t)]$ relative to the number needed to evaluate $H_{\text{eff}}(t)\ket{\psi(t)}$ differs by a factor of $N$, so that $\beta \approx \alpha + 1$, and Eq.~\eqref{eq:timecost} becomes
\bes
\begin{align}
T_{\text{AME}} &= k_1 s_{1}(N) N^{\alpha + 1} \,,\\
T_{\text{StS}} &=
\begin{dcases} 
      k_2' s_{2}(N) N^{\alpha - x} &\text{\space  for } N < N^{*}  \\
      k_{2} s_{2}(N) N^{\alpha} &\text{\space  for }  N\geq N^{*} 
   \end{dcases}
\end{align}
\ees
where 
\begin{equation}\label{eqt:k2prime}
k_2' = k_2\left(\frac{\Lambda_{B}}{\hat{\sigma}_{t}^2}\right) \,,
\end{equation}
and $k_2'$ hence depends on the required accuracy as well. In many situation $s_{1}(N)$ and $s_{2}(N)$ grow with $N$, but they are roughly equal or grow in same manner with $N$. By dividing these two expressions, we can obtain the ratio of $T_{\text{AME}}/T_{\text{StS}}$, 
\begin{equation}
\frac{T_{\text{AME}}} {T_{\text{StS}}} =
\begin{dcases} 
      \frac{k_1  }{k_2' }N^{1 + x} & \text{\space  for }N < N^{*} \\
      \frac{k_1  }{k_2 }N & \text{\space  for } N\geq N^{*} 
   \end{dcases}  \ . 
\end{equation}
Since $0\leq x \leq 1$, we can write
\begin{equation} 
\label{eqt:timeComparison}
\begin{aligned}
       \frac{k_1}{k_2'}N &\leq \frac{T_{\text{AME}}}{T_{\text{StS}}} \leq  \frac{k_1}{k_2'}N^2 & \text{\space  for } N < N^{*} \,,\\
     &  \frac{T_{\text{AME}}}{T_{\text{StS}}} = \frac{k_1}{k_2}N & \text{\space for } N\geq N^{*} \,. 
\end{aligned} 
\end{equation}

We say that the trajectories method has an advantage over the direct master equation solution if $\frac{T_{\text{AME}}}{T_{\text{StS}}}> 1$. The constant factor $\frac{k_1}{k_2'}$ is typically a small number because it is proportional to the required standard error squared [Eq.~\eqref{eqt:k2prime}]. 
Therefore there is an advantage for the trajectories method when either $N$ is sufficiently large or when a sufficient number of CPU cores $C$ is available (see the next subsection). 
Equation~\eqref{eqt:timeComparison} shows that an advantage
beyond linear in $N$ is attainable for $N<N^\ast$ on a single CPU. The reason is that the number of trajectories needed to achieve a fixed accuracy decreases with increasing system dimension.  For $N>N^\ast$, only one trajectory is required, and the advantage scales as $O(N)$.   

We note that the larger-than-linear advantage only holds if we are interested in estimating operators with the same self-averaging property. This is in contrast to evolving the entire density matrix, as in the AME, which allows the expectation value of any observable to be calculated.  If we demand this same capability from the trajectories approach, then only the linear advantage holds.
 
\subsection{Parallel implementation}
\label{sec:parallel}

The stochastic wave function method is very well-suited for parallel computing implementations. The communication needed between each core is minimal since each trajectory is simulated independently. Assuming $C$ CPU cores are used, where $C \leq R(N)$, we can adjust the time cost [Eq.~\eqref{eq:T-StS}] for the stochastic method to
\begin{equation}
T_{\text{StS}} = k_{2} \frac{R(N)}{C} s_{2}(N) N^{\alpha} \,.
\end{equation}
Note that the number of cores $C$ is held constant, i.e., is independent of the system dimension $N$. Therefore, we can update Eq.~\eqref{eqt:timeComparison} to:
\begin{equation}
\begin{aligned}
       \frac{k_1}{k_2'}C N &\leq \frac{T_{\text{AME}}}{T_{\text{StS}}} \leq  \frac{k_1}{k_2'}C N^2 & \text{\space  for } N < N^{\star} \,,\\
     &  \frac{T_{\text{AME}}}{T_{\text{StS}}} = \frac{k_1}{k_2}  N & \text{\space for } N\geq N^{\star} \,.
      \end{aligned} 
\end{equation}
where $N^{\star} =  \lceil\left({\Lambda_{B}}/\left({C \hat{\sigma}_{t}^2} \right) \right)^\frac{1}{x}\rceil$.  Here $N^{\star}$ is the system dimension where $R(N^{\star}) = C$, and one execution of the $C$ parallel CPU cores is enough to obtain the desired standard error.  

Again, the larger-than-linear advantage in $N$ only holds if we are interested in estimating operators with the same self-averaging property, and otherwise we can only expect a linear advantage in $N$.

\section{Conclusion of this chapter}
\label{secconclude: qt}
In this chapter, we have shown how quantum trajectories (in the form of quantum jumps) can be unravelled from the adiabatic master equation. We have described and demonstrated a simulation procedure in terms of the waiting time distribution that reproduces the results of the master equation for examples involving $8$ and $16$-qubit systems. 
Direct master equation simulations for the $16$-qubit example would take a long time, but the simulation of the quantum trajectories remains computationally feasible for larger system dimensions by allowing us to simulate many trajectories in parallel. 
A scaling cost comparison of the two methods shows that, generically, the quantum trajectories method yields an improvement by a factor linear in the system dimension $N$ over directly solving the adiabatic master equation. However, the trajectories method can be expected to be up to a factor $c N^2$ faster than a direct simulation of the master equation if only the expectation value of specific self-averaging observables is desired. Here $c$ is a constant proportional to the number of parallel processes and the target standard error.
 
We therefore believe this approach will be particularly useful in enabling the study of larger systems than has been possible using a direct simulation of the AME. 

In addition, the quantum trajectories method offers fresh physical insight into the nature of individual trajectories and their statistics, which may become a helpful tool in interpreting computational bottlenecks in quantum annealing and adiabatic quantum computing.

Finally, while we did not address this in the present chapter, the quantum trajectories approach is well known to be a convenient path towards continuous measurement and the inclusion of quantum feedback control~\cite{Wiseman:book}. This approach might in the future provide a path towards error correction of adiabatic quantum computing, e.g., by formulating control targets that push the system back to the ground state after diabatic or thermal transitions. We address this in Chap.~\ref{chap: feedback}.

\section{Acknowledgments of this chapter}
The research is based upon work (partially) supported by the Office of
the Director of National Intelligence (ODNI), Intelligence Advanced
Research Projects Activity (IARPA), via the U.S. Army Research Office
contract W911NF-17-C-0050. The views and conclusions contained herein are
those of the authors and should not be interpreted as necessarily
representing the official policies or endorsements, either expressed or
implied, of the ODNI, IARPA, or the U.S. Government. The U.S. Government
is authorized to reproduce and distribute reprints for Governmental
purposes notwithstanding any copyright annotation thereon. Computation for the work described in this chapter was supported by the University of Southern California's Center for High-Performance Computing. The codes in this chapter are available at:

\url{https://github.com/USCqserver/Adiabatic-master-equation-and-quantum-trajectories}

%% file: chapter2.tex
\chapter{$1/f$ noise}
\label{chap: 1f}
\section{Introduction}
The trajectories method in the previous Chapter can be applied to non-Markovian noise, in particular $1/f$ (telegraph) noise. $1/f$ noise means that the spectrum of the environment $S(f) \propto 1/f$. To model $1/f$ (telegraph) noise, one needs to add stochastic elements to trajectories simulations and perform ensemble averaging. We start with the known Bloch vector-magnetic field approach~\cite{1f, Bergli2009, paladino2003decoherence, faorodynamical}, which in many ways is similar to  spin vector dynamics~\cite{Radcliffe_1971, klauder1979path}. Then we formulate a stochastic Hamiltonian approach for $1/f$ noise. The latter approach  allows an easier adaption to the time-dependent annealing Hamiltonian, and also to  high energy-level modeling of superconducting qubits.

There are also experimental studies of $1/f$ noise on superconducting qubits~\cite{bylander2011noise,quintana2017observation}. Note that in~\cite{bylander2011noise}, a stochastic Hamiltonian has been considered, but only one stochastic Hamiltonian is included in the model. In this thesis we formulate a stochastic Hamiltonian approach for $1/f$ noise, that includes a series of fluctuator terms with different flipping frequencies and couplings, which can be further added to the annealing system Hamiltonian. The new and original contributions of this chapter include the followings: 
\begin{itemize}
    \item We derive from the Bloch vector-magnetic field approach a stochastic Hamiltonian approach that includes a series of fluctuator terms with different flipping frequencies and Gaussian couplings that altogether constitute the $1/f$ power spectrum;
    \item We, for the first time, show how to incorporate the many fluctuators into the inherently time-dependent annealing Hamiltonians, and construct an overall time-dependent stochastic Hamiltonian. We thoroughly analyze how noise with $1/f$ power spectrum affects quantum annealing;
    \item We introduce flexibility into the description of the noise fluctuators. For example, we allow the noise fluctuators to take any form of multi-level operators and any noise-axis direction, depending on the superconducting qubits and circuit decoherence process.
\end{itemize}

\section{$1/f$ noise: Bloch vector-Magnetic field approach}
\label{sec:bbfield}
\subsection{One qubit and one fluctuator}
\begin{figure}[h!]
\centering
\includegraphics[width=4.5cm]{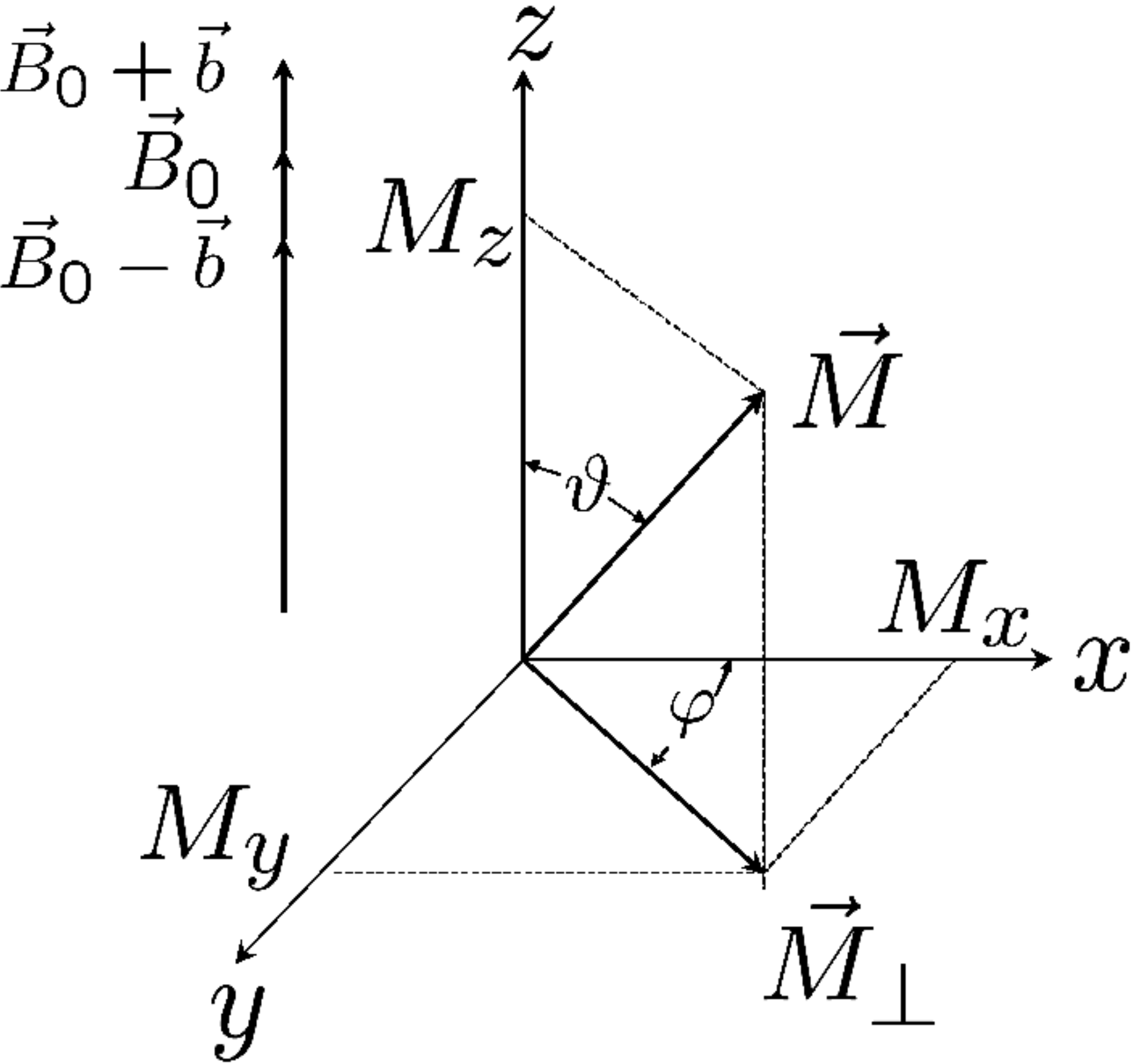}
\caption{Bloch vector representing the state of a qubit in the rotating 
(with angular frequency $B_0$) frame of reference.  In the laboratory frame of reference it precesses around the $z$-axis
 with (time-dependent) angular velocity $B$. Fluctuations of 
 this velocity is $b(t)$  Photo credit:~\cite{1f}.}
\label{figAGB1}
\end{figure}

\subsubsection{Qubit}
One can express the qubit under Hamiltonian drive in the form of a Bloch vector $\mathbf{M} = (M_x, M_y, M_z)$ under the influence of a sum of magnetic fields $\mathbf{B}$.

The dynamics of the qubit is described by the precession equation (Schrodinger equation)
\begin{equation} 
\label{bv1}
\dot{\mathbf{M}} = \mathbf{B} \times \mathbf{M} \, 
\end{equation}

\subsubsection{Fluctuator}
\label{sec:fluctuatorbfield}
Assume for simplicity that the qubit is subjected to a time-independent magnetic field pointing in the $z$-direction $\mathbf{B} = B = B_z$. However, a stochastic magnetic field $b(t)$ exists and is pointing parallel to $B$ such that the total magnetic field is $B+b(t)$. In this section we focus on one stochastic magnetic field $b(t)$. 

The stochastic magnetic field $b(t)$ originates from a so-called fluctuator, which can find itself in one of two (meta)stable states, $1$ and $-1$, and once in a while makes a switch between them. We consider the symmetric random telegraph (RT) process: $\gamma_{1\to -1}=\gamma_{-1\to 1}=\gamma/2$. When the fluctuator is in state 1, the stochastic magnetic field is $b(t) = +b$; similarly when it is in state $-1$, the stochastic magnetic field is $b(t) = -b$. 

Define $m_{+}(t) = M_{x}(t) + iM_{y}(t)$. Its ensemble average $\braket{m_{+}(t)}$ is called free induction decay (FID)~\cite{1f}. The decay constant of the ensemble average $\braket{m_{+}(t)}$ gives $T^{*}_2$. The differential equation satisfied by $\braket{m_{+}(t)}$ and its analytical solution is given in~\cite{1f} as:
\begin{equation} 
\label{FID01}
\braket{\ddot{m}_+} + \gamma \braket{\dot{m_+}} = -b^2  \braket{m_+} 
\end{equation}
with initial conditions $  \braket{ m_+ (0)}=1$, $\braket{\dot{m}_+(0)}=0$ (The initial state of the qubit is $(\ket{0}+\ket{1})/\sqrt{2}$, such that the initial Bloch vector pointing in the $x$-direction: $M(0) = (1, 0, 0)$). The solution of Eq.~(\ref{FID01}) with these 
initial conditions is
\begin{eqnarray} 
\label{FID02}
\braket{m_+}&=&(2\mu)^{-1}e^{-\gamma t/2}\left[(\mu +1)e^{\gamma \mu t/2}+ (\mu -1)e^{-\gamma \mu t/2}\right]\, , \nonumber \\
\mu &\equiv& \sqrt{1-(2b/\gamma)^2}\, .
\end{eqnarray}

\subsubsection{Comparison of analytical solutions and stochastic simulations}
We develop new and efficient parallelizable simulation techniques based on the algorithm developed in the previous chapter on quantum trajectories (Chap.~\ref{chap: qt}). Using this algorithm, one does not need to draw a random number at each timestep, but instead calculate the waiting times for each flipping of the stochastic $b$-field.
We compare the analytical solution of Eq.~\eqref{FID02} with the stochastic random field process simulation results we produced, which are averaged over many trajectories. The simulation procedure is detailed next.

Define the ratio of the strength of the fluctuator over the rate of the fluctuator as:
\begin{equation}
    g =b/(\gamma/2) =  \frac{2b}{\gamma} \,.
\end{equation}
Using Gillispie's algorithm~\cite{gillespie1977exact} in the previous chapter of quantum trajectories (Chap.~\ref{chap: qt}) to determine the flipping (waiting) times of fluctuator (thus the sign of the stochastic magnetic field), and averaging over many realizations, we plot the time evolution of the ensemble average $\braket{m_{+}(t)}$ with $g = 0.8$ and $g = 5$. Although in each realization the initial state of the qubit is fixed, the initial state of the fluctuator is not. If the fluctuator starts in either state with equal probability, the initial condition $\braket{\dot{m}_+(0)}=0$ above is fulfilled.

\begin{figure}[h!]
\centering
\includegraphics[width=0.75\textwidth]{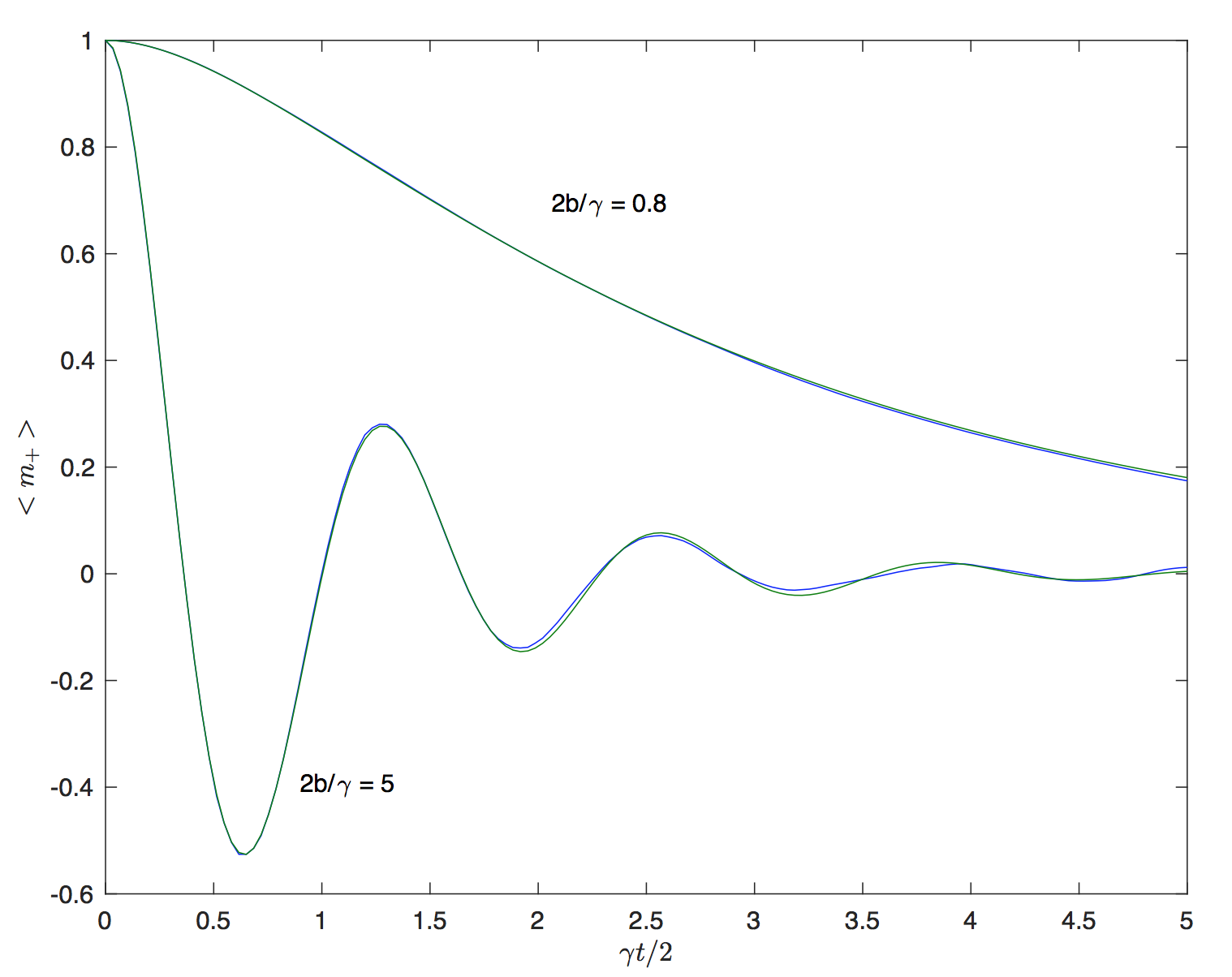}
\caption{
Blue lines: results reproduced from averaging 8000 realizations. The initial state of the fluctuator is either state 1 or $-1$ with equal probability. Green lines: analytical solutions. Note that $b = 5$ for both curves.}
\label{fig:decayfactorlook0}
\end{figure}

We numerically simulate $\braket{m_{+}(t)}$s with different $g$ and plot the results as blue lines in Fig.~\ref{fig:decayfactorlook0}. In the same figure, we also plot the analytical solutions of Eq.~\eqref{FID02} as green lines. Note that the unit of $\gamma$ is (number of switches)/(time unit)~\cite{Bergli2009}. 

\begin{figure}[h!]
\centering
\includegraphics[width=0.75\textwidth]{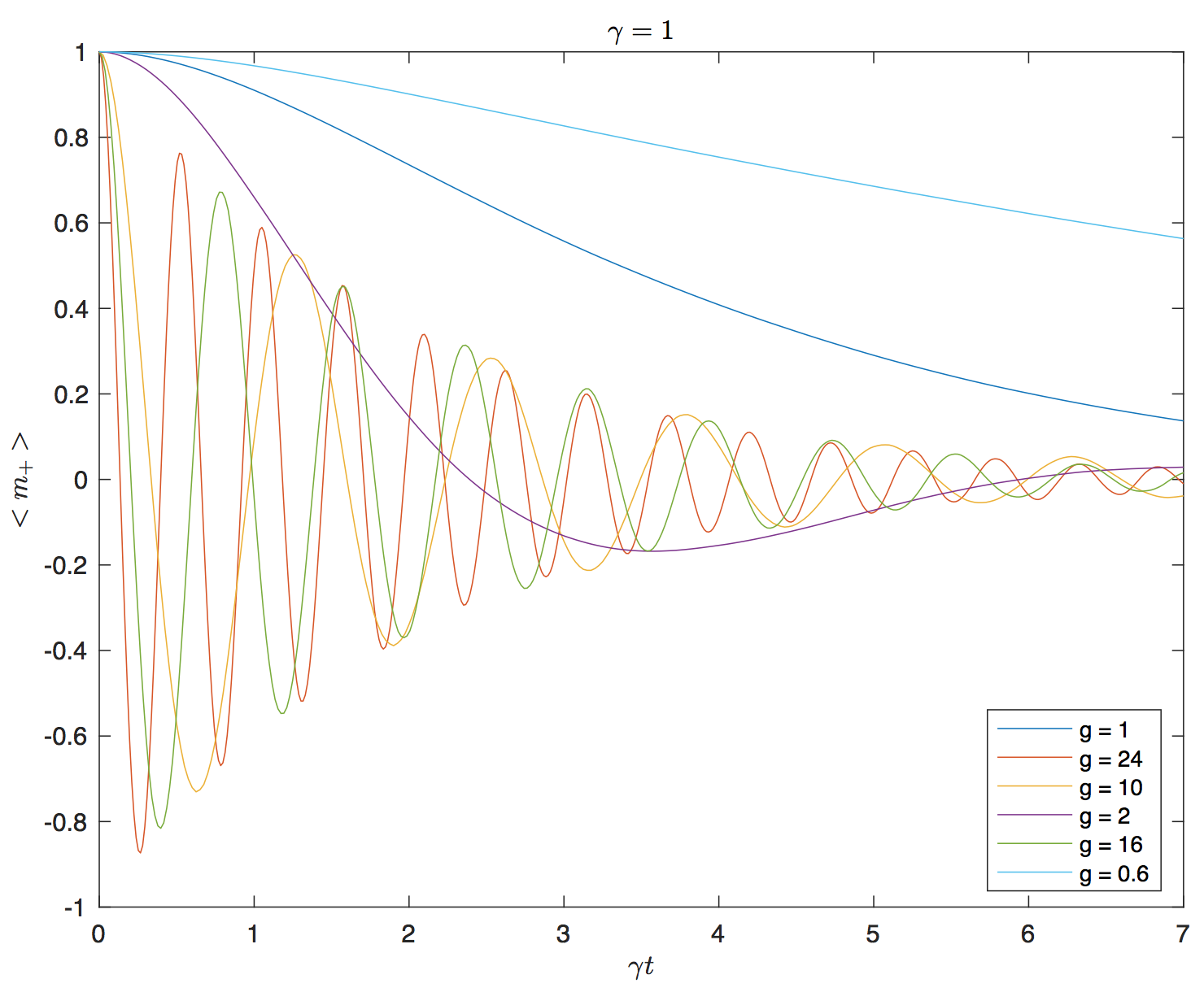}
\caption{Saturation of decay as $b$ further increases. Note that $\gamma$ = 1 switch/(time unit) for all curves. The decay rate is saturated (cannot decay faster) after $g = 2b/\gamma$ reaches 1. The curves are produced from averaging 10000 realizations with a random initial fluctuator state for each realization. }
\label{fig:decayfactorlook}
\end{figure}

We perform simulations for a wider range of $g$ and plot the results in Fig.~\ref{fig:decayfactorlook}. We have several observations. The resulting ensemble average $\braket{m_{+}}$ of the qubit decays faster (or $T^{*}_2$ is shorter) when $g$ is larger (the stronger the fluctuator is or the slower the fluctuator switches).
However, its decay rate saturates as the ratio $g$ reaches $1$, where the behavior of $\braket{m_{+}(t)}$ transitions from strictly decay to damped oscillation. 

\subsection{One qubit and an ensemble of fluctuators}
\subsubsection{An ensemble of fluctuators to generate an overall $1/\omega$ power spectrum: $S(\omega) = \mathcal A /\omega$.}
\label{sec:1fspectrum}
The following terminology is summarized from~\cite{1f, Bergli2009, paladino2003decoherence, faorodynamical}.
A single two-state fluctuator is described by a stochastic variable $\chi (t)= \pm 1$ with rates $\gamma_{1\to 2}=\gamma_{2\to 1}=\gamma/2$. The power spectrum of the noise generated by the $i$-th fluctuator is:
\begin{equation} \label{ns2}
S_i(\omega)=b_i^2 \cL_{\gamma_i} (\omega)\, ,
\end{equation}
where 
$\cL_{\gamma_i} (\omega)$ is a Lorentzian function of frequency, 
\begin{equation} \label{Lorentzian}
\mathcal{L}_{\gamma_i} (\omega) \equiv \frac{1}{\pi} \frac{\gamma_i}{\omega^2 +\gamma_i^2}.
\end{equation}
The couplings $b_i$ are distributed with a  small dispersion around an average value $\langle b \rangle$. 

If a distribution $P(\gamma) 
\propto 1/\gamma$ is assumed for the switching rates $\gamma_i \in 
[\gamma_{\text{min}}, \gamma_{\text{max}}]$, 
the total fluctuation after summing up all the fluctuating fields $\Xi (t) = \sum_i \chi_i(t)$ exhibits a 1/$\omega$ total power 
spectrum of the form $S(\omega) = \mathcal A/\omega$, $\mathcal A >0$. In this case, let the constant of proportionality in the distribution function $P(\gamma)$ be
\begin{equation}
\label{eq:1fdist}
    P(\gamma) = \frac{P_0}{2 \gamma}\,.
\end{equation}

Under these conditions the total power spectrum $S(\omega)$ reads
\begin{equation} 
S(\omega) = \langle b^2 \rangle \int_{\gamma_{\text{min}}}^{\gamma_{\text{max}}} d\gamma \, \frac{P_0}{2 \gamma} \,  \cL_{\gamma} (\omega) 
\approx    \frac{{\mathcal A}}{\omega}\, .
\label{eq:totalpowerspectrum}
\end{equation}
 The amplitude  $\mathcal A$ can be expressed in terms of the number of \fs per noise
decade, $n_d= {\mathcal N}_T \ln{(10)}/\ln{(\gamma_{\text{max}}/\gamma_{\text{min}})}$, as follows 
${\mathcal A} = \langle b^2 \rangle P_0/4 = \langle b^2 \rangle  n_d/(2 \ln(10))$.

Other notes:
\begin{itemize}
    \item $n_d$ means the number of fluctuators per decade.
    \item The strength $b_i$ of the fluctuators has Gaussian distribution with average as $\langle b \rangle$, and standard deviation $\Delta b$.
    \item The initial value of each fluctuator is randomly sampled according to thermal equilibrium distribution $\delta p_{eq}$, with $\langle \delta p_0 \rangle > 0$ means more fluctuators start at state $1$ than in $-1$.
\end{itemize}
\subsubsection{Distributions of the fluctuators}
\label{sec:distoff}
\begin{figure}[h!]
	\subfigure[]{\includegraphics[width = 0.5\columnwidth]{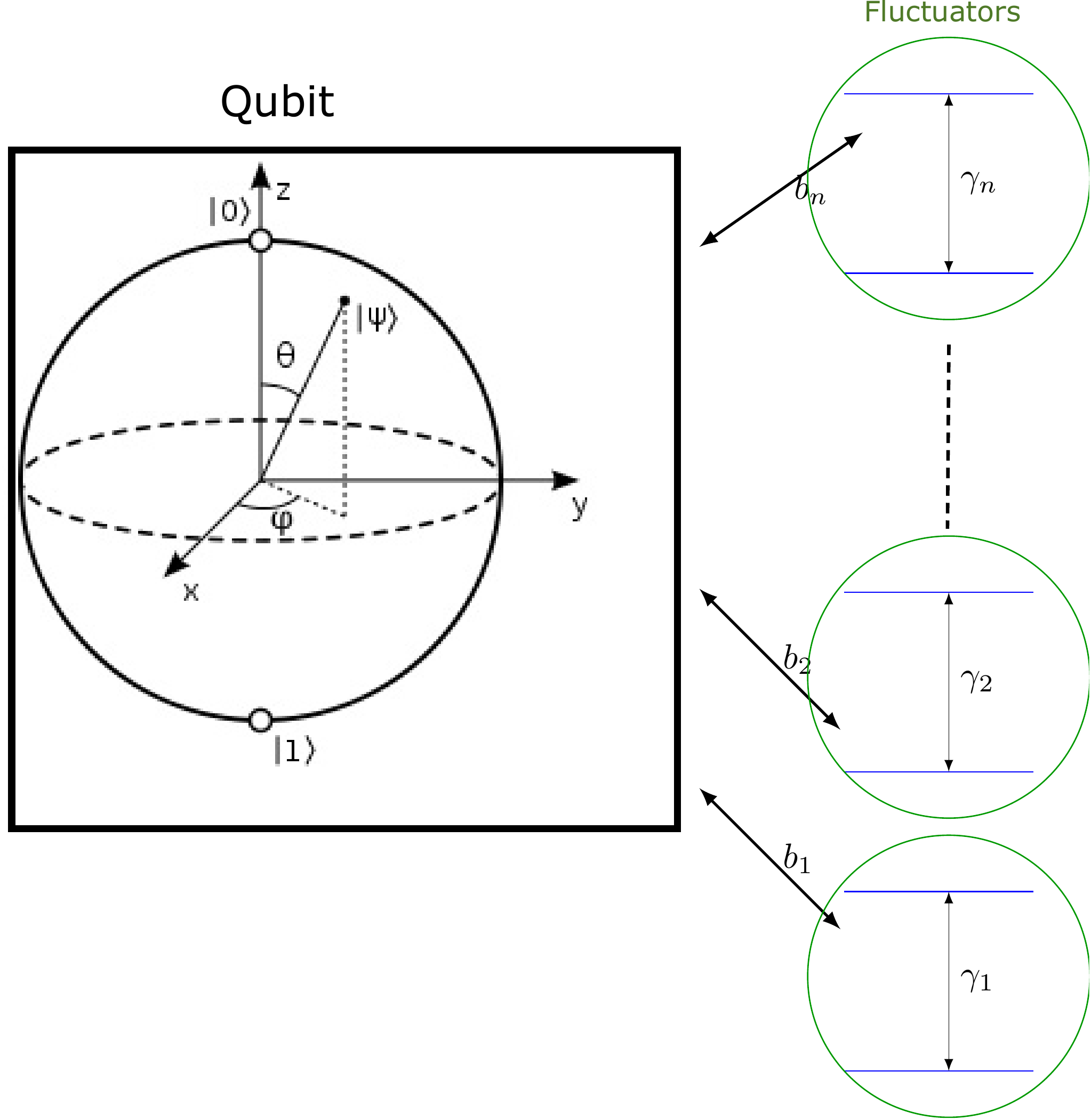}\label{fig:1fcartoon}}
	\subfigure[]{\includegraphics[width = 0.5\columnwidth]{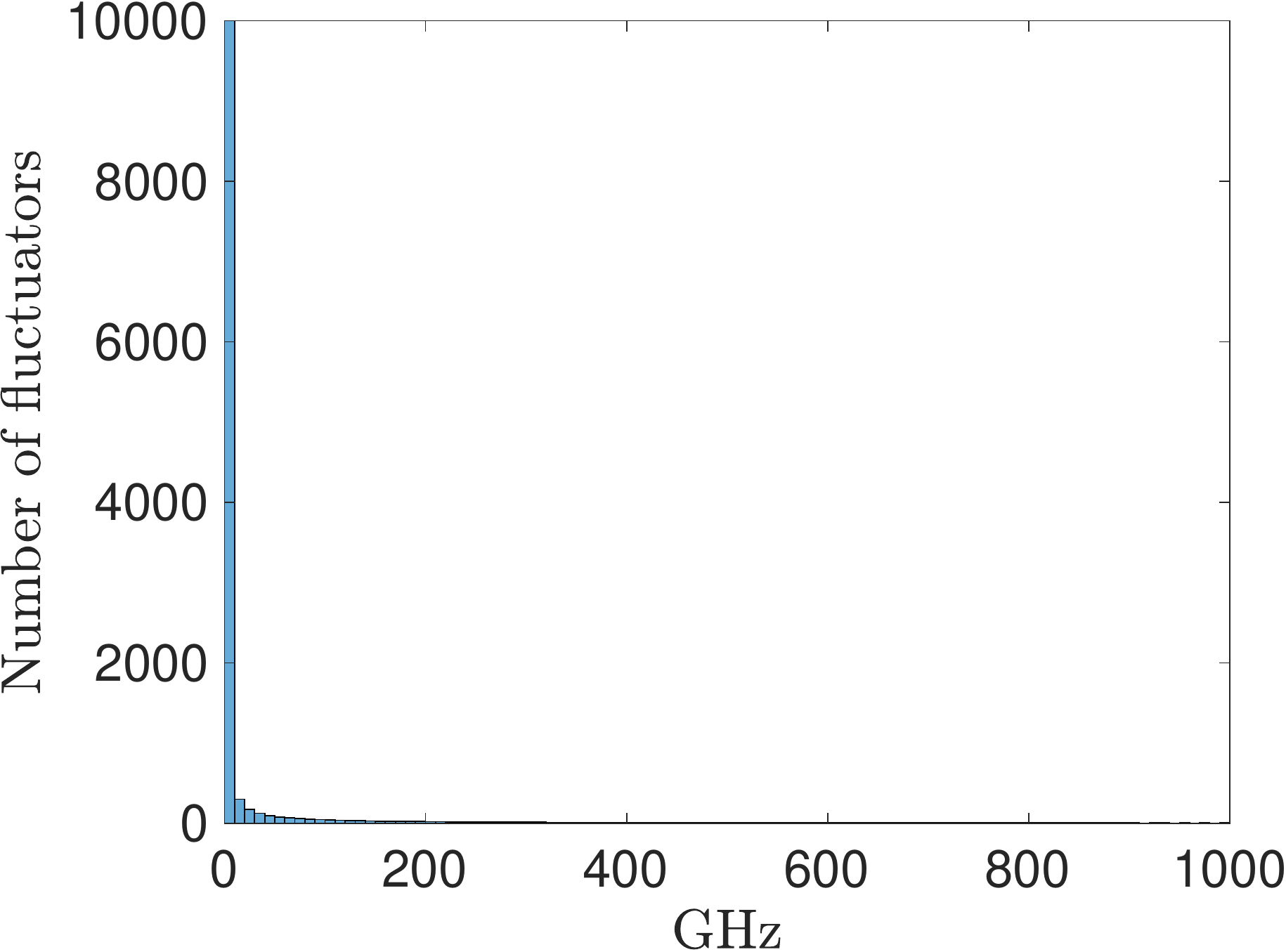}\label{fig:distribution12000}}
	\caption{(a) A cartoon illustrating a single qubit coupled to a series of fluctuators with different individual flipping frequencies
	(b) Distribution of frequencies of $12000$ fluctuators. 
	}
\end{figure}

Fig.~\ref{fig:1fcartoon} illustrates the couplings between the qubit and fluctuators with strength $b_i$ and flipping frequency $\gamma_i$.
The ($1/\gamma$) distribution function (Eq.~\eqref{eq:1fdist}) of the frequencies of the fluctuators corresponds to a uniform distribution of $\ln\gamma$.
Fig.~\ref{fig:distribution12000} shows the frequency distribution of $12000$ fluctuators, where there are $12$ decades
of noise ($[10^{0}, 10^{12}]$ Hz) and the number of \fs per decade is $n_d = 100$. 

\subsubsection{Simulation results: The importance of the number of decades and $\gamma_{\text{min}}$}
Unlike the case of a single fluctuator, there are no analytical solutions for an ensemble of fluctuators. We have to perform numerical simulations. Similar to the case of a single fluctuator, we use Gillispie's algorithm to determine the flipping times of fluctuators. At each iteration, we randomly select a fluctuator based on the uniform distribution of $\ln\gamma$ (Sec.~\ref{sec:distoff}).
Again one can measure how fast the coherence is lost by how fast the FID $\braket{m_{+}(t)}$ decays (or oscillates). 

\begin{figure}[h!]
\centering
\includegraphics[width=0.55\textwidth]{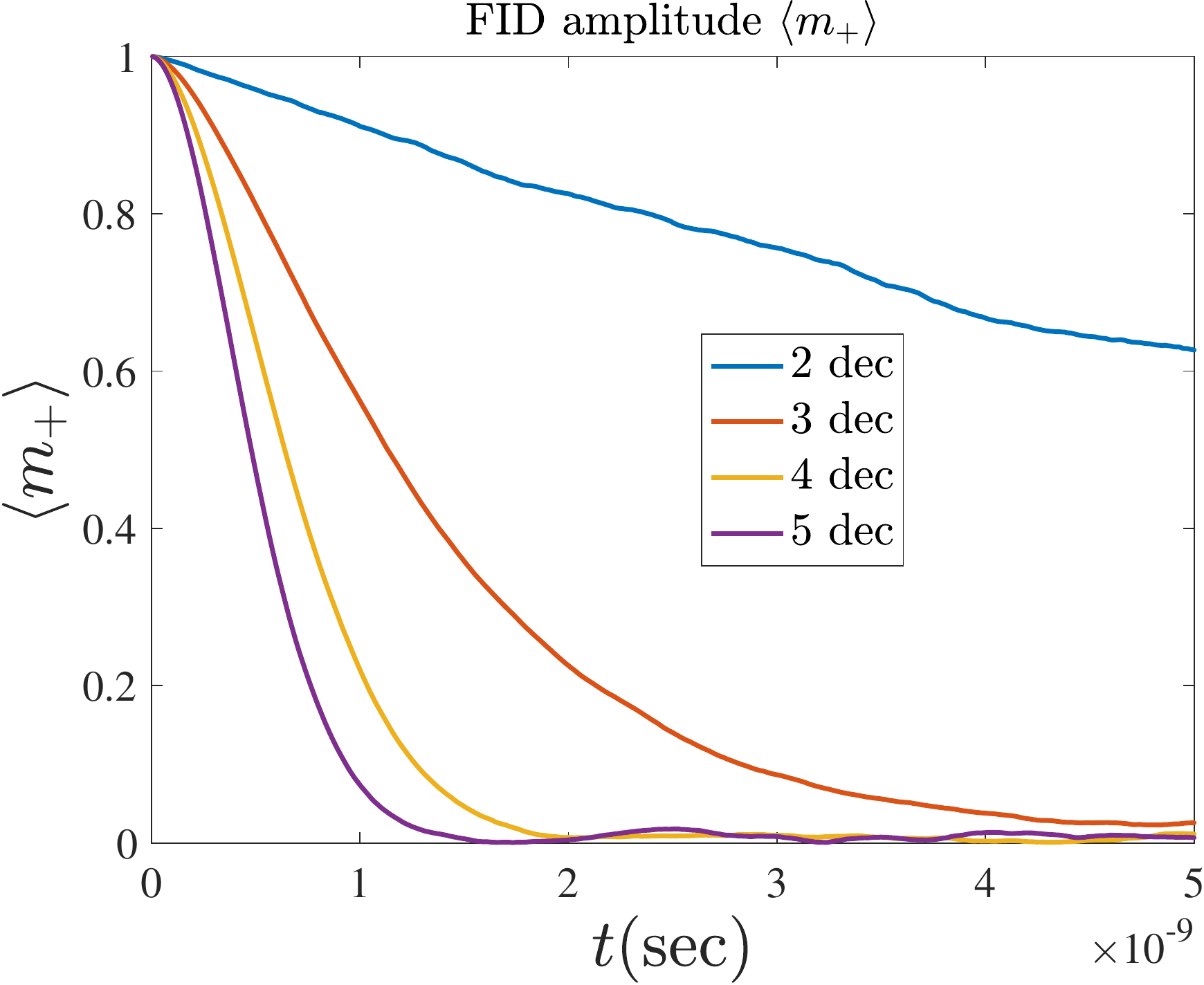}
\caption{Reproduced figures of 2 to 5 decades. Couplings distributed with $\langle \Delta b \rangle/\langle 
b \rangle =0.2$ and around $\langle b \rangle=  4.6 \times 10^7$~Hz. $\gamma_{\text{max}} = 10$~THz with different $\gamma_{\text{min}}$.  $n_d = 1000$. 
$dp_{0j}= \pm 1$ are distributed according to $\langle dp_{0j} \rangle= dp_{\text{eq}} = 0.08$.}
\label{fig:decadessensitity}
\end{figure}

Due to the $1/\gamma$ distribution function, most of the fluctuators have their flipping frequency close to $\gamma_{\text{min}}$. By fixing $\gamma_{\text{max}}$, the more decades of noise, the smaller the $\gamma_{\text{min}}$. We perform simulations to investigate the sensitivity of qubit coherence to the number of decades of fluctuator frequencies (and thus $\gamma_{\text{min}}$) and plot the results in Fig.~\ref{fig:decadessensitity}. For example, if $\gamma_{\text{max}} = 10$ THz, $2$ decades of noise means $\gamma_{\text{min}} = 100$ GHz and $4$ decades of noise means $\gamma_{\text{min}} = 1$ GHz. As we can see in Fig.~\ref{fig:decadessensitity}, the decay of $\braket{m_{+}(t)}$ is faster with more decades of noise (and thus a smaller $\gamma_{\text{min}}$).

\subsubsection{Simulation results: Dependence on $\Delta b /\langle b \rangle$} 

\begin{figure}[h!]
\centering
\includegraphics[width=0.7\textwidth]{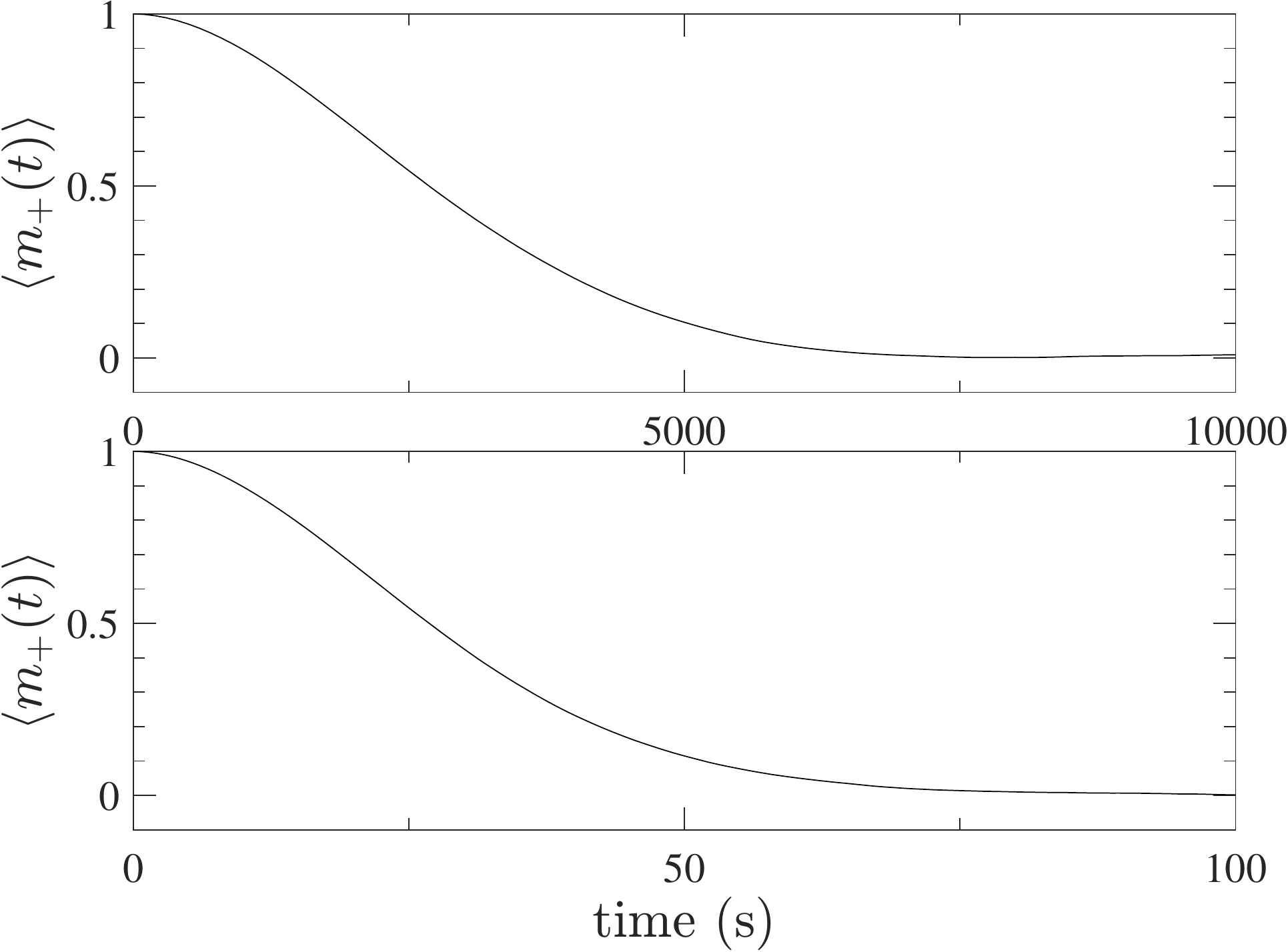}
\caption{Top: $\gamma_{\text{min}}=10^{-6}$ GHz, $\gamma_{\text{max}}=100$ GHz, $n_d=100$, $\langle b\rangle = 10^{-4}/4$;
Bottom: $\gamma_{\text{min}}=10^{-4}$ GHz, $\gamma_{\text{max}}=100$ GHz, $n_d=100$, $\langle b\rangle = 0.01/4$.
Results are averaged over $8k$ realizations.  Each fluctuator is initially in a pure 
state randomly sampled according to 
$\langle \delta p_0 \rangle= \delta p_{eq}=0.08$. $\Delta b /\langle b \rangle \sim 0.2$.} 
\label{fig:decayfactorlook2}
\end{figure}

We plot in Fig.~\ref{fig:decayfactorlook2} another case where the decay time is sensitive to the average strength of fluctuators $\gamma_{\text{min}}$. In general, the stronger the fluctuators are the faster  $\braket{m_{+}(t)}$ decays. We observe two interesting points in Fig.~\ref{fig:decayfactorlook2}. Comparing the top subfigure with the bottom subfigure, the bottom subfigure has $\Delta b /\langle b \rangle$ increased $100$ times.
\begin{itemize}
    \item The decay in the bottom subfigure is $100$ times faster.
    \item to obtain a similar \textbf{decay envelope} as in the top subfigure,  $\gamma_{\text{min}}$ also has to increase $100$ times.
\end{itemize}

\subsubsection{Simulation results: Non-exponential decay}
\begin{figure}[h!]
\centering
\includegraphics[width=0.77\textwidth]{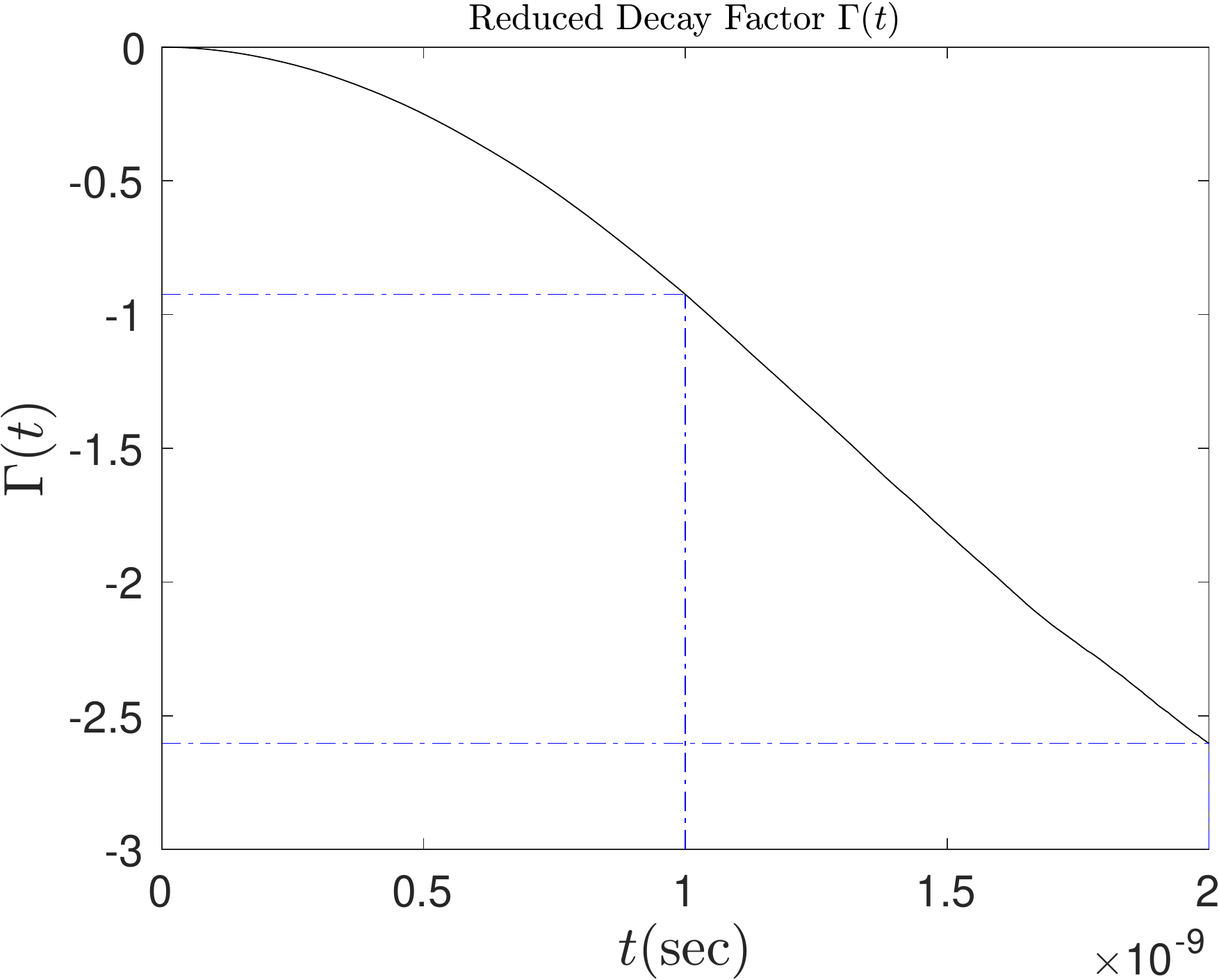}
\caption{$12$ decades of noise: $\gamma_{\text{max}} = 10^{12}$~Hz with $\gamma_{\text{min}}= 10^{0}=1$~Hz. Couplings distributed with $\langle \Delta b \rangle/\langle 
b \rangle =0.2$ and around $\langle b \rangle =  (4.6/\pi) \times 10^7 $~Hz.   $n_d = 1000$. $dp_{0j}= \pm 1$ are distributed according to $\langle dp_{0j} \rangle= dp_{\text{eq}}$. As time doubles, the reduced decay factor drops more than $2$ times.}
\label{fig:nonexponential}
\end{figure}

We have seen in Fig.~\ref{fig:decadessensitity} that the more decades of noise are included, the more rapid the coherence is lost. For exponential decay, $\ln{\braket{m_{+}(t)}}$ would decrease linearly. We can define the reduced Decay factor of the qubit as $\Gamma(t) = \ln{\braket{m_{+}(t)}}$~\cite{1f}. We plot in Fig.~\ref{fig:nonexponential} the effect of the $12000$ fluctuators on the $\Gamma(t)$ evolution with frequencies distributed according to Fig.~\ref{fig:distribution12000}.

As seen in Fig.~\ref{fig:nonexponential}, the reduced decay factor does not decrease in a linear manner. As time doubles from $1$ns to $2$ns, the reduced decay factor drops more than $2$ times.

\section{$1/f$ noise: Stochastic Hamiltonian formulation}
We consider the stochastic Hamiltonian formulation of $1/f$ noise, which can also be applied to the case where the Hamiltonian drive of the qubit is time-dependent.

A Hamiltonian is only defined up to a unitary transformation (change of basis). Therefore we can define the system Hamiltonian with $\sigma_x$ and $\sigma_z $ components only, and with real $A(t)$ and $B(t)$ (for example, they can be the DWave NASA schedule in Fig.~\ref{fig: DWavenasaschedule}):

\begin{equation}\label {qb1}
\hat{H}_{\text{sys}}=\frac{1}{2}\left(A(t) \,  \sigma_x + B(t) \, \sigma_z \right) \, .
\end{equation}
It can be cast into the spin-(time dependent) magnetic field description in Sec.~\ref{sec:bbfield} as

\begin{equation} \label{qb2}
\hat{H}_{\text{sys}}=(1/2)\mathbf{B}\cdot \sigma\, , \ B_z\equiv  B(t), \ B_x \equiv A(t)\, .
\end{equation}

The Schr{\"o}dinger equation with this system Hamiltonian drive turns out to be equivalent to the precession equation for the
Bloch vector:
\begin{equation} \label{bv1:H}
\dot{\mathbf{M}} = \mathbf{B} \times \mathbf{M} \, .
\end{equation}


\subsection{Single fluctuator}
\label{sec:qubitfluctuator}
A single fluctuator induces an extra stochastic magnetic field $b_1(t) = b_1 \chi_1(t)$ along the $B_z$ axis. Its effect can be included in the $\hat{H}_{\text{sys}}$, as:
\begin{equation}
\hat H_{\text{sys}} = \frac{1}{2}\left(A(t) \,  \sigma_x + B(t) \, \sigma_z \right)
+ \frac{1}{2} \,  b_1 \chi_1(t) \sigma_z \,,
\label{H_SF_general1}
\end{equation}
where $\chi_1(t)$ switches with rate $\gamma_1/2$. The $\frac{1}{2}$ convention follows from Sec.~\ref{sec:1fspectrum}.

\subsection{Many fluctuators}
For many fluctuators,
\begin{equation}
\hat H_{\text{sys}} = \frac{1}{2}\left(A(t) \,  \sigma_x + B(t) \, \sigma_z \right)
+ \frac{1}{2} \,  \sum_{i}b_i \chi_i(t) \sigma_z \,,
\label{H_SF_general_many}
\end{equation}
where $\chi_i(t)$ switches with rate $\gamma_i/2$. 

The flipping frequencies  $\chi_i(t)$ of the fluctuator-$i$ is randomly drawn from the distribution of $1/\gamma_i$. The initial thermal distribution of $\chi_i(t)$, and Gaussian distribution of $b_i$ follows from Sec.~\ref{sec:1fspectrum}.

\subsection{Case studies}
\subsubsection{Single fluctuator: $A(t) = 0$ and $B(t) = 0$ for all $t$}
Consider the following system Hamiltonian, where the system drive is inherently zero ($A(t) = 0$ and $B(t) = 0$). Therefore the system Hamiltonian only has a single fluctuator pointing along the $z$-axis.
\begin{equation}
\hat H_{\text{sys}} =  b_1 \chi_1(t) \sigma_z \,,
\label{H_SF_general0}
\end{equation}
is the case where there is no system driving. (where $\chi_1(t)$ switches with $\gamma_1/2$. )

The qubit's initial state is $\ket{+} = \frac{\ket{0}+\ket{1}}{\sqrt{2}}$. Denote $\ket{\psi(t)} = c(t)\ket{0} + d(t)\ket{1}$ the qubit state at time $t$. With this initial condition, the Bloch vector can be expressed in terms of the $\ket{+}$'s coefficients as:
\begin{align}
M_{x}(t) &= cd^{*} + c^{*}d,\\
M_{y}(t) &= icd^{*} - ic^{*}d \,.
\end{align}

\begin{figure}[h!]
\centerline{\includegraphics[width=10.5cm]{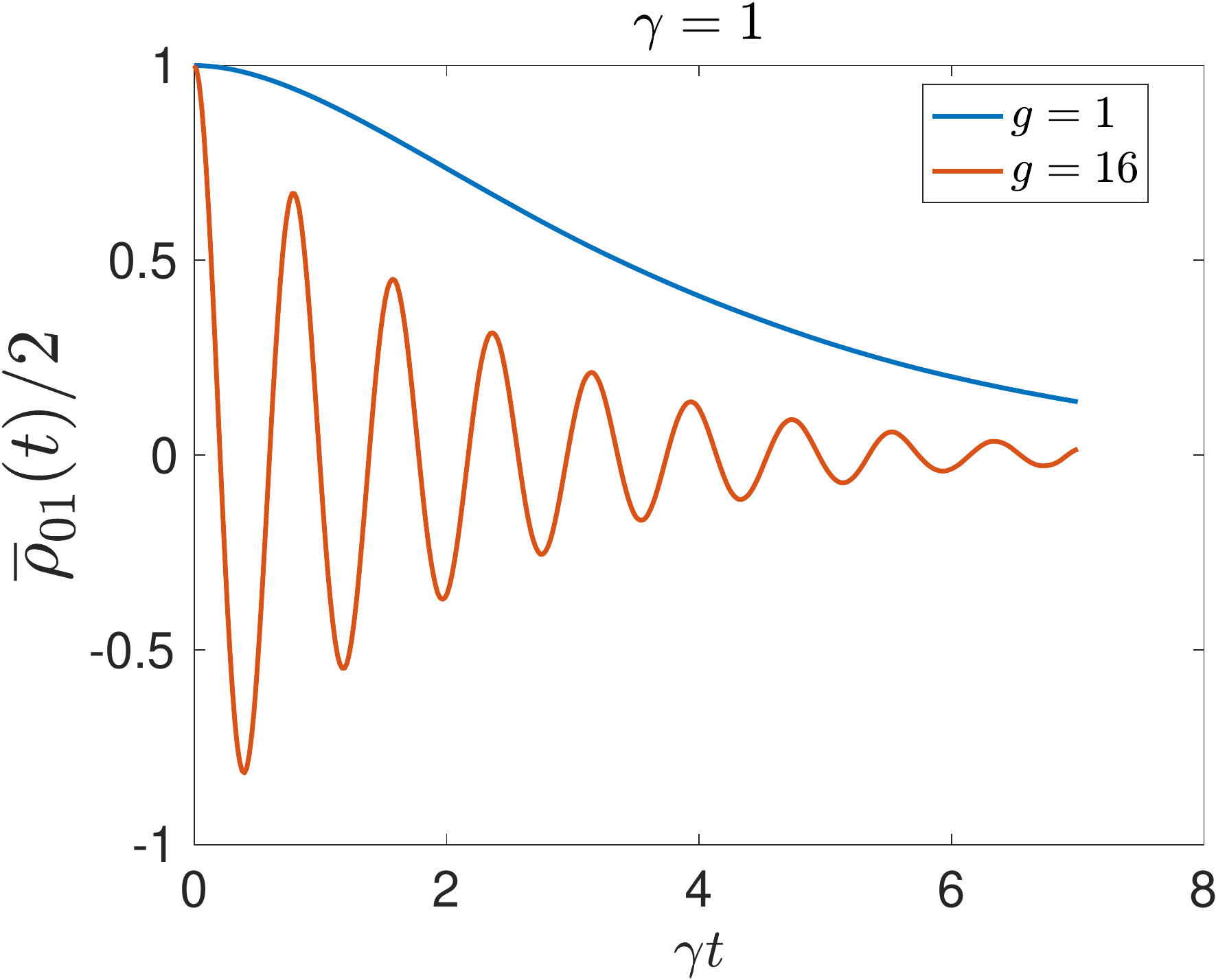}}
\caption{$g = 2b/\gamma$. Averaged over $10000$ trajectories.}
\label{fig:hamdecayfactor0}
\end{figure}

Denote $|\psi(t)\rangle$ the state vector obtained in one trajectory of the stochastic Schr\"{o}dinger equation:
\begin{equation}
    i\hbar\frac{d}{dt}|\psi(t)\rangle = \hat H_{\text{sys}}(t)|\psi(t)\rangle \, .
\end{equation}
After averaging many trajectories, we recover the density matrix evolution. Define $\overline{\rho}(t)$ as
\begin{align}
\overline{\rho}(t) = \braket{|\psi(t)\rangle\langle \psi(t)|} &= \frac {1}{2}{\begin{pmatrix}1+\braket{M_{z}(t)} & \braket{M_{x}(t)}-i\braket{M_{y}(t)}\\\braket{M_{x}(t)}+i\braket{M_{y}(t)} & 1-\braket{M_{z}(t)}\end{pmatrix}} \nonumber\\
&= \frac {1}{2}{\begin{pmatrix}1+\braket{M_{z}(t)} & \braket{M_{x}(t)}-i\braket{M_{y}(t)}\\\mathbf{\braket{m_{+}(t)}} & 1-\braket{M_{z}(t)}\end{pmatrix}} \,.
\label{eq:fiddm}
\end{align}
The off-diagonal element of the density matrix coincides with the definition of $\braket{m_{+}(t)}$ in Sec.~\ref{sec:fluctuatorbfield}. We plot in Fig.~\ref{fig:hamdecayfactor0} the evolution the off-diagonal element $\overline{\rho}_{01}(t)/2$ of the density matrix, by performing a trajectories simulation of the stochastic  Schr{\"o}dinger equation, for two values of $g= 2b_1/\gamma_1$. Note that in this time-independent drive (Eq.~\eqref{H_SF_general0}) the $\overline{\rho}_{01}(t)/2$ is a real number ($\braket{M_{y}(t)} = 0$). We see in Fig.~\ref{fig:hamdecayfactor0} that the simulation data agrees with Fig.~\ref{fig:decayfactorlook} obtained by the Bloch vector-Magnetic field approach.

\subsubsection{Multiple fluctuators with a time-dependent system Hamiltonian}
Consider the following example.
\begin{equation}
\hat H_{\text{sys}} = -\frac{A}{2}\left(\left(1-\frac{t}{t_f}\right) \,  \sigma_x  + \frac{t}{t_f} \, \sigma_z \right)
+ \frac{1}{2} \,  \sum_{i = 1}^{10}b_i \chi_i(t) \sigma_z \, .
\label{H_SF_general3}
\end{equation}
where: 
\\ $A = 1$ GHz; The sum over fluctuator-$i$ with the fluctuator $\chi_i(t)$ distributed in $1/\gamma_i$ with  $\gamma_i \in [\gamma_{\text{min}}, \gamma_{\text{max}}]$; \\
$\gamma_{\text{min}} = 0.1$ GHz, $\gamma_{\text{max}} = 10$ GHz. \\
$\chi_i(t)$ switches with $\gamma_i/2$ and takes value of $\{-1, 1\}$;\\
The strengths of the fluctuators has mean $\langle b\rangle = 0.2$ GHz
and standard deviation $\Delta b /\langle b \rangle = 0.2$.

\begin{figure}[h!]
	\subfigure[]{\includegraphics[width = 0.5\columnwidth]{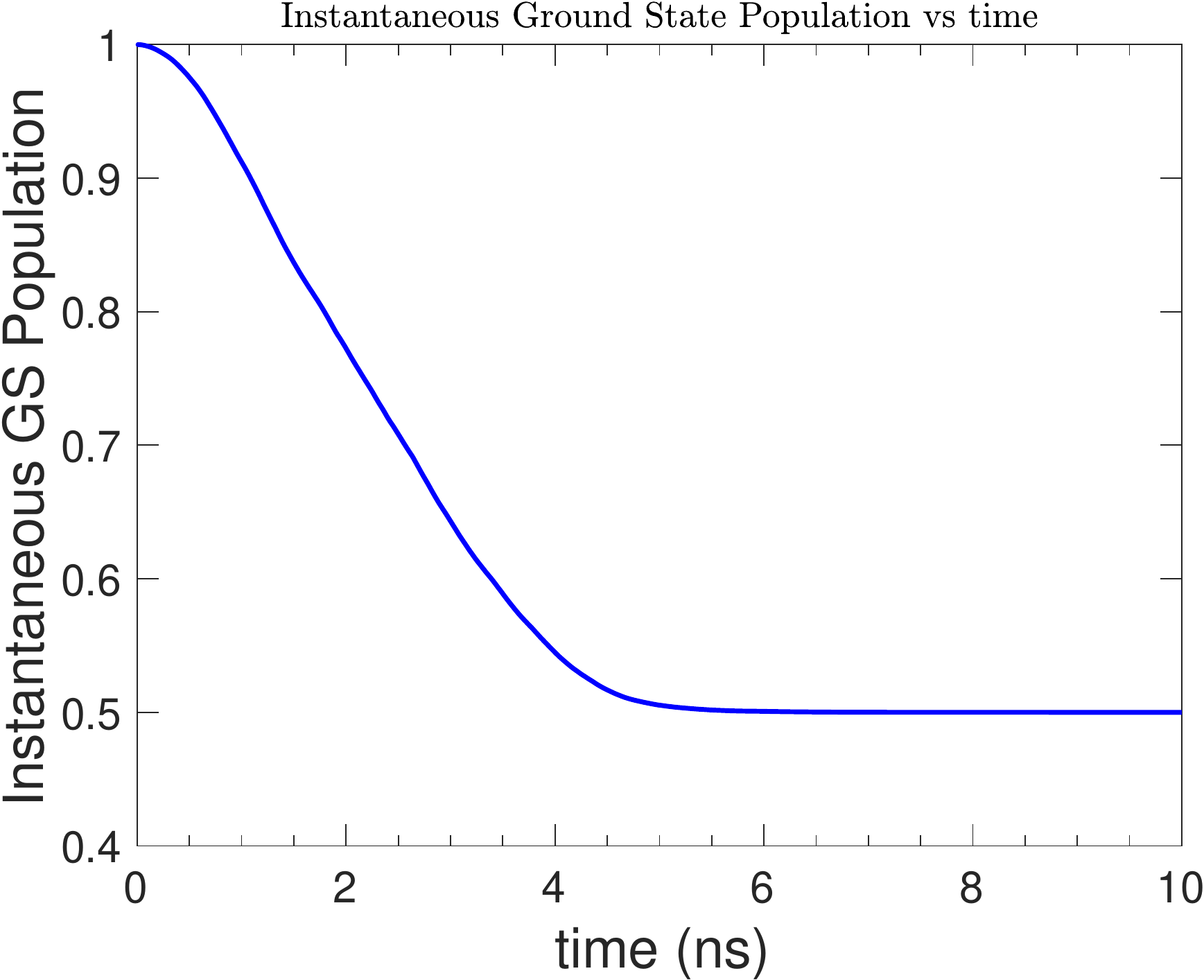}\label{fig:GS}}
	\subfigure[]{\includegraphics[width = 0.5\columnwidth]{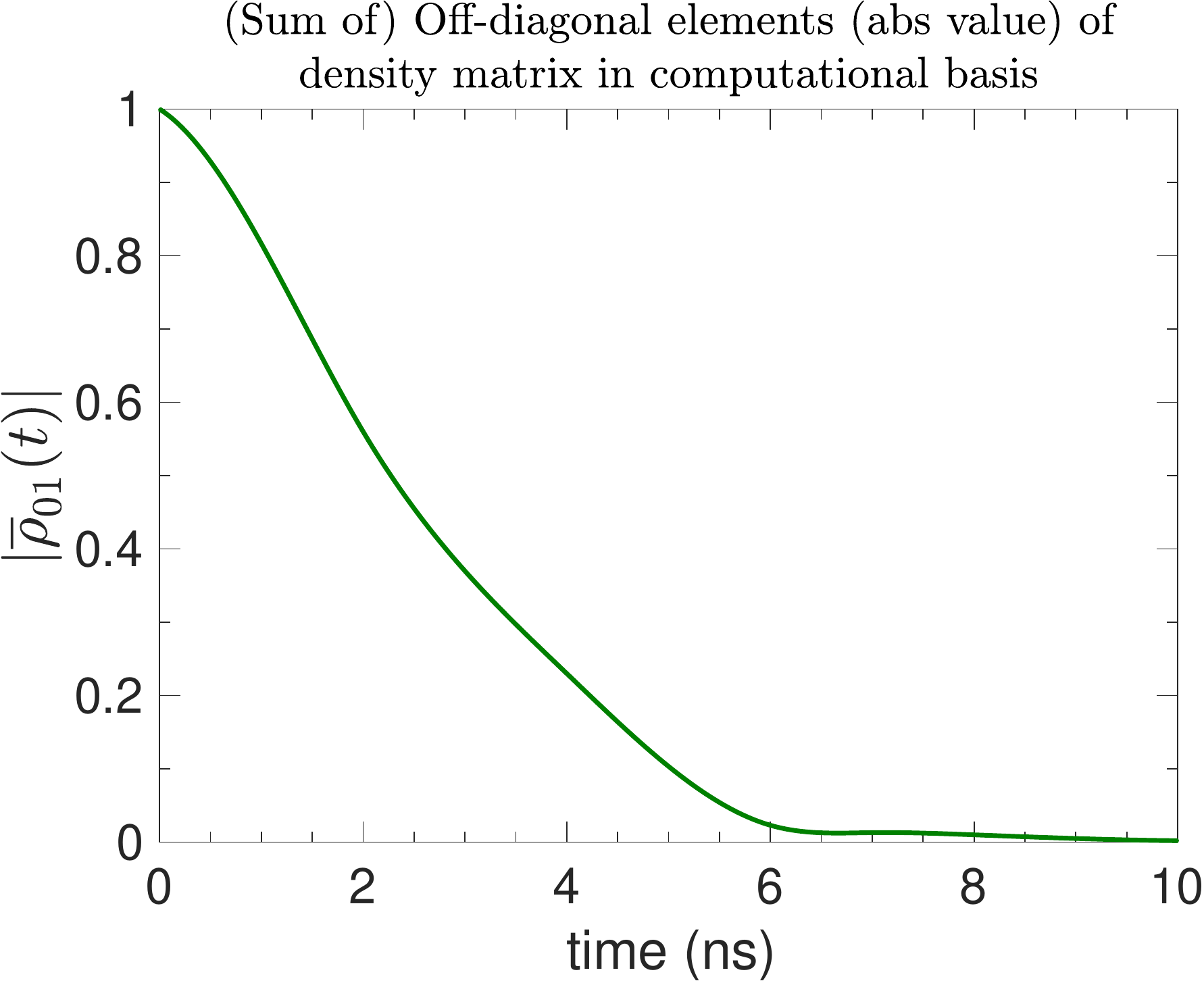}\label{fig:coherence}}
	\caption{Pure dephasing noise: $\gamma_{\text{min}}=0.1$ GHz, $\gamma_{\text{max}}=10$ GHz. 
Number of \fs is $10$. $\langle b\rangle = 0.2$ GHz,
$\Delta b /\langle b \rangle = 0.2$. Results are averaged over $30k$ trajectories.  The initial value of each fluctuator is randomly sampled according to the thermal equilibrium distribution $\delta p_{eq}= 0.08$ (i.e. each fluctuator is $0.08$ more likely to be initialized in state $1$ than state $-1$). (a) Instantaneous Ground state population, (b) (Absolute value) of off-diagonal elements in the computational basis.
	}
\end{figure}

For the ground state population (where the ground state is defined w.r.t the system Hamiltonian without the stochastic noise), its evolution with an annealing time $t_f = 1 \mu s$ is plotted in Fig.~\ref{fig:GS}.

For the coherence, because of the time-dependent system Hamiltonian, the off-diagonal part of the density matrix (in the computational basis) now has a nonzero imaginary component. Therefore, the absolute value 
$|\overline{\rho}_{01}(t)| = \frac{1}{2}|\braket{M_{+}(t)}|$
is plotted. Its evolution with an annealing time $t_f = 1 \mu s$ is plotted in Fig.~\ref{fig:coherence}.

\section{Effect of temperature, $T_1$, $T_2^{*}$ decay time, and strength of fluctuators}
We investigate the temperature dependence of $1/f$ telegraph noise simulation on an annealing Hamiltonian; and look into the $T_1$, $T_2^{*}$ decay due to the $1/f$ telegraph noise.
\subsection{Two temperature limits}
\label{sec:2Tl}
The initial distribution of fluctuators depends on temperature. The initial value of each fluctuator is randomly sampled according to the thermal equilibrium distribution $\delta p_{eq}$, which is related to the temperature $T$ by~\cite{1f}: 
\begin{equation}
\label{eq:fluctuatorequilibrium}
    \delta p_{eq} = 1 - (e^{-\epsilon/2 k_B T} )\simeq  \epsilon/2 k_B T \,.
\end{equation}
We now perform simulations in the two temperature limits ($T\rightarrow \infty$ and $T\rightarrow 0$).
Here we study the system evolution with the following annealing Hamiltonianm whose 
initial state is also its ground state:
\begin{equation}
\hat H_{\text{sys}} = -\frac{A}{2}\left(\left(1-\frac{t}{t_f}\right) \,  \sigma_x  + \frac{t}{t_f} \, \sigma_z \right)
+ \frac{1}{2} \,  \sum_{i}b_i \chi_i(t) \sigma_z \, .
\label{H_SF_general2}
\end{equation}
where: 
\\ $A = 1$ GHz;
\\The sum over fluctuator-$i$ with the fluctuator $\chi_i(t)$ distributed in $1/\gamma_i$ with  $\gamma_i \in [\gamma_{\text{min}}, \gamma_{\text{max}}]$; \\
$\gamma_{\text{min}} = 0.01$ GHz, $\gamma_{\text{max}} = 1$ GHz. \\
$\chi_i(t)$ switches with $\gamma_i/2$ and takes value of $\{-1, 1\}$;\\
The strengths of the fluctuators has mean $\langle b\rangle = 0.2$ GHz and standard deviation $\Delta b /\langle b \rangle = 0.2$.

\subsubsection{Infinite temperature $T \rightarrow \infty$} 
\begin{figure}[h!]
\subfigure[]{\includegraphics[width = 0.5\columnwidth]{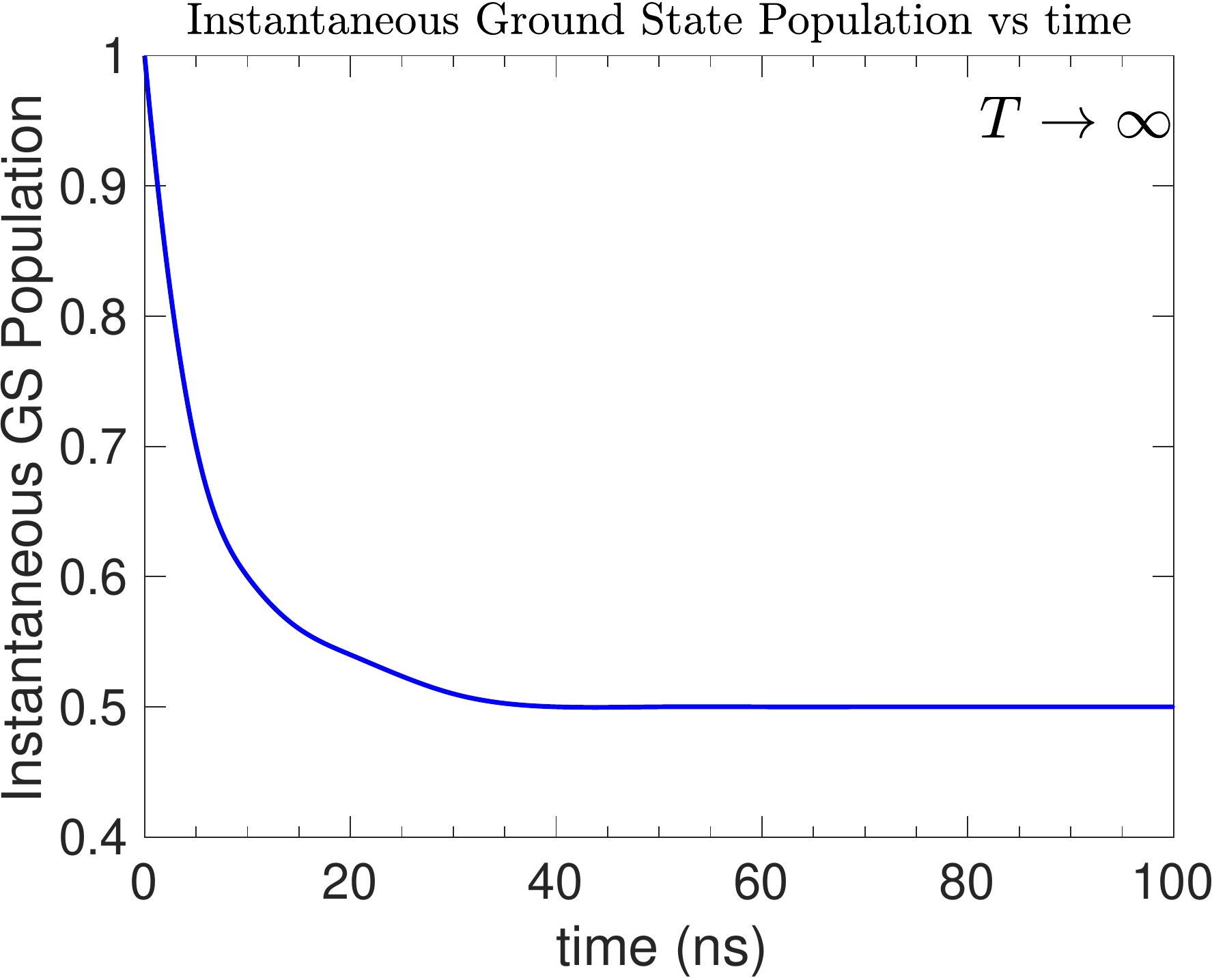}}
\subfigure[]{\includegraphics[width = 0.5\columnwidth]{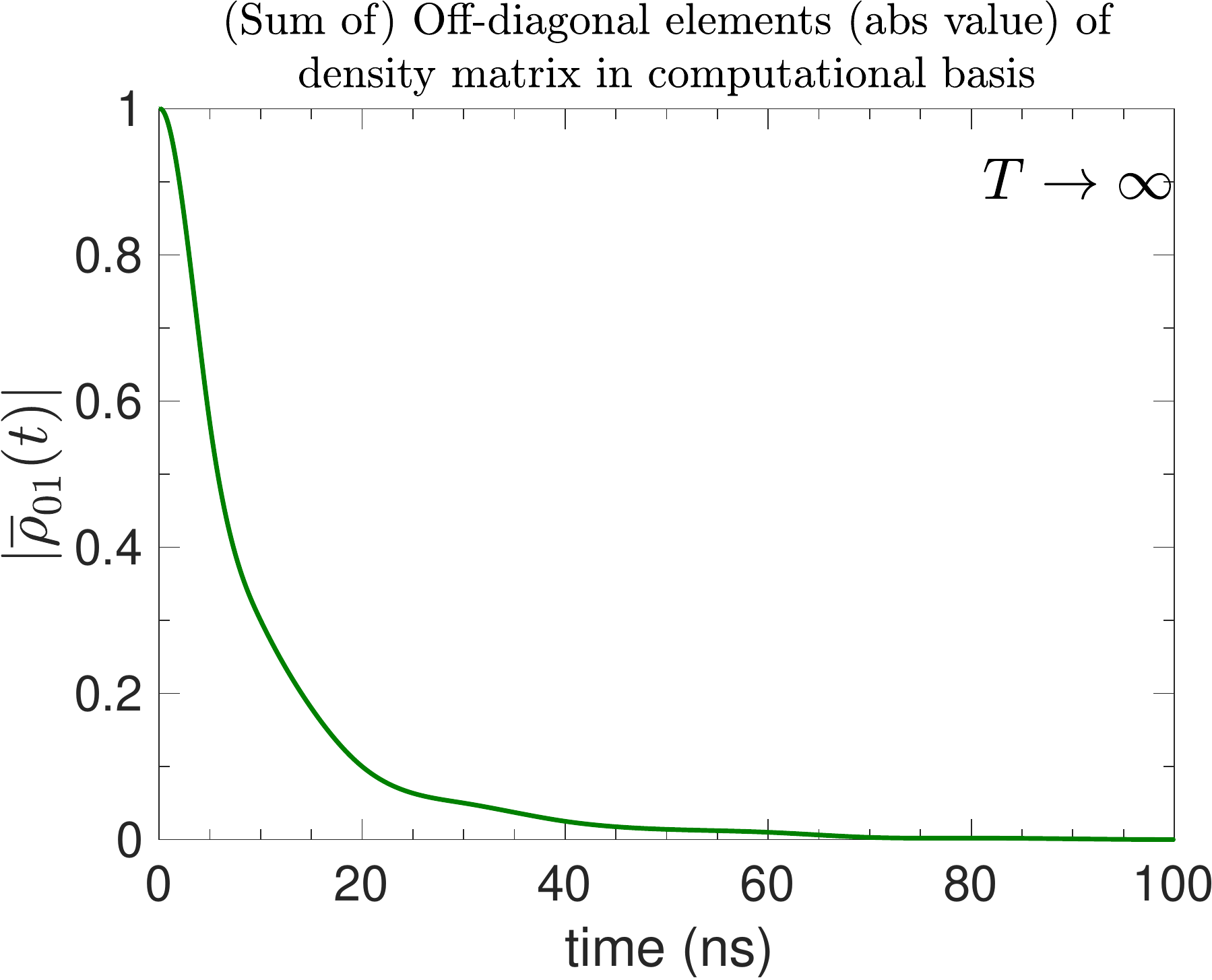}}
\caption{High temperature limit, pure dephasing noise: $\gamma_{\text{min}}=0.01$ GHz, $\gamma_{\text{max}}=1$ GHz. 
Number of \fs is $20$. $\langle b\rangle = 0.2$ GHz,
$\Delta b /\langle b \rangle = 0.2$. Results are averaged over $30k$ trajectories.  The initial value of each fluctuator is randomly sampled according to thermal equilibrium distribution $\delta p_{eq}= 0$ (i.e. each fluctuator is equally likely to be initialized in state $1$ than state $-1$), since $T \rightarrow \infty$. (a) Instantaneous Ground state population, (b) (Absolute value) of off-diagonal elements in the computational basis.}
\label{fig:highTlimit}
\end{figure}

Take $T \rightarrow \infty$, the thermal equilibrium distribution of $\delta p_{eq} \rightarrow 0$. In this case, the initial distribution of fluctuators has equal probability of being in spin up or spin down, i.e. $P(\chi_i(0) = 1) = P(\chi_i(0) = -1) = \frac{1}{2}$.
Fig.~\ref{fig:highTlimit} shows that the steady state is a maximally mixed state, with the (instantaneous) ground state approaching $1/2$, and off-diagonal elements in computational basis decaying to $0$.

\subsubsection{Zero temperature $T \rightarrow 0$}
\begin{figure}[h!]
\subfigure[]{\includegraphics[width = 0.5\columnwidth]{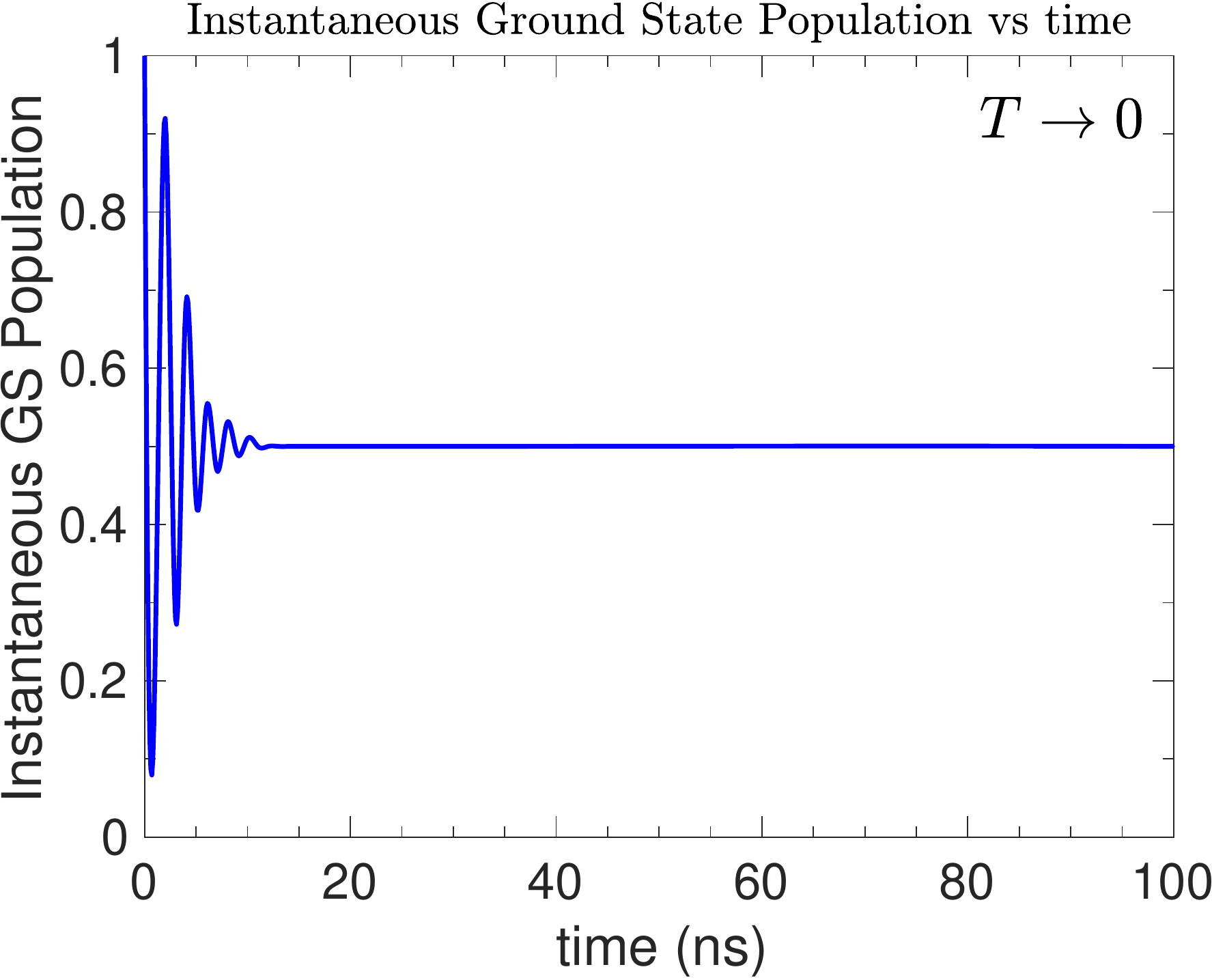}}
\subfigure[]{\includegraphics[width = 0.5\columnwidth]{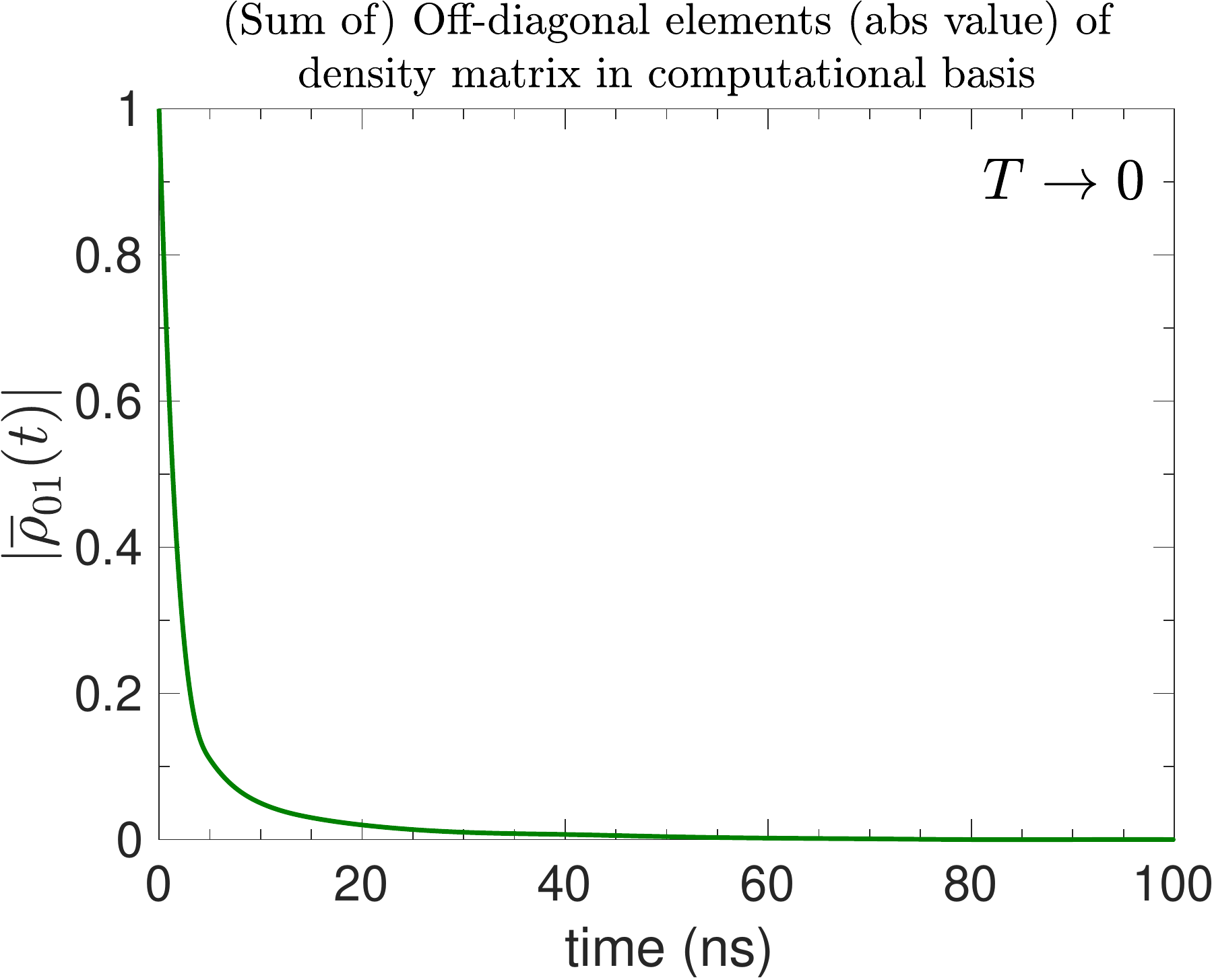}}
\caption{
Zero temperature limit, pure dephasing noise. Same fluctuator parameters as in Fig.~\ref{fig:highTlimit}. Results are averaged over $30k$ trajectories.  $\delta p_{eq}= 1$, since $T \rightarrow 0$. (a) Instantaneous Ground state population, (b) (Absolute value) of off-diagonal elements in the computational basis.
}
\label{fig:zeroTlimit}
\end{figure}

For $T \rightarrow 0$, the thermal equilibrium distribution of $\delta p_{eq} \rightarrow 1$. In this case, the initial distribution of fluctuators are all collectively spin up, i.e. $P(\chi_i(0) = 1) = 1$. 
Fig.~\ref{fig:zeroTlimit} shows that the steady state is also a maximally mixed state, with (instantaneous) ground state approaching $1/2$, and off-diagonal elements in computational basis decaying decaying to $0$.
Assuming the strengths and frequencies of the fluctuators do not depend on temperature, with temperature $T \rightarrow 0$, we can see that the fluctuators destroy the coherence faster and lead to oscillations in the (instantaneous) GS population early in the anneal.

\subsection{$T_{1}$ and $T_{2}^{*}$ time}
\subsubsection{$T_1$ time}
To measure the $T_1$ time, we initialize the state in the $\ket{1}$ state and evolve it with the following Stochastic Hamiltonian with fluctuators in the transverse direction $\sigma_x$ (Note that with fluctuator as $\sigma_z$, the $T_{1}$ time is essentially infinity since it commutes with the system Hamiltonian):
\begin{equation}
    \hat H_{\text{sys}} = \frac{E}{2}\sigma_z + \frac{1}{2} \,  \sum_{i}b_i \chi_i(t) \sigma_x \,,
\end{equation}
where $\chi_i(t)$ switches with $\gamma_i/2$. The flipping frequencies  $\chi_i(t)$ of the fluctuator-$i$ is randomly drawn from the distribution of $1/\gamma_i$ At large time $t \gg T_1$, after averaging many trajectories, the resulting density matrix under this initial condition and Hamiltonian can be expressed as:
\begin{equation}
\overline{\rho}(t)  = {\begin{pmatrix}-\frac{1}{2}e^{-\frac{t}{T_1}} + \frac{1}{2} & 0\\
0 & \frac{1}{2}e^{-\frac{t}{T_1}} + \frac{1}{2}\end{pmatrix}}
\label{eq:fiddm_form}
\end{equation}

\begin{figure}[h!]
\centerline{\includegraphics[width=12.5cm]{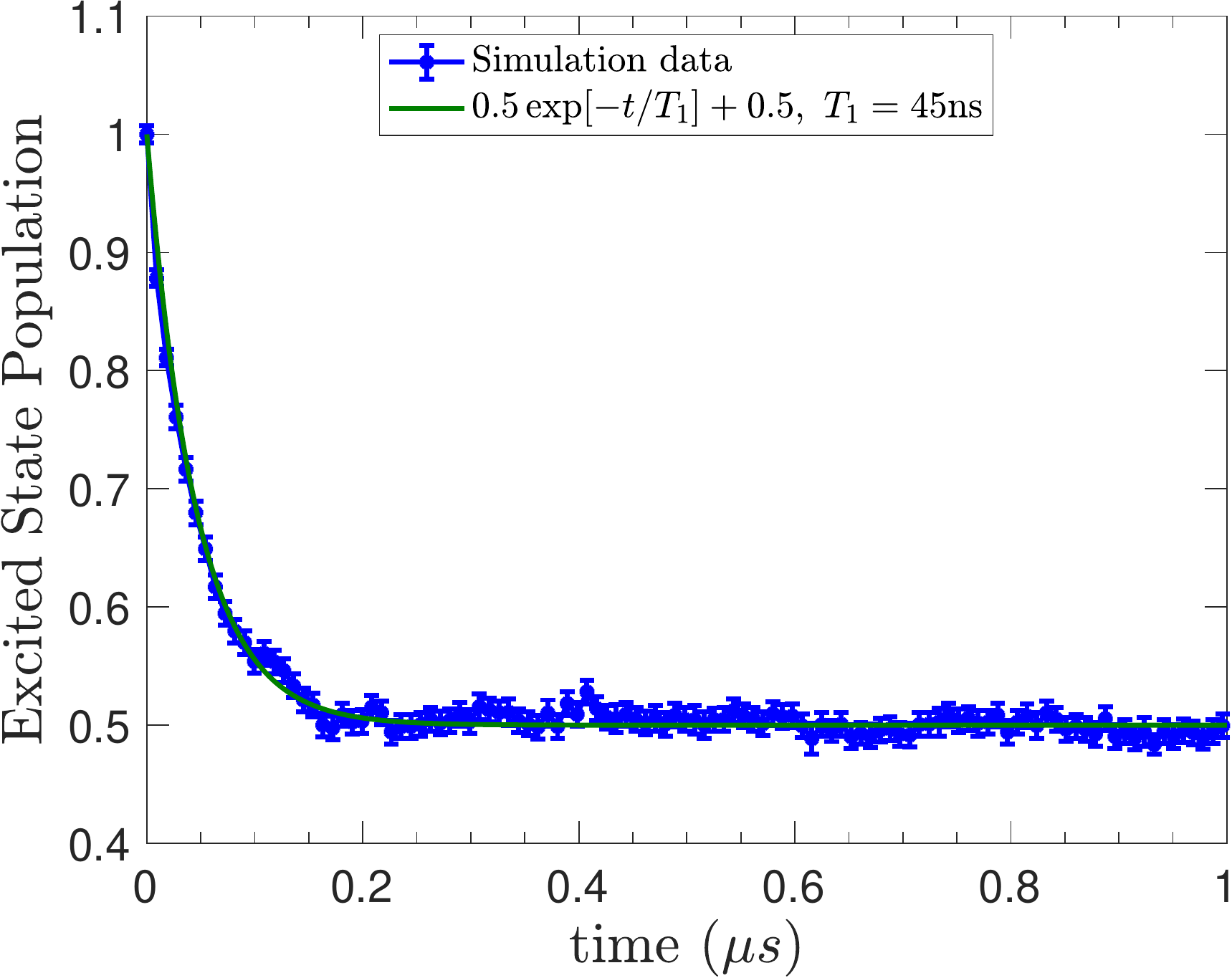}}
\caption{Transverse noise: $\gamma_{\text{min}}=0.01$ GHz, $\gamma_{\text{max}}=1$ GHz. 
Number of \fs is $200$. $\langle b\rangle = 0.046$ GHz,
$\Delta b /\langle b \rangle = 0.2$. Results are averaged over $1k$ trajectories.  The initial value of each fluctuator is randomly sampled according to thermal equilibrium distribution $\delta p_{eq}= 0.08$ (i.e. each fluctuator is $0.08$ more likely to be initialized in state $1$ than state $-1$). Error bars are $2\sigma$ over $1k$ trajectories.}
\label{fig:T1}
\end{figure}

The $T_1$ time can then be found by finding the decay profile of the diagonal elements of Eq.~\eqref{eq:fiddm_form}. For the example of a qubit with energy of $E/2 = 1$ GHz and the fluctuator strength being $\langle b\rangle = 0.046$ GHz, the $T_1$ time and decay profile is plotted in Fig.~\ref{fig:T1}.

\subsubsection{$T_2^{*}$ time}
To measure the $T_2^{*}$ time (without presence of Longitudinal noise), we initialize the state in the $\ket{+}$ state and evolve it with the following Stochastic Hamiltonian with fluctuators in the longitudinal direction $\sigma_z$:
\begin{equation}
    \hat H_{\text{sys}} = \frac{E}{2}\sigma_z + \frac{1}{2} \,  \sum_{i}b_i \chi_i(t) \sigma_z \,,
\end{equation}
where $\chi_i(t)$ switches with $\gamma_i/2$. After averaging many trajectories, the resulting density matrix under this initial condition and Hamiltonian can be expressed as:

\begin{equation}
\overline{\rho}(t)  = {\begin{pmatrix}\frac{1}{2} & \frac{1}{2}e^{-i\omega t} e^{-\frac{t}{T_2^{*}}}\\
\frac{1}{2}e^{i\omega t} e^{-\frac{t}{T_2^{*}}} & \frac{1}{2}\end{pmatrix}}
\, \text{       with       }
\rho(0)  = \begin{pmatrix}\frac{1}{2}\,\,\, & \frac{1}{2}\\
\frac{1}{2}\,\,\, & \frac{1}{2}\end{pmatrix} \,.
\label{eq:fiddm2}
\end{equation}

The $T_2^{*}$ time can then be found by finding the decay profile of expectation value of the $\ket{+}$ state. Its time evolution according to Eq.~\eqref{eq:fiddm2} is:
\begin{equation}
    \left<+|\overline{\rho}(t)|+\right> = \frac{1}{2} + \frac{1}{2}e^{-\frac{t}{T_2^{*}}} \cos{\omega t} \,.
\end{equation}

\begin{figure}[h!]
\centerline{\includegraphics[width=12.5cm]{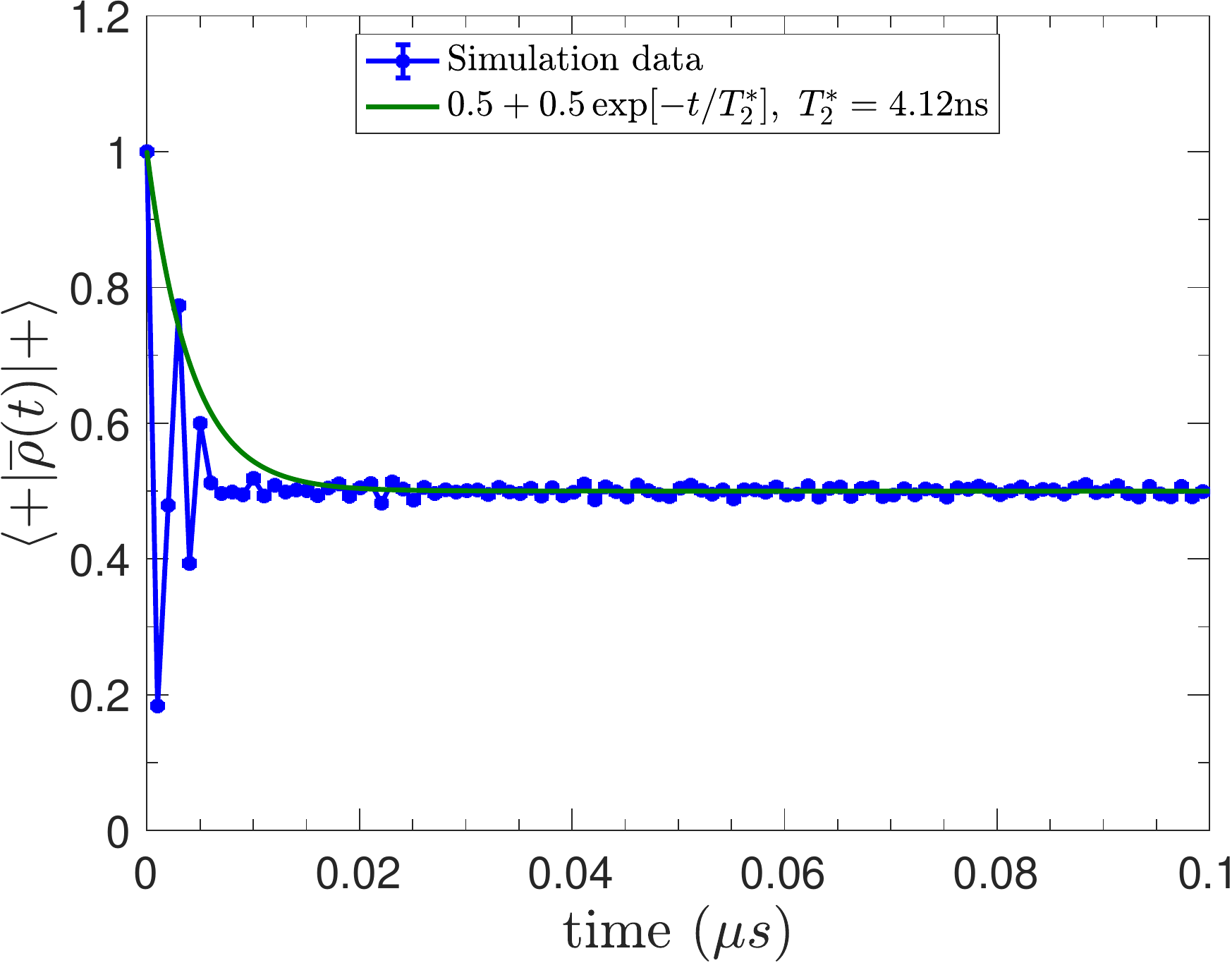}}
\caption{Longitudinal noise: $\gamma_{\text{min}}=0.01$ GHz, $\gamma_{\text{max}}=1$ GHz. 
Number of \fs is $\mathbf{100}$. $\langle b\rangle = 0.046$ GHz,
$\Delta b /\langle b \rangle = 0.2$. Results are averaged over $1k$ trajectories.  The initial value of each fluctuator is randomly sampled according to the thermal equilibrium distribution $\delta p_{eq}= 0.08$ (i.e. each fluctuator is $0.08$ more likely to be initialized in state $1$ than state $-1$). Here we denote the decay profile as $e^{-t/T_2^{*}}$ since there is no longitudinal noise (i.e. $T_1 = \infty$). Error bars are $2\sigma$ over $1k$ trajectories.}
\label{fig:T2}
\end{figure}

As $t \gg T^*_2$, the qubit goes to the maximally mixed state. For the example of a qubit with energy of $E/2 = 1$ GHz and fluctuator strength of $\langle b\rangle = 0.046$ GHz, the expectation value and decay profile is plotted in Fig.~\ref{fig:T2}.

\subsection{Different fluctuator strengths}
We proceed to study the dependence on qubit evolution with the average fluctuator strength $\langle b\rangle$. Consider the following system Hamiltonian and fluctuator parameters.
\begin{equation}
    \hat H_{\text{sys}} = \frac{E}{2}\sigma_z + \frac{1}{2} \,  \sum_{i}b_i \chi_i(t) \sigma_x\,,
\end{equation}
where: 
\\ $E = 1$ GHz;
$\gamma_{\text{min}} = 0.01$ GHz, $\gamma_{\text{max}} = 1$ GHz. \\
$\chi_i(t)$ switches with $\gamma_i/2$ and takes value of $\{-1, 1\}$;\\

\begin{figure}[h!]
\centerline{\includegraphics[width=12.5cm]{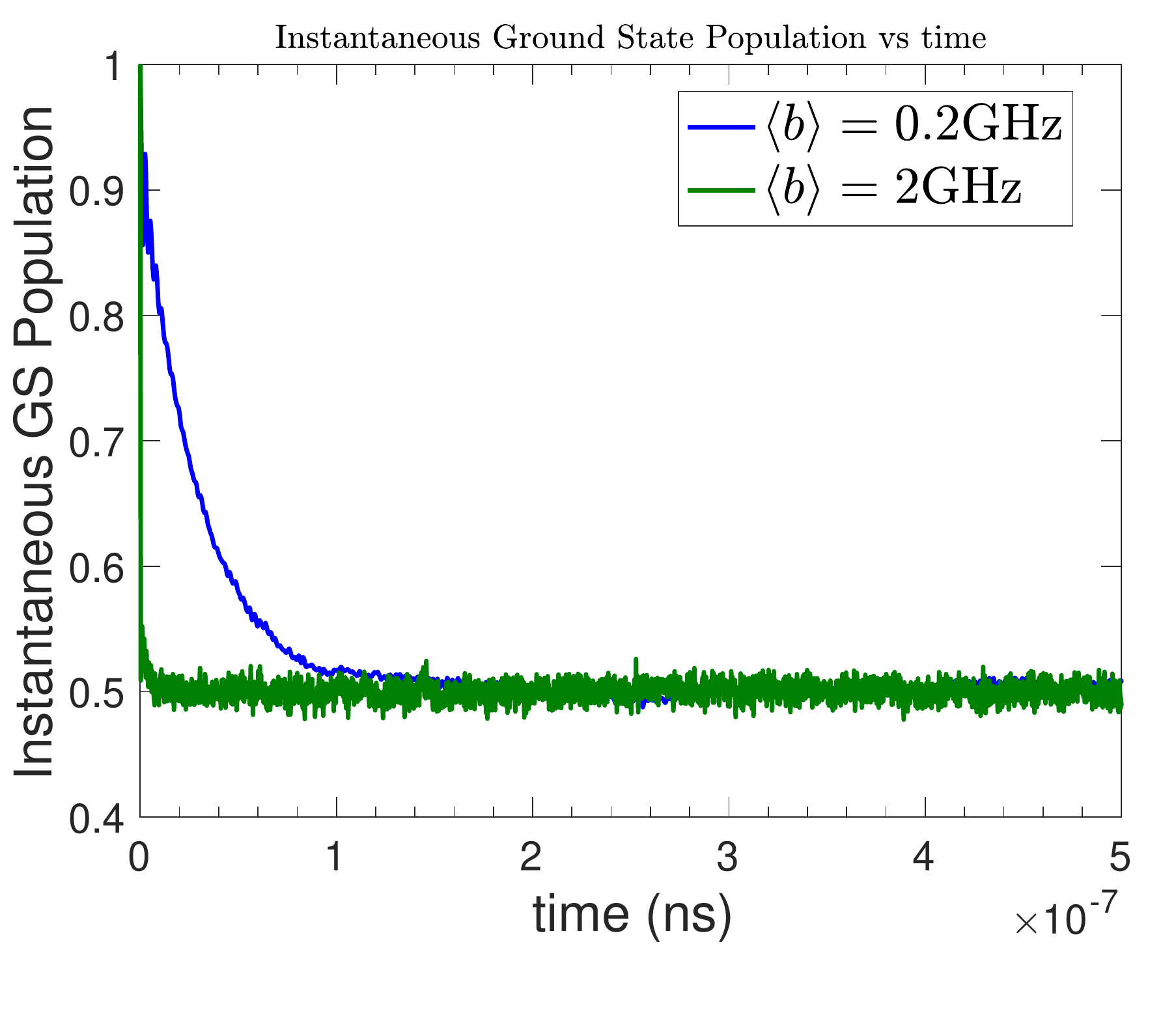}}
\caption{Longitudinal noise: $\gamma_{\text{min}}=0.01$ GHz, $\gamma_{\text{max}}=1$ GHz. 
Number of \fs is $20$. Results are averaged over $3k$ trajectories.  The initial value of each fluctuator is randomly sampled according to the thermal equilibrium distribution $\delta p_{eq}= 0.08$ (i.e., each fluctuator is $0.08$ more likely to be initialized in state $1$ than state $-1$). Blue line: $\langle b\rangle = 0.2$ GHz; Green line: $\langle b\rangle = 2$ GHz. $\Delta b /\langle b \rangle = 0.2$.}
\label{fig:gspplot_ssb}
\end{figure}

We compare the two cases where $\langle b\rangle = 2$ GHz and $\langle b\rangle = 0.2$ GHz. Fig.~\ref{fig:gspplot_ssb}, we see that a higher coupling strength of the fluctuators speeds up the evolution to a maximally mixed state.

\section{Modeling $1/f$ noise for superconducting qubit: Capacitively shunted flux qubit (CSFQ)}
In this and the next section, we study how to model $1/f$ noise for a superconducting circuit Hamiltonian, and how the open-system dynamics caused by $1/f$ noise is different from that of the weak-coupling limit. We study two kinds of superconducting circuit Hamiltonians, namely Capacitively shunted flux qubit (CSFQ) and Transmon.
\subsection{Capacitively shunted flux qubit (CSFQ) circuit Hamiltonian}
\begin{figure}
    \centering
    \includegraphics[width=5.5cm]{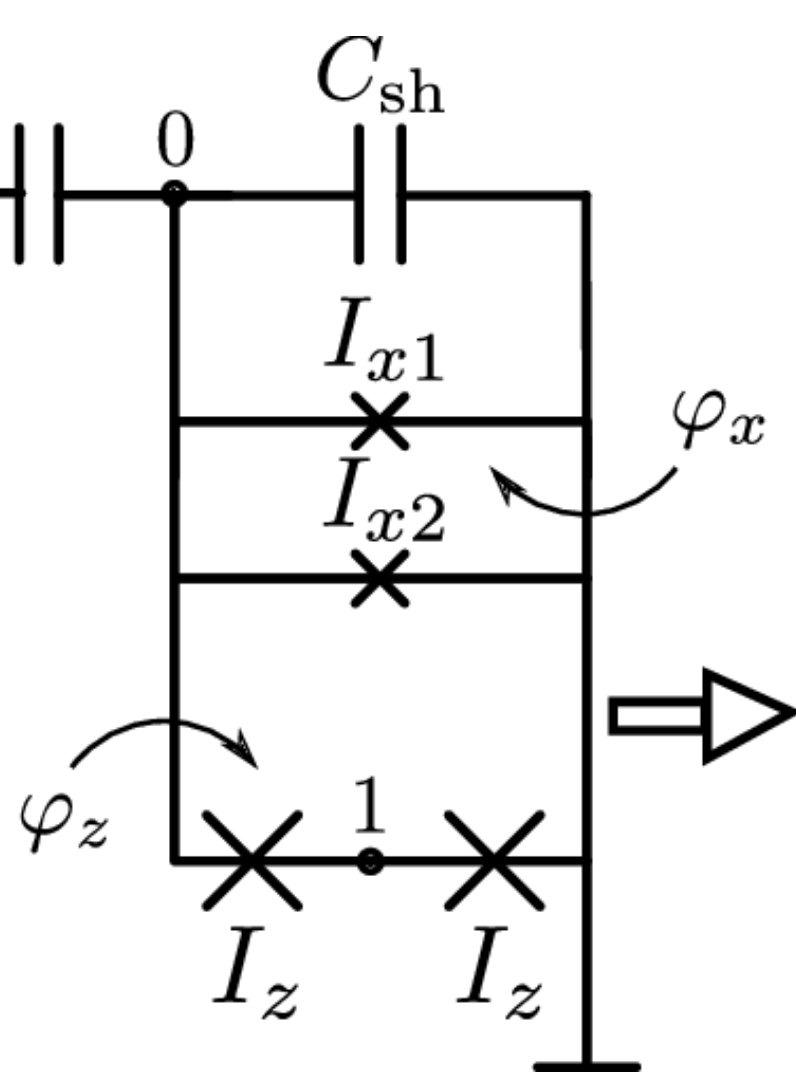}
    \caption{Schematic of the experimental set up. Capacitively shunted flux qubit (CSFQ) in the center, with curved arrows showing x and z fluxes threading their corresponding loops. Source: Ref.~\cite{khezri2021anneal}.}
    \label{fig:CSFQ_layout}
\end{figure}

The capacitively shunted flux qubit (CSFQ) circuit Hamiltonian $H_{S}(s)$ takes the form of~\cite{khezri2021anneal}:
\begin{equation*}
H_S(s)= \frac{E_C}{2}{\hat{n}}^2-2E_J\cos{(\hat{\varphi})+2\alpha E_J\sqrt{1+d^2\tan^2{\left(\frac{\phi_x(s)}{2}\right)}}\cos{\left(\frac{\phi_x(s)}{2}\right)}}\cos\left(2\hat{\varphi}+\varphi_z(s)-\varphi_0(s)\right) \,,
\end{equation*}
which can be rewritten as displacement operators in the charge basis:
\begin{eqnarray*}
\label{Eq:CH}
    H_{S}(s) &=& E_{J}\left(\frac{E_C}{2E_J} \hat{n}^2 - (\hat{d}_{1} + \hat{d}_{1}^{\dagger}) + \right. \notag \\
&& \hspace{-0.5cm} \left. \qquad\qquad \alpha
    \cos{\left(\frac{\phi_x(s)}{2}\right)}\sqrt{1 + d^2 \tan^2{\frac{\phi_x(s)}{2}}}\left(e^{i(\phi_z(s)-\phi_0(s))}\hat{d}_2 + e^{-i(\phi_z(s)-\phi_0(s))}\hat{d}_2^{\dagger}\right) \right)\,,
\end{eqnarray*}
where (as in Fig.~\ref{fig:CSFQ_layout}): \\$E_J = I_z\frac{\Phi_0}{2\pi}$ is the Josephson junction energy; \\
$E_C$ is the mean Capacitive energy $= \frac{e^2}{C_{\text{sh}}}$; \\ 
$\hat{n}$ is the charge operator (with the same dimension as $H_{S}$); \\ $\hat{d}_{1}$ is the charge displacement operator by 1 cooper pair (with the same matrix dimension as $H_{S}$); \\
$\alpha < 1$ is ratio of the current in one of the small $x$-loop junctions to the current in the large $z$-loop junction. \\
$d = \frac{E_{J1} - E_{J2}}{E_{J1} + E_{J2}}$ is the $x$-loop junction asymmetry; \\ $\phi_x(s)$ is the time-dependent $x$ (barrier) bias phase; \\
$\phi_z(s)$ is the time-dependent $z$ (tilt) bias phase; \\
$\phi_0 = \tan^{-1}{\left(d\tan{\left(\frac{\phi_x(s)}{2}\right)}\right)}$; \\
$\hat{d}_{2}$ is the charge displacement operator by 2 cooper pairs (with the same dimension as $H_{S}$).

\subsection{CSFQ: Persistent current operator as fluctuator}
One important question is how to include the fluctuators and form the overall stochastic Hamiltonian for $1/f$ noise simulations. The total Stochastic Hamiltonian $H(s)$ can be formulated as fluctuations in the persistent current operator, i.e., the fluctuators are all proportional to the persistent current operator.
\begin{equation}
\label{eq:stocHCSFQ}
    H(s) =  H_{S}(\phi_x(s), \phi_z(s)) + \frac{1}{2} \,  \sum_{i}b_i \chi_i(s) I_p(s) \,,
\end{equation}
where $H_S(s)$ is the time-dependent circuit Hamiltonian; $I_p(s) = I_p((\phi_x(s), \phi_z(s)))$ is the persistent current operator.

\begin{table}[h!]
\centering
\begin{tabular}{ |c|c|c| } 
 \hline
 & Qubit & CSFQ  \\ 
 \hline
 fluctuation & $\sigma_z$ &$I_{p}(s)$   \\ 
 \hline
\end{tabular}
\caption{Form of fluctuation (Qubit vs CSFQ)}
\label{tab:qubitandfluctuation}
\end{table}

Table.~\ref{tab:qubitandfluctuation} compares the form of fluctuators for qubit and superconducting circuits.


\subsection{Case studies}
\begin{figure}[h!]
  \begin{minipage}{.5\textwidth}
\begin{align}
    E_J &= 592.9434/2\pi \text{ GHz} \,,\notag\\
    E_c &= 5.4092 \text{GHz} \,,\notag\\
    \alpha &= 0.46 \,,\notag\\
    d &= 0.1 \,,\notag\\
    \phi_x(s) &= 0.96s+1.04\pi \,,\notag\\
    \phi_z(s) &= (0.326s + \phi_z(0))\pi \,.
    \label{eq:CSFQ-params}
\end{align}
  \end{minipage}%
  \begin{minipage}{.5\textwidth}
  \vspace{2cm}
    \centering
    \includegraphics[width=8cm]{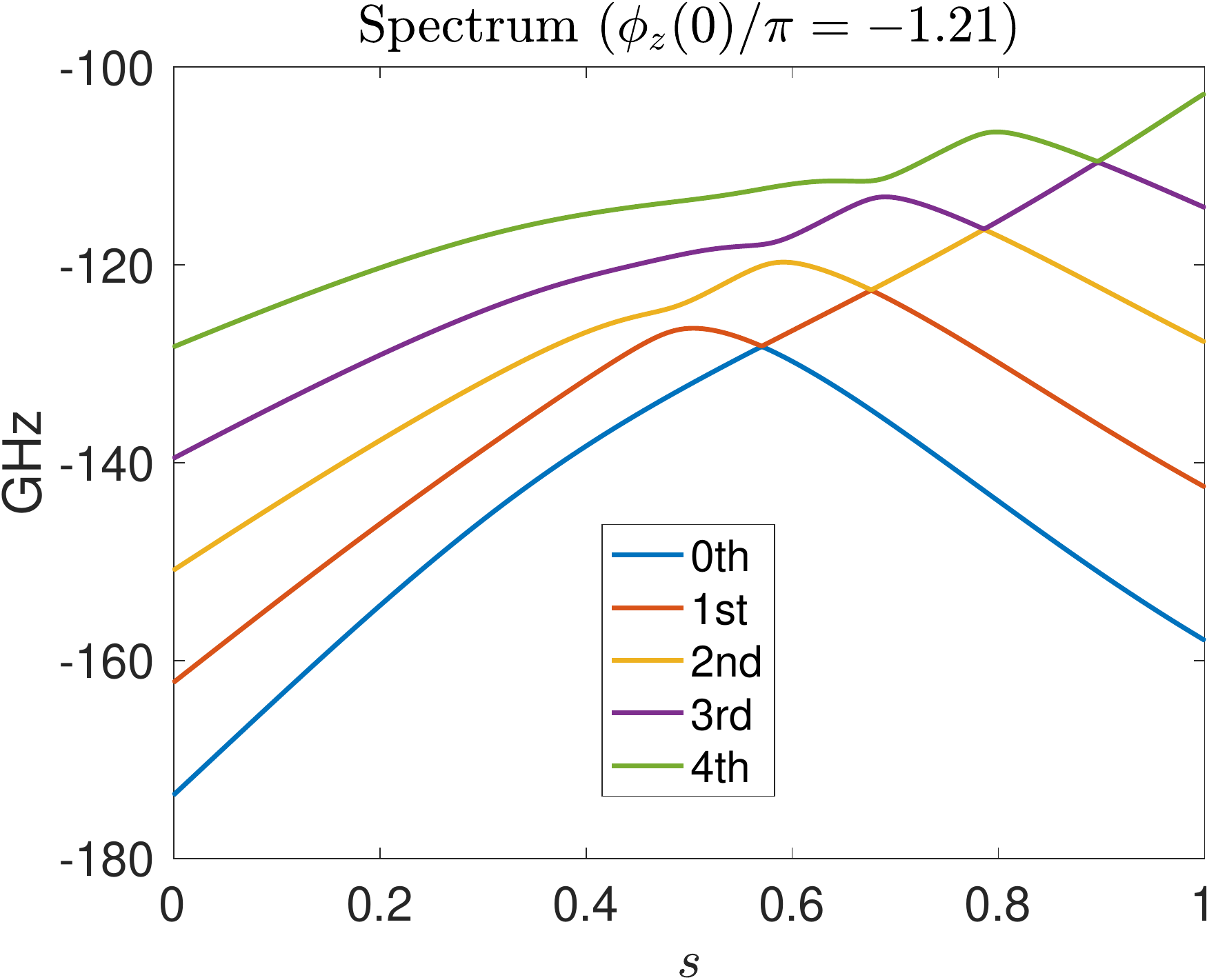}
    \caption{Spectrum with  $\phi_z(0)/\pi = -1.21$.}
    \label{fig:spectrumcsfq}
  \end{minipage}
\end{figure}

We study the time evolution under the circuit Hamiltonian, with the parameters (4 decimal figures) and flux schedules given in Eq.~\eqref{eq:CSFQ-params} (which also appears in the $s$-curve studies in~\cite{khezri2021anneal}).

The initial tilt bias $\phi_z(0)$ is an external parameter whose value we can set. Here we focus on $\phi_z(0)/\pi = -1.21$ (the spectrum of the first $5$ levels is plotted in Fig.~\ref{fig:spectrumcsfq}). We consider the cases of closed-system evolution, open-system evolution from the adiabatic master equation, and open-system evolution from the fluctuator $1/f$ simulation. For closed system evolution and the adiabatic master equation, we truncate the energy levels of the system to $10$, and plot the populations of the first $5$ levels. For fluctuator $1/f$ simulation, again benefiting from the fact that in the trajectories approach we simulate the state vector evolution rather than the density matrix's , we can efficiently simulate the populations of all $121$ levels of the CSFQ circuit Hamiltonian.

\subsection{Closed-system annealing evolution}
We simulate the closed-system annealing evolution for $t_f = 6$ ns with value of $\phi_z(0)/\pi = -1.21$, and plot the results in Fig.~\ref{fig:scurveclosed}. In the later stage of the anneal, there is a sequence of diabatic transitions of population to excited states due to the small gaps at that point. The population first transfers from the ground state to the first excited state, then to the $2$-nd excited state, then to the $3$-rd excited state, and ultimately ends up in the $4$-th excited state.

\subsection{Adiabatic master equation simulation}
\begin{figure}[h!]
	\subfigure[]{\includegraphics[width = 0.5\columnwidth]{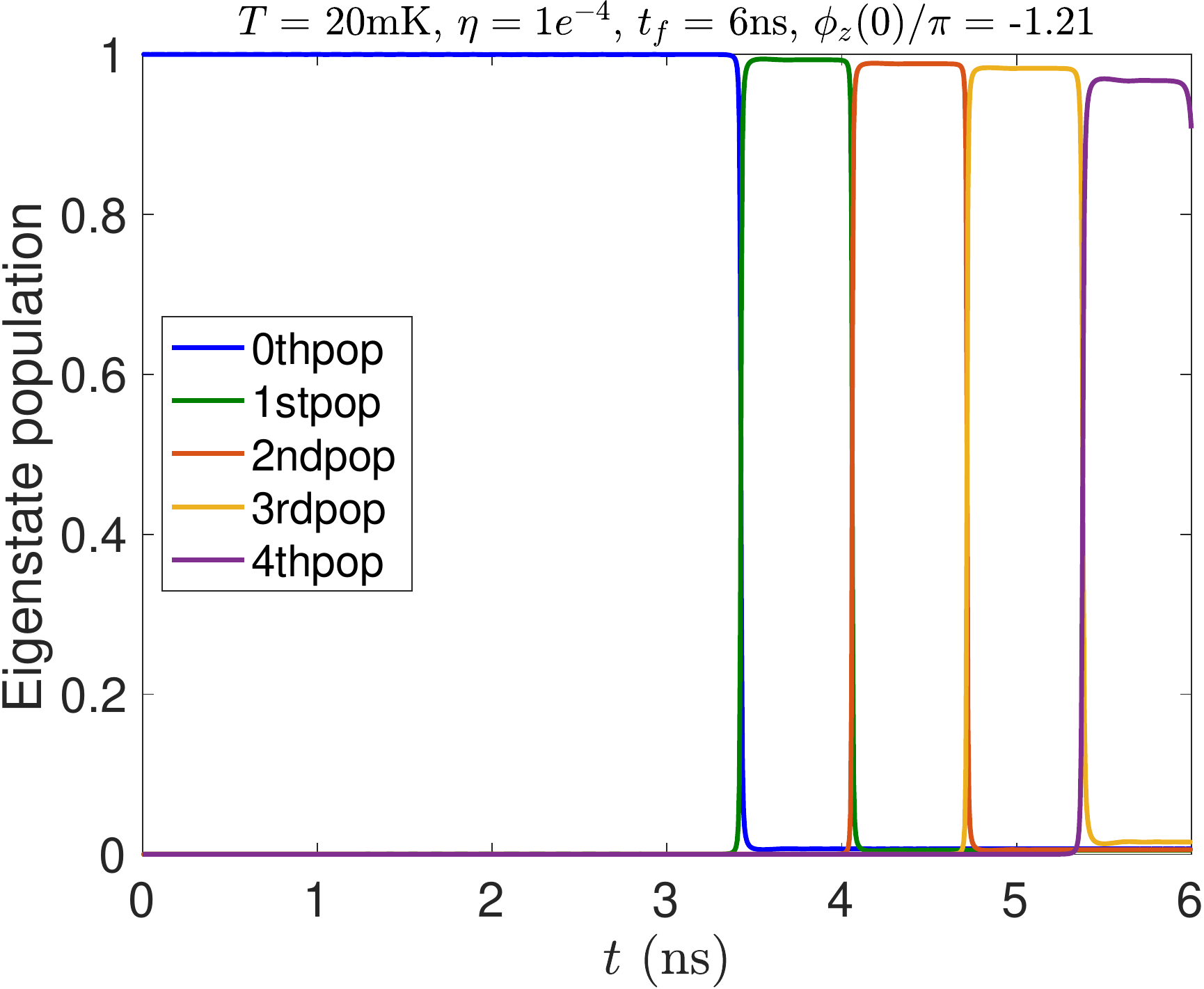}\label{fig:scurveclosed}}
	\subfigure[]{\includegraphics[width = 0.5\columnwidth]{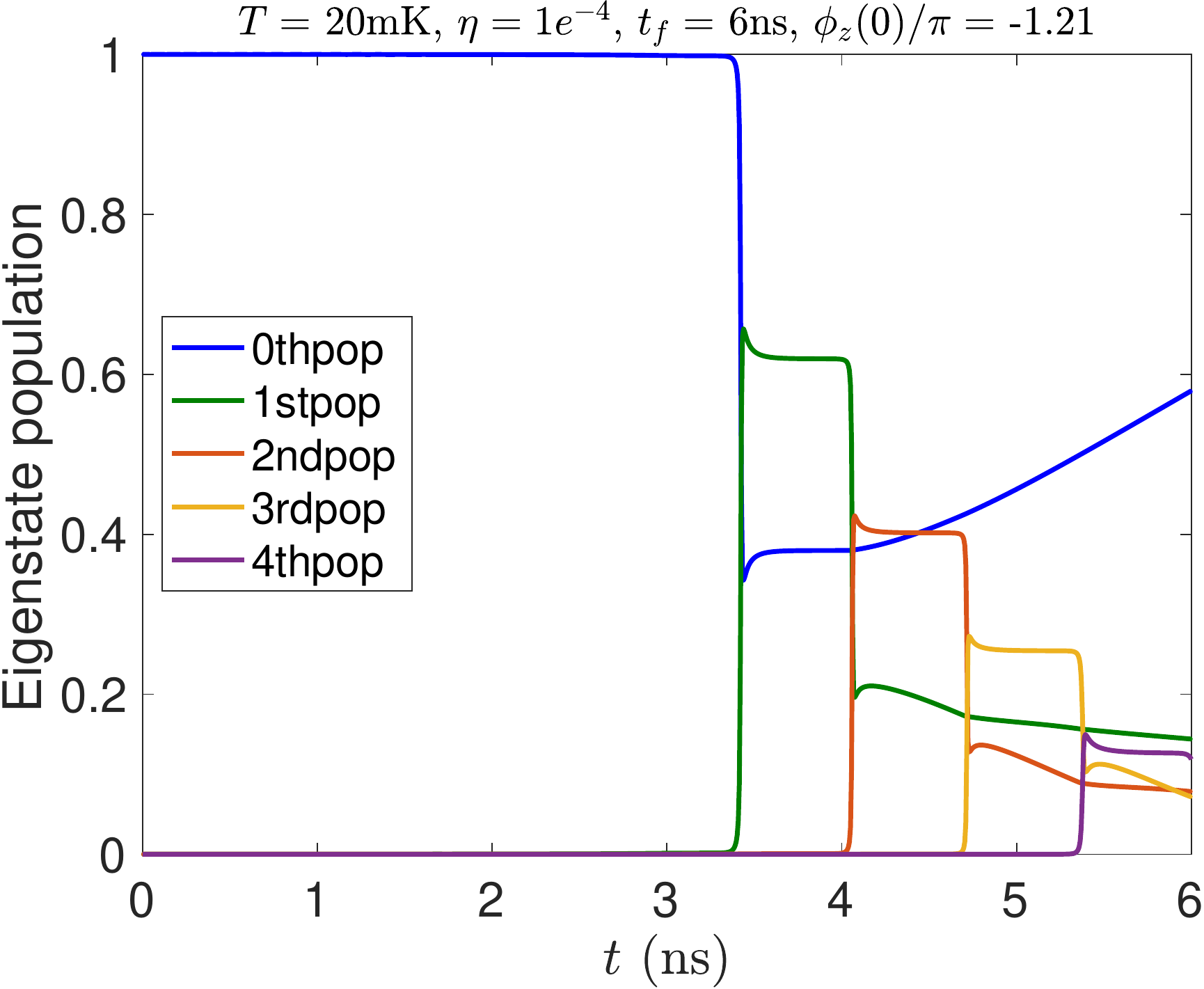}\label{fig:scurveame}}
	\begin{center}
	    \subfigure[]{\includegraphics[width = 0.5\columnwidth]{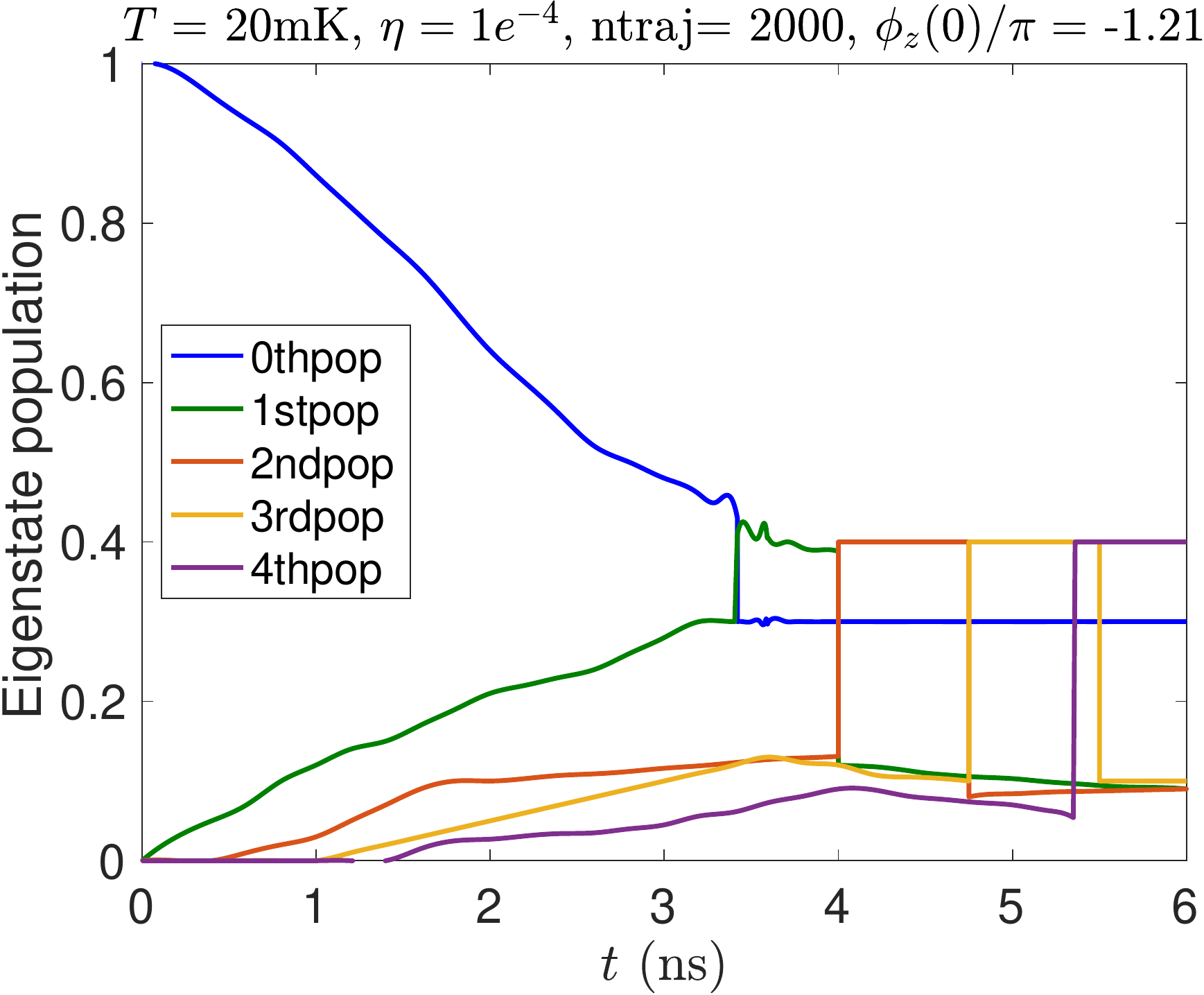}\label{fig:scurve1f}}
	\end{center}
	\caption{The eigenstate populations for $\phi_z(0)/\pi = -1.21$. (a) Closed-system evolution, (b) Evolution under the adiabatic master equation. (c) $1/f$ noise: $\gamma_{\text{min}}=0.01$ GHz, $\gamma_{\text{max}}=1$ GHz. 
Number of \fs is $100$. $\langle b\rangle = 0.46$ GHz,
$\Delta b /\langle b \rangle = 0.2$. Results are averaged over $1k$ trajectories.  The initial value of each fluctuator is randomly sampled according to the thermal equilibrium distribution $\delta p_{eq}= 0.08$. Some of the population is lost to higher excited states.
	}
\end{figure}

We perform the adiabatic master equation simulations for $t_f = 6$ ns with $\phi_z(0)/\pi = -1.21$. We set the temperature to be 20mK, $g^2$ to be $1e^{-4}$ and the cutoff frequency $\omega_c$ to be $4$GHz. Since the Lindblad operator is defined based on energy differences, we also additionally set the absolute tolerance of energy difference ($\omega$) to be $\frac{0.1}{2\pi}$Hz for the grouping of Lindblad operators. For example, if $|\omega_{ab} - \omega_{cd}| < \frac{0.1}{2\pi}$Hz, then $L_{ab}$ and $L_{cd}$ are combined into one Lindblad operator $L(\omega)$. The results are plotted in Fig.~\ref{fig:scurveame}. We observe the same sequence of diabatic transitions to high excited states, but they are relatively suppressed. At the end of the anneal there is still have more than half of the population in the ground state.

\subsection{$1/f$ noise simulation}
We perform the $1/f$ noise simulation with the stochastic Hamiltonian Eq.~\eqref{eq:stocHCSFQ}. We use $10$ fluctuators in the form of $I_{p}(s)$. The flipping frequency is sampled from $1/\gamma$ distribution with $\gamma \in \left[\gamma_{\text{min}}=0.01, \gamma_{\text{max}}=1\right]$ GHz, $\langle b\rangle = 0.46$ GHz and
$\Delta b /\langle b \rangle = 0.2$. $\phi_z(0)/\pi = -1.21$. The results are shown in Fig.~\ref{fig:scurve1f}. Compared to closed system evolution and AME simulations, not only are diabatic transitions further suppressed, but there are also population transfers during the annealing. The $1/f$ noise is also more detrimental to the qubit's lower-level populations. In fact, at the end of the anneal, the sum of the first $5$ levels is no longer $1$, since some of the population is lost to even higher excited states.

\section{Modeling $1/f$ noise for superconducting qubit: Transmon}
\subsection{Transmon Circuit Hamiltonian}
The transmon circuit Hamiltonian $H_{S}(s)$ takes the form of:
\begin{equation*}
H_S = 4E_C{\hat{n}}^2-E_{J_{+}}\,\sqrt{1+d^2\tan^2{\left(\frac{\phi_x}{2}\right)}}\cos{\left(\frac{\phi_x}{2}\right)}\cos\left(\hat{\varphi}-\phi_0\right) \,,
\end{equation*}
which can be rewritten as:
\begin{equation*}
H_S = 4E_C{\hat{n}}^2-E_{J_{+}}\,\sqrt{1+d^2\tan^2{\left(\frac{\phi_x}{2}\right)}}\cos{\left(\frac{\phi_x}{2}\right)}\frac{1}{2} \left(e^{-i\phi_0}\hat{d}_1 + e^{i\phi_0}\hat{d}_1^{\dagger}\right) \,,
\end{equation*}
where: \\$E_{J_{+}}$ is the total Josephson junction energy $= (E_{J1} + E_{J2})$; \\
$E_J$ is the Josephson junction energy $= \frac{(E_{J1} + E_{J2})}{2\alpha }$; \\
$E_C$ is the mean Capacitive energy; \\ $\hat{n}$ is the charge operator (with the same dimension as $H_{S}$); \\ $\hat{d}_{1}$ is the charge displacement operator by 1 cooper pair (with the same dimension as $H_{S}$); \\
$d = \frac{E_{J1} - E_{J2}}{E_{J1} + E_{J2}}$ is the $x$-loop junction asymmetry; \\ 
$\phi_x$ is the time-dependent $x$ (barrier) bias phase; \\
$\phi_0 = \tan^{-1}{\left(d\tan{\left(\frac{\phi_x}{2}\right)}\right)}$; 

\subsection{Transmon: Form of fluctuators}
The transmon circuit Hamiltonian $H_{S}(s)$ is time-independent. Particularly, the first two energy eigenstates $\ket{\overline{0}}$ and $\ket{\overline{1}}$ do not evolve with time. Therefore, in addition to using the persistent current operator as the fluctuator, one can also define the fluctuator to be a logical $\overline{\sigma_z} = \ket{\overline{0}}\bra{\overline{0}} - \ket{\overline{1}}\bra{\overline{1}}$ operator as in the longitudinal noise for qubit (Sec.~\ref{sec:qubitfluctuator})
Denote the form of fluctuator by $A$. The total Stochastic Hamiltonian $H(s)$ can then be formulated as fluctuation of the longitudinal noise or persistent current operator (See table.~\ref{tab:Asigmazpersistent}). In general, 
\begin{equation}
\label{eq:stocH_tran}
    H(t) =  H_{S}(\phi_x) + \frac{1}{2} \,  \sum_{i}b_i \chi_i(t) A \,,
\end{equation}
where $\chi_i(t)$ switches with $\gamma_i/2$, with $\gamma_i$ sampled from $1/\gamma$ distribution in the range of $\{\gamma_{\text{min}}, \gamma_{\text{max}}\}$. $H_S$ is the time-independent, transmon circuit Hamiltonian.

\begin{table}[h!]
\centering
\begin{tabular}{ |c|c|c| } 
 \hline
 & Longitudinal noise & Persistent current operator  \\ 
 \hline
 fluctuation (A) & $\overline{\sigma_z}$ &$I_{p}$   \\ 
 \hline
\end{tabular}
\caption{Form of fluctuation. $\overline{\sigma_z} = \ket{\overline{0}}\bra{\overline{0}} - \ket{\overline{1}}\bra{\overline{1}}$. ; $I_p = I_p(\phi_x)$ is the persistent current operator.}
\label{tab:Asigmazpersistent}
\end{table}

\subsection{Case studies}
We study the time evolution under the circuit Hamiltonian, with the following parameters (4 decimal figures) and flux schedules:
\begin{figure}[h!]
  \begin{minipage}{.5\textwidth}
\begin{align}
    E_J &= 15.3 \text{ GHz} \,,\notag\\
    E_c &= 0.25 \text{ GHz} \,,\notag\\
    d &= 0.03 \,,\notag\\
    \phi_x &= 0.5\pi  \,.
\end{align}
  \end{minipage}%
  \begin{minipage}{.5\textwidth}
  \vspace{2cm}
    \centering
    \includegraphics[width=8cm]{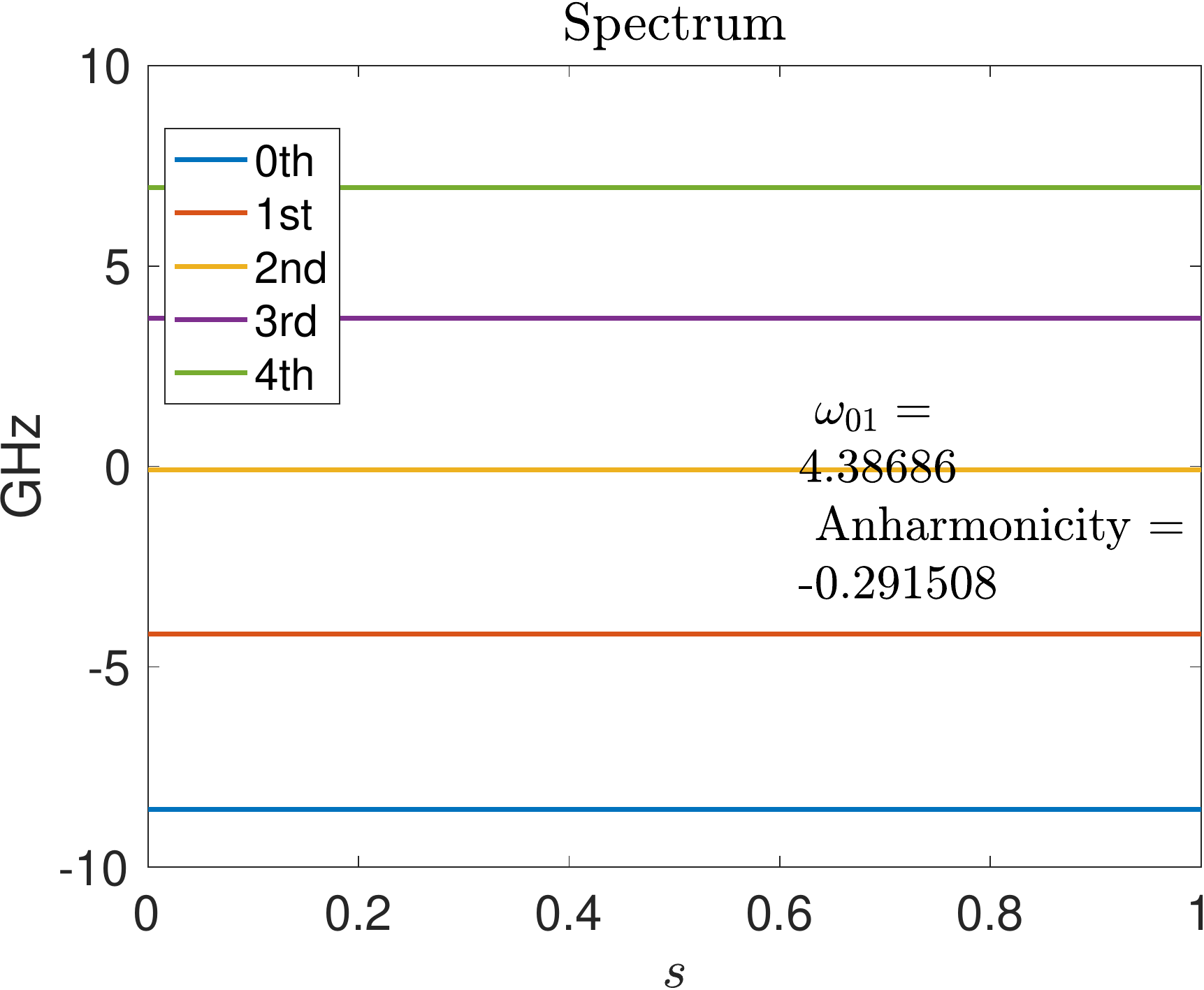}
    \caption{Spectrum with  $\phi_x = 0.5\pi$.}
    \label{fig:spectrum}
  \end{minipage}
\end{figure}

Here we focus on $\phi_x/\pi = -0.5$ (the spectrum of the first $5$ levels is plotted in Fig.~\ref{fig:spectrum}). We look into the cases of open-system evolution from adiabatic master equation, and open-system evolution from the fluctuator $1/f$ simulation.



\subsection{Adiabatic master equation simulation}
\begin{figure}[h!]
\subfigure[]{\includegraphics[width = 0.5\columnwidth]{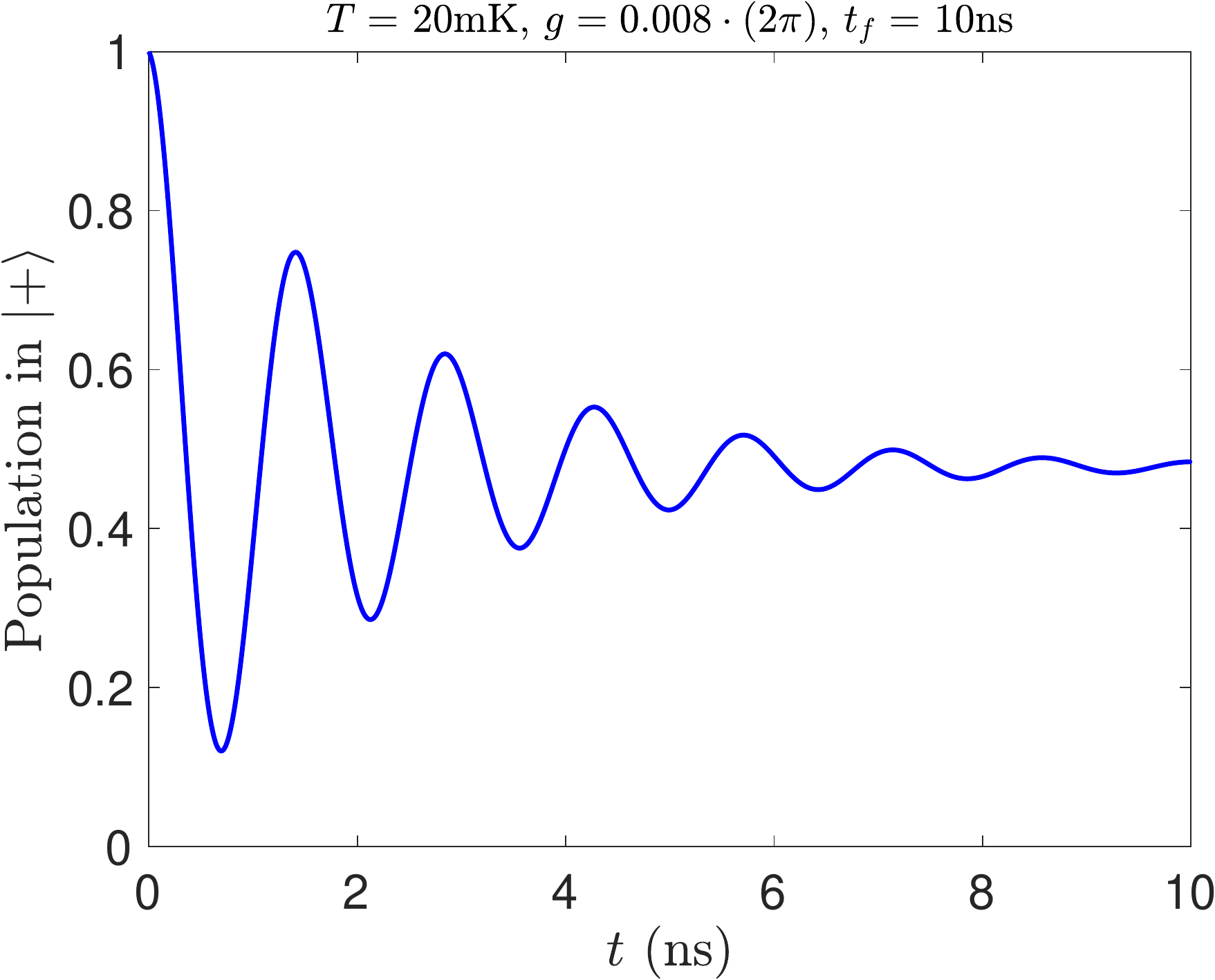}\label{fig:ametran_1}}
\subfigure[]{\includegraphics[width = 0.5\columnwidth]{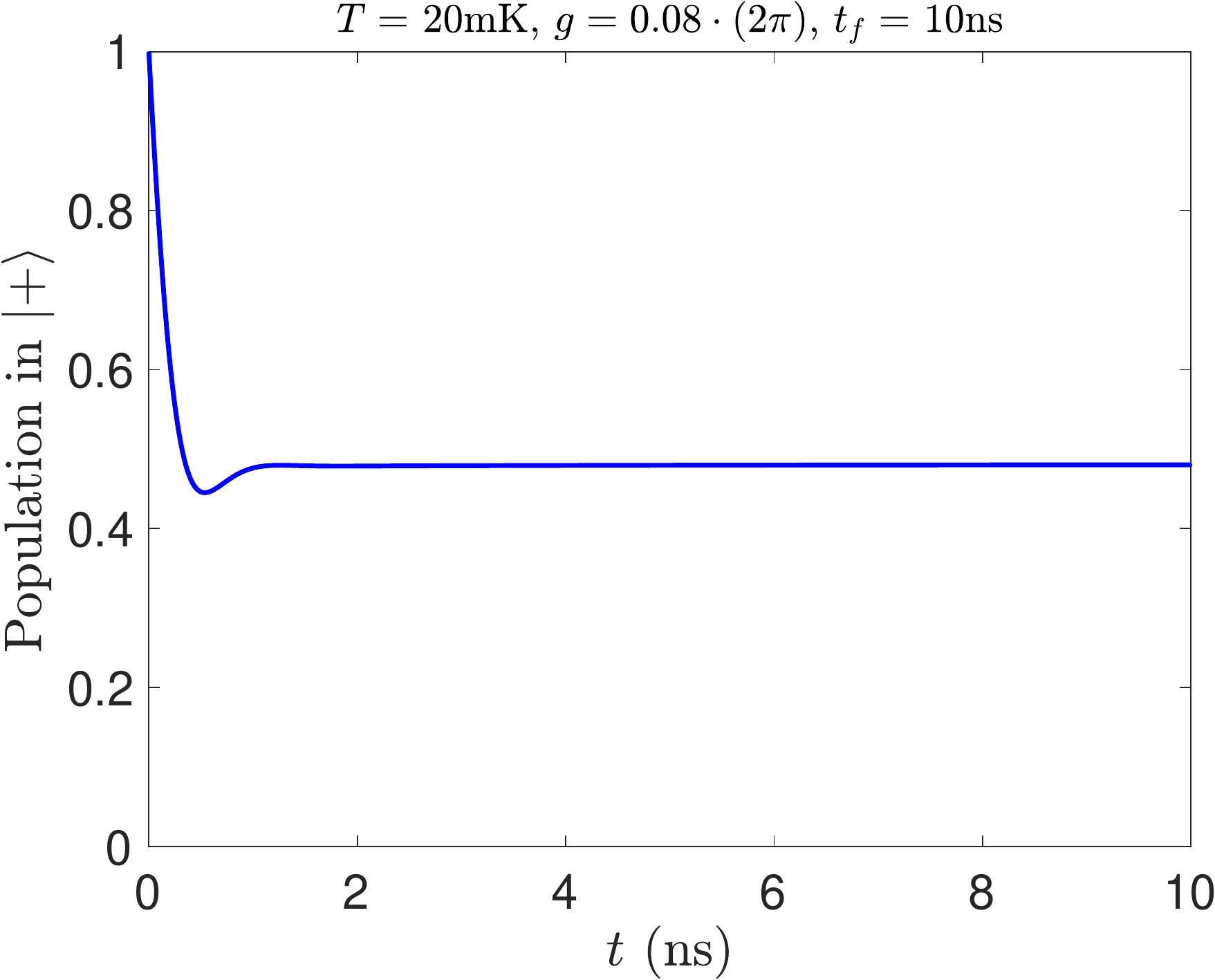}\label{fig:ametran_2}}
	
\caption{The population in $\ket{+}$ for coupling (a) $g= 0.008(2\pi)$. (b) $g= 0.08(2\pi)$. Evolution under (adiabatic) master equation.}
\label{fig:ame}
\end{figure}

We perform the adiabatic master equation simulations for $t_f = 10$ ns with value of $\phi_x/\pi =  -0.5$. Set the temperature to be 20mK and $\omega_c$ cutoff frequency to be $8$GHz. Since the Lindblad operator is defined based on energy difference, We also additionally set the absolute tolerance of energy difference ($\omega$) to be $\frac{0.1}{2\pi}$Hz for the grouping of Lindblad operator. For example, if $|\omega_{ab} - \omega_{cd}| < \frac{0.1}{2\pi}$Hz, then $L_{ab}$ and $L_{cd}$ are combined into one Lindblad operator $L(\omega)$. We truncate the energy levels of the system to $10$. The results of coupling $g= 0.008(2\pi)$ and $g= 0.08(2\pi)$ are plotted in Fig.~\ref{fig:ametran_1} and Fig.~\ref{fig:ametran_2}. As expected, the stronger the coupling to the environment, the faster the coherence is lost.


\subsection{$1/f$ noise simulation}

\begin{figure}[h!]
\centerline{\includegraphics[width=10.5cm]{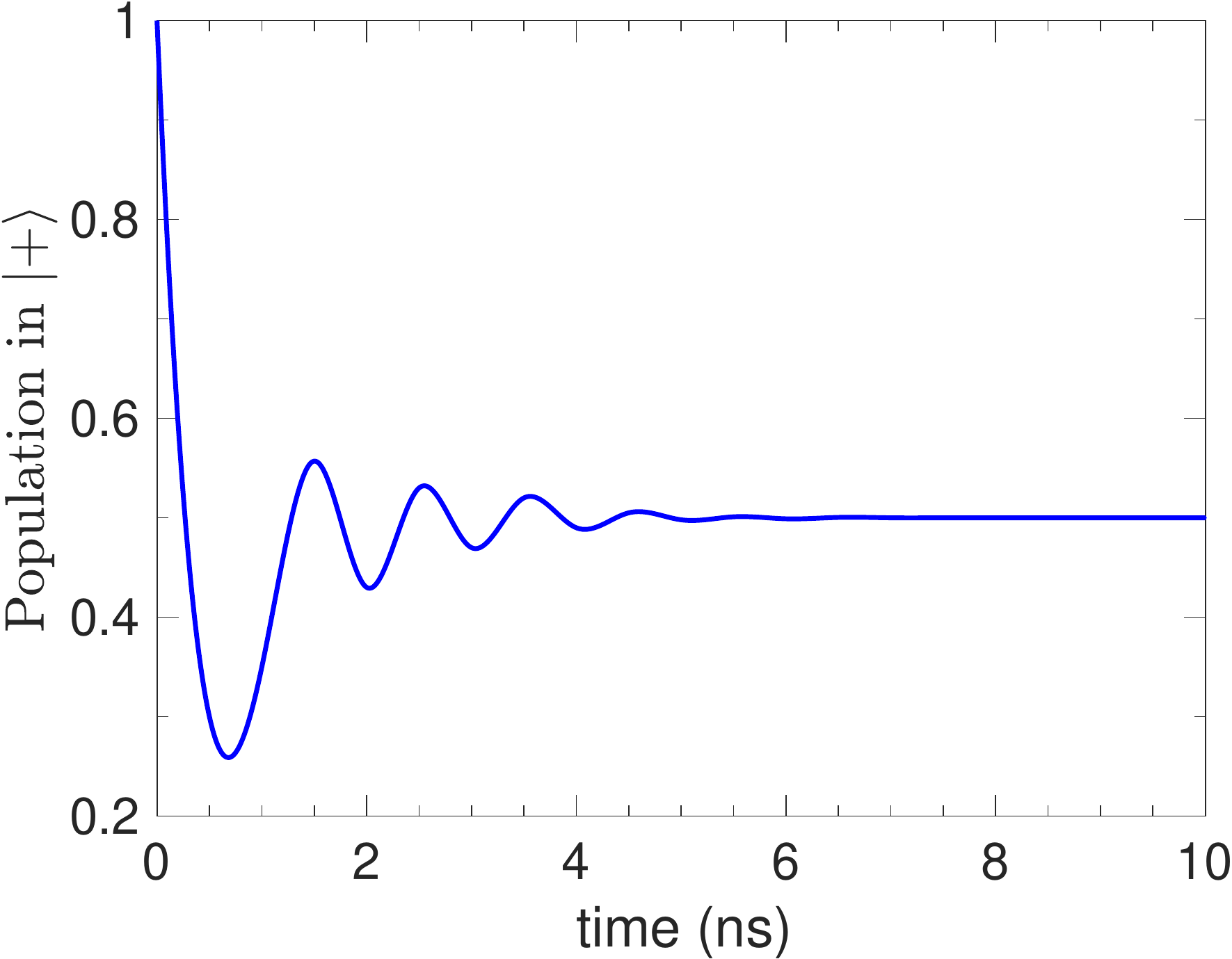}}
\caption{$A = \overline{\sigma_z}$. The population of $\ket{+}$ for  $\phi_x/\pi = -0.5$. $\gamma_{\text{min}}=1$ GHz, $\gamma_{\text{max}}=10$ GHz. 
Number of \fs is $20$. $\langle b\rangle = 0.4600$ GHz,
$\Delta b /\langle b \rangle = 0.2$. Results are averaged over $300$ trajectories. $\delta p_{eq}= 0.08$.}
\label{fig:1f_transmonlogicalz}
\end{figure}
\begin{figure}[h!]
\centerline{\includegraphics[width=10.5cm]{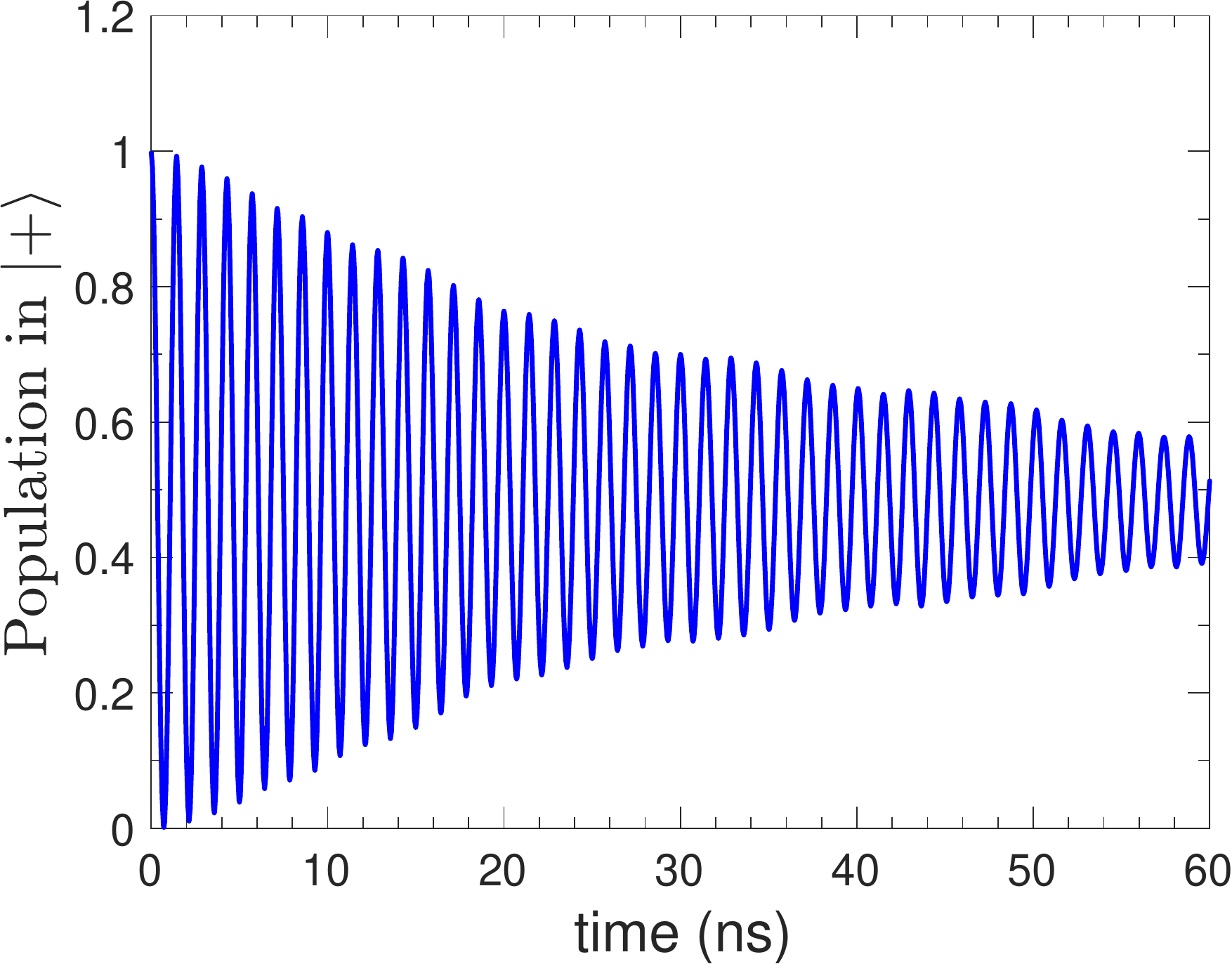}}
\caption{$A = I_p$. The population of $\ket{+}$ for  $\phi_x/\pi = -0.5$. $\gamma_{\text{min}}=0.1$ GHz, $\gamma_{\text{max}}=10$ GHz. 
Number of \fs is $20$. $\langle b\rangle = 0.0092$ GHz,
$\Delta b /\langle b \rangle = 0.2$. Results are averaged over $300$ trajectories. $\delta p_{eq}= 0.08$.  }
\label{fig:1f_transmonpc}
\end{figure}

For fluctuator $1/f$ simulation, benefited from the fact that in trajectories approach we simulate the state vector evolution rather than the density matrix's , we can efficiently simulate the populations of all the $21$ levels, defined by the Transmon circuit Hamiltonian.
We perform the $1/f$ noise simulation with the stochastic Hamiltonian Eq.~\eqref{eq:stocH_tran}. We consider both form of fluctuators: $A = \overline{\sigma_z}$ and $A = I_p(\phi_x)$. $\phi_x/\pi = -0.5$. We use $20$ fluctuators in both cases. The flipping frquency is sampled from $1/\gamma$ distribution with $\gamma \in \left[\gamma_{\text{min}}, \gamma_{\text{max}}\right]$. The result of $A = \overline{\sigma_z}$ is shown in Fig.~\ref{fig:1f_transmonlogicalz} and that of $A = I_p(\phi_x)$ is shown in Fig.~\ref{fig:1f_transmonpc}. In both cases we observe non-exponential decay in the population in $\ket{+}$ (off-diagonal elements of density matrix, after averaging over many trajectories)


\section{Fitting experimental data and reverse engineering of the spectrum}
\label{page:56start} 
The goal of this section is to try to evaluate the effectiveness of $1/f$ noise simulation in the fittings of Transmon qubits (IBM Quantum) experimental data and how dynamical decoupling can eliminate part of the $1/f$ noise. 

There are two directions. Assume that one is given either experimental data of free evolution and dynamical decoupling from a superconducting qubit experiment. One direction is reconstructing the $1/\omega^{\alpha}$ spectrum by fitting the experimental data of $T_1$ and $T_2$ times of free evolution and dynamical decoupling. This is what~\cite{bylander2011noise} did: find out the parameters that fit with the experimental data from free evolutions and dynamical decoupling, and reconstruct the $1/\omega^{\alpha}$ spectrum. 
Recall the form of $S(\omega)$ in Eq.~\eqref{eq:totalpowerspectrum}.
By finding the correct parameters like $\overline{b}$, $\gamma_M$, and $\gamma_m$ to best fit with the experimental data, one can find out the spectrum $S(\omega)$. This is what we do in this section. We do not know what the power spectrum of the IBM quantum computer is.

\begin{figure}[h!]
    \centering
    \includegraphics[width=17.5cm]{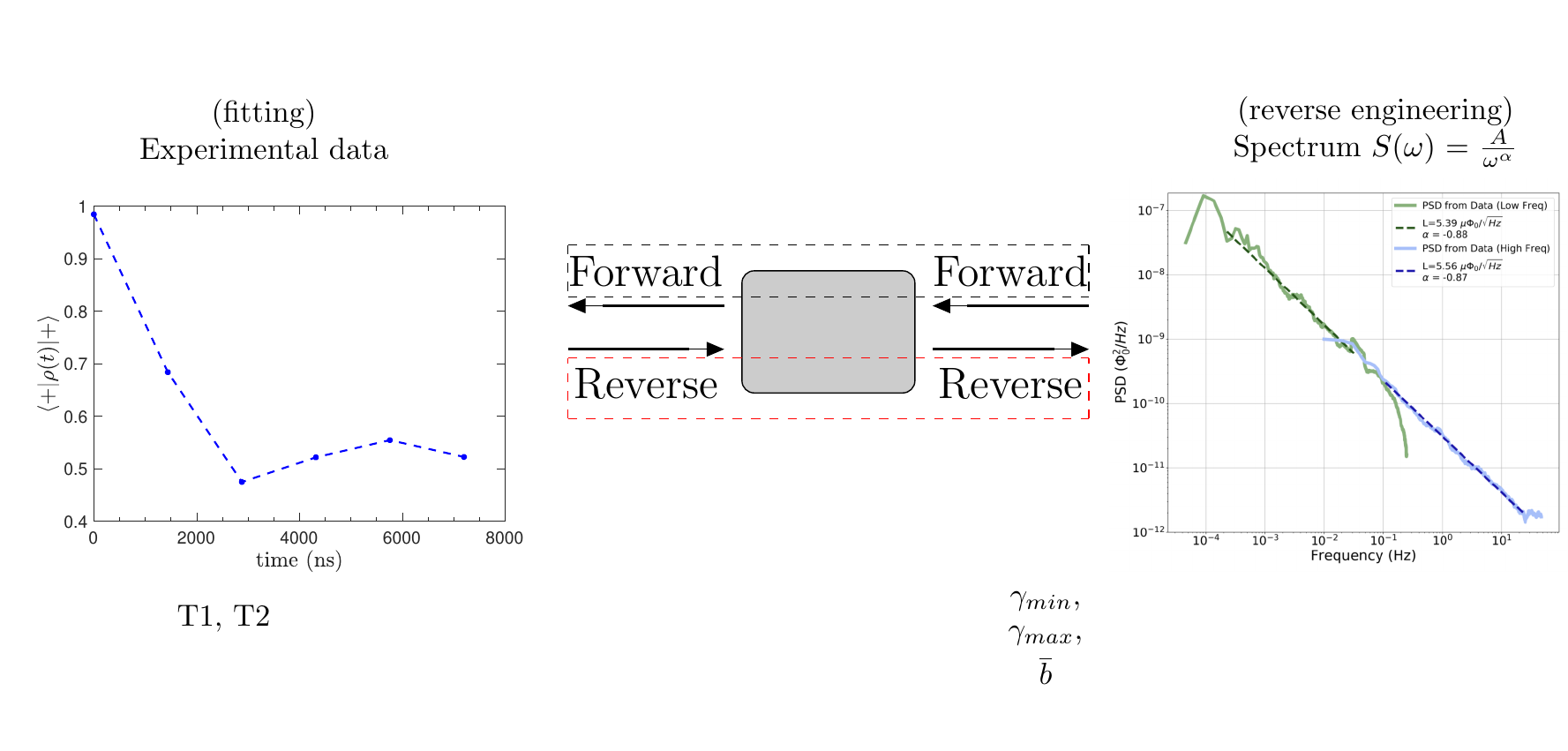}
    \caption{The processes of determining the spectrum~\cite{bylander2011noise} and performing simulations based on the spectrum.}
    \label{fig:ibmqeo}
\end{figure}

Another direction is, given the power spectrum, simulate the qubit free evolutions and evolutions under dynamical decoupling. In this case, we know exactly what $S(\omega)$ is in a given range of $\omega$. With this information, one can deduce the parameters like $\overline{b}$, $\gamma_M$, and $\gamma_m$, and simulate the free evolutions and evolutions under dynamical decoupling with such parameters, etc. 
Fig.~\ref{fig:ibmqeo} illustrates both directions.



\subsection{Fitting results with IBM machine}
\label{sec:fitIBM}
We try to fit the IBM experimental data with our $1/f$ noise simulation. The goal is to reduce the sum of L2 norm of the difference of the simulation data and the experimental data, i.e. $\sum_i \norm{s_i-e_i}_2$, and extract the best fluctuator parameters. ($\overline{b}, \gamma_{\min}, \gamma_{\max}$, etc)

We perform the simulations in a rotated frame, where the frame of reference is the original system Hamiltonian.

\subsubsection{Free induction decay: Fluctuation $A= I_p$}
We try to fit the simulations with the IBM experimental data of free induction decay, where the gates along evolution are the identity gates. 

\begin{figure}[h!]
    \centering
    \subfigure[]{\includegraphics[width = 0.493\columnwidth]{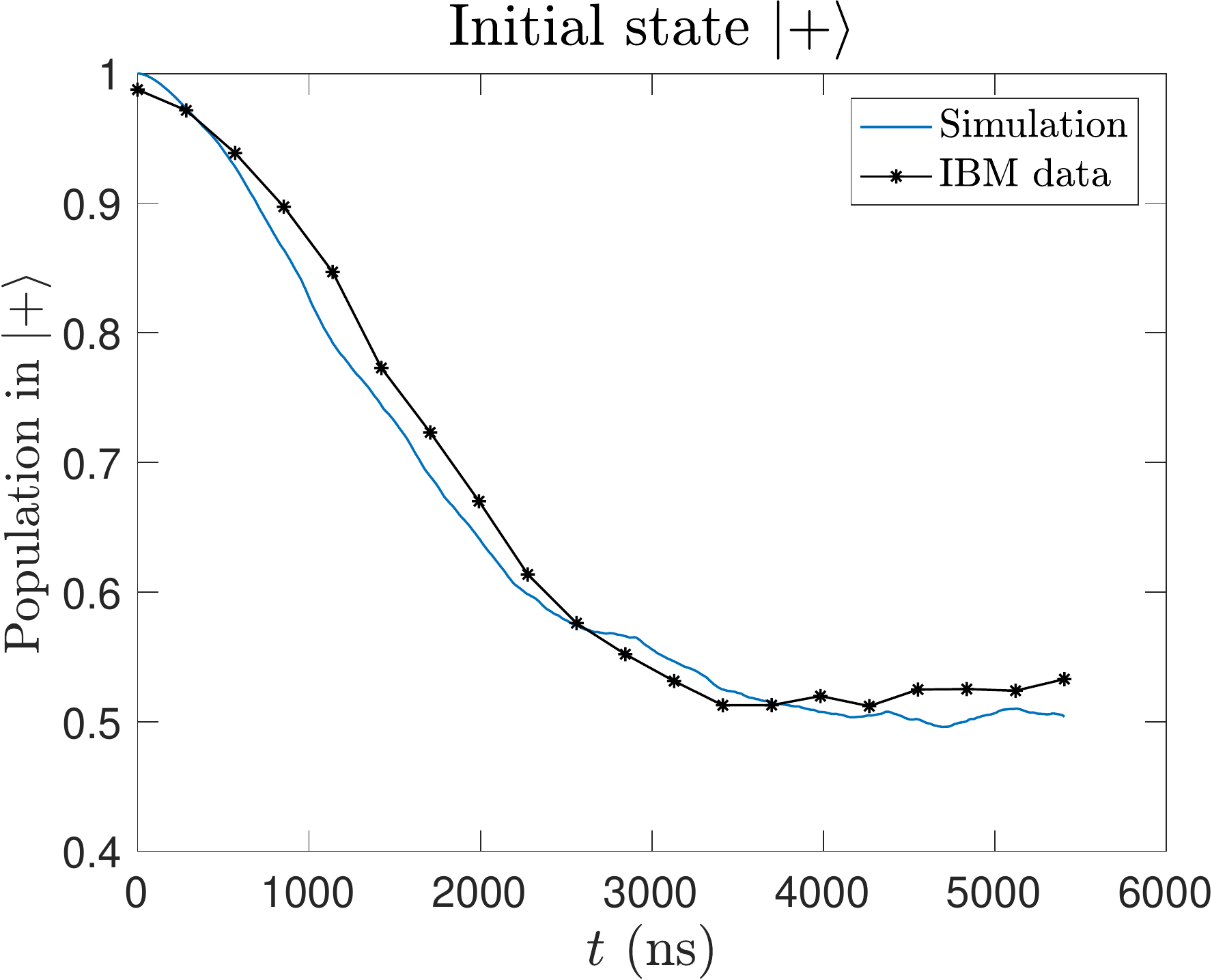}\label{fig:free_fit1}}
	\subfigure[]{\includegraphics[width = 0.493\columnwidth]{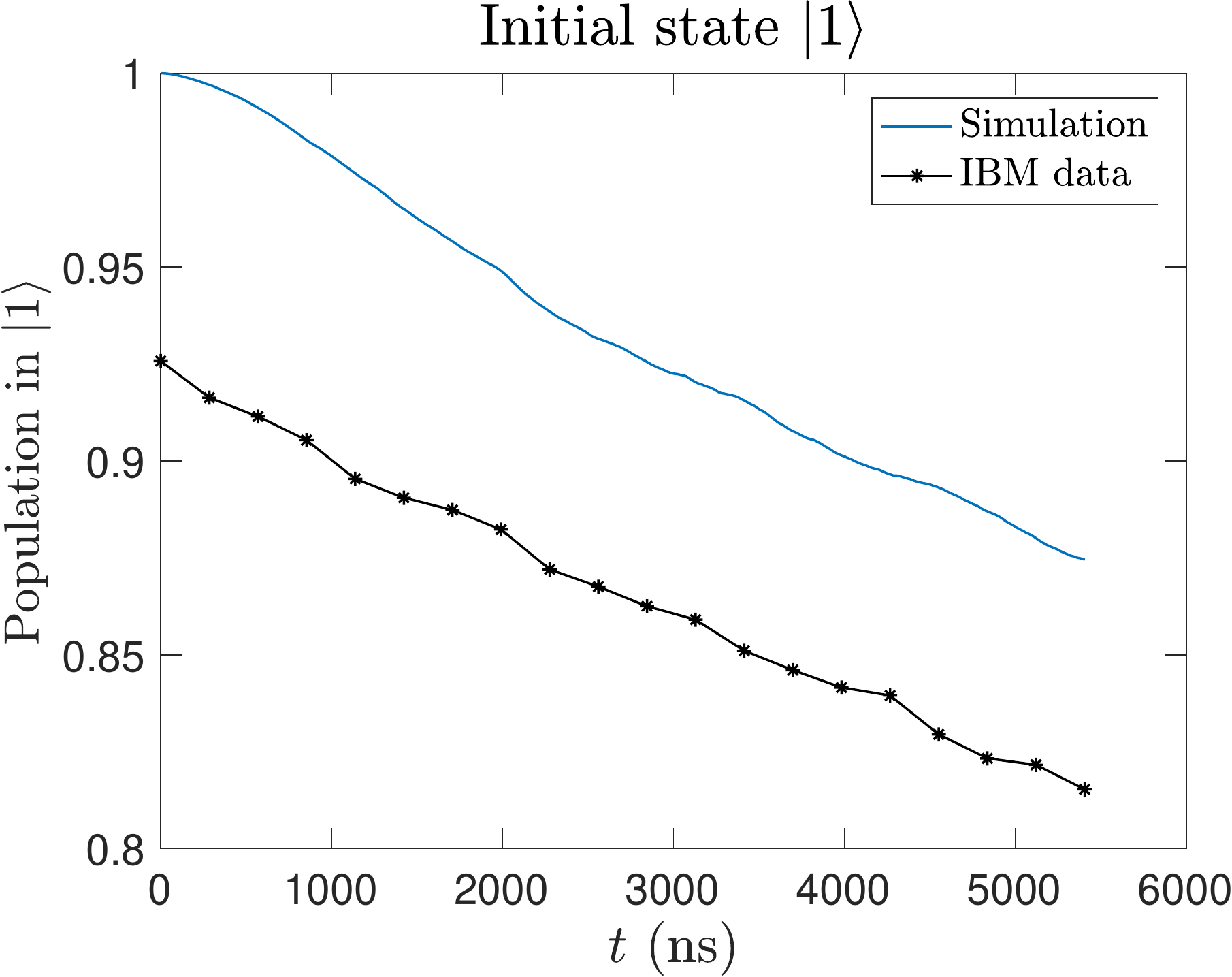}\label{fig:free_fit2}}
	\subfigure[]{\includegraphics[width = 0.493\columnwidth]{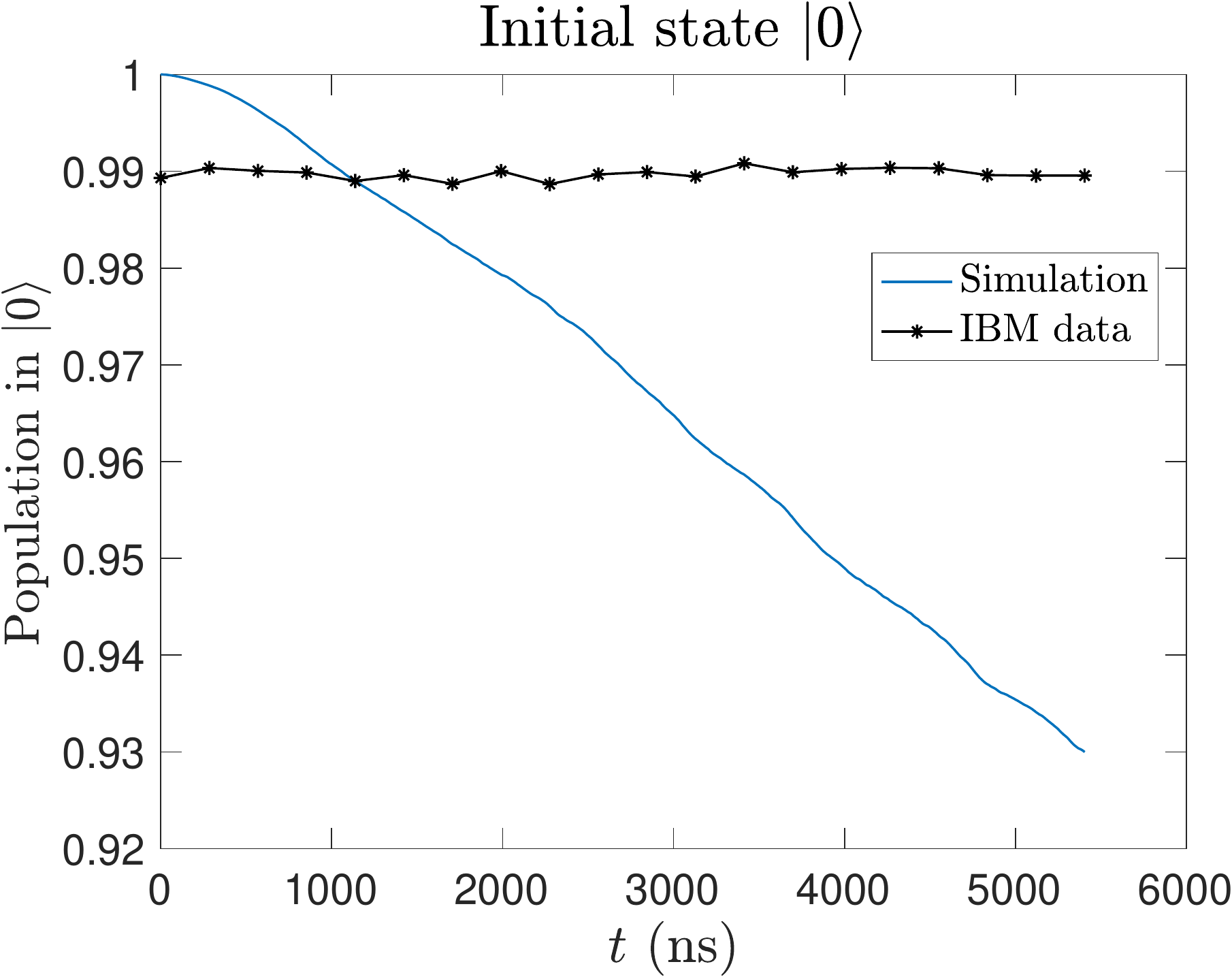}\label{fig:free_fit3}}
    \caption{$t_f = 5405$ns, $\overline{b} = 0.12(9.2) \times 10^{-3}, \gamma_{\min} = 0.05~\text{GHz}, \gamma_{\max} = 5~\text{GHz}$, fluctuations in $I_p$, $20$ fluctuators, $200$ trajectories. Free induction decay. Rotated frame.}
\end{figure}

The fluctuation chosen in our simulations is $A= I_p$. We fix the standard  deviation of fluctuator strength to be $\Delta b /\langle b \rangle = 0.2$. The initial value of each fluctuator is randomly sampled according to the thermal equilibrium distribution $\delta p_{eq}= 0$ (i.e., each fluctuator is equally likely to be initialized in the state $1$ and state $-1$). The number of fluctuators is $20$.
We first output the best-fit parameters ($\overline{b}, \gamma_{\min}, \gamma_{\max}$, etc) to the free induction decay curve with duration of $t_f = 5405$ns, with the initial state being $\ket{+}$. The best fit parameters of fluctuators are $\overline{b} = 0.12(9.2) \times 10^{-3}, \gamma_{\min} = 0.05~\text{GHz}, \gamma_{\max} = 5~\text{GHz}$ (Fig.~\ref{fig:free_fit1}). 
With these best-fit parameters, we can perform simulations with a new initial state. The simulation data of initial state $\ket{1}$ also compares well with the free induction decay experimental data (Fig.~\ref{fig:free_fit2}). However, the simulation data of initial state $\ket{0}$ does not compare well experiments (Fig.~\ref{fig:free_fit3}). If the initial state of the transmon qubit is the ground state, the experimental data suggests that it stays in the ground state, while the simulation data from $1/f$ noise suggests it goes to the maximally mixed state.

\subsubsection{Dynamical decouplings: Fluctuation $A= [I_{p}]_{2\times 2}$}
We try to fit the simulations with the IBM experimental data of dynamical decoupling. 

We fix the standard deviation of fluctuator strength to be $\Delta b /\langle b \rangle = 0.2$. The initial value of each fluctuator is randomly sampled according to thermal equilibrium distribution $\delta p_{eq}= 0$. The number of fluctuators is $20$. 
Based on the fitting with free evolution, we perform simulations with the $XYXY$ dynamical decoupling sequence.
We model the dynamical decoupling as ideal pulses in terms of Pauli two-level $\sigma_x$ and $\sigma_y$ matrices. Therefore, our system Hamiltonian and fluctuators are also truncated to the first two energy levels. $A = [I_{p}]_{2\times 2}$.

\begin{figure}[h!]
    \centering
    \subfigure[]{\includegraphics[width = 0.493\columnwidth]{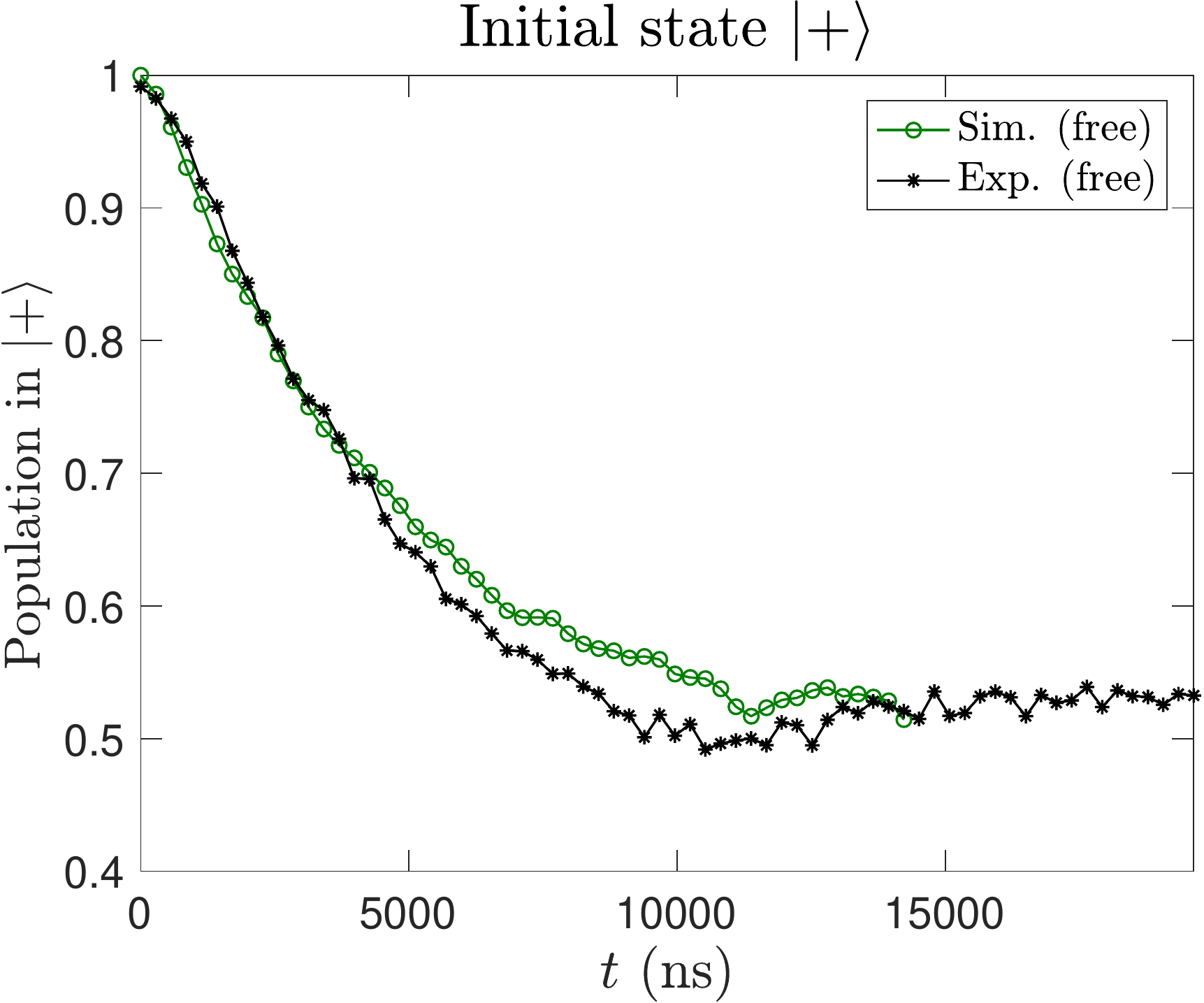}\label{fig:dd_fit1}}
	\subfigure[]{\includegraphics[width = 0.493\columnwidth]{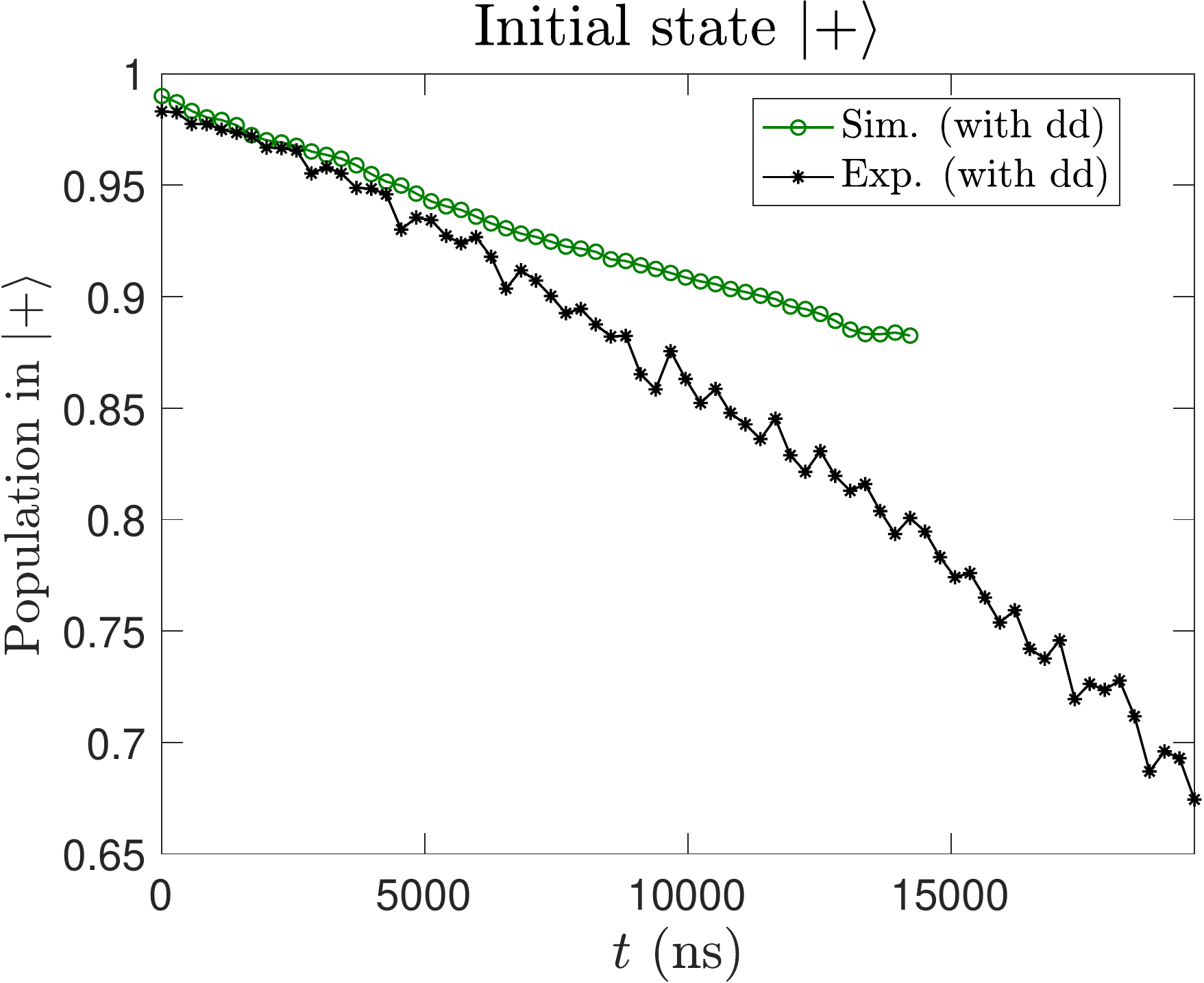}\label{fig:dd_fit2}}
	\subfigure[]{\includegraphics[width = 0.493\columnwidth]{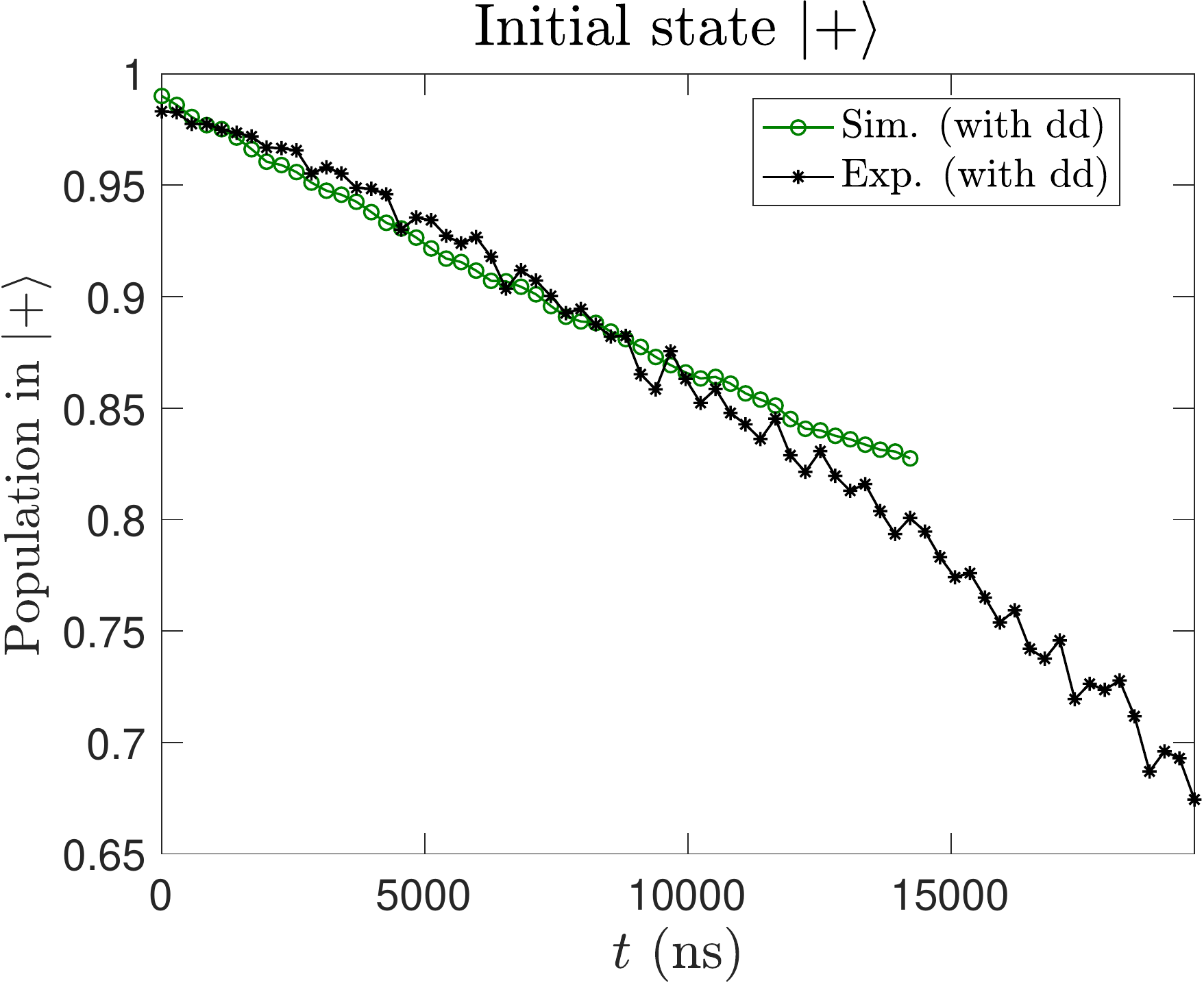}\label{fig:dd_fit3}}
    \caption{Fluctuations in $[I_{p}]_{2\times 2}$, $20$ fluctuators, $300$ trajectories. Rotated frame. $\gamma_{\min} = 0.01~\text{GHz}, \gamma_{\max} = 0.1~\text{GHz}$. (a), (b): $\overline{b} = 9.2929 \times 10^{-4}~\text{GHz}$, (c) Modified averaged fluctuator strength: $\overline{b} = 1.05 \times 10^{-3}~\text{GHz}$.}
\end{figure}

We first find out the best parameters in ($\overline{b}, \gamma_{\min}, \gamma_{\max}$, etc) to the free induction decay curve with duration of $t_f = 15000$ns, with initial state $\ket{+}$. The best fit parameters of fluctuators are $\overline{b} = 0.12(9.2) \times 10^{-3}~\text{GHz}, \gamma_{\min} = 0.05~\text{GHz}, \gamma_{\max} = 5~\text{GHz}$ (Fig.~\ref{fig:dd_fit1}).
With these best-fit parameters, we can perform simulations of dynamical decoupling (XYXY sequence between each data point) under $1/f$ noise, also with initial state $\ket{+}$. However, as we can see in Fig.~\ref{fig:dd_fit2}, the distance between the experimental data and simulation data diverges at long times. Fast noises, which cannot be removed by dynamical decoupling, are present in Transmon qubit, and thus the experimental dynamical decoupling curve decays faster than pure $1/f$ noise simulation.

Nevertheless, we increase the average fluctuator strength by a bit, and can obtain a better fit (Fig.~\ref{fig:dd_fit3}).

\section{Conclusion of this chapter}
\label{secconclude: 1f}
In this chapter, based on the previously studied Bloch vector-magnetic field approach, we have formulated a stochastic fluctuator Hamiltonian simulation method that includes a series of fluctuator terms with different flipping frequencies and Gaussian couplings that altogether constitute the $1/f$ power spectrum. We, for the first time, show how to incorporate them into the inherently time-dependent annealing Hamiltonians, and construct an overall time-dependent stochastic Hamiltonian to perform parallelizable simulations. We thoroughly analyzed how noise with $1/f$ power spectrum affects the quantum annealing, and we also studied  temperature effects, loss of coherence ($T_1$ and $T_2$) in QA under different fluctuator parameters. We developed a flexible approach, allowing the noise fluctuators to take any form of multi-level operators and any noise-axis direction depending on the superconducting qubits and circuit decoherence process. We showed how to append these operators to two different superconducting circuit Hamiltonians. We fit our simulation results with experimental data, and showed how dynamical decoupling can extend the coherence of qubits under $1/f$ noise.
    
\section{Acknowledgments of this chapter}
Sec.~\ref{sec:fitIBM} would be impossible without Vinay Tripathi's experimental data and discussions. 
The codes in this chapter are available at:
\url{https://github.com/USCqserver/1fnoise}




%% file: chapter3.tex
\chapter{Quantum feedback control}
\label{chap: feedback}
\paragraph{}
As mentioned in the conclusion~\ref{secconclude: qt} of Chapter.~\ref{chap: qt}, an extension of the trajectories method is the study of weak measurement and feedback in annealing system. We devise a quantum feedback error correction method to reverse the effect (error) of thermal excitations during quantum annealing.

\section{Introduction}
In QA, the solution of the computational problem is encoded as the ground state of the final Hamiltonian. In an ideal situation, a quantum state is initiated as the ground state of an initial Hamiltonian, and is evolved to the ground state of the final Hamiltonian. It is desirable is to maintain the state during the anneal as the ground state of the time-dependent Hamiltonian. However, due to diabatic or thermal transitions, this is not always successful, i.e., the quantum state at the end of the process is not the ground state of the final Hamiltonian. 

Quantum error correction (QEC) aims to correct such quantum errors. There were previous extensive studies on quantum error correction in QA~\cite{young2013error, lidar2008towards, mishra2016performance}, mostly centered on using error correcting codes to perform majority voting at the end of the anneal. This chapter is about a particular active quantum error correction approach that can used \emph{during} an anneal: feedback-based error correction for QA.

Quantum feedback-based error correction aims to manipulate the system's quantum state or trajectory to evolve the system towards some desired outcome. Depending on the delay between detection and feedback, the effect of the feedback can be Markovian or non-Markovian. We study the following topics about feedback-based error correction on QA.

\subsection{Feedback and control}
In active quantum error correction, one may want to know when and what types of corrections to apply to the system qubit(s) during the quantum operation/process. One way to achieve this goal is to perform measurement on the quantum system at a current time-step and use the measurement result as feedback to determine/improve the error correction for future time-steps. 


\subsection{Weak measurement}
\begin{figure}
    \centering
    \includegraphics[scale=0.5]{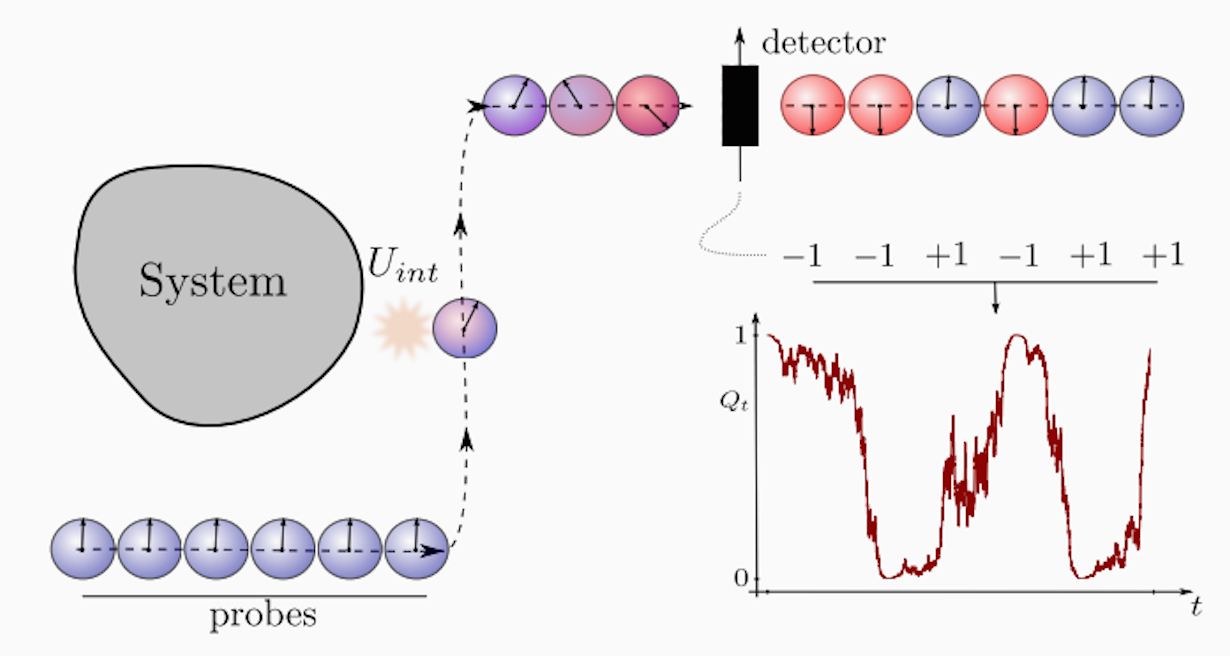}
    \caption{By successively interacting the system with environmental (probe) qubits which are subsequently measured, we can obtain the weak measurement trajectories.}
    \label{fig:weakmeasurement}    
\end{figure}

One way to obtain information about quantum system is by weak measurement. Weak measurement is a type of quantum measurement that results in an observer obtaining very little information about the system on average, but also disturbs the state very little~\cite{brun2002simple}. Weak measurement is also often referred to as continuous measurement, where (weak) measurement is performed continuously in successive tiny time-steps. Weak measurement arises from the system qubit successively interacting with environmental (probe) qubits which are subsequently measured (See Fig.~\ref{fig:weakmeasurement}). Depending on the basis in which the environmental qubits is measured, one can obtain different types (jumpy/diffusive) of measurement trajectories for the system qubit being measured.

When little information is gathered during the measurement, the quantum state is almost entirely preserved. Nevertheless, taking a sufficiently large sample of data still permits the same average information to be extracted as a textbook projective measurement that fully collapses the state. By continuously monitoring partial information about a quantum system, one can control experimental parameters in real time for a variety of purposes, such as error correction, coherence stabilization, state purification.

\section{Thermal excitation as error: detection}
In QA, thermal excitation is considered an error since we want to preserve the ground state of the system Hamiltonian. To do that in feedback-based error correction, the system evolution is continuously monitored and feedback is applied conditioned on each weak measurement result. A photo-detection current $I(t)$ of thermal excitation is obtained, $0<t<t_f$. 

With the assumption of weak-coupling limit, one can derive a master equation describing the evolution of system qubit as:
\begin{align}
\frac{d}{dt} \rho(t) =  -i \left[H_{\textrm{S}}(t) + H_{\textrm{LS}}(t), \rho(t) \right] &+ D[A_{-\omega}(t)]\rho(t) + D[A_{0}(t)]\rho(t) + D[A_{\omega}(t)]\rho(t) \,.
\end{align}

Specifically,
\begin{equation}
\label{eq:exciteop}
    A_{-\omega}(t) \propto |\eps_1(t) \rangle  \langle \eps_0(t)| 
\end{equation} 
is the Lindblad operator of thermal excitation in the instantaneous energy eigenbasis. 
$D[\,\cdot\,]$ refers to the decoherence channel (excitation/dephasing/damping). This master equation can be unraveled into a stochastic Schr\"{o}dinger equation describing the weak measurement. 

\begin{figure}[h!]
\centering
\includegraphics[width=0.35\textwidth]{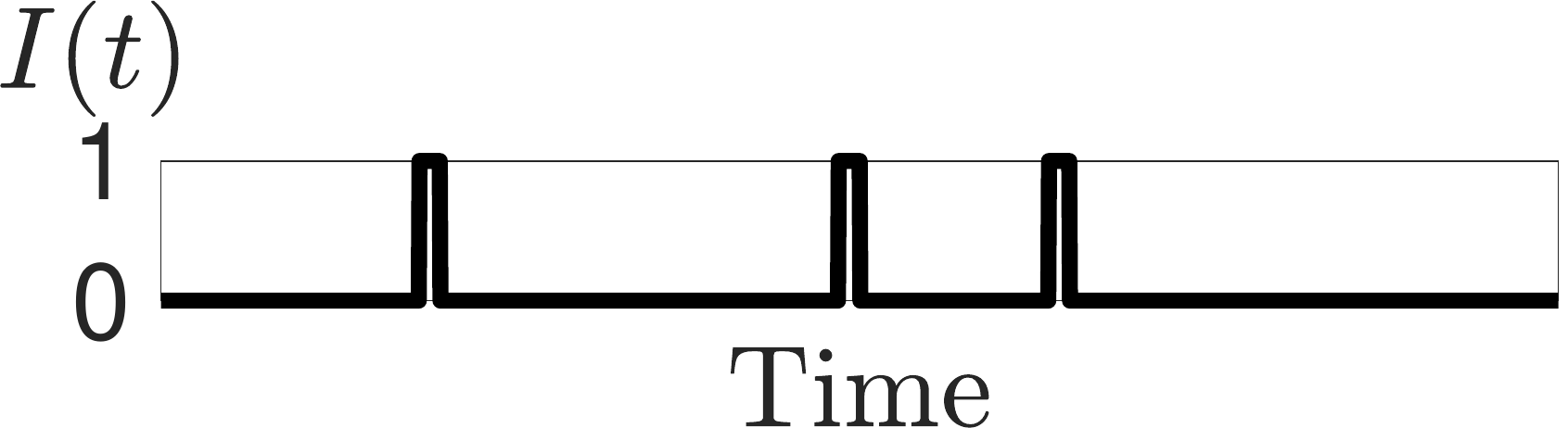}
\caption{Photodetection current $I(t)$.}
\label{fig:photocurrent}
\end{figure}

The system evolution under annealing is continuously monitored and feedback is applied conditioned on each continuous measurement result. The continuous measurement results are in the form of photodetection current $I(t)$, $0\leq t \leq t_f$. The photodetection current $I(t)$ corresponds to the detection of excitation of the system qubit. $I(t) = 1$ when excitation is detected, $=0$ otherwise  (See Fig.~\ref{fig:photocurrent}).

Conditioned on this current, we can obtain an evolution (unravelling in the quantum jump picture) of the \textit{conditioned} state $\rho_{I}(t)$.

\section{Delay between detection and feedback}
After the detection of a thermal excitation ($I(t) = 1$) from ground to excited state, we then want to apply feedback to correct such errors. However, there usually is a delay between detection and feedback. The most general expression for the effect of the feedback on the conditioned state $\rho_{I}(t)$ is~\cite{Wiseman-Milburn-book}:
\begin{equation}
[\dot{\rho}_{\text{I}}(t)]_{\text{fb}} = \mathcal{F}[t, \mathbf{I}_{[0, t)}]\rho_{\text{I}}(t)   \,,
\end{equation}
where $ \mathbf{I}_{[0, t)}$ is the complete photocurrent record from time $0$ to $t$, and $\mathcal{F}[t, \mathbf{I}_{[0, t)}]$ is a superoperator.
Assuming that the feedback is of linear functional form, the feedback evolution can be expressed as:
\begin{equation}
[\dot{\rho}_{\text{I}}(t)]_{\text{fb}} = \int_{0}^{\infty} h(s) I(t-s) \mathcal{L}_f \rho_{\text{I}}(t) ds\,,
\end{equation}
where $\mathcal{L}_f$ is the Liouville superoperator representing the feedback control.
If the delay is fixed as $\tau$, an equation incorporating the effect of the feedback is:
\begin{equation}
\label{Eq:implicitfeedbackevolution}
[\dot{\rho}_{I}(t)]_{\text{fb}} = I(t-\tau)\mathcal{L}_f\rho_{I}(t) \,.
\end{equation}
Note that Eq.~\eqref{Eq:implicitfeedbackevolution} is non-Markovian for finite $\tau$. An analytical solution of Eq.~\eqref{Eq:implicitfeedbackevolution} is hard to obtain and numerical simulation is needed. 

\section{Feedback superoperator $\mathcal{L}_f$: Lindbladian}
The feedback superoperator $\mathcal{L}_f$ can take the form of a Lindbladian or a Hamiltonian. In the form of a Lindbladian:
\begin{equation}
\mathcal{L}_f[\,\cdot\,] = F(t)\,\cdot\,F^{\dagger}(t) - \frac{1}{2}\left\{F^{\dagger}(t)F(t), \,\cdot\, \right\} \,.
\end{equation}
Physically, for the purpose of QA, this Lindbladian represents a cooling/damping channel in which the qubit is relaxed from an excited state to the ground state. 

$F(t)$ can take many forms. For example, $F(t) = A^{\dagger}_{-\omega}(t-\tau) = | \eps_{0}(t-\tau)   \rangle\langle \eps_{1}(t-\tau)  |$, the complex conjugate of the thermal excitation operator [Eq.~\eqref{eq:exciteop}] a time $\tau$ ago.
This exactly cancels the error caused by thermal excitation if $\tau = 0$. With this form of $F(t)$, we can devise a quantum feedback algorithm -- see Algorithm.~\ref{alg:feedbackalgo}. We perform such trajectories feedback simulation in Section~\ref{sec:feedback case studies}.

\begin{algorithm}[H]
\caption{Quantum feedback algorithm with delay}\label{alg:feedbackalgo}
\begin{algorithmic}
\Procedure{Feedback with delay}{$t_f, t_j = 0, \tau$}\Comment{Initialize $t_j = 0$.}
\While{$t_j + \tau \leq t_f$}
\State Evolve according to quantum trajectory method.
    \If {Excitation ${A}_{-\omega}(t)$ occurs at $t = t_j$ to $\ket{\eps(t_j)}$ (Indicated by $I(t) = 1$)}
        \State Apply feedback $F(t_j+\tau) = A^{\dagger}_{-\omega}(t_j)$ at $t_j + \tau$ to $\ket{\eps(t_j +\tau)}$.
    \EndIf
\EndWhile
\EndProcedure
\end{algorithmic}
\end{algorithm}

\section{Case studies}
\label{sec:feedback case studies}
We consider the case of a single qubit with annealing Hamiltonian, i.e.,
${H_S(t)} = A(t) \sigma_x + B(t) \sigma_z$. The $A(t) = A(1-t/t_f)$ and $B(t) = B(t/t_f)$ schedules are the linear schedules, with $A = 1.57$ GHz and $B = 2.01$ GHz (in units $\hbar = 1$). 

\begin{figure}[h!]
\centering
\includegraphics[width=0.7\textwidth]{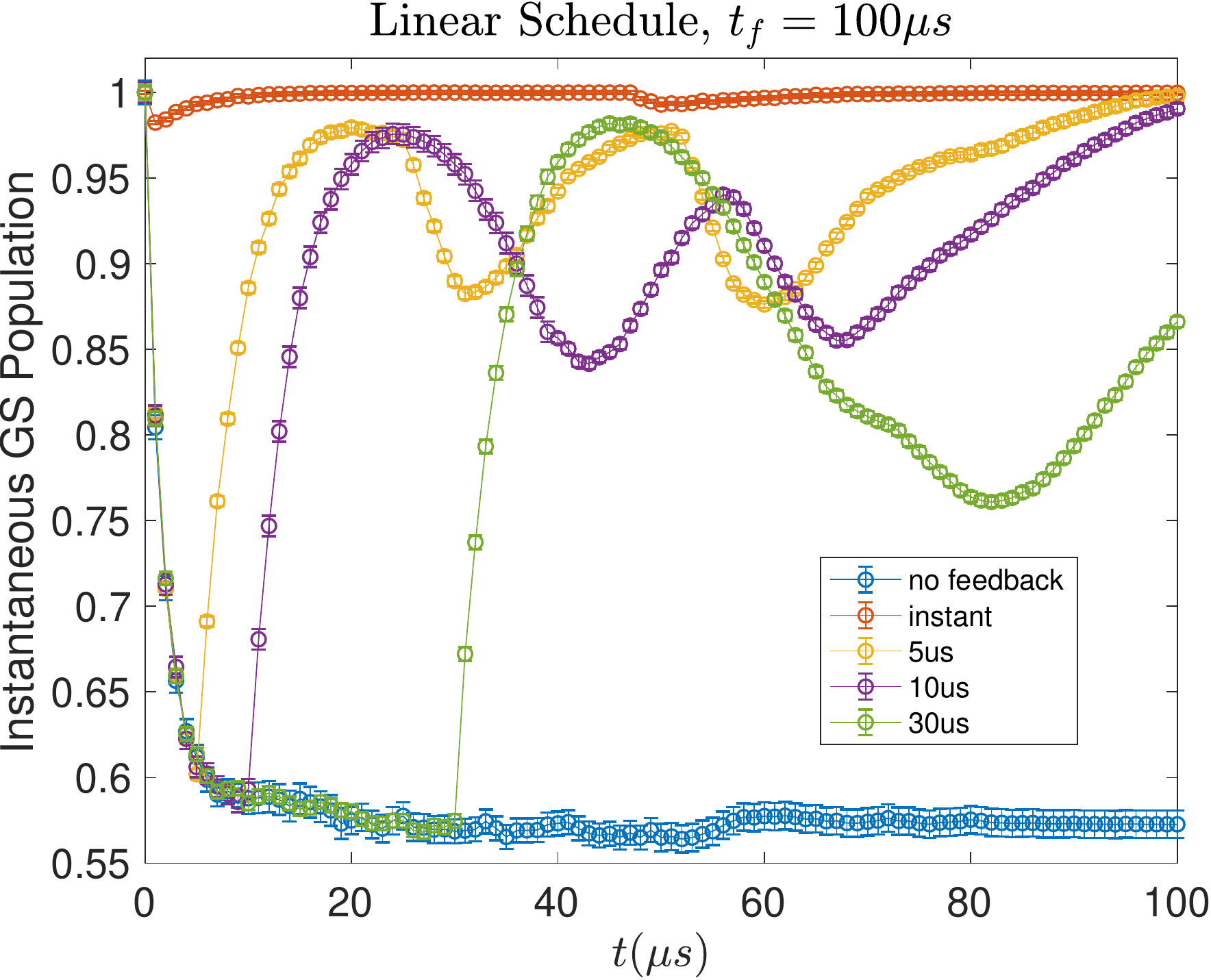}
\caption{Evolution of instantaneous ground state population for a total time of $t_{f} = 100 \mu s$ and temperature $2.6$~GHz, with feedback applied with some different delays $\tau$ ($0,5,10,30 \mu s$). Results are obtained from averaging 5$k$ quantum trajectories of the quantum feedback algorithm. Error bars are $2\sigma$ over $5k$ trajectories.}
\label{fig:feedbackplotlinearsimple}
\end{figure}

We use $F(t) = A^{\dagger}_{-\omega}(t-\tau) = | \eps_{0}(t-\tau)   \rangle\langle \eps_{1}(t-\tau)  |$, with different delays $0\leq \tau \leq t_{f}$.
In Fig.~\ref{fig:feedbackplotlinearsimple}, we simulate the performance of the feedback error protocol for different delay $\tau$ and plot the ground state populations along the anneal.

\begin{figure}[h!]
\centering
\includegraphics[width=0.7\textwidth]{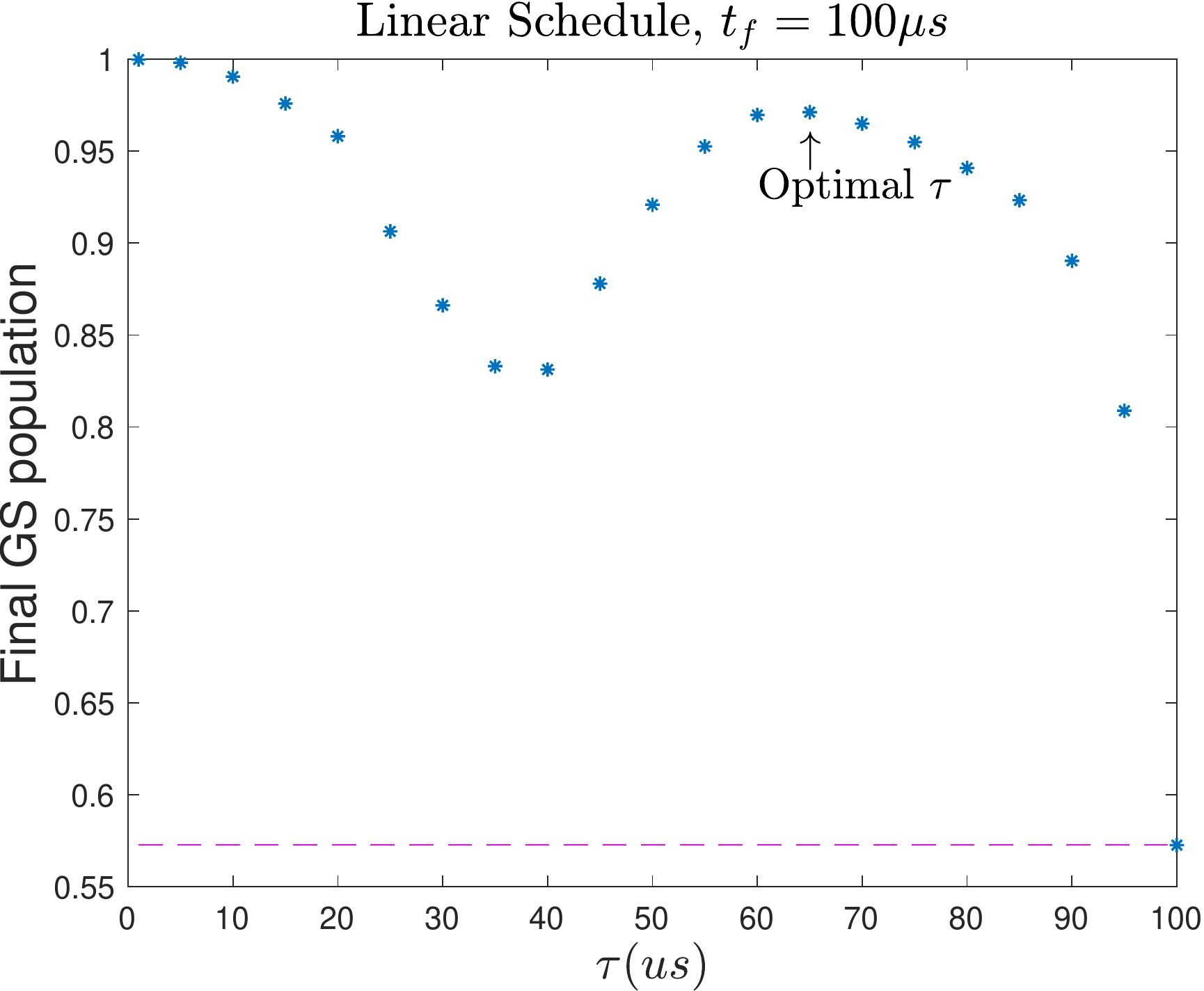}
\caption{Linear Schedule. Final ground state population. Single qubit. Total annealing time $t_f = 100\mu s$. Dashed purple line is the final ground state population without feedback. Average over $5k$ trajectories with Lindbladian feedback. There is a non-small (thus experimentally allowed) $\tau \approx 60us$ where the Final GS population is maximized.}
\label{fig:feedbackdeltalinear_new}
\end{figure}

For the purpose of QA we want to maximize the final ground state population, i.e. ground state population at $t_f$. This serves as a metric to measure the effectiveness of the quantum feedback error correction. Note that the final ground state population need not decrease as delay increases. There is an optimal $\tau$ we can search for. In Fig.~\ref{fig:feedbackdeltalinear_new}, we can see that there is a nonzero delay $\tau$ where final ground state population reaches a local maximum. Thus there is an \textit{experimentally allowed} delay time $\tau$ where the final ground state population is maximized.

\section{Feedback superoperator $\mathcal{L}_f$: Hamiltonian and basis of feedback}
$\mathcal{L}_f$ can also take the form of Hamiltonian~\cite{Wiseman-Milburn-book}.
\begin{equation}
\mathcal{L}_f[\,\cdot\,] = -i\left[H_{\text{fb}}(t), \,\cdot\,\right] \,,
\end{equation}
where $H_{\text{fb}}(t)$ is called the feedback Hamiltonian. Some examples of $H_{\text{fb}}(t)$ are
\begin{itemize}
\item $\sigma^x$ in the energy eigenbasis
\begin{equation}
H_{\text{fb}}(t) =  \frac{\pi}{2}\left(| \eps_{0}(t-\tau)   \rangle\langle \eps_{1}(t-\tau)  | + | \eps_{1}(t-\tau)   \rangle\langle \eps_{0}(t-\tau)  |\right) \,,
\label{eq:feedbackHenergy}
\end{equation}
\item $\sigma^x$ in the computational basis
\begin{equation}
H_{\text{fb}}(t) = H_{\text{fb}} = \frac{\pi}{2} \sigma^x \,,
\label{eq:feedbackHsigmax}
\end{equation}
\item $\sigma^z$ in the computational basis
 \begin{equation}
H_{\text{fb}}(t) = H_{\text{fb}} =  \frac{\pi}{2} \sigma^z \,.
\label{eq:feedbackHsigmaz}
\end{equation}
\end{itemize}

\begin{figure}[h!]
\centering
\includegraphics[width=0.82\textwidth]{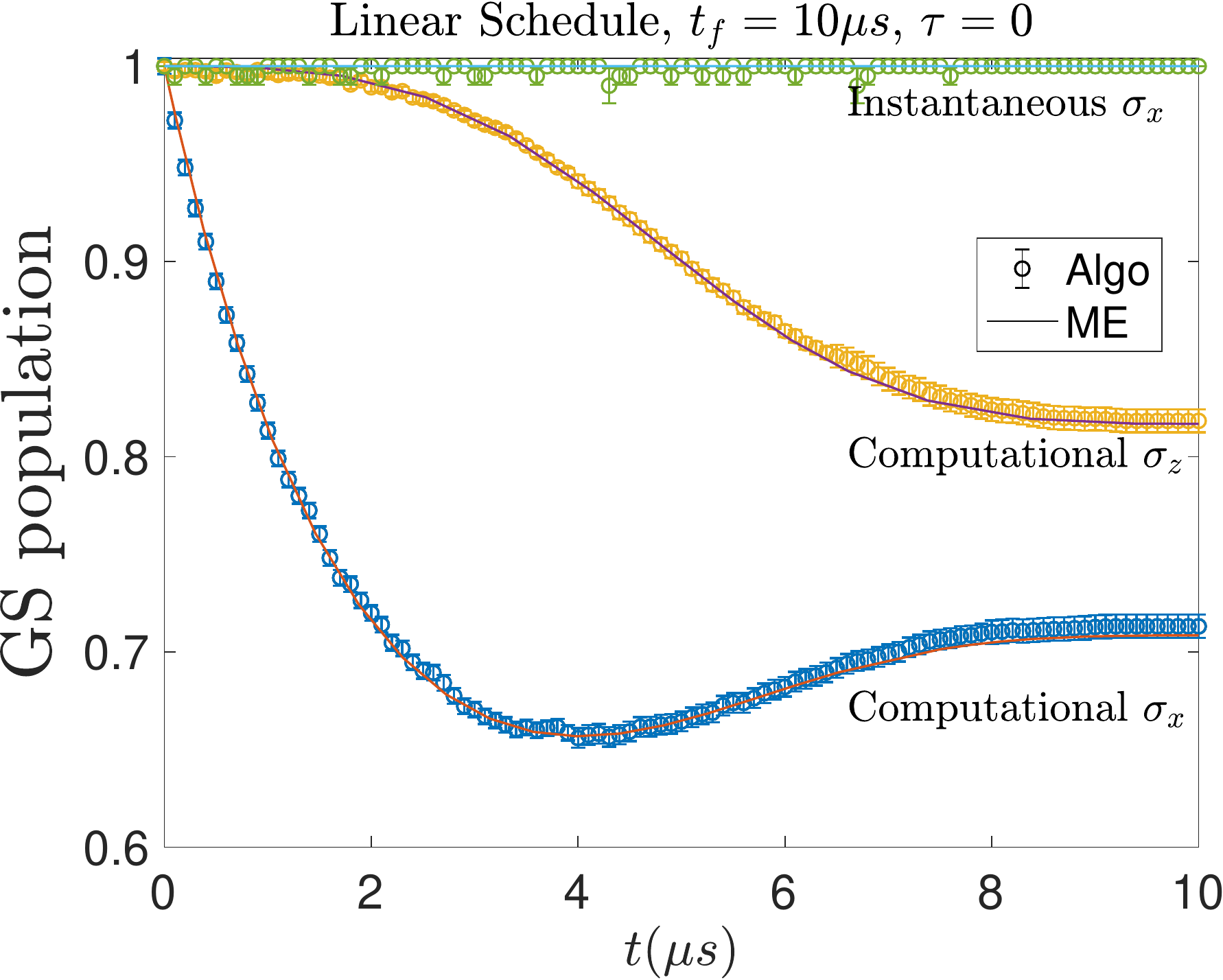}
\caption{Evolution of ground state population under feedback error correction of different feedback Hamiltonians ($\tau = 0$). The results of quantum jump trajectories (averaged over $5k$ trajectories) of quantum feedback algorithm agree with the feedback master equation solutions. Error bars are $2\sigma$ over $5k$ trajectories.}
\label{fig:Hfeedall}
\end{figure}

These are Pauli operators in different bases. Note that in the decoherence model of the weak coupling limit, any excitation/relaxation takes place in the energy eigen-basis~\cite{ABLZ:12-SI}, thus one would expect that the feedback Hamiltonian $\sigma^x$ in the energy eigenbasis performs better than $\sigma^x$ in the computational eigenbasis (see simulations in Fig.~\ref{fig:Hfeedall}). Note also that the significance of the $\frac{\pi}{2}$ implies that the feedback Hamiltonian acts instantly (pulse/gate-like). 

Another advantage of the feedback Hamiltonian formalism is that for delay $\tau \rightarrow 0$, one can obtain the feedback master equation that explains the ensemble average of the effect of feedback. The feedback master equation is 
\begin{align}
\frac{d}{dt} \rho(t) &= -i \left[H_{\textrm{S}}(t) + H_{\textrm{LS}}(t), \rho(t) \right] \notag\\
&+ \left(
e^{-iH_{\text{fb}}(t)}A_{-\omega}(t)\rho(t) A^\dagger_{-\omega}(t)e^{iH_{\text{fb}}(t)}  - \frac{1}{2} \left\{ A_{-\omega}^{\dagger}(t) A_{-\omega}(t) , \rho(t) \right\} \right) \notag\\
&+ D[A_{0}(t)]\rho(t) + D[A_{\omega}(t)]\rho(t)
\label{Eq:feedbackME_H}
\end{align}
This also allows direct master equation simulation, which can be faster than trajectories simulations for a small number of qubits. Fig.~\ref{fig:Hfeedall} below compares the two approaches of master equation simulation and the quantum feedback trajectories algorithm, over the three feedback bases mentioned in Eqs.~\eqref{eq:feedbackHenergy},~\eqref{eq:feedbackHsigmax} and~\eqref{eq:feedbackHsigmaz}.


%% file: chapter4.tex
\chapter{Reverse Annealing: Iterative reverse annealing}
\label{chap:ira}
We want to apply our open-system simulation techniques the study of open-system effects on a particular QA application called reverse annealing (RA). Reverse annealing has been predicted to have better performance than forward annealing in several systems~\cite{perdomo:sombrero,chancellor:reverse,King:2018aa,Ottaviani2018,nishimori:reverse-pspin, nishimori:reverse-pspin-2,Venturelli2019,marshall}.

Reverse annealing can be generally separated into two formalisms: iterated reverse annealing (IRA) and adiabatic reverse annealing (ARA). ARA is similar to standard quantum annealing, but with an extra initialization Hamiltonian. IRA does not have such term but starts at $s=1$ with a random Initial state that is usually an eigenstate of the problem Hamiltonian. 

Iterated reverse annealing (IRA) involves (1) annealing backward from a known classical state to a mid-anneal state of quantum superposition, (2) searching for optimum solutions at this mid-anneal point while in the presence of an increased transverse field (quantum state), and then (3) proceeding forward to a new classical state at the end of the anneal.

In this chapter, we study the open-system effects on iterative reverse annealing and perform simulations of IRA in the weak coupling limits to bath, for $p$-spin model with $p=3$. This chapter is based on~\cite{Passarelli2019}. In Chapter.~\ref{chap: IRA_exp}, we perform the IRA simulations with more realistic experimental settings of D-Wave annealers, for $p$-spin model with $p=2$. In Chapter.~\ref{chap: ara}, we study the open-system effects on adiabatic reverse annealing (ARA), for $p$-spin model with $p=3$.


\section{Introduction}
In iterative reverse annealing, the initial state is an eigenstate of the final problem Hamiltonian and the transverse field is cycled rather than strictly decreased as in standard (forward) quantum annealing. We present a numerical study of the reverse quantum annealing protocol applied to the $p$-spin model ($p=3$), including pausing, in an open system setting accounting for dephasing in the energy eigenbasis, which results in thermal relaxation. We consider both independent and collective dephasing and demonstrate that in both cases the open system dynamics substantially enhances the performance of reverse annealing. Namely, including dephasing overcomes the failure of purely closed system reverse annealing to converge to the ground state of the $p$-spin model. We demonstrate that pausing further improves the success probability. The collective dephasing model leads to somewhat better performance than independent dephasing. The protocol we consider corresponds closely to the one implemented in the current generation of commercial quantum annealers, and our results help to explain why recent experiments demonstrated enhanced success probabilities under reverse annealing and pausing.

\section{Iterative reverse annealing: the protocol}
\label{sec:reverse}
\begin{figure*}[t]
	\subfigure[]{\includegraphics[width = 0.5\columnwidth]{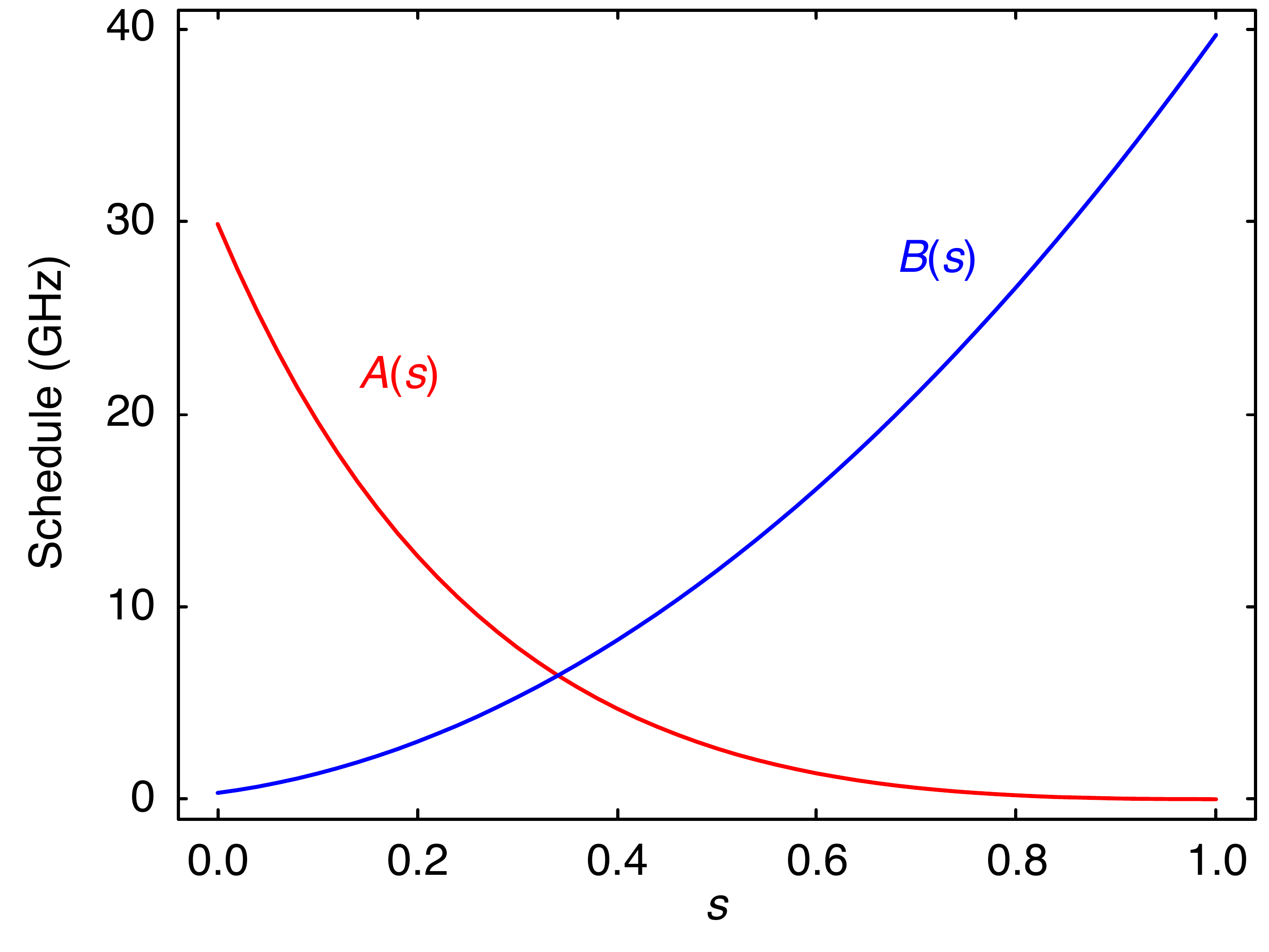}\label{fig:schedules}}
	\subfigure[]{\includegraphics[width = 0.5\columnwidth]{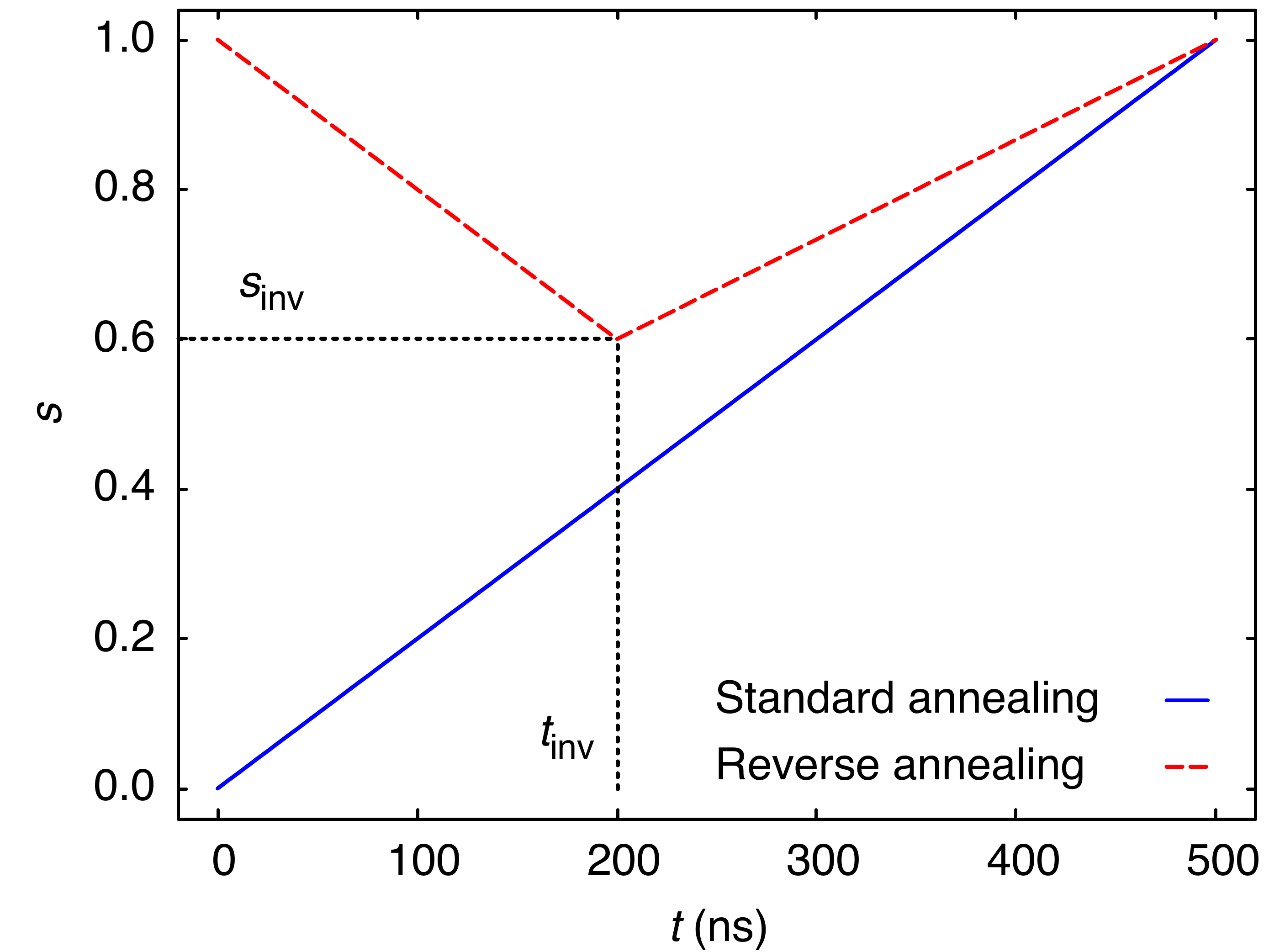}\label{fig:s-of-t}}
	\caption{(a) Annealing schedules (in units such that $\hbar=1$) as a function of the annealing fraction $s(t)$, chosen to be similar to the schedules of the D-Wave processors.
	(b) Annealing fraction $ s(t) $. The blue solid curve represents standard forward quantum annealing of total annealing time $\tau = 500$ns. 
	}
\end{figure*}	
	
Standard quantum annealing aims at solving optimization problems by employing quantum fluctuations that are slowly decreased to zero, to efficiently explore the solution space~\cite{kadowaki:qa, Santoro, albash:review-aqc}. A system of $ \nspin $ qubits is prepared at $ t = 0 $ in the ground state of a transverse field Hamiltonian, \ie, the uniform superposition over the $\n = 2^\nspin $ computational basis states $ \{\ket{0}, \ket{1}, \dots, \ket{\n - 1}\} $. 
The magnitude of transverse field
is then slowly decreased, while the magnitude of the Hamiltonian $ \hamtarget $, encoding the optimization problem, is simultaneously increased. The adiabatic theorem guarantees that if the evolution is long on the timescale set by the inverse of the minimal gap $ \mingap = \min_t [E_1(t) - E_0(t)] $ between the ground state and the first excited state (we set $ \hslash = 1 $ henceforth), then at the end of the anneal $ t = \tf $ the system populates the target ground state of $ \hamtarget $ with a probability $ \pgs $ that approaches unity~\cite{Jansen:07,lidar:102106}. However, any finite sweep rate leads to diabatic Landau-Zener transitions at the avoided crossings~\cite{Joye:LZ}, thus reducing the success probability $ \pgs $ of the adiabatic algorithm. Therefore, the output state of quantum annealing is in general a trial solution of the optimization problem, ideally having a large overlap with the target ground state.
	
Reverse annealing instead aims at refining an already available trial solution~\cite{perdomo:sombrero}. For instance, NP-hard optimization tasks are solved using heuristics, whose output is often an approximation of the true global minimum. The algorithm of reverse annealing is the following.
\begin{enumerate}
	\item At $ t = 0 $, the system is prepared in the trial solution state.
	\item Quantum fluctuations are increased, until a turning point $ \tinv $ is reached during the dynamics. This ends the reversed part of the dynamics.
	\item After the turning point, the dynamics follows the standard quantum annealing schedule; quantum fluctuations are decreased until $ t = \tf $, when the state is eventually measured.
\end{enumerate}
Careful choices of the turning point, and of the initial state, can lead to an improvement in the solution. Moreover, this scheme can also be repeated multiple times, each time starting from the output of the previous stage; hence the terminology of iterated reverse annealing. Assuming that each iteration improves the success probability $ \pgs $, this procedure can systematically improve the quality of the solution. 

We consider the following time-dependent Hamiltonian, suitable for IRA but not ARA (which includes another term):
\begin{equation}
\label{eq:time-dependent-hamiltonian}
	\ham(t) = A\bigl(s(t)\bigr) \hamtf + B\bigl(s(t)\bigr) \hamtarget, \quad  \hamtf = -\sum_{i=1}^{\nspin} \sigma_i^x .
\end{equation}
Since we only consider the IRA protocol, henceforth we simply refer to it as reverse annealing. The function of time $ s(t) $ in Eq.~\eqref{eq:time-dependent-hamiltonian} is the annealing fraction (or dimensionless time) and satisfies $ 0 \le s(t) \le 1 $ for all $ t $. The two functions $ A(s) $ and $ B(s) $ determine the annealing schedule, which we choose to match the annealing schedule of the D-Wave processors [see Fig.~\ref{fig:schedules}]. They satisfy $ A(0) \gg B(0) $ and $ B(1) \gg A(1) $.
	
The functional form of $ s(t) $ distinguishes between standard (forward) and reverse annealing. In standard quantum annealing the dimensionless time is defined as $ s(t) = t / \tf $, $ \tf $ being the annealing time. Thus, $ s(t) $ is a monotonic function of $ t $, and is represented in the plane $ (t, s) $ by a straight line going from $ (0, s_0) $ to $ (\tau, s_1) $, where $ s_0 = s(0) = 0 $ and $ s_1 = s(\tau) = 1 $. This is shown in Fig.~\ref{fig:s-of-t} using a blue solid line, for $ \tau = 500$ ns. During standard quantum annealing, quantum fluctuations are very large at $ t = 0 $, and decrease monotonically until $ t = \tau $.
	
In contrast, in reverse annealing $ s_0 = s_1 = 1 $. Starting from $ s = s_0 $, where quantum fluctuations are zero, the annealing fraction is first decreased until it reaches the inversion point, $ s = \sinv $, at a time $ t = \tinv $. In this first branch, quantum fluctuations are increased. At $ s = \sinv $, the annealing fraction is then increased towards $ s = s_1 $, and quantum fluctuations are decreased again to zero. In Fig.~\ref{fig:s-of-t}, we show a typical function $ s(t) $ for a reverse annealing of annealing time $ \tf = \SI{500}{\nano\second} $ with an inversion point $ (\tinv = \SI{200}{\nano\second}, \sinv = 0.6) $, using a red dashed line.
		
In general, $ \sinv $ and $ \tinv $ can be chosen independently of each other. However, in this work we choose the following linear relation, in order to have only one free parameter:
\begin{equation}
\label{eq:inversion-time}
	\tinv = \tf (1 - \sinv)\ , \ \ \sinv \ne 0, 1.
\end{equation}
In this way, we have that
\begin{equation}\label{eq:schedule}
	s(t) =
	\begin{cases}
		1-t / \tf & \text{for $ t \le \tinv $},\\
		\frac{1 - \sinv}{\tf \sinv} t + \frac{2 \sinv - 1}{\sinv} & \text{for $ t > \tinv $}.
	\end{cases}
\end{equation}
	
Another possible choice would be to fix $ \tinv $ so that the two slopes are the same, \ie, $ \tinv = \tf / 2 $ for all choices of $ \sinv $. This is similar in spirit to what is discussed in Ref.~\cite{nishimori:reverse-pspin-2}. 
	
In what follows, we will focus on the fully-connected ferromagnetic $ p $-spin model, a model with a permutationally invariant Hamiltonian and a nontrivial phase diagram, often used as a benchmark for the performance of quantum annealing~\cite{gross:p-spin, derrida:p-spin}. 
	
\section{Ferromagnetic $p$-spin model}\label{sec:pspin}

The Hamiltonian of the ferromagnetic $ p $-spin model, in dimensionless units, reads
\begin{equation}\label{eq:pspin}
	\hamtarget = -\frac{\nspin}{2}{\left(\frac{1}{\nspin} \sum_{i = 1}^{\nspin} \sigma^z_i\right)}^p,
\end{equation}
with $ p \ge 2 $. For even $ p $, there are two degenerate ferromagnetic ground states, whereas for odd $ p $ the ferromagnetic ground state is nondegenerate. For $ p = 2 $, the Hamiltonian of Eq.~\eqref{eq:time-dependent-hamiltonian} is subject to a second-order QPT in the thermodynamic limit at the critical point $ \scrit $ (or, equivalently, $ \tcrit = \tf \scrit $), separating a para- and a ferromagnetic phase. For $ p > 2 $, the QPT is first-order~\cite{bapst:p-spin}. The presence of QPTs affects also the finite-size behavior of the system, as the minimal spectral gap $ \mingap $, found at $ \sgap $ ($ \tgap = \tf \sgap $), closes as $ \nspin^{-1/3} $ for $ p = 2 $ or exponentially in $ \nspin $ for $ p > 2 $. $ \sgap $ (\ie, $ \tgap $) approaches $ \scrit $ (\ie, $ \tcrit $) as $ \nspin \to \infty $. First-order QPTs are especially detrimental for quantum annealing, as the annealing time has to grow exponentially with the system size to compensate the closure of the gap at $ s = \sgap $. In the following, we will focus on the case $ p = 3 $.
	
We can define the total spin operator $\mathbf{S} = (S_x,S_y,S_z)$ and the dimensionless magnetization operators
\begin{equation}
	m_{\alpha} = \frac{1}{\nspin} S_{\alpha} ,\quad S_{\alpha} = \sum_{i = 1}^{\nspin} \sigma_i^{\alpha}, \quad \alpha\in\{x,y,z\}
\end{equation}
that allow the rewriting of the time-dependent Hamiltonian of Eq.~\eqref{eq:time-dependent-hamiltonian} as
\begin{equation}
	 \ham(t) = -\frac{A(t)}{2} S_x - \frac{B(t)N}{2} m_z^p .
\end{equation}
Since $[\mathbf{S}^2,S_z]=0$ and both are Hermitian, they share an orthonormal basis $\{\ket{S,\mu_S}\}$ such that the eigenvalues of $\mathbf{S}^2$ are $S(S+1)$ with $S\in\{0,1/2,1,\dots,\nspin/2\}$ for even $ \nspin $ and $S\in\{1/2,1,\dots,\nspin/2\}$ for odd $ \nspin $, and the eigenvalues of $S_z$ are $\mu_S \in \{-S,-S+1,\dots,S\}$.
In the subspace with maximum spin $S=\nspin/2$, we instead label the basis states as $ \ket{w} \equiv \ket{\nspin/2 - w} $, with $ w \in \{0, 1, \dots, \nspin\} $. These are the eigenstates of $ m_z $ with eigenvalues $ m = 1 - 2 w / \nspin $. The target state is the ferromagnetic ground state $ \ket{0} $, \ie, the eigenstate of $ m_z $ with eigenvalue $ m = 1 $. 

The Hamiltonian of the $ p $-spin model commutes with $\mathbf{S}^2$. Hence sectors differing by $S$ do not become coupled under the dynamics generated by the $ p $-spin model Hamiltonian. Since the ferromagnetic ground state and the initial one belong to the subspace with maximum spin $ S = \nspin / 2 $, the interesting dynamics occurs in this subspace, whose dimension scales \emph{linearly} with the number of qubits: $ \n = \nspin + 1 $. This fact enables us to perform numerical calculations with relative large numbers of qubits $\nspin$.

In the following, we will start reverse annealing in each of the $ \nspin $ excited states $ \{\ket{1}, \dots, \ket{\nspin} \}$ of $ \hamtarget $ in the symmetric sector with $ S = \nspin/2 $. The similarity to the ferromagnetic ground state is quantified by the corresponding starting eigenvalue of $ m_z $, denoted $ m_0 $. Note that the $ w $th excited state differs from the ferromagnetic ground state by $ w $ spin flips. Therefore, the initial state and the target solution differ by a fraction $ c = \nspin_\uparrow / \nspin = 1 - w /\nspin $ of up-aligned qubits. These parameters are also related to the Hamming distance $ d\ped{H} $, 
via $ d\ped{H} = \nspin - \nspin_\uparrow = \nspin (1 - c) $.

\section{Unitary dynamics}
\label{sec:unitary}

\begin{figure*}[t]
	\subfigure[]{\includegraphics[width = 0.48\columnwidth]{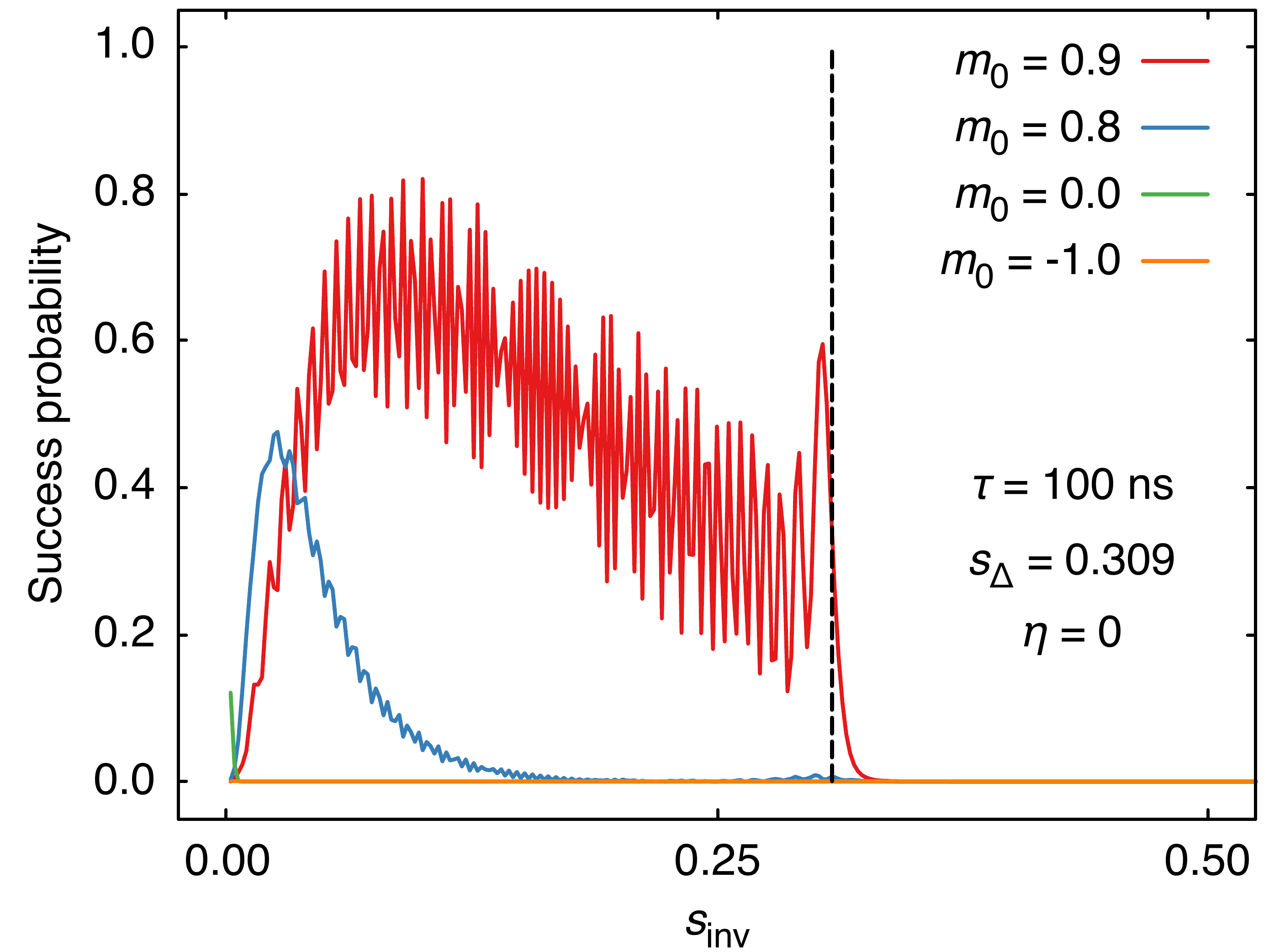}\label{fig:unitary-tf-100}}
	\subfigure[]{\includegraphics[width = 0.48\columnwidth]{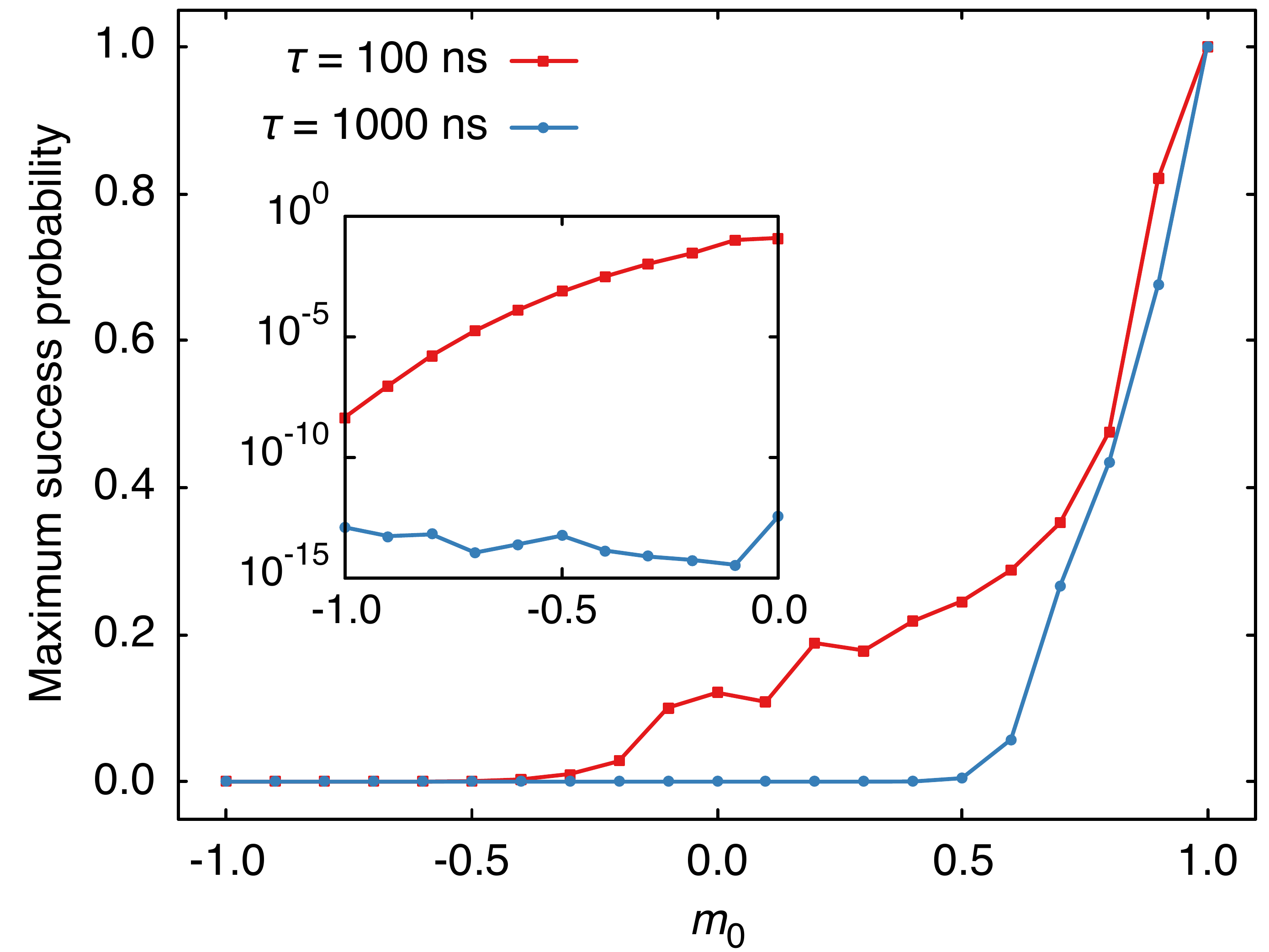}\label{fig:unitary-scaling}}
	\caption{(a) Success probability in unitary reverse annealing as a function of the inversion point $ \sinv $, for several values of the magnetization of the initial state. The dashed vertical line indicates $ \sinv = \sgap \approx 0.309 $. The annealing time is $ \tf = \SI{100}{\nano\second} $. We sampled the interval $ \sinv \in \rngopen{0}{1} $ using a step size of $ \Delta s = 0.002 $, and repeated the dynamics for each choice of $ \sinv $. (b) Maximum success probability achievable with unitary reverse annealing, as a function of the magnetization of the initial state, for two annealing times: $ \tf = \SI{100}{\nano\second} $ and $ \tf = \SI{1000}{\nano\second} $. The inset zooms in on the region $ m\in\rng{-1.0}{0.0} $. Note the logarithmic scale on the vertical axis.}
\end{figure*}

In this section we study the closed system case of a system of $ \nspin = 20 $ qubits, with $ p = 3 $. For our choice of parameters and in terms of the annealing schedules shown in Fig.~\ref{fig:schedules}, the $ p $-spin system has a minimal gap $ \mingap \approx \SI{2.45}{\giga\hertz} $ at $ \sgap \approx 0.309 $. The annealing time is $ \tf = \SI{100}{\nano\second} $. 
	
In Fig.~\ref{fig:unitary-tf-100}, we report the ground state population $ \pgs $  at $ t = \tf $, as a function of the inversion point $ \sinv $, for several initial states: $ m_0 = \text{\numlist{0.9;0.8;0;-1}} $. Recall that the target ground state $ \ket{0} $ has $m=1$. We focus on the region $ \sinv \in \rnglopen{0.0}{0.5}$.
	
The rightmost part of Fig.~\ref{fig:unitary-tf-100} corresponds to cases in which the anneal is reversed too early, \ie, for $ \tinv < \tgap $ and $ \sinv > \sgap $. The system does not cross its quantum critical point, and the success probability is zero. Therefore, no effects on the outcome of the procedure are visible, as the dynamics is slow compared with the minimal inverse level spacing and diabatic transitions are exponentially suppressed. Thus, the system is forced to stay in its initial state, or transition to other excited states. In fact, avoided crossings between pairs of excited eigenstates occur at $ s > \sgap $ for this model, and Landau-Zener processes can further excite the $ p $-spin system.
	
On the other hand, if $ \sinv < \sgap $ the system crosses the minimal gap twice. Here, the success probability benefits from Landau-Zener processes, inducing transitions towards the ground state. In this region, we also note some non-adiabatic oscillations of the success probability, due to the finite annealing time. These oscillations are more evident for large $ m_0 $. As expected from the adiabatic theorem, they are suppressed for longer annealing times. For instance, we verified that they are no longer visible for $ \tf = \SI{1000}{\nano\second} $ (not shown). The sharp rise of the success probability for $ m_0 = 0.9 $ occurs exactly at $ \sinv = \sgap $. For smaller values of $ m_0 $, the success probability rises more smoothly, as the ground state is reached after a preliminary sequence of Landau-Zener transitions between pairs of excited states, whose corresponding avoided crossings occur at $ s > \sgap $. For $ m_0 = 0.8 $, a very small rise of the success probability can still be observed around $ \sinv = \sgap $. This is due to the fact that during the reverse annealing, the system prepared in the second excited state first encounters an avoided crossing with the first excited state, where part of the population is transferred to the latter, and then the avoided crossing with the ground state, where the system populates its ground state. After reversing the dynamics, the two avoided crossings are encountered again (in the reverse order) and part of the population gets excited, thus reducing the success probability $ \pgs $.

As expected, reverse annealing is more effective when the initial state is close to the correct ground state. Moreover, as is also clear from Fig.~\ref{fig:unitary-tf-100}, the inversion time $ \sinv $ must be increasingly close to $0$ for decreasing $ m_0 $, in order to obtain a nonzero success probability at $ t = \tf $. This means that almost the entire dynamics is spent in the reverse part of the annealing, and the system is eventually quenched towards $ s = s_1 $ for $ t \approx \tf $. Even so, if the initial state is too far in energy from the correct solution, the success probability of reverse annealing is always close to zero, as evident from the curves for $ m_0 = 0 $ and $ m_0 = -1 $ in Fig.~\ref{fig:unitary-tf-100}. 
	
The maximum success probability decreases rapidly as a function of $ m_0 $. This is clearly seen in Fig.~\ref{fig:unitary-scaling}, where we report the maximum attainable success probability as a function of $ m_0 $, for annealing times $ \tf = \text{\SIlist{100;1000}{\nano\second}} $. 
Increasing the annealing time reduces non-adiabaticity and results in a lower success probability, compared with that at the end of a faster reverse anneal. As shown in Fig.~\ref{fig:unitary-scaling}, which zooms in on the region $ m_0 \in \rng{-1}{0} $, this decrease can be of several orders of magnitude for poorly chosen trial solutions.  The influence of the annealing time is less pronounced close to $ m_0 = 1 $, and more evident for intermediate and lower values of $ m_0 $. This is consistent with the adiabatic theorem, since a longer anneal time guarantees that the system will have a higher probability of remaining close to the initial eigenstate it has the largest overlap with (not necessarily the ground state)~\cite{Jansen:07}. 

Adopting a conventional quantum annealing procedure, the success probability in the case of $ \tf = \SI{100}{\nano\second} $ would be $ P_0 = 0.96 $. This value is larger than any $ P_0 $ achievable using reverse annealing. However, this argument cannot be used to discredit reverse annealing for two main reasons. First, it is clear that, in the analyzed case, we are very close to adiabaticity, where conventional annealing is efficient. Second, as clarified in the next Sections, the role of dissipation may strongly affect this scenario.
	
The results of this Section are in agreement with those reported in Ref.~\cite{nishimori:reverse-pspin-2}. Namely, as is clear from Fig.~\ref{fig:unitary-scaling}, upon iteration the IRA protocol will only decrease the success probability under unitary, closed system dynamics, unless the initial state was already chosen as the solution of the optimization problem.
	
\begin{figure*}[t]
	\subfigure[]{\includegraphics[width = 0.48 \columnwidth]{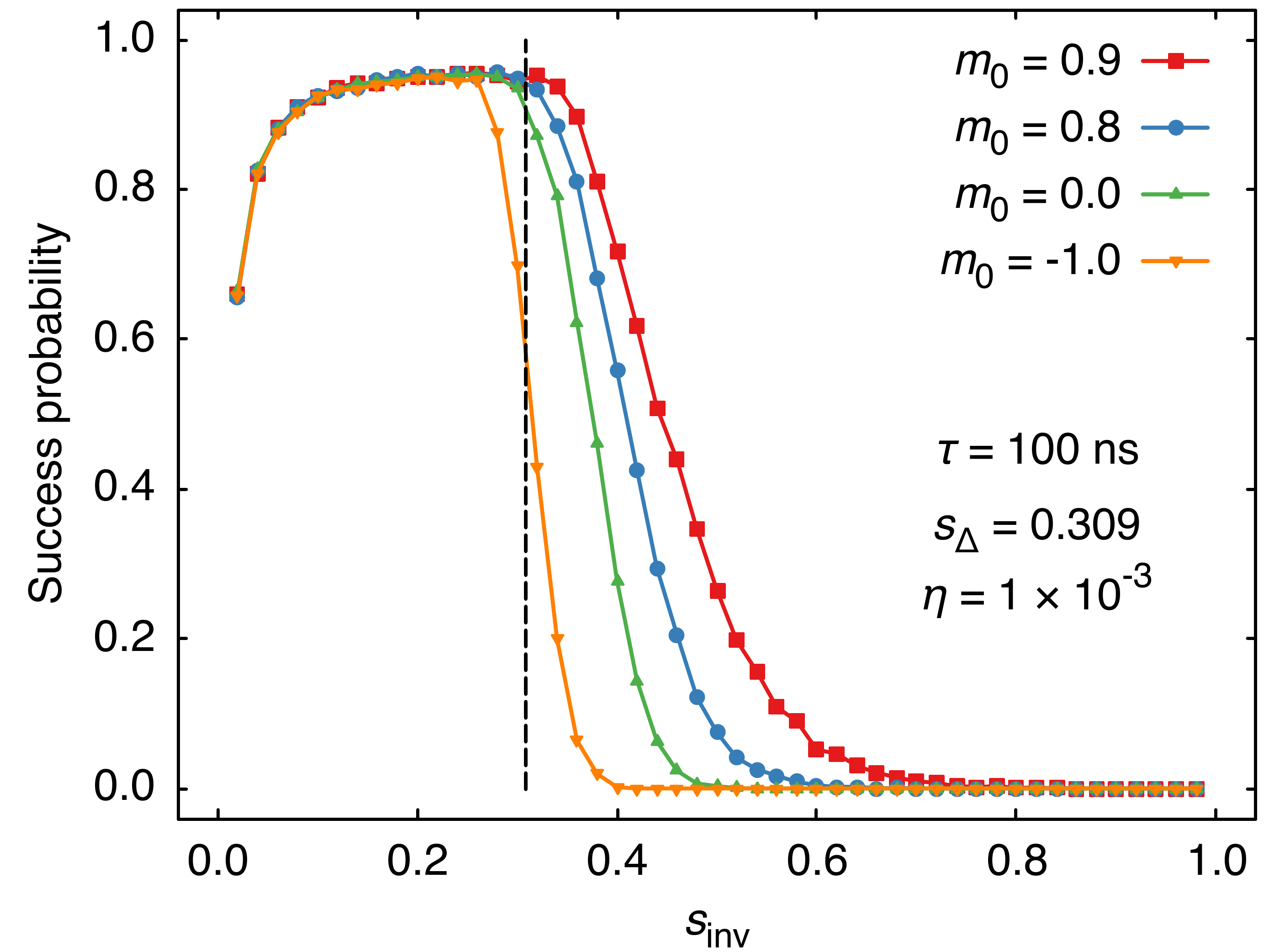}\label{fig:dissipative-tf-100}}
	\subfigure[]{\includegraphics[width = 0.48 \columnwidth]{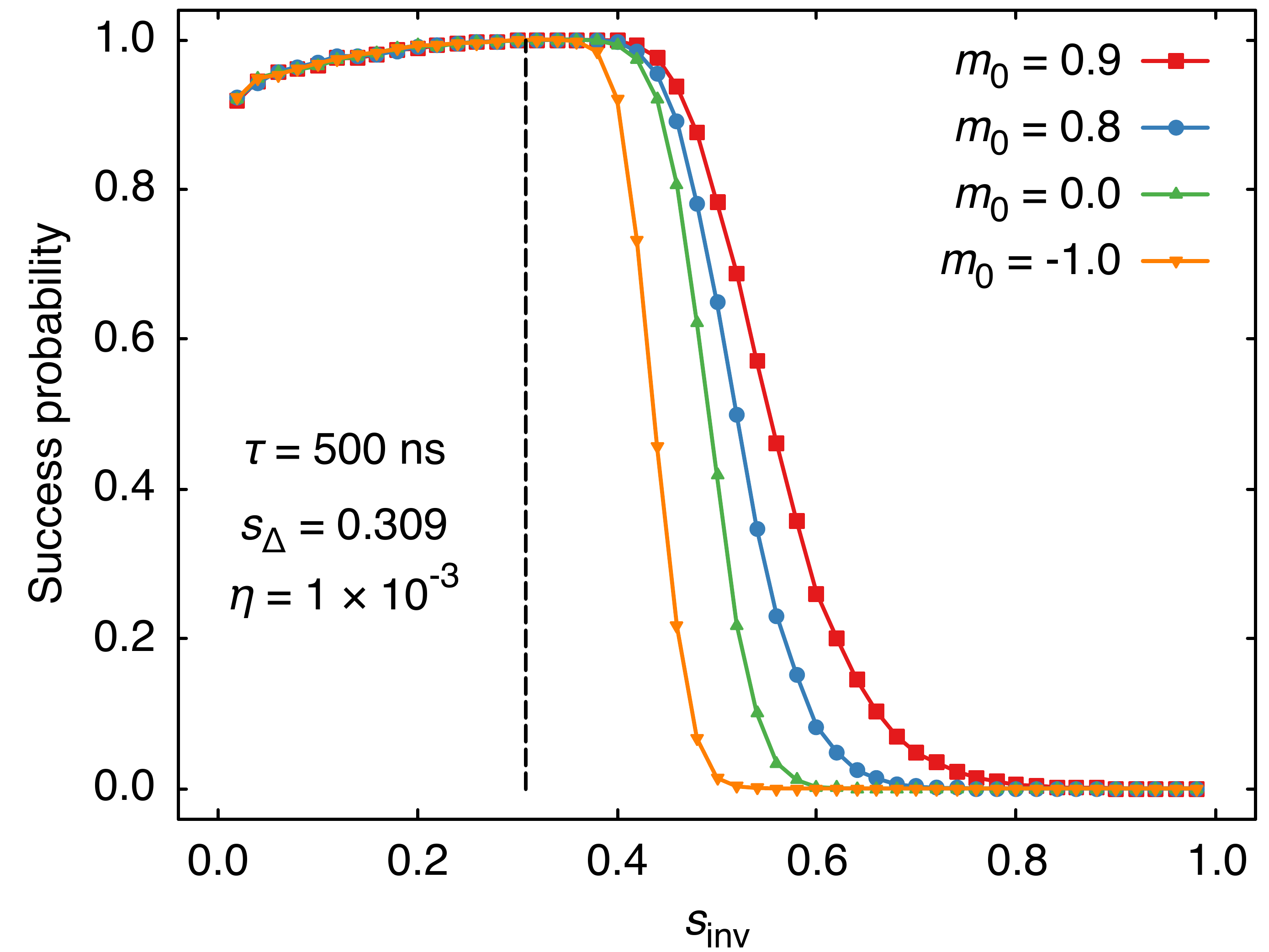}\label{fig:dissipative-tf-500}}
	\caption{Success probability in reverse annealing as a function of the inversion point $ \sinv $, for several values of the magnetization of the initial state. The $ p $-spin system is coupled to a collective dephasing bosonic environment as in Eq.~\eqref{eq:system-bath-collective}, and the coupling strength is $ \eta = \num{1e-3} $. The dashed vertical line denotes the time $ \sgap $ of the avoided crossing between the ground state and the first excited state. In (a) the annealing time is $ \tf = \SI{100}{\nano\second} $, in (b) $ \tf = \SI{500}{\nano\second} $. We sampled the interval $ \sinv \in \rngopen{0}{1} $ using a step $ \Delta s = 0.02 $. All other parameters are given in the main text.}
	\label{fig:dissipative}
\end{figure*}

\section{Open system dynamics subject to dephasing-induced relaxation}\label{sec:dissipative}
	
Physical quantum processors always interact with the surrounding environment, which induces decoherence and thermal excitation/relaxation, which in turn impacts the performance of quantum annealing~\cite{childs:robustness,amin_decoherence_2009,Albash:2015nx}. We 
assume weak coupling between the qubit system and the environment. 
It can then be shown that the reduced system density matrix evolves according to a quantum master equation in time-dependent Lindblad form~\cite{zanardi:master-equations}, known as the adiabatic master equation:
\begin{equation}\label{eq:lindblad}
	\frac{d\rho(t)}{dt} = \iu \bigl[\rho(t), \ham(t) + \ham\ped{LS}(t)\bigr] + \diss\bigl[\rho(t)\bigr].
\end{equation} 
In Eq.~\eqref{eq:lindblad}, $ \ham\ped{LS}(t) $ is a Lamb shift term and $ \diss $ is the dissipator superoperator, which makes the dynamics non-unitary and irreversible. They are expressed in terms of Lindblad operators, inducing dephasing or quantum jumps (pumps and decays) between pairs of adiabatic energy eigenstates~\cite{yip:mcwf}:
\begin{align}
	\ham\ped{LS} &= \sum_{ab} L_{ab}^\dagger(t) L_{ab}(t) S_{ab}(\omega_{ab});\\
	\diss[\rho(t)] &= \sum_{ab} \gamma_{ab} \biggl( L_{ab}(t) \rho(t) L_{ab}^\dagger(t) \notag\\
	&\quad- \frac{1}{2} \left\lbrace L_{ab}^\dagger(t) L_{ab}(t), \rho(t) \right\rbrace \biggr).
\end{align}
The Lindblad operators are determined by the instantaneous eigenbasis of the system Hamiltonian $H(t)$ and the system-bath interaction Hamiltonian $ \hamsysbath $. In general, this coupling Hamiltonian involves local operators
that break the spin symmetry of the $ p $-spin model, as each qubit is then coupled to its own bath. We study both this independent decoherence model and the collective decoherence model, wherein  
all the qubits are coupled to a collective bath with the same coupling energy $ g $, in order to preserve the spin symmetry. More specifically, we first consider collective dephasing, for which the system-bath coupling Hamiltonian is
\begin{equation}\label{eq:system-bath-collective}
	\hamsysbath\api{col} = g S_z 
	\otimes B ,
\end{equation}
where $B$ is a bath operator [e.g., $B = \sum_k (a_k + a_k^\dagger)$ for an oscillator bath with annihilation operators $a_k$ for the $k$th bosonic mode].
The Lindblad operators are represented in the instantaneous energy eigenbasis of $H(t)$ as~\cite{zanardi:master-equations}:
\begin{equation}\label{eq:Lab(t)}
	L_{ab}(t) = \braket{E_a(t)| S_z |E_b(t)} \ket{E_a(t)}\bra{E_b(t)}.
\end{equation}
This represents collective dephasing in the energy eigenbasis, wherein the dephasing process randomizes the relative phase between eigenstates of the system Hamiltonian. Thus, this model does not support phase coherence between energy eigenstates.\footnote{It is worth pointing out a caveat. Namely, the collective dephasing model in general supports decoherence free subspaces (DFSs), \ie, subspaces that evolve unitarily despite the coupling to the bath~\cite{Zanardi:97c,Lidar:1998fk}. For instance, the $ S = 0 $ subspace (for even $ \nspin $) is a DFS of the $ p $-spin model. However, the $p$-spin model is unsuitable for performing quantum annealing inside a DFS, since its Hamiltonian consists of operators that preserve the DFS, so that no dynamics would take place if we were to try to encode a computation using states inside the DFS. Instead, to obtain meaningful dynamics (performing a computation) subject to the collective dephasing model, we would need to add Heisenberg exchange terms to the system Hamiltonian~\cite{Kempe:2001uq}.} As a consequence, thermal relaxation tends to equilibrate the system towards its Gibbs state, with a characteristic timescale set by the inverse of the bath spectral density at the gap frequency~\cite{Albash:2015nx}.
	
If instead the qubit system is coupled to independent, identical baths, the system-bath coupling operator becomes
\begin{equation}\label{eq:system-bath-independent}
	\hamsysbath\api{ind} = g \sum_{i} \sigma_i^z \otimes B_i, 
\end{equation}
where, e.g., in the bosonic case $B_i = \sum_k (a_{k, i} + a_{k, i}^\dagger)$. Thermal relaxation effects occur here similarly to the collective dephasing case.
However, simulations in this case are more demanding due to the fact that the spin symmetry is broken and that we have $\nspin$ times as many Lindblad operators, \ie,
\begin{equation}
	L_{ab, i}(t) = \braket{E_a(t)| \sigma_i^z |E_b(t)} \ket{E_a(t)}\bra{E_b(t)}.
\end{equation}
Therefore, in this case we will only investigate reverse annealing starting from the first excited state in the symmetric subspace with maximum spin, \ie, $ \ket{w = 1} $, for $ \nspin \in \{3,\dots,8\}$. Moreover, for the particular cases of $ \nspin = 7 $ and $ \nspin = 8 $, we truncate our system to the lowest $ 29 $ and $ 37 $ eigenstates, respectively, to speed up the numerics. This choice is made since the first three levels of the maximum spin subspace at $ s = 1 $ are spanned by $ \sum_{i=0}^2 \binom{7}{i} = 29 $ (for $ \nspin = 7 $) and $ \sum_{i=0}^{2} \binom{8}{i} = 37 $ (for $ \nspin = 8 $) energy eigenstates. We confirm that this is a good approximation by checking that the total population among these levels is close to $1$ during the reverse annealing when additional levels are included in the simulation.
	
\begin{figure}[t]
	\centering
	\includegraphics[width = 0.9 \linewidth]{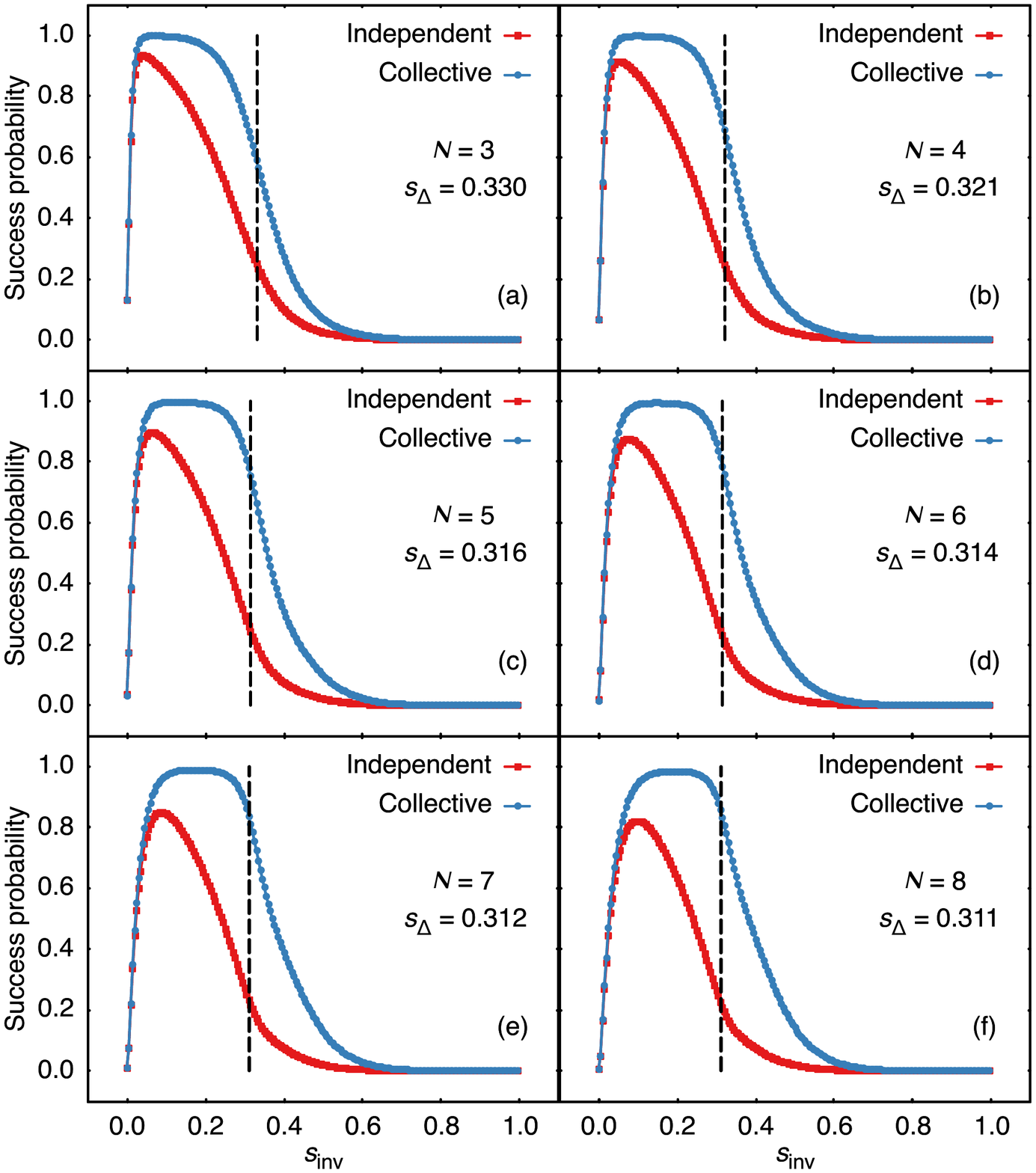}
	\caption{Success probability in reverse annealing as a function of the inversion point $ \sinv $, for $\nspin \in \{3,\dots,8\}$. The initial state is the first excited state of the maximum spin subspace ($ m_0 = 1 - 2/\nspin $). The dashed vertical line denotes the time $ \sgap $ of the avoided crossing between the ground state and the first excited state. The annealing time is $ \tf = \SI{100}{\nano\second} $. We sampled the interval $ \sinv \in (0, 1) $ using a step size of $ \Delta s = 0.005 $.}
	\label{fig:withoutpausing}
\end{figure}

\begin{figure}[h!]
	\centering
	\includegraphics[width=0.5\linewidth]{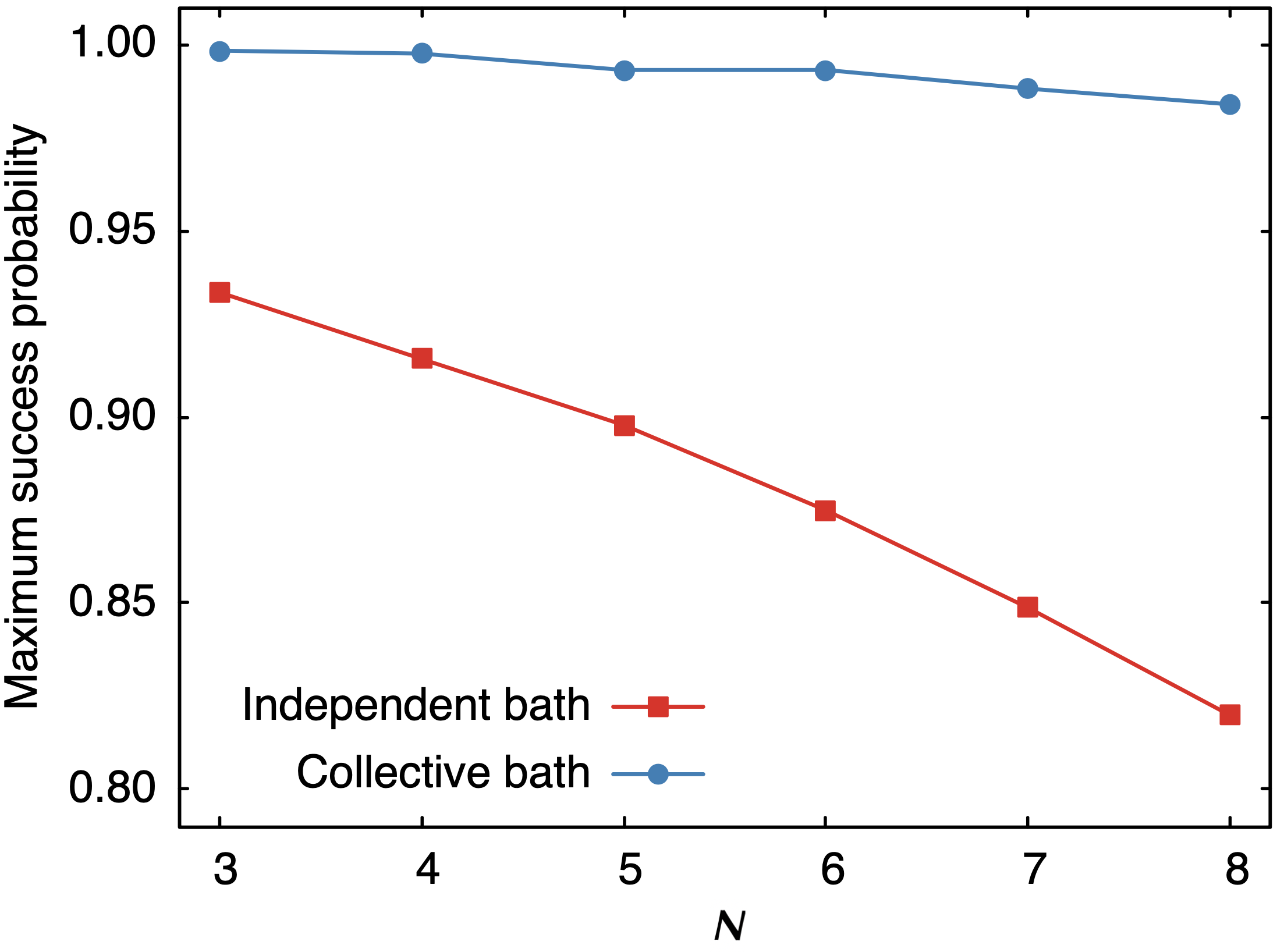}
	\caption{Maximum success probability achievable with reverse annealing as a function of the number of qubits $ \nspin $, using the collective and the independent dephasing models of Eqs.~\eqref{eq:system-bath-collective} and~\eqref{eq:system-bath-independent}, respectively. The annealing time is $ \tf = \SI{100}{\nano\second} $.}
	\label{fig:fiddependenceonn}
\end{figure}

\begin{figure*}[h!]
	\subfigure[]{\includegraphics[width = 0.49 \columnwidth]{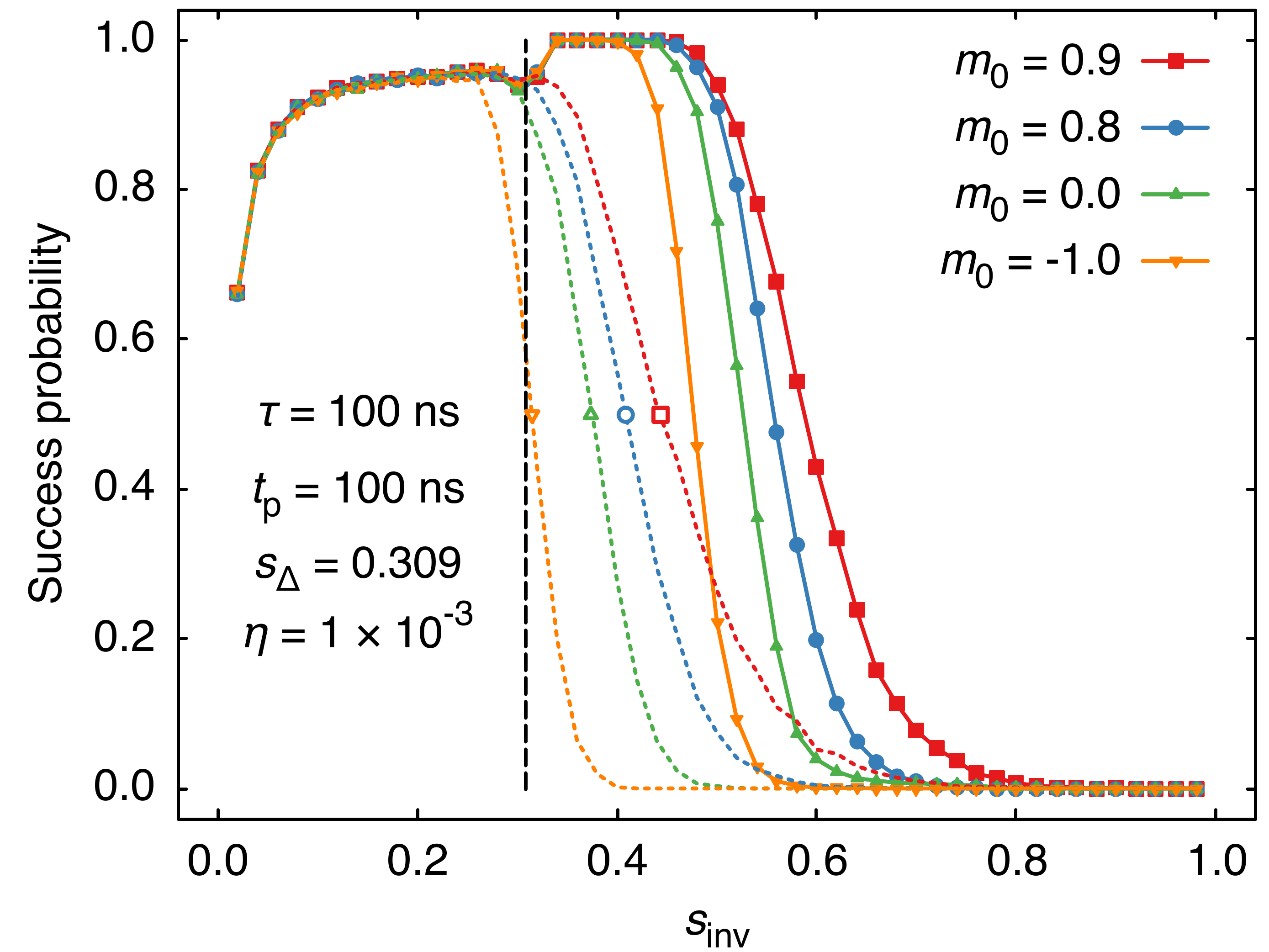}	\label{fig:dissipative-tf-100-pause-100}}
	\subfigure[]{\includegraphics[width = 0.49 \columnwidth]{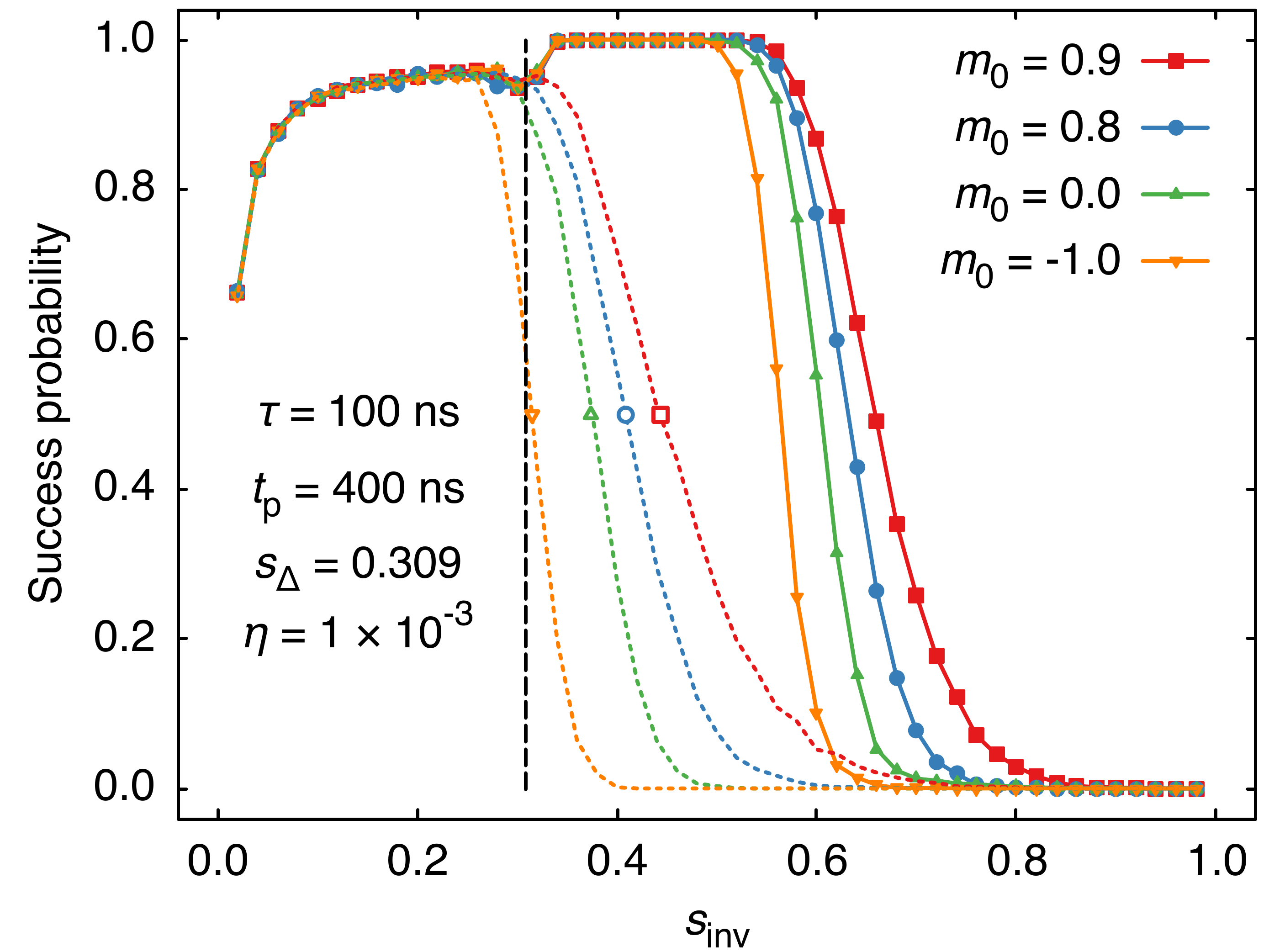}\label{fig:dissipative-tf-100-pause-400}}
	\caption{Success probability in paused reverse annealing for the collective dephasing model, as a function of the inversion point $ \sinv $, for several values of the magnetization of the initial state. The coupling strength is $ \eta = \num{1e-3} $. The annealing time is $ \tf = \SI{100}{\nano\second} $. In (a) a pause of duration $ \lpause = \SI{100}{\nano\second} $ is inserted at the inversion point, while in (b) $ \lpause = \SI{400}{\nano\second} $. All other parameters are given in the main text. Dotted lines with empty symbols refer to open system reverse annealing of time $ \tf = \SI{100}{\nano\second} $ and no pauses. The dashed vertical line denotes $ \sinv = \sgap $. The interval $ \sinv \in \rngopen{0}{1} $ is sampled using a step size $ \Delta s = 0.02 $.}
	\label{fig:dissipative-tf-100-pause}
\end{figure*}		

\begin{figure}[tb]
	\includegraphics[width = 0.9 \linewidth]{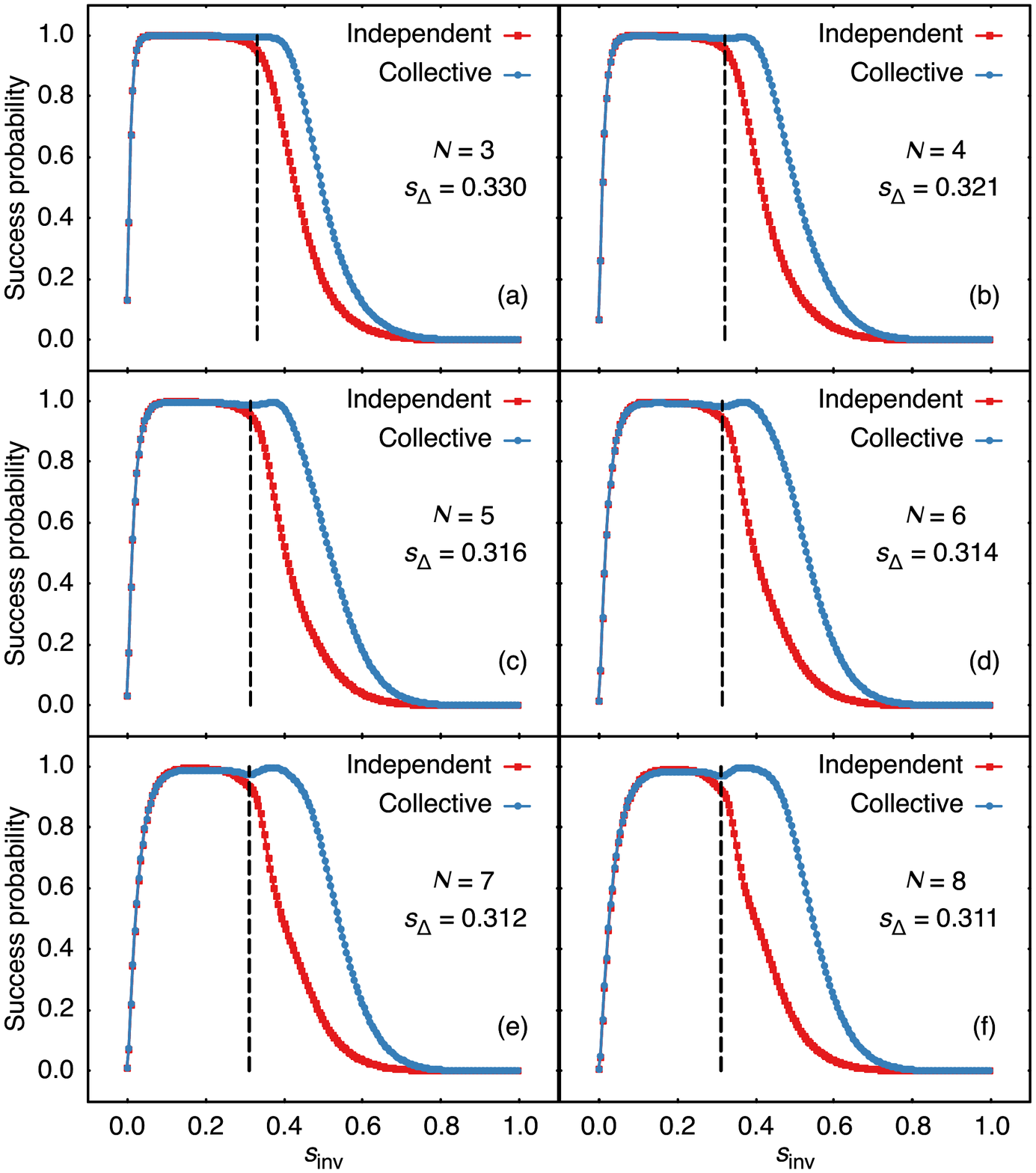}
	\caption{Comparison of the success probability in reverse annealing for the collective and independent dephasing models, as a function of the inversion point $ \sinv $, for $\nspin \in\{3,\dots,8\}$. The initial state is the first excited state of the maximum spin subspace ($m_0 = 1 - 2/\nspin$). The dashed vertical line denotes the time $\sgap$ of the avoided crossing between the ground state and the first excited state. The annealing time is $\tf = \SI{100}{\nano\second}$, and a pause of duration $ \lpause = \SI{100}{\nano\second}$ is inserted at the inversion point.}
	\label{fig:withpausing}
\end{figure}

The adiabatic master equation in Eq.~\eqref{eq:lindblad} is unraveled using a time-dependent Monte Carlo wavefunction (MCWF) approach~\cite{yip:mcwf}. The advantage of MCWF is that it allows to work with wavefunctions rather than density matrices, thus saving quadratically in the dimension of the objects we need to store for numerical calculations. The tradeoff is that to recover the statistical properties of the density operator we need to average over a large number $ K $ of independent trajectories. For the collective system-bath coupling of Eq.~\eqref{eq:system-bath-collective}, the time evolution operator of each trajectory is generated by the effective non-Hermitian Hamiltonian
\begin{align}\label{eq:non-hermitian-hamiltonian}
	\ham\ped{eff}(t) &= \ham(t) + \ham\ped{LS}(t) - \frac{\iu}{2} \sum_{a \ne b} \gamma_{ab} L_{ab}^\dagger(t) L_{ab}(t) \notag \\ & \quad-\gamma_0 \frac{\iu}{2} \sum_{ab}  L_{aa}^\dagger(t) L_{bb}(t),
\end{align}
where $ \gamma_{ab} $ and $ \gamma_0 $ are the rates for jumps and dephasing, respectively. They are related to the temperature and to the spectral density of the bosonic bath,
\begin{equation}\label{eq:spectral-density}
	J(\omega) = g^2 \sum_k \delta(\omega -  \omega_k) = 2\pi \eta \omega e^{-\omega / \omegac},
\end{equation}
where $ \omega_k $ are the bath eigenfrequencies, $ \omegac $ is a high-frequency cutoff and $ \eta $ is the dimensionless coupling strength. We fix $ \omegac = \SI{1}{\tera\hertz} $ and $ \eta = \num{1e-3} $. The working temperature is chosen to be $ T = \SI{12.1}{\milli\kelvin} = \SI{1.57}{\giga\hertz} $, as in experimental quantum annealing systems~\cite{dwave-site}. Eq.~\eqref{eq:non-hermitian-hamiltonian} is easily extended to the independent dephasing case by including a summation over $ i $ for the Lindblad operators and in the Lamb shift term.
	
The time evolution operator generated by the non-Hermitian Hamiltonian of Eq.~\eqref{eq:non-hermitian-hamiltonian} is not unitary. Therefore, the norm of the wavefunction decays in time. Whenever the squared norm decreases below a randomly extracted threshold $ r \in \rng{0}{1} $, a quantum jump occurs, projecting the wavefunction on one of the eigenstates of the Hamiltonian $ \ham(t) $.
The adiabatic master equation dynamics is found after averaging over all stochastic trajectories. In this work, we fix $ K = 5000 $ trajectories and consequently find relative Monte Carlo errors $ \delta\ev{O} / \ev{O} $ of the order of $ \SI{1.5}{\percent} $ over all observables $ O $.
	
We repeat the simulations we reported in Sec.~\ref{sec:unitary} for $ \nspin = 20 $, $ p = 3 $ and $ \tf = \SI{100}{\nano\second} $, but now include the role of the environment. We consider both the collective and independent dephasing models. 
	
In Fig.~\ref{fig:dissipative-tf-100}, we show the success probability as a function of the inversion time $ \sinv $, for the four initial magnetizations $ m_0 = \text{\numlist{0.9;0.8;0;-1}} $. Monte Carlo errors are of the order of the point size in all cases and are invisible. 
	
As in the unitary case, if the inversion occurs too early (\ie, for $ \tinv \ll \tgap $, or, equivalently, $ \sinv \gg \sgap $), the reverse annealing protocol fails to find the ferromagnetic ground state. In fact, thermal excitations are suppressed, as well as Landau-Zener transitions, due to the large level spacing, compared with the temperature and the inverse of the annealing time. For $ \tinv \approx \tgap $ ($ \sinv \approx \sgap $), however, the scenario is drastically different from the unitary case of Fig.~\ref{fig:unitary-tf-100}.

The first difference is that the success probability can be nonzero even if the inversion occurs for $ \sinv \gtrsim \sgap $, especially for $ m_0 = 0.9 $, where the tail of the curve extends to $ \sinv \approx 0.75 $. When the instantaneous gap is of the same order of magnitude as the temperature, thermal processes influence reverse annealing even before crossing the minimal gap. Second, for all $ m_0 $ we observe a sudden increase in the success probability around $ \sinv \approx \sgap $, that eventually brings all curves to an almost flat region at $ \sinv < \sgap $, where the success probability reaches the large value $ \pgs \approx 0.957 $. The value of the maximum success probability at the plateau is $ m_0 $-independent within Monte Carlo errors. The time at which the success probability starts to increase with respect to the baseline depends on $ m_0 $. Moreover, the flat region is wider for larger $ m_0 $, although it has a finite width for all $ m_0 $.
	
These results show that even trial solutions far in Hamming distance from the ferromagnetic ground state can result in a large success probability at the end of a reverse anneal. Moreover, the time window in which inverting the annealing favors the ferromagnetic ordering is relatively large. 
	
We also studied a longer annealing time, $ \tf = \SI{500}{\nano\second} $, as shown in Fig.~\ref{fig:dissipative-tf-500}.
Here, we note that the onset of the success probability plateau shifts towards longer values of $ \sinv $, compared with the $ \tf = \SI{100}{\nano\second} $ case. Therefore, the plateau is wider, and the maximum success probability at the plateau is $ \pgs \approx 1 $ within Monte Carlo errors for all $ m_0 $ we considered. This is in contrast with the unitary case of Fig.~\ref{fig:unitary-scaling}, where increasing the annealing time had detrimental effects on the algorithm. This evidence supports the notion that the success probability enhancement is due to thermal effects, rather than due to purely unitary quantum dynamics~\cite{passarelli:pspin}. Moreover, the adiabatic theorem for open quantum system guarantees convergence to the steady state of the superoperator generator of the dynamics in the large $\tf$ limit~\cite{Venuti:2015kq,Venuti:2018aa}. This too helps to explain our observations: the steady state of the Davies-Lindblad generator of the open system dynamics we considered here is the Gibbs distribution of the final Hamiltonian, which at sufficiently low temperature relative to the gap is the ferromagnetic ground state. Recall that in our case $ \mingap \approx \SI{2.45}{\giga\hertz} $ (at $ \sgap \approx 0.309 $) and $ T = \SI{1.57}{\giga\hertz} $.

Comparing these results with conventional forward annealing in the presence of dissipation, we note that the success probability at the plateau is similar to that of standard quantum annealing for $ \eta = \num{1e-3} $ ($ P_0 = 0.98 $). The reason is that collective dephasing favors the ferromagnetic ordering in the $p$-spin model and its induced relaxation increases the success probability compared to the isolated case, in agreement with previous findings~\cite{passarelli:pspin}. We also observe that when the annealing is far from adiabaticity (e.g. $\tf = \SI{1}{\nano\second}$), reverse annealing becomes a more favorable approach to forward annealing for the $p$-spin model.
				
We also compare the collective and independent dephasing models of Eqs.~\eqref{eq:system-bath-collective} and~\eqref{eq:system-bath-independent}.
Fig.~\ref{fig:withoutpausing} shows the simulation results for the two models using the adiabatic master equation of Eq.~\eqref{eq:lindblad} for $ \nspin \in \{3,\dots,8\} $. As shown in the figure, simulations using the collective dephasing model have larger success probabilities for almost every $ \sinv$. This is because, in the independent dephasing model, other states not in the subspace of maximum spin become accessible by thermal excitation or diabatic transition during the reverse anneal. For all of the system sizes we simulated, we had to reverse anneal to a smaller inversion point $ \sinv $ for the independent dephasing model to achieve the same success probability as the collective dephasing model. Moreover, the maximum success probability achievable is always smaller for the independent dephasing model. The success probabilities from both models, however, are very similar as $ \sinv \to 0 $, \ie, in the quench limit of the direct part of the evolution.

Figure~\ref{fig:fiddependenceonn} shows how the maximum success probability (over $ \sinv $) of both bath models depends on the number of qubits. As $ \nspin $ increases, the maximum success probability of the independent dephasing model decreases more rapidly than that of the collective dephasing model. While we can infer that if we modeled independent dephasing for $ \nspin = 20 $ we would not observe as large success probabilities as in Fig.~\ref{fig:dissipative}, we stress that  reverse annealing in the independent dephasing model still yields a significantly larger success probability  (for the same $\nspin$ values) than the unitary dynamics case described in Section~\ref{sec:unitary}.

\section{Open system dynamics with a pause}\label{sec:pausing}

Quantum annealing in the presence of a low temperature bath can benefit from pauses inserted at certain times during the dynamics~\cite{marshall, passarelli:pausing}. During a pause, $ s(t) = \text{constant} $, and the system evolves with a time independent Hamiltonian, subject to dephasing. When a pause is inserted some time after $ \sgap $, the environment favors a redistribution of the repopulation according to the Gibbs state at the pause point; at sufficiently low temperature (relative to the gap at this point), this can result in a repopulation of the instantaneous ground state. In this section, we show that pauses at the inversion point can further improve the performance of reverse annealing of the $ p $-spin model.

We repeat the simulations for $ \nspin = 20 $, $ p = 3 $ and $ \tf = \SI{100}{\nano\second} $, using the collective dephasing model. A pause of duration $ \lpause = \tf $ is inserted at $ t = \tinv $, so that the total annealing time, including the pause, is $ \tf' = \tf + \lpause = \SI{200}{\nano\second} $. 
	
In Fig.~\ref{fig:dissipative-tf-100-pause-100}, we report the success probability as a function of the inversion point, for starting magnetizations $ m_0 = \text{\numlist{0.9;0.8;0;-1}} $. We compare the paused case with the unpaused case, for which $ \tf = \SI{100}{\nano\second} $. 
As can be seen in the figure, if the dynamics is reversed too early ($ \sinv \gg \sgap $), the success probability at the end of the anneal vanishes. The level spacing is large compared with the temperature. The relaxation rate is small and the pause is too short to have impact on the dynamics. 
	
However, the presence of a pause significantly changes the outcome of the annealing around $ \sinv \approx \sgap $.  In fact, when a pause is inserted at $ \sinv \gtrsim \sgap $, the success probability reaches $ \pgs \approx 1 $ for a wide range of inversion points and for all $ m_0 $, within Monte Carlo errors. Here, the ground state is completely repopulated by thermal relaxation. This is in contrast with conventional quantum annealing, where the success probability exhibits a peak as a function of the pausing time, when the pause is inserted about $ \SI{20}{\percent} $ later than $ \sgap $, and then rapidly returns to its baseline value~\cite{marshall, passarelli:pausing}.
In contrast, for $ \sinv < \sgap $, the effect of the pause is negligible; the solid (with pause) and dotted (no pause) lines in Fig.~\ref{fig:dissipative-tf-100-pause-100} overlap in this region.

We repeated our analysis for a pause duration $ \lpause = \SI{400}{\nano\second} $, with total annealing time $ \tf' = \SI{500}{\nano\second} $. As shown in Fig.~\ref{fig:dissipative-tf-100-pause-400}, the longer pause duration affects the results only marginally. Comparing with Fig.~\ref{fig:dissipative-tf-100-pause-100}, we note that the qualitative behavior of the curves is the same in the two cases. The pause duration affects mostly the region $ \sinv \gtrsim \sgap $. A longer pause enhances thermal relaxation, thus the success probability starts to increase from its baseline earlier than for shorter $ \lpause $. This results in a wider plateau where the success probability is large, compared with Fig.~\ref{fig:dissipative-tf-100-pause-100}.
	
Finally, we compare the collective and independent dephasing models while including pausing,
starting from the first excited state of the maximal spin sector.
The results are shown in Fig.~\ref{fig:withpausing}, for a pause of duration $ \lpause = \SI{100}{\nano\second} $ inserted at the inversion point. The collective dephasing model continues to exhibit higher success probabilities than the independent dephasing model, as in the case discussed in the previous section, but the results of the two models coincide when $ \sinv < \sgap $. 
Thus, relaxation to the ground state during the pause improves performance for both dephasing models. 
Note that, as $ \nspin $ increases, the maximum success probability of the collective dephasing model is achieved at $ \sinv > \sgap $, while it is achieved at $ \sinv < \sgap $ in the independent dephasing model. This is in agreement with the $ \nspin = 20 $ result shown in Fig.~\ref{fig:dissipative-tf-100}.

\section{Conclusion of this chapter}
\label{sec:conclusions}

Earlier work revealed an intriguing tension between experimental results demonstrating a substantial enhancement in success probabilities for random spin glass instances under reverse annealing compared to standard (forward) annealing~\cite{marshall}, and theoretical results finding that reverse annealing adversely affects performance for the $p$-spin model, in a closed system setting~\cite{nishimori:reverse-pspin-2}. In this work we resolved this tension by performing a numerical study of reverse annealing of the $p$-spin model in an open system setting, where we included dephasing in the instantaneous energy eigenbasis. We found that the associated thermal relaxation results in significant increase in the success probabilities, as long as the inversion point of the reverse annealing protocol is chosen to be close to the avoided crossing point, or before it. Pausing at the inversion point further improves performance. 

Since closed-system, unitary dynamics predicts that reverse annealing fails, yet its open system analogue succeeds, it follows that thermal relaxation is the mechanism responsible for the success. Reverse annealing is thus an example of a family of protocols that strictly benefit from thermal effects~\cite{verstraete2009quantum,Venuti:2017aa}.
It may be worth noting that quantum effects are likely to play an important role in thermal relaxation because the success probability is very small for $s_{\rm inv}$ close to $1$, i.e. when the Hamiltonian stays classical during the anneal, even with a pause.
Whether this can lead to any quantum speedups is an interesting problem worthy of future investigations.

\section{Acknowledgments of this chapter}
This chapter was originally published in~\cite{Passarelli2019}. The simulation data of collective coupling model is provided by Gianluca Passarelli.
The research of KY, DL, and HN is based upon work (partially) supported by the Office of
the Director of National Intelligence (ODNI), Intelligence Advanced
Research Projects Activity (IARPA), via the U.S. Army Research Office
contract W911NF-17-C-0050. The views and conclusions contained herein are
those of the authors and should not be interpreted as necessarily
representing the official policies or endorsements, either expressed or
implied, of the ODNI, IARPA, or the U.S. Government. The U.S. Government
is authorized to reproduce and distribute reprints for Governmental
purposes notwithstanding any copyright annotation thereon. Computation for some of the work described in this chapter was supported by the University of Southern California Center for High-Performance Computing and Communications (hpcc.usc.edu).

%% file: chapter5.tex
\chapter{Experimental iterative reverse annealing}
\label{chap: IRA_exp}
In this chapter, we continue to study the open-system effects on iterative reverse annealing. We perform the IRA simulations with more realistic experimental settings of D-Wave annealers, for $p$-spin model with $p=2$. We explore both quantum and semiclassical simulation methods, and also compare our simulation results to the experiments.
\section{Introduction}
Iterated reverse annealing (IRA) is a variant of quantum annealing, in which the system is prepared in a trial solution state, reverse-annealed to an inversion point, and then forward-annealed. It may also be iterated with the last output state as the input of the next iteration.
We perform experiments of reverse annealing on the D-Wave 2000Q device, with a focus on the $p$-spin problem with $p=2$. We examine the dependence of the performance on parameters such as problem size, annealing time and inversion points, as well as pausing and the number of iterations. The performance is evaluated in terms of total and partial success probabilities, the latter being the probabilities for each of two degenerate ground states. To explain the experimental results, we perform open-system simulations with decoherence in the instantaneous energy eigenbasis. The experimental total success probabilities  match the simulation data well. We find a difference between the independent and collective dephasing models, which is explained by non-conservation of spin symmetry of classical inputs. Simulations show that iterated reverse annealing fails in a closed system, suggesting the importance of thermal relaxation in the convergence to the correct solution. We also perform semi-classical simulations of spin-vector Monte Carlo to supplement  our understanding of the experimental data.


Recently, D-Wave devices that actually realize quantum annealing on a hardware have been developed. Not only can the D-Wave device be used to solve optimization problems by following quantum annealing  physically, but it is also able to perform physics experiments, for example, spin-glass phase transitions~\cite{Harris2018}, the Kosterlitz-Thouless phase transition~\cite{King:2018aa, King:2019aa}, the Kibble-Zurek mechanism~\cite{Gardas2018}, the Griffiths-McCoy singularity \cite{Nishimura2020}, spin ice \cite{King2020}, and the Shastry-Southernland model \cite{Kairys2020}.

As in the case of simulated annealing~\cite{kirkpatrick:sa}, which is a classical counterpart of quantum annealing, one of the simple ways to enhance the performance of quantum annealing is to control the ``annealing schedule''. For conventional quantum annealing~\cite{kadowaki:qa, Santoro}, we choose the ground state of the transverse field term as the initial state and define forward annealing as the temporal change of the system from the transverse field Hamiltonian to the target Hamiltonian. Traditionally, this transition process is supposed to be adiabatic. Diabatic transition processes have also been proposed to overcome the difficulties of adiabatic computing, see e.g., ~\cite{Brady2020,Mbeng2019,Crosson2020}. On the other hand, reverse annealing is defined as the process of choosing an appropriate classical state as the initial state, starting from the target Hamiltonian and mixing it with the transverse field Hamiltonian to return to the original target Hamiltonian~\cite{perdomo:sombrero, nishimori:reverse-pspin, nishimori:reverse-pspin-2}. \par

In chapter.~\ref{chap:ira} (Ref.~\cite{Passarelli2019}) we tested the effectiveness of reverse annealing on the $p$-spin models with $p=3$. Numerical calculations using the time-dependent Lindblad equation~\cite{ABLZ:12-SI, yip:mcwf} showed that the presence of coupling with the boson field significantly changes the performance of reverse annealing. This result shows that the interaction with the environment can lead to an improvement in performance. Since qubits and the environment should always interact with each other on the hardware, it is likely that this result reflects real physical processes in the D-Wave device. However, it is non-trivial if the coupling model  to the environment of chapter.~\ref{chap:ira} is good enough to explain experimental data. 

In chapter.~\ref{chap:ira},  the $p$-spin model with $p=3$  is discussed because it has a first-order phase transition in the thermodynamic limit~\cite{derrida:p-spin, Gross1984, Bapst2012} and thus is a difficult problem for simple quantum annealing. 
In the present chapter, we instead choose the $p$-spin model with $p=2$. This is because the $p$-spin model with $p=2$ can be more easily embedded in the hardware graph and thus may avoid possible errors by minor embedding of three-body interactions, and also larger system sizes can be implemented up to about 40. Another reason is the existence of ground-state degeneracy, which leads to additional information as will be described below. Our goal is not to study if a high success probability can be achieved but to understand the mechanism of physical processes working in the device.
\par

In this chapter, we perform reverse annealing on the $p$-spin model with $p=2$ experimentally and numerically. We compare the success probability with numerical results to verify how the performance of reverse annealing on D-Wave 2000Q is enhanced by  coupling to baths of boson field.\par

The structure of this chapter is as follows.  We first define the $p$-spin model with $p=2$ in Sec.~\ref{sec:problem} and describe the sampling details in this Chapter. In Sec.~\ref{sec:exp}, we discuss the sampling results from the D-Wave device. We show and examine the simulation results based on closed system Schr\"{o}dinger equation and quantum adiabatic master equation with two system-bath coupling models in Sec.~\ref{sec:simulations}. In Sec.~\ref{sec:classical}, we perform semi-classical simulations of spin-vector Monte Carlo to further understand the experimental data. The chapter is concluded in Sec.~\ref{section:Conc}.

\section{\label{sec:problem}Problem definition and protocols of experiment}
In this section we describe the problem Hamiltonian and the protocol of reverse annealing we used both in our experiments on the D-Wave device and in numerical simulations.
\subsection{\label{sec:p-spin}$p$-spin model with $p=2$}
\begin{figure}[h!]
\centering
\includegraphics[width=0.85\columnwidth]{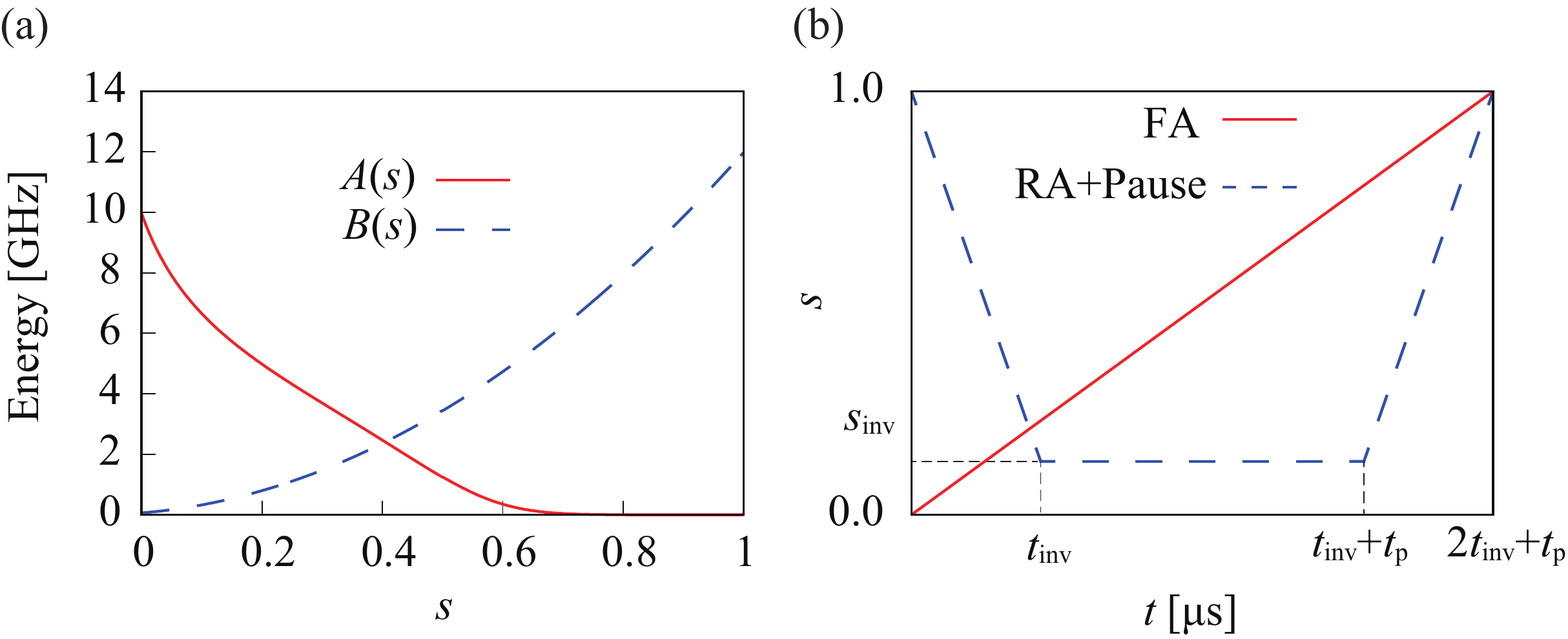}%
\caption{(a) Annealing schedule of D-Wave 2000Q with the ``DW\_2000Q\_6`` solver. (b) Forward (FA) and reverse annealing (RA) protocols are defined in Eq.~\eqref{eq:FA}) and Eq.~\eqref{eq:RA}), respectively. The latter incorporates a pause of duration $t_p$. The parameter values for this plot are $t_a=1.2~\mathrm{[\mu s]}$ for forward annealing, and $\tau=1~\mathrm{[\mu s]}$ and $s_\mathrm{{inv}}=0.2$ for reverse annealing. These symbols are defined below Eq.~\eqref{eq:RA}.}
\label{fig:schedule}
\end{figure}

The total Hamiltonian of quantum annealing is composed of the driver Hamiltonian $H_{D}$ and the target Hamiltonian $H_{0}$, the latter encoding the combinatorial optimization problem represented as an Ising model, as a function of dimensionless time $0\le s(t)\le 1$,
\begin{equation}
\label{eq:H}
H(s) = \frac{A(s)}{2} H_{D} + \frac{B(s)}{2}H_{0}.
\end{equation}
Here, $A(s)$ and $B(s)$ are device-dependent annealing schedules. The D-Wave device used in the present experiment has $A(s)$ and $B(s)$ as shown in Fig.~\ref{fig:schedule}(a).

The final values of the schedule functions are $A(s=1)\approx0$ and $B(s=1)>0$ so that ideally the ground state of $H_{0}$ is realized as the final state, irrespective of the type of annealing protocol, i.e., the time dependence of $s(t)$ as discussed in the next subsection.
The driver Hamiltonian $H_{D}$ is usually chosen as
\begin{equation}\label{eq:QA}
H_{D}=-\sum_{i=1}^{N}\sigma_{i}^{x}
\end{equation}
where $N$ is the number of qubits and $\sigma_{i}^{x}$ is the $x$ component of the Pauli matrix acting on the $i$th qubit. 

We study the $p$-spin model,
\begin{equation}
H_{0}=-N\left(\frac{1}{N}\sum_{i=1}^{N}\sigma_{i}^{z}\right)^p
\end{equation}
with $p=2$, in contrast to $p=3$ used in chapter.~\ref{chap:ira}. The main reason for the choice of $p=2$ is that we can study the effects of ground-state degeneracy in $p=2$ in contrast to the non-degenerate ground state of $p=3$. Another technical reason is that two-body interactions can be directly embedded on the D-Wave device, which would reduce possible minor errors coming from reduction of three-body interactions to two-body representations.

\subsection{Reverse annealing protocol}

The traditional protocol of forward  annealing has
\begin{equation}
\label{eq:FA}
s(t)=\frac{t}{t_{a}}\ \ \ t\in[0,t_{a}],
\end{equation}
where $t_a$ is the total annealing time (see Fig.~\ref{fig:schedule}(b), the red full line). The initial state of forward annealing is the ground state of the driver Hamiltonian $H_D$.

In reverse annealing, we start from $s=1$ and decrease $s$ to an intermediate value $s_{\mathrm{inv}}$, pause, then increase $s$ to finish the process at $s=1$. The explicit time dependence $s(t)$ realized on the D-Wave device is
\begin{equation}
\label{eq:RA}
s(t)=\begin{cases}
1-\,\displaystyle\frac{\,t}{\tau} & 0 \leq t \leq t_{\rm inv} \\
1-\displaystyle\frac{t_{\rm inv}}{\tau} = s_{\mathrm{inv}} & t_{\rm inv}\le t \leq t_{\rm inv}+t_p \\
2s_{\mathrm{inv}} - 1 + \displaystyle\frac{t}{\tau} & t_{\rm inv} +t_p< t \leq 2t_{\rm inv}+t_p,
\end{cases}
\end{equation} 
where $1/\tau$ is the annealing rate,  $t_{\rm inv}$ is the inversion time defined as $t_{\rm inv}=\tau(1-s_{\mathrm{inv}})$, and $t_p$ is the length of an intermediate pause. Our reverse annealing begins from $s(t=0)=1$ and ends at $s(t=t_a)=1$, where
\begin{equation}
t_a = 2t_{\rm{inv}}+t_p = 2 \tau(1-s_{\mathrm{inv}}) + t_p  ,
\label{eq:t_a}
\end{equation}
passing through the inversion point $s_{\mathrm{inv}}$; see the blue dashed line in Fig.~\ref{fig:schedule}(b). To distinguish between $\tau$ and $t_a$, we henceforth refer to the former as the ``annealing time" and the latter as the ``total annealing time". The initial state of reverse annealing is a classical state, usually a candidate solution to the combinatorial optimization problem, i.e, a state that is supposed to be close to the solution.

We can iteratively repeat the process of reverse annealing with the final state of a process fed into the system as the initial condition of the next iteration. We call this protocol iterated reverse annealing and the denote the iteration number as $r$.
In our experiments for single-run reverse annealing, we constructed 15 instances (embeddings on the Chimera graph of the device) and generated 10 random gauges for each set of parameter values, and 1000 annealing runs were repeated for a given random gauge. Thus, the total number of runs for each set of parameter values is 150,000. In iterated reverse annealing, we constructed 15 instances and generated 90 random gauges for each $r$. Thus, the total number of samples at each $r$ and each $s_{\rm inv}$ is 1350.\par
 
Notice that the present protocol of reverse annealing on the D-Wave device is different from the one adopted in Chapter.~\ref{chap:ira} ( Ref.~\cite{Passarelli2019}) for numerical computation. In particular, when $s_{\rm inv}$ is close to 0, the second part to increase $s$ of their protocol has a very sharp increase of $s$ as a function of $t$, which seems to have lead to a significant drop of success probability for $s_{\rm inv}\approx 0$. This is not the case in the present protocol, where the time derivative of $s(t)$ is a constant $\pm 1/\tau$ (or 0 in pausing) irrespective of $s_{\rm inv}$.

\section{Experimental results}
\label{sec:exp}

In this section we report on the performance of reverse annealing for the $p=2$ $p$-spin model on the D-Wave 2000Q device. 

\subsection{Dependence on initial conditions}
\label{sec:initcond}

Figure~\ref{fig:RA_20_initial} shows the experimental success probability, i.e., the ground-state probability of the final state for $N=20$ and no pause. The initial condition is specified by the initial value of the normalized total magnetization,
\begin{align}
\label{eq:mo}
    m_0=\frac{1}{N}\sum_{i=1}^N \langle \psi_0 |\sigma_i^z |\psi_0\rangle, 
\end{align}
where $|\psi_0\rangle$ denotes the initial wave function (a classical state). Since the doubly degenerate ground state of the problem Hamiltonian $H_0$ has $\pm 1$ as the normalized magnetization, a value of $m_0$ closer to 1 (or $-1$) represents an initial condition exhibiting higher overlap with the ground state. In Fig.~\ref{fig:RA_20_initial} we present results for a completely unbiased initial condition ($m_0=0$) and initial conditions strongly biased toward the all-up state $\sigma_i^z=1~\forall i$ ($m_0=0.8,0.9$). The state with $m_0=0$ is the highest excited state, and the state with $m_0=0.9$ is the first excited state.

\subsubsection{$m_0=0$}

\begin{figure}[h!]
\subfigure[]{\includegraphics[width = 0.4939\columnwidth]{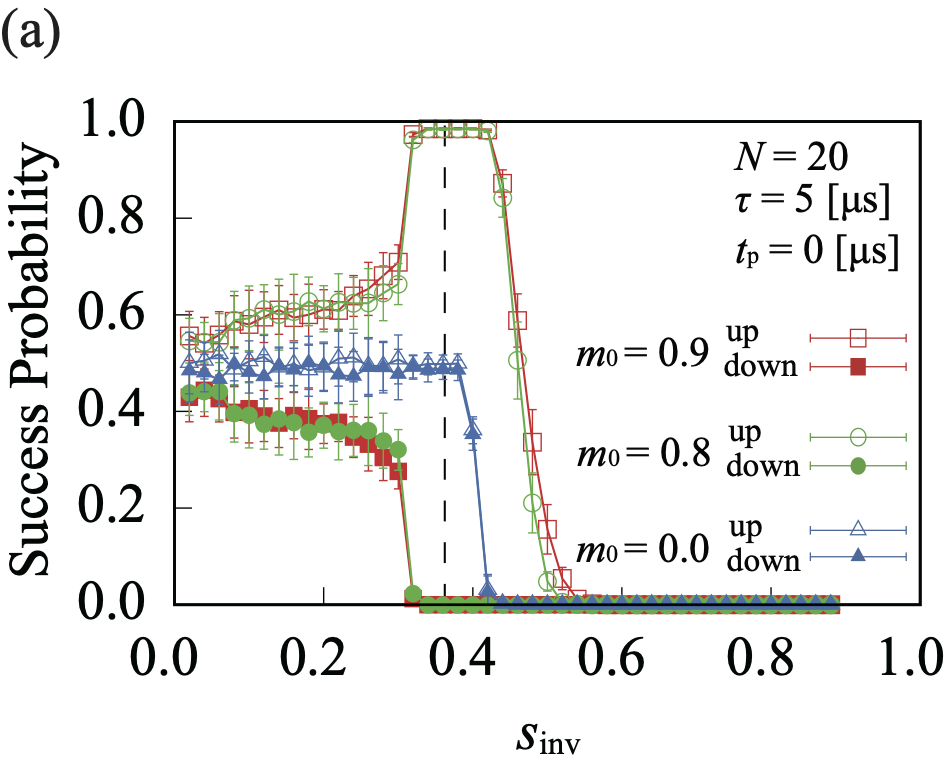}}
\subfigure[]{\includegraphics[width = 0.4939\columnwidth]{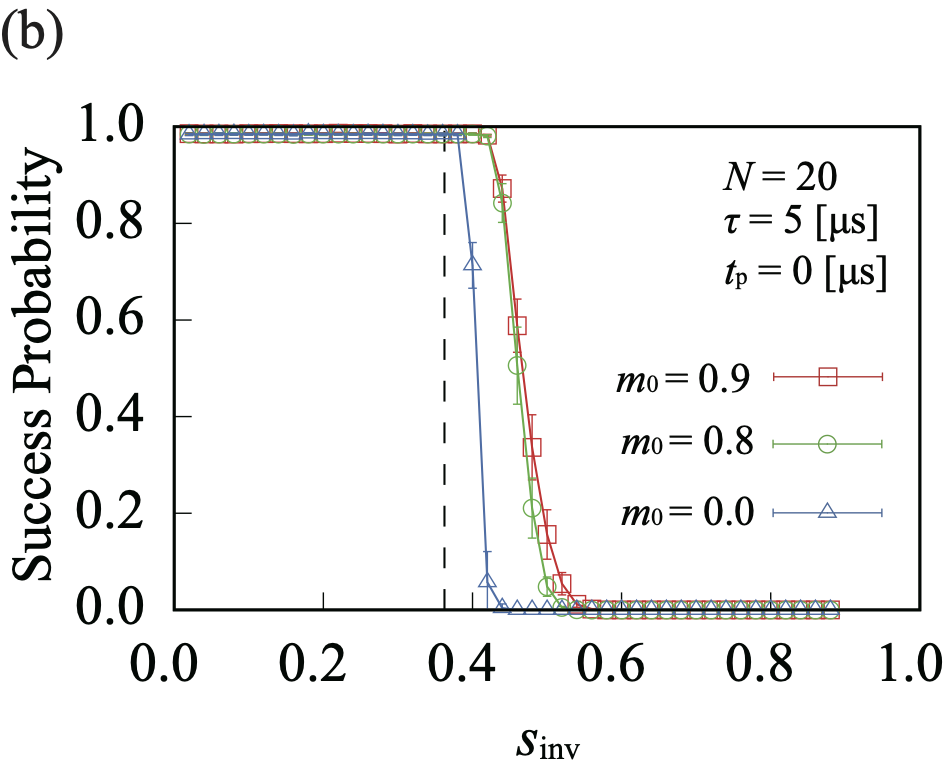}}
\caption{Success probability for different initial conditions $m_0$ (magnetization) in reverse annealing on the D-Wave 2000Q device as a function of the inversion point $s_{\mathrm {inv}}$. The dashed line is the minimum-gap point at $s_{\Delta}\approx 0.36$ for $N=20$. When $s_{\mathrm {inv}} < s_{\Delta}$ the reverse direction of the anneal goes through and past the minimum gap point, and then crosses it again during the forward anneal. When $s_{\mathrm {inv}} > s_{\Delta}$ there is no crossing of the minimum gap point.
(a) Partial success probability for $m_0=0, 0.8$, and $0.9$. (b) Total success probability for the same set of $m_0$ as in (a). Here and in all subsequent figures the labels ``up" and ``down" mean the populations of the all-up state and the all-down state, respectively. Error bars denote standard deviations.}
\label{fig:RA_20_initial}
\end{figure}

Figure~\ref{fig:RA_20_initial}(a) shows the probability that the system reaches the all-up state (denoted ``up") and the all-down state ($\sigma_i^z=-1,~\forall i$, denoted ``down"). We call these the partial success probabilities.

When the initial condition is $m_0=0$, the up and down probabilities are equal to within the error bars for all $s_{\rm {inv}}$, as expected because the initial state is unbiased. Note that both success probabilities are close to $0.5$ for $s_{\mathrm{inv}}<0.4$ whereas they are zero for $s_{\mathrm{inv}}>0.5$.  The vertical dashed line at $s_{\mathrm{inv}}\approx 0.36$ indicates the minimum-gap point, i.e., the point where the energy gap between the ground and the second excited state becomes minimum, which we denote by $s_{\Delta}$ (the first excited state becomes degenerate with the ground state at the end of the anneal, hence it is the second excited state that is relevant). It is likely that the system remains close to its initial state for $s_{\mathrm{inv}}>0.5$, where the transverse field and the associated quantum fluctuations are small, and the true final ground state with $m_0=\pm 1$ is hard to reach, since the initial state with $m_0=0$ has small overlap with the latter. The situation is different when $s_{\mathrm{inv}}<0.4$, i.e., when the system traverses the minimum-gap region. In this case the system enters the paramagnetic (quantum-disordered) phase dominated by the driver $H_D$, so that the initial condition is effectively erased. This renders the process similar to conventional forward quantum annealing, by which the two true ground states are reached with nearly equal probability $0.5$. Recall that the phase transition around $s_{\Delta}$ is of second order in the present problem, and therefore the system finds the true final ground state relatively easily by forward annealing because of the mild, polynomial, closing of the energy gap, as we discuss in more detail later (see Fig.~\ref{fig:RA_20_size}(c)).

\subsubsection{$m_0=0.8,0.9$: up/down symmetry breaking for $0.3 \lesssim s_{\mathrm{inv}} \lesssim 0.5$}

For the initial states with $m_0=0.8$ and $m_0=0.9$, the experimental results shown in Fig.~\ref{fig:RA_20_initial}(a) reveal significant differences between the probabilities of the final all-up and all-down states. Of course, the initial conditions $m_0=0.8,0.9$ have much larger overlap with the all-up state, and this fact alone suggest a mechanism for breaking the symmetry between the probabilities of the final state being the all-up or all-down state. However, 
it is remarkable that the success probability for the all-up state is almost 1 in a region $0.3 \lesssim s_{\mathrm{inv}}\lesssim 0.5 $ both to the left and to the right of $s_{\Delta}\approx 0.36$ (we discuss the additional asymmetry for $s_{\mathrm{inv}} \lesssim 0.3$ below). As we explain in Sec.~\ref{sec:AME}, this behavior is inconsistent with a model of an open quantum system that is weakly coupled to its environment and is thus described by the adiabatic master equation~\cite{ABLZ:12-SI}. However, it is consistent with a simple semiclassical model captured by the spin-vector Monte Carlo algorithm~\cite{shin2014quantum,Albash2020ComparingRM}, as explained in Sec.~\ref{sec:PSP-SVMC}.

\subsubsection{$s_{\mathrm{inv}} \gtrsim 0.5 > s_{\Delta}$: freezing}
\label{sec:freezeout}

Note that for all three initial conditions the success probability eventually vanishes for $s_{\mathrm{inv}} \gtrsim 0.5 > s_{\Delta}$. The reason is that in all cases the system is initialized in an excited state, and remains in an excited state at the end of the anneal, since there is no mechanism for thermal relaxation to a lower energy state when $s_{\mathrm{inv}}$ is large. This is a manifestation of the phenomenon of freezing~\cite{ABLZ:12-SI,amin:freeze-out}, i.e., the extreme slowdown of relaxation due to the fact that the system-bath interaction (nearly) commutes with the system Hamiltonian when the transverse field is very small. In addition, the annealing timescale is manifestly too slow for downward diabatic transitions. This is true even in the absence of pausing, which suggests a discontinuity in the derivative of the schedule depicted in Fig.~\ref{fig:schedule}(b). Despite this, the reversal of the anneal direction is apparently too slow in practice to have a non-adiabatic effect, or if diabatic transitions do occur, then they exclusively populate higher excited states.

\subsubsection{$s_{\mathrm{inv}} \lesssim 0.3 < s_{\Delta}$: spin-bath polarization}
\label{sec:spinbathpol}

It is also noteworthy from Fig.~\ref{fig:RA_20_initial}(a) that the initial condition results in different probabilities for the all-up and all-down states even in the paramagnetic region $s_{\mathrm{inv}} \lesssim 0.3 < s_{\Delta}$, where quantum fluctuations are large and the all-up and all-down states are expected to have the same probability in equilibrium. The system ``remembers" the initial condition to a certain extent, an anomaly that may be attributable to spin-bath polarization~\cite{spin-bath-polarization}. Namely, the persistent current flowing in the qubit body during the anneal produces a magnetic field that can partially align or polarize an ensemble of environmental spins local to the qubit wiring, with a much slower relaxation time than the anneal duration. Given the polarized initial condition $m_0=0.8$ or $0.9$, this polarized spin-bath will be aligned with the all-up state even after the system crosses into the paramagnetic phase, thus preventing the system from equilibrating and explaining the observed memory effect.

\subsubsection{Total success probability}

Figure~\ref{fig:RA_20_initial}(b) shows the total success probability, the sum of the final up and down probabilities. The resulting curve stays almost flat and close to $1$ for $s_{\mathrm{inv}}<0.4$.  This constant $1$ is a mixture of the two effects manifest in Fig.~\ref{fig:RA_20_initial}(a), the genuine effects of reverse annealing around $s_{\mathrm{inv}}\approx 0.4$ for $m_0=0.9$ and 0.8 and the effectively-forward-annealing-like behavior for $m_0=0$. We shall see in Sec.~\ref{sec:simulations-init} that all the salient features seen in Fig.~\ref{fig:RA_20_initial}(b), such as the shift to the left with decreasing $m_0$, are captured by our open system simulations.

The results seen in Fig.~\ref{fig:RA_20_initial}(b) are also consistent with the trend found numerically in Ref.~\cite{Passarelli2019} for the $p=3$ $p$-spin model except for the very small values of success probability near $s_{\mathrm{inv}}=0$ seen in the latter's numerical simulations. This may be explained by the very fast forward annealing in this parameter region in Ref.~\cite{Passarelli2019}, which keeps the system almost unchanged from the quantum-disordered state at $s=s_{\mathrm{inv}}$.
Aside from this subtlety, our experimental data are consistent with the result of Ref.~\cite{Passarelli2019}, supporting the latter's expectation that the system-environment interaction (through spin-boson coupling) prompts relaxation of the system toward the ground state around the minimum-gap region.

\subsection{Dependence on annealing time}
\begin{figure}[h!]
\subfigure[]{\includegraphics[width = 0.4939\columnwidth]{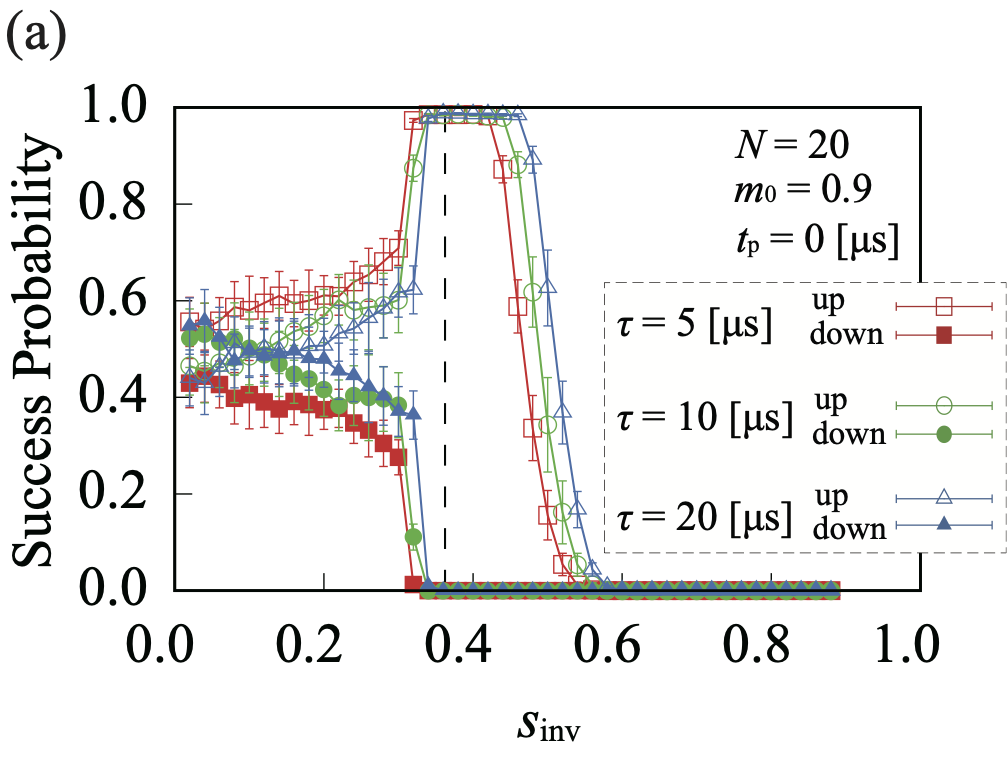}}
\subfigure[]{\includegraphics[width = 0.4939\columnwidth]{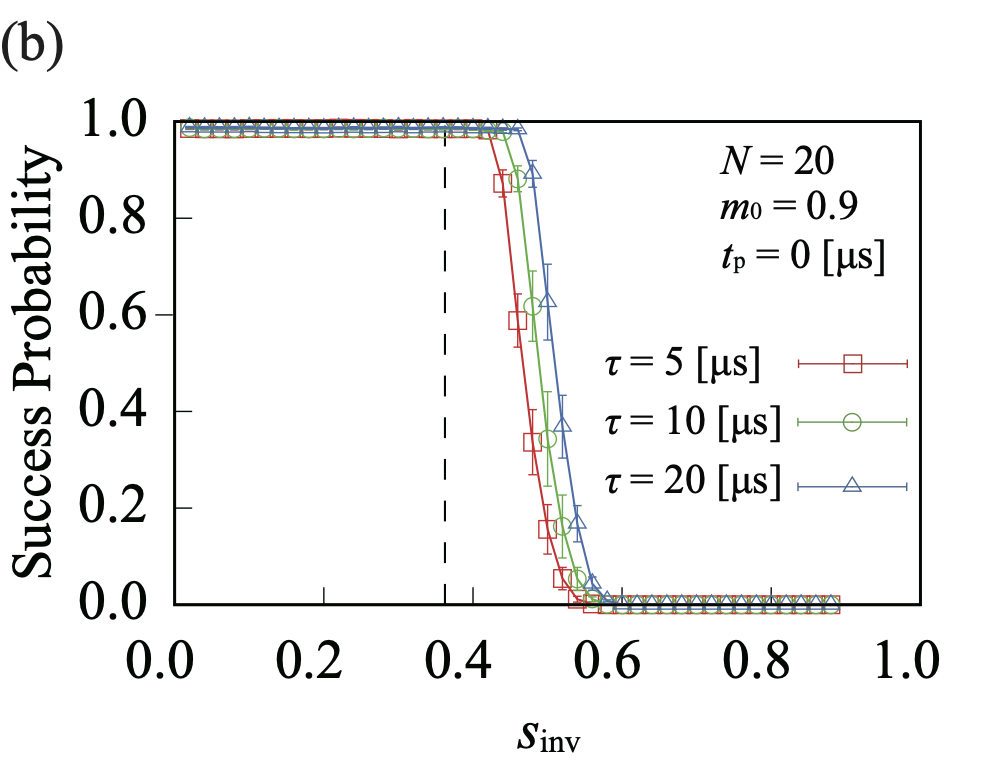}}
\caption{Success probability for different annealing times $\tau$ in reverse annealing on the D-Wave 2000Q device as a function of $s_{\mathrm{inv}}$.
(a) Partial success probability for $\tau=5, 10$, and $20[\mu s]$. (b) Total success probability for the same set of $\tau$ as in (a). The dashed line is the minimum-gap point at $N=20$, $s_{\Delta}\approx 0.36$.}
\label{fig:RA_20_diff_tau}
\end{figure}

As seen from Eq.~\eqref{eq:t_a}, the total annealing time $t_a$ is linearly dependent on the annealing time $\tau$, and proportional to $\tau$ when $t_p=0$.
Figure~\ref{fig:RA_20_diff_tau} shows the success probability for different $\tau$ with $N=20$, $m_0=0.9$, and no pausing.  The overall trend is similar to Fig.~\ref{fig:RA_20_initial}.

As shown in Fig.~\ref{fig:RA_20_diff_tau}(a), an increase of $\tau$ leads to slightly higher all-up success probabilities for $s_{\mathrm{inv}}> s_{\Delta}$ but the other way around for $s_{\mathrm{inv}} < s_{\Delta}$. This can be explained in terms  of an increased relaxation to the ferromagnetic ground state near the minimum energy gap for larger $\tau$ and an enhanced relaxation to the paramagnetic ground state for $s_{\mathrm{inv}}<s_{\Delta}$.

The total success probability stays close to $1$ for $s_{\mathrm{inv}}<s_{\Delta}$ and benefits slightly from increasing $\tau$ for $s_{\mathrm{inv}}>s_{\Delta}$ (Fig.~\ref{fig:RA_20_diff_tau}(b)).

\subsection{Effects of pausing}

\begin{figure}[h!]
\subfigure[]{\includegraphics[width = 0.4939\columnwidth]{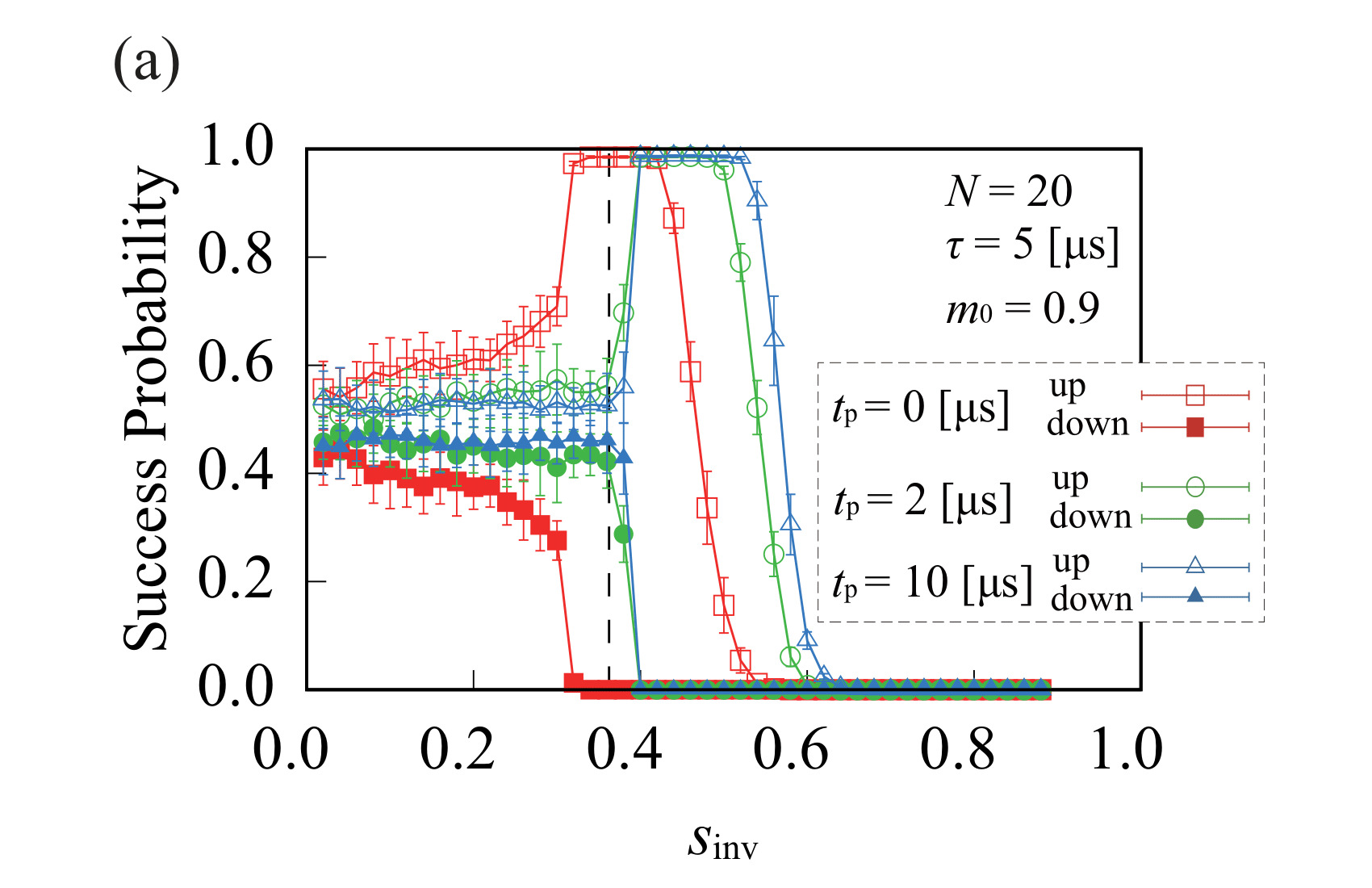}}
\subfigure[]{\includegraphics[width = 0.4939\columnwidth]{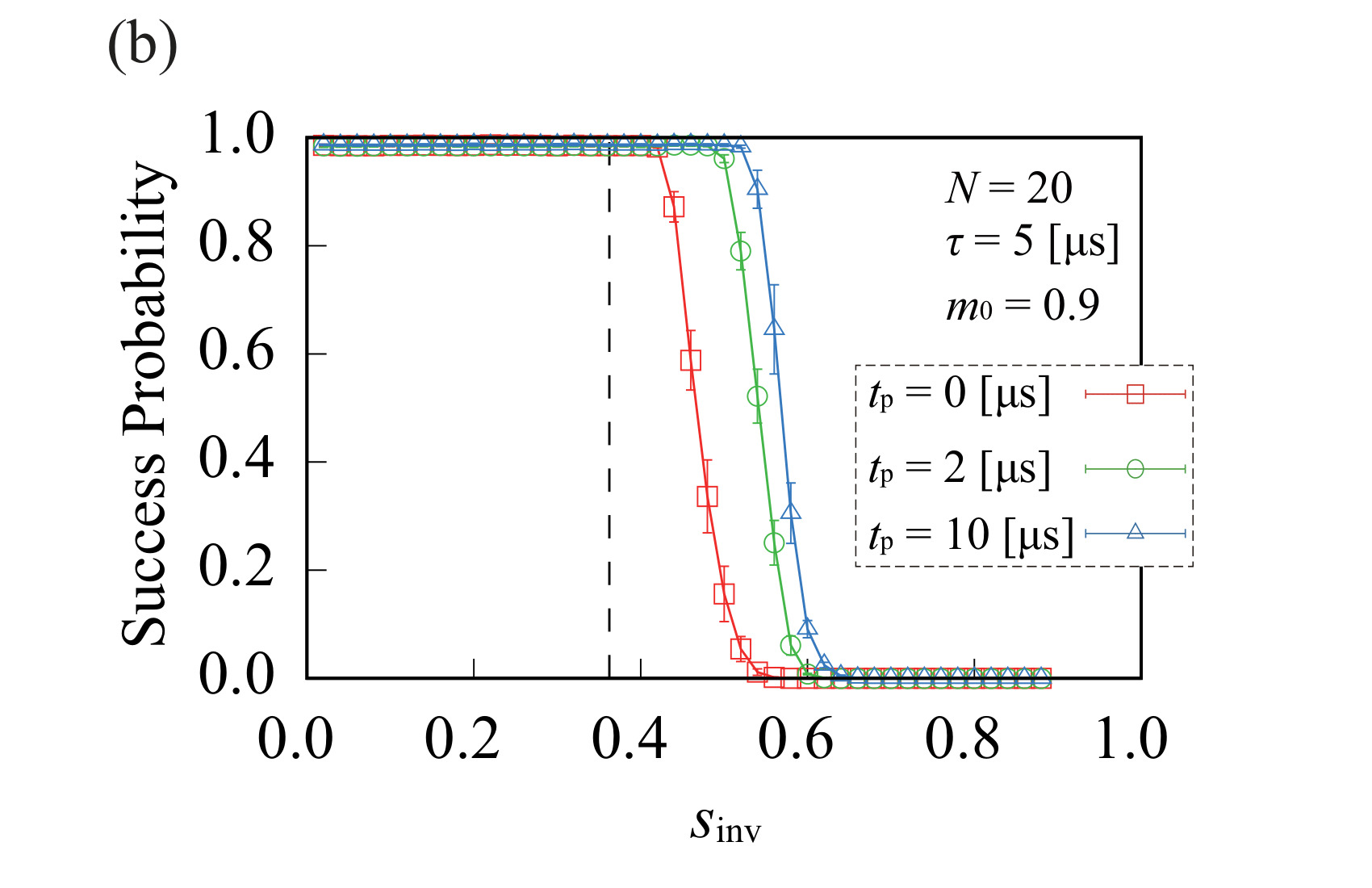}}
\caption{Success probability for different pausing times $t_p$ in reverse annealing as a function of $s_{\mathrm{inv}}$. 
(a) Partial success probability for pause $t_p=0, 2$, and $10~[\mu s]$. (b) Total success probability for the same set of $t_p$ as in (a). The dashed line is the minimum-gap point at $N=20$, $s_{\Delta}\sim0.36$.}
\label{fig:RA_20_pause}
\end{figure}

Figure~\ref{fig:RA_20_pause} shows success probabilities for pausing time $t_p\in\{0, 2, 10\}~[\mu s]$ as a function of $s_{\mathrm{inv}}$  with $N=20$, $m_0=0.9$ and $\tau=5~[\mu s]$. 

The overall trend is similar to Fig.~\ref{fig:RA_20_diff_tau}, i.e., a longer pause leads to an increased relaxation, but the effects are more significant in the present case. Pausing at $0.5<s_{\mathrm{inv}}<0.6$ greatly improves the success probability from nearly 0 ($t_p= 0~[\mu s]$) to nearly 1 ($t_p= 2~[\mu s]$ and $10~[\mu s]$), implying that pausing in a relatively narrow region slightly past the minimum gap  point is most effective. This is in line with previous findings~\cite{marshall,chen2020pausing,Albash2020ComparingRM}. 

Another significant difference between pausing and not pausing is that the partial success probabilities shown in Fig.~\ref{fig:RA_20_pause}(a) are flat for $t_p>0$, showing that the effect responsible for the anomaly disappears under pausing. This is consistent with the spin-bath polarization effect discussed in Sec.~\ref{sec:initcond}, in that the pause provides the time needed for this polarization to relax to equilibrium, on a timescale of a few $\mu s$.

\begin{figure}[h!]
\subfigure[]{\includegraphics[width = 0.4939\columnwidth]{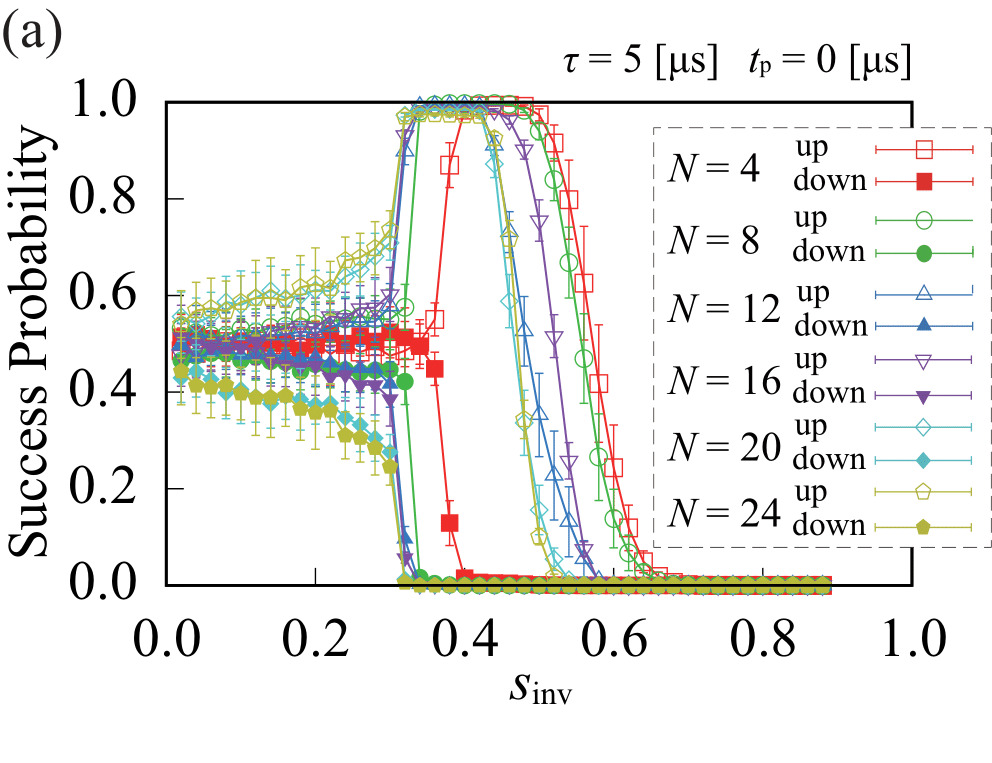}}
\subfigure[]{\includegraphics[width = 0.4939\columnwidth]{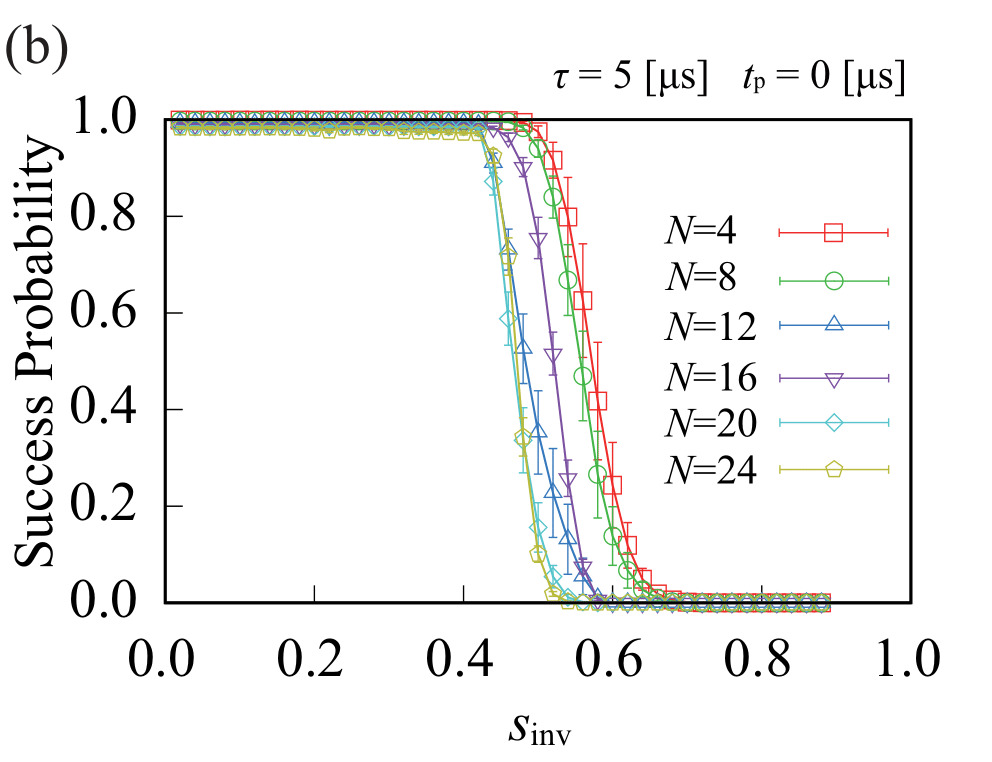}}
\subfigure[]{\includegraphics[width=0.4739\linewidth]{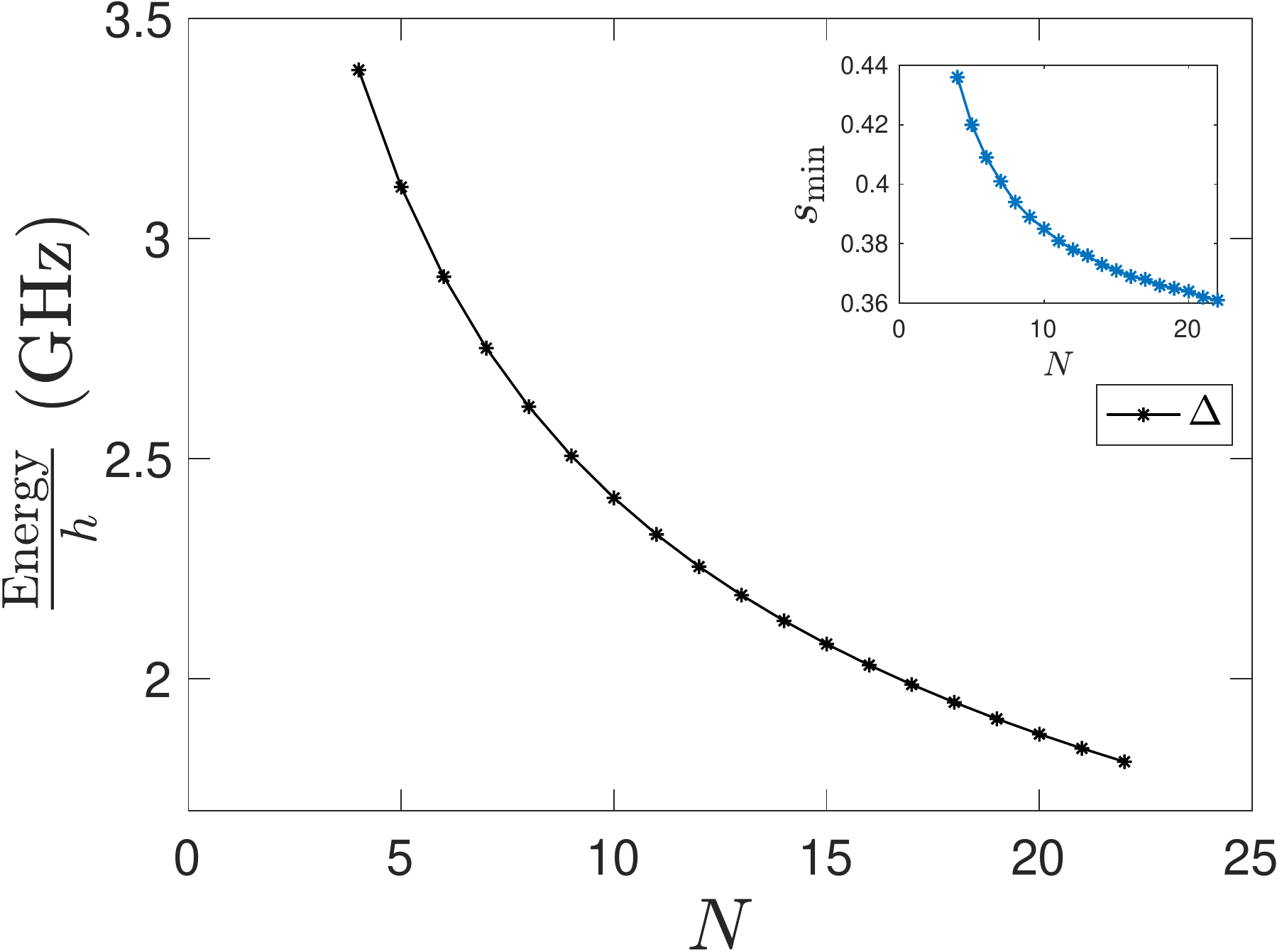}}
\caption{Success probability for different system sizes $N$ in reverse annealing  as a function of $s_{\mathrm{inv}}$. 
(a) Partial success probability for system size $N=4, 8, 12, 16, 20$, and $24$. (b) Total success probability for the same set of $N$ as in (a). (c) The position of the minimum gap $\Delta$ for $4 \le N \le 22$, along with the value of $s_{\mathrm{inv}}$ at which the success probability $=.5$ for each $N$. Inset: the minimum gap for each $N$.} 
\label{fig:RA_20_size}
\end{figure}

\subsection{Size dependence}


Figures~\ref{fig:RA_20_size}(a) and (b), respectively, show the partial and total success probability for different system sizes $N$ as a function of $s_{\mathrm{inv}}$ with $\tau=5~[\mu s]$. The initial state for each $N$ is the first excited state with $m_0<1.0$ (closest to the ground state for which $m_0=1.0$), i.e., a state with one spin flipped. Although we observe a slight non-monotonicity with $N$ in the region $s_{\mathrm{inv}}\geq s_{\Delta}$ in Fig.~\ref{fig:RA_20_size}(a), the overall trend is clear, namely, the drop-off to zero success probability occurs at smaller $s_{\mathrm{inv}}$ as $N$ is increased. This is consistent with the reduction in $s_\Delta$ as a function of $N$, shown in Fig.~\ref{fig:RA_20_size}(c), which is tracked by the value of $s_{\mathrm{inv}}$ at which the success probability equals $0.5$.

Additionally, the asymmetry between the all-up and all-down states for $s_{\mathrm{inv}}<s_\Delta$ is enhanced with increasing $N$. This is consistent with the decreasing gap seen in Fig.~\ref{fig:RA_20_size}(c), which should lead to slower relaxation in thermal equilibrium: at inverse temperature $\beta$, the probability of an excitation is $\exp(-\beta \Delta)$ times the probability of relaxation between the same pair of states, separated by a gap $\Delta$. I.e., the asymmetry induced by the spin-bath polarization effect is expected to take longer to dissipate for larger $N$ due to the slower relaxation of the system.

Finally, the overlap of data for $N=20$ and $24$ 
suggests that large-$N$ effects have already converged at these sizes, i.e., that $N\sim20$ is  sufficiently large to infer the behavior at larger $N$. Also, even the smallest systems with $N=4$ and $8$ already share qualitative features with larger systems.

\subsection{Effects of iteration}

\begin{figure}[h!]
\centering
\includegraphics[width=0.95\columnwidth]{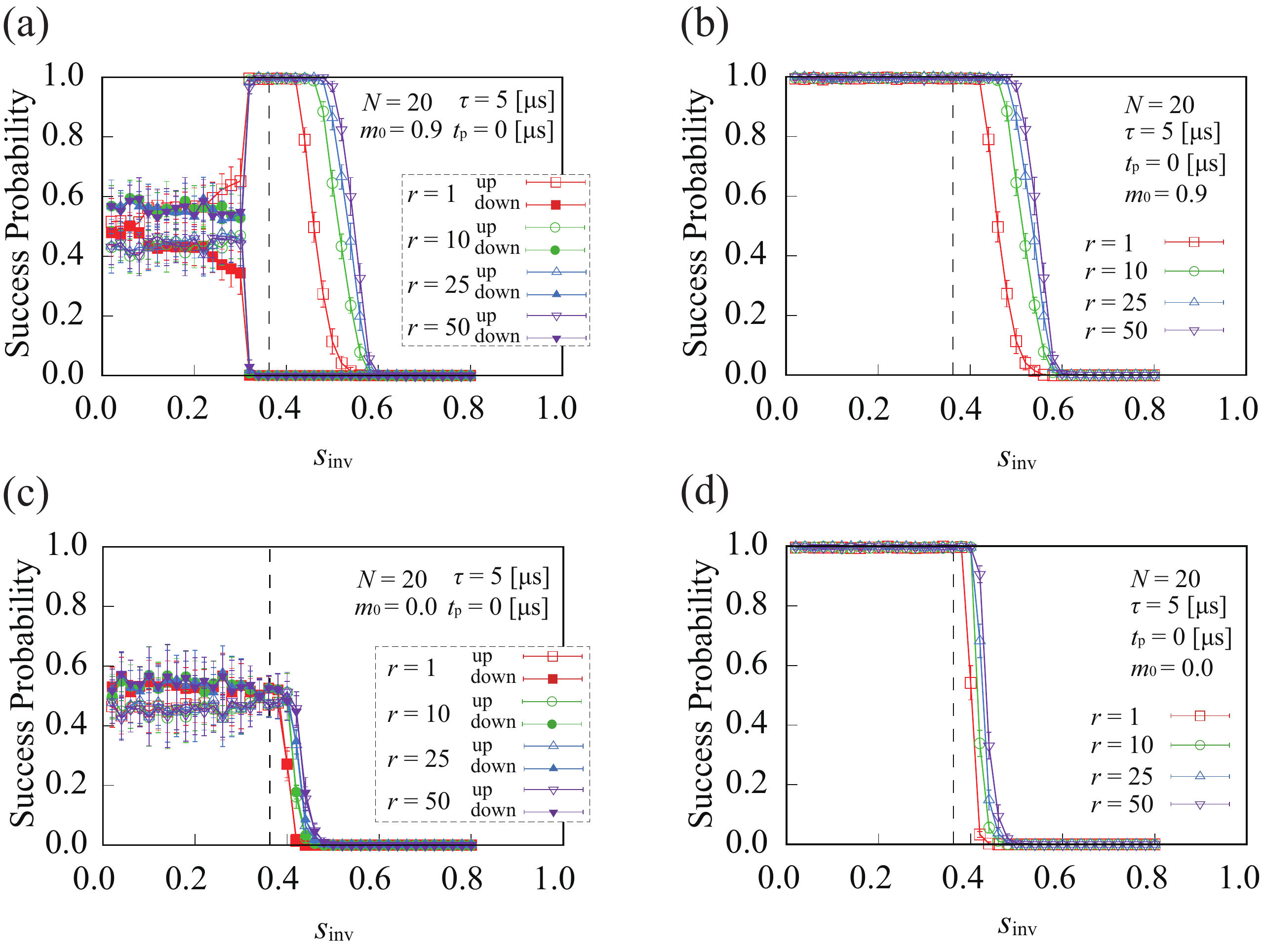}%
\caption{Success probability for different number of iterations $r$ in iterated reverse annealing  as a function of $s_{\mathrm{inv}}$. (a) and (b) are partial and total success probabilities for number of iterations $r=1, 10, 25$, and $50$ with the initial state $m_0=0.9$ (the first excited state), respectively. (c) and (d) are partial and total success probabilities with the initial state $m_0=0$ (the highest excited state). The dashed line is the minimum-gap point at $N=20$, $s_{\Delta}\sim0.36$.}
\label{fig:IRA}
\end{figure}

We next study the effects of iteration on reverse annealing, i.e., how the success probability depends on the number of iterations $r$. Figure~\ref{fig:IRA} shows the result for $r=1, 10, 25$ and $50$ at $s_{\mathrm{inv}}$ with $N=20$, $m_0=0.9$, $t_p=0~[\mu s]$, and $\tau=5~[\mu s]$.
In Fig.~\ref{fig:IRA}, the success probabilities improve in the region $s_{\mathrm{inv}}\geq s_{\rm \Delta}$ as the number of iterations $r$ increases, regardless of the initial state $m_0$. The relaxation to the low-energy state due to coupling to the environment should be successively induced by iterations, and the occupation of the ground state should increase. The results in Fig.~\ref{fig:IRA} confirm this expectation.

Comparing Figs.~\ref{fig:IRA}(a) and (b) with $m_0=0.9$ and Figs.~\ref{fig:IRA}(c) and (d) with $m_0=0$, we can see that the initial state with $m_0=0.9$ has a larger improvement in success probability with fewer iterations $r$ in the region where $s_{\mathrm{inv}} \geq s_{\Delta}$. The initial state with $m_0 = 0$ deviates greatly from the ground state, and there are many excited states between it and the ground state. Therefore, when the system state transitions from the initial state of $m_0 = 0$ to other states by relaxation due to coupling to the environment, those excited states become candidates. Many iterations are expected to be necessary to obtain a significant improvement in the success probability. On the other hand, there are only a few excited states between the initial state $m_0\lesssim 1.0$ and the ground state. Therefore, it is expected that only a few iterations will be sufficient to make the transition to the ground state. Our results support this picture of improved performance under iterated reverse annealing.

\section{Numerical simulations}
\label{sec:simulations}

We next present closed and open-systems simulations and compare the results with the data presented in the previous section. We choose relatively small system sizes $N=4$ and $8$ to facilitate numerical computations. We note that while we can thus simulate the initial condition $m_0=0$ (equal number of spins up and down), we are limited to $m_0=0.5$ for $N=4$ and $m_0=0.75$ for $N=8$ (a single spin down). However, our experimental data do not show a strong size dependence, as discussed in the previous section, so the qualitative comparison our numerical results will enable should suffice to draw relevant physical conclusions.

For simplicity, we focus on reverse annealing without pausing (i.e., $t_p=0$) in this section. We expect that in general pausing will lead to an overall success probability  in the results of open-system simulation.

\subsection{Closed system model}
An analysis of the closed system case, while being unrealistic due to the absence of thermal effects, is instrumental in isolating the effect and importance of diabatic transitions in explaining our experimental results.

The time evolution of reverse annealing in a closed system is described as follows.
The state after a single iteration (cycle) is
\begin{equation}
\ket{\psi(2t_{\text{inv}})} = U(2t_{\text{inv}},0)\ket{\psi(0)}\,,
\end{equation}
where $\ket{\psi(0)}$ is the initial state and
\begin{equation}
    U(2t_{\text{inv}}, 0) = {\cal T}\exp\left[-i\int_0^{2t_{\text{inv}}}H(t')dt'\right]
\end{equation}
is the unitary time-evolution operator and ${\cal T}$ denotes forward time ordering. 
At the end of $r$ cycles, the final state $\ket{\psi(2rt_{\text{inv}})}$ can be expressed as:
\begin{equation}
\ket{\psi(2rt_{\text{inv}})} = U(2rt_{\text{inv}}, 0)\ket{\psi(0)}\,,
\end{equation}
with
\begin{equation}
    U(2rt_{\text{inv}}, 0) = \prod_{i=0}^{r-1}U(2(i+1)t_{\text{inv}}, 2it_{\text{inv}}) \,.
\end{equation}
Here, the final state of the $r$th cycle is the initial state of the next cycle. This condition is shared by the experiments in the previous section. The solution states are doubly degenerate, and the total success probability at the end of $r$ cycles is
\begin{equation}
p(r)=|\langle\psi(2rt_{\text{inv}})\ket{\text{up}}|^2 + |\langle\psi(2rt_{\text{inv}})|\ket{\text{down}}|^2 \,,
\end{equation}
where $\ket{\text{up}} = |\uparrow\rangle^{\otimes N}$ and $\ket{\text{down}} = |\downarrow\rangle^{\otimes N}$.
For higher efficiency without loss of accuracy, we rotate the state vector into the instantaneous energy eigenbasis representation at each time step in our numerical simulations. 

\subsubsection{Dependence on system size and annealing time}

We initialize the state to have a single spin down, i.e., as the computational basis state $\ket{0001}$ ($\ket{0} \equiv \ket{\uparrow}$, $\ket{1} \equiv \ket{\downarrow}~(m_0=0.5)$) for $N=4$ and $\ket{0000001}~(m_0=0.75)$ for $N=8$, respectively.
Note that in our simulations these are not exact eigenstates of $H(1)$, due to a small residual transverse field at $s=1$, as in the D-Wave annealing schedule shown in Fig.~\ref{fig:schedule}.
Note further that our chosen computational basis initial states are not eigenstates of $\bold{S}^2 = \sum_{\alpha\in\{x,y,z\}}(S^{\alpha})^2$, where $S^\alpha=\frac{1}{2}\sum_{i=1}^{N}\sigma_i^\alpha$. Therefore the computational basis states do not lie in the subspace of any fixed value of $S$, where $S$ represents the total spin quantum number of $N$ qubits.
When the initial state $\ket{\psi(0)}$ is a computational basis state, there is a simple upper bound on the success probability achievable in closed system reverse annealing: the population of the initial state in the maximum-spin sector (see App.~\ref{append:successbound}). This upper bound explains why the following closed system results have relatively low success probabilities.

\begin{figure}[h!]
\subfigure[]{\includegraphics[width=0.49\columnwidth]{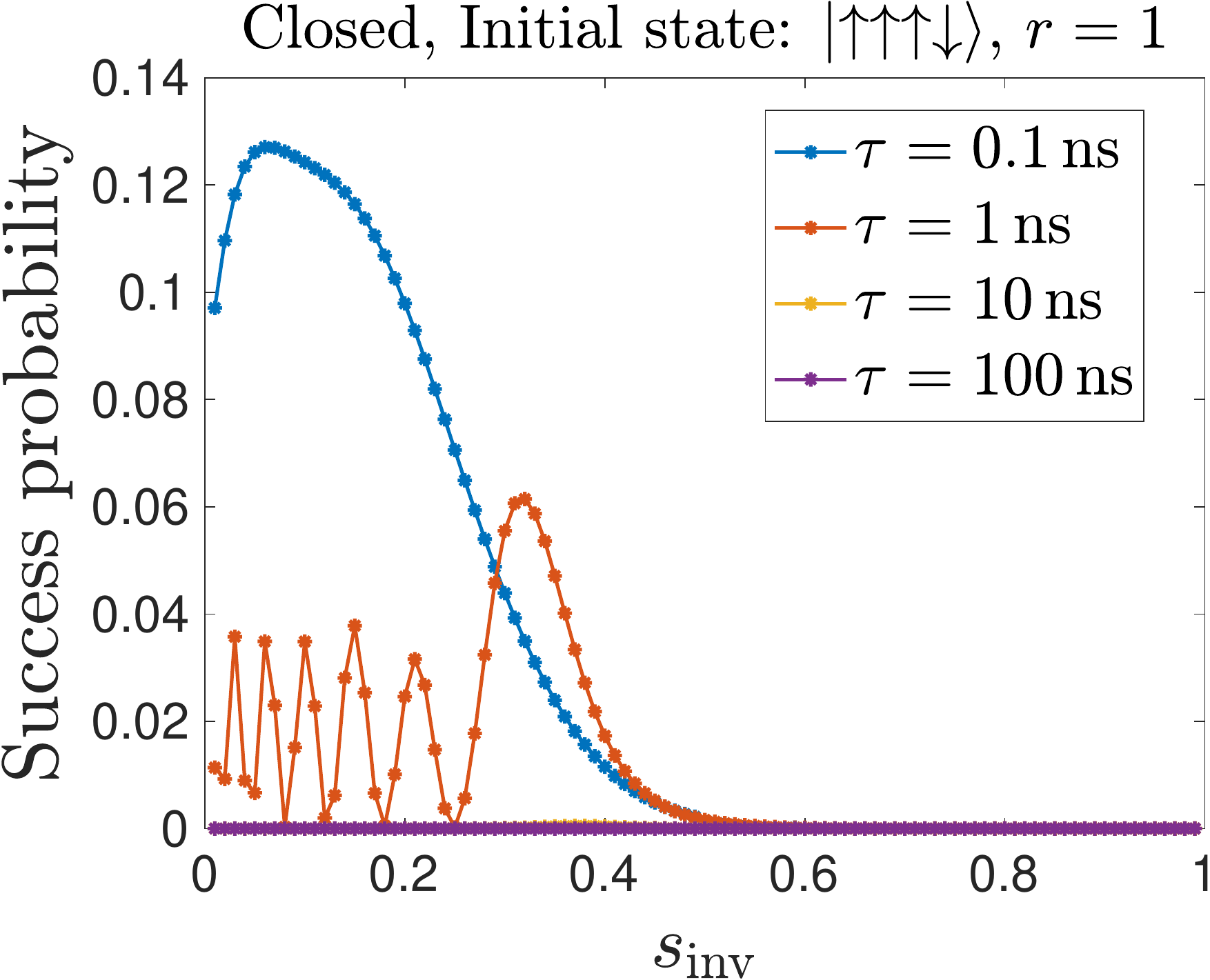}\label{fig:4qubitsunitary}}
\subfigure[]{\includegraphics[width=0.4939\columnwidth]{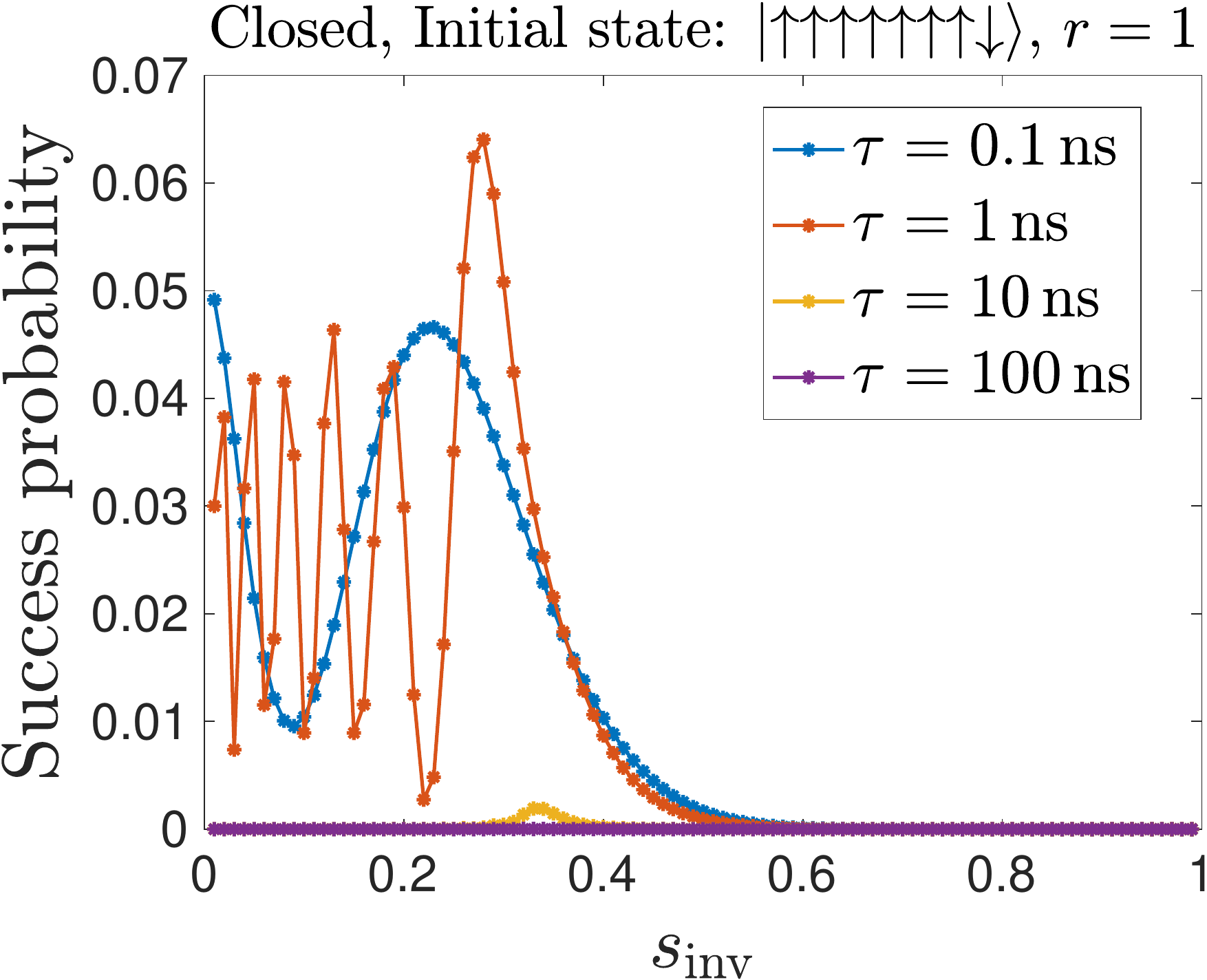}\label{fig:8qubitsunitary}}
\caption{Total success probability for closed system with $r=1$ (single cycle) and a single-spin down as the initial state. (a) $N=4$, (b) $N=8$. Here and in the subsequent figures the nanoscecond timescale is set by the energy scale of the D-Wave annealing schedule shown in Fig.~\ref{fig:schedule}.}
\label{fig:close0001_r1}
\end{figure}

We plot in Fig.~\ref{fig:close0001_r1} the simulation results for the total success probability for various annealing rates, subject to the D-Wave annealing schedule shown in Fig.~\ref{fig:schedule}, in the same units. We see that the success probability after a single cycle is non-negligible only when $\tau$ is small enough ($\tau < 1$ [ns]), in which case diabatic transitions to states with lower energies take place for $s_{\mathrm{in}}< s_\Delta$. However, the success probability remains small even in this case, in any case much smaller than in our experimental results where thermal relaxation plays the dominant role.

\subsubsection{Dependence on initial state and number of iterations}
\label{sec:closed2}

We next choose the initial state $\ket{\psi(0)}$ with two spins down $\ket{0011}~(m_0=0)$ for $N=4$ and  $\ket{00000011}~(m_0=0.5)$ for $N=8$, respectively, i.e., the second excited states of $H_{0}$ for these respective system sizes.

\begin{figure}[h!]
\subfigure[]{\includegraphics[width=0.4939\columnwidth]{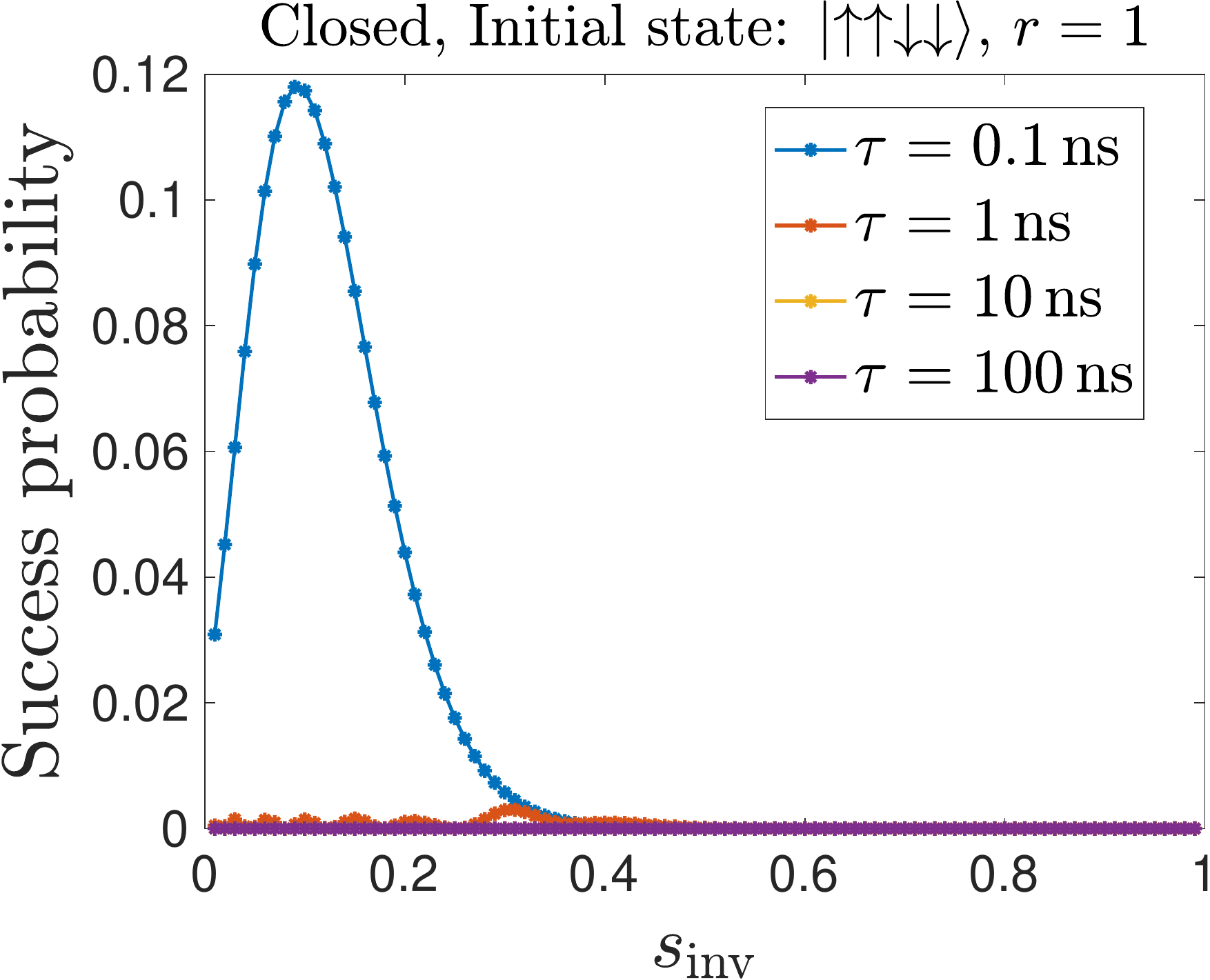}\label{fig:4qubitsunitary2}}
\subfigure[]{\includegraphics[width=0.4939\columnwidth]{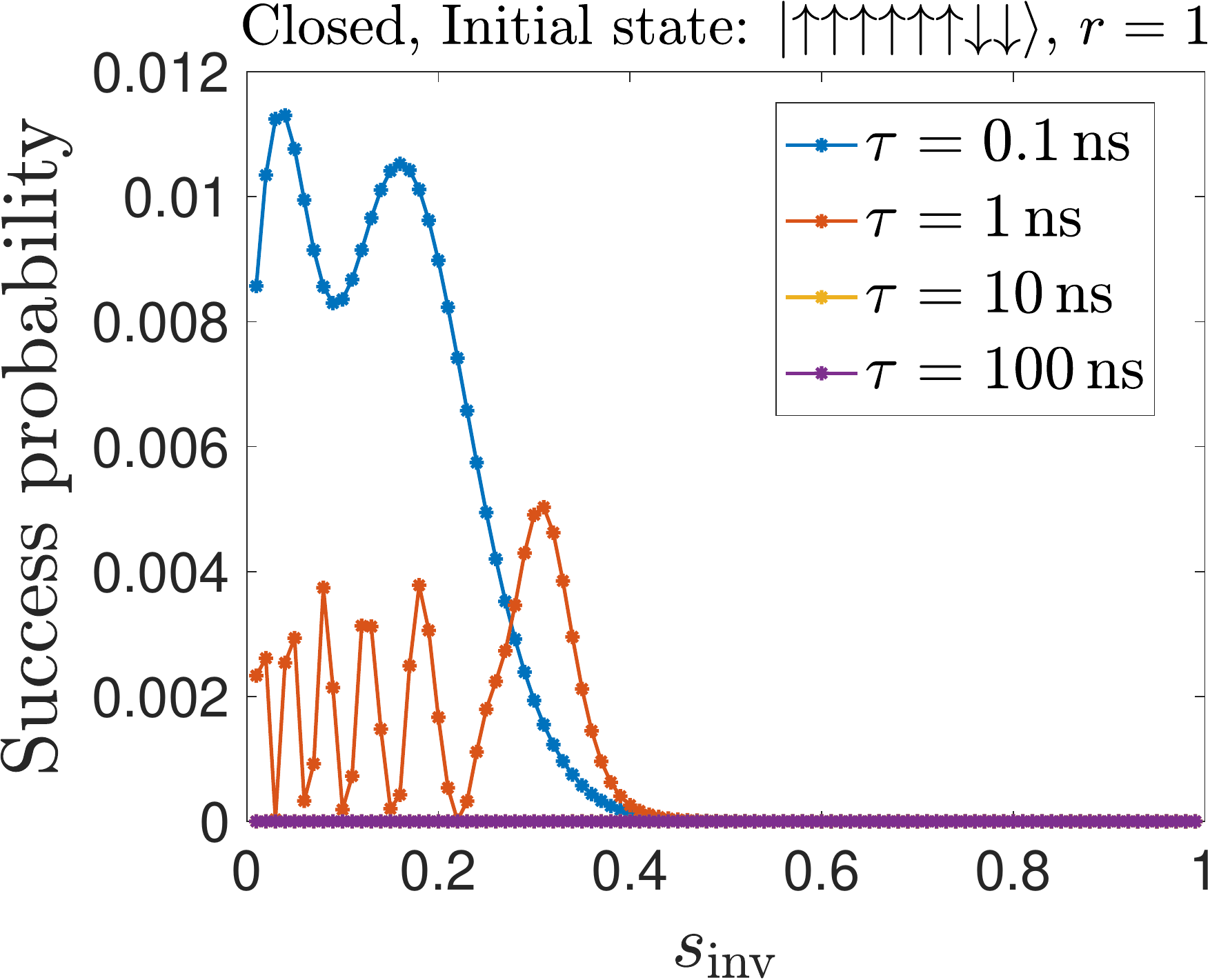}\label{fig:8qubitsunitary2}}
\subfigure[]{\includegraphics[width=0.4939\columnwidth]{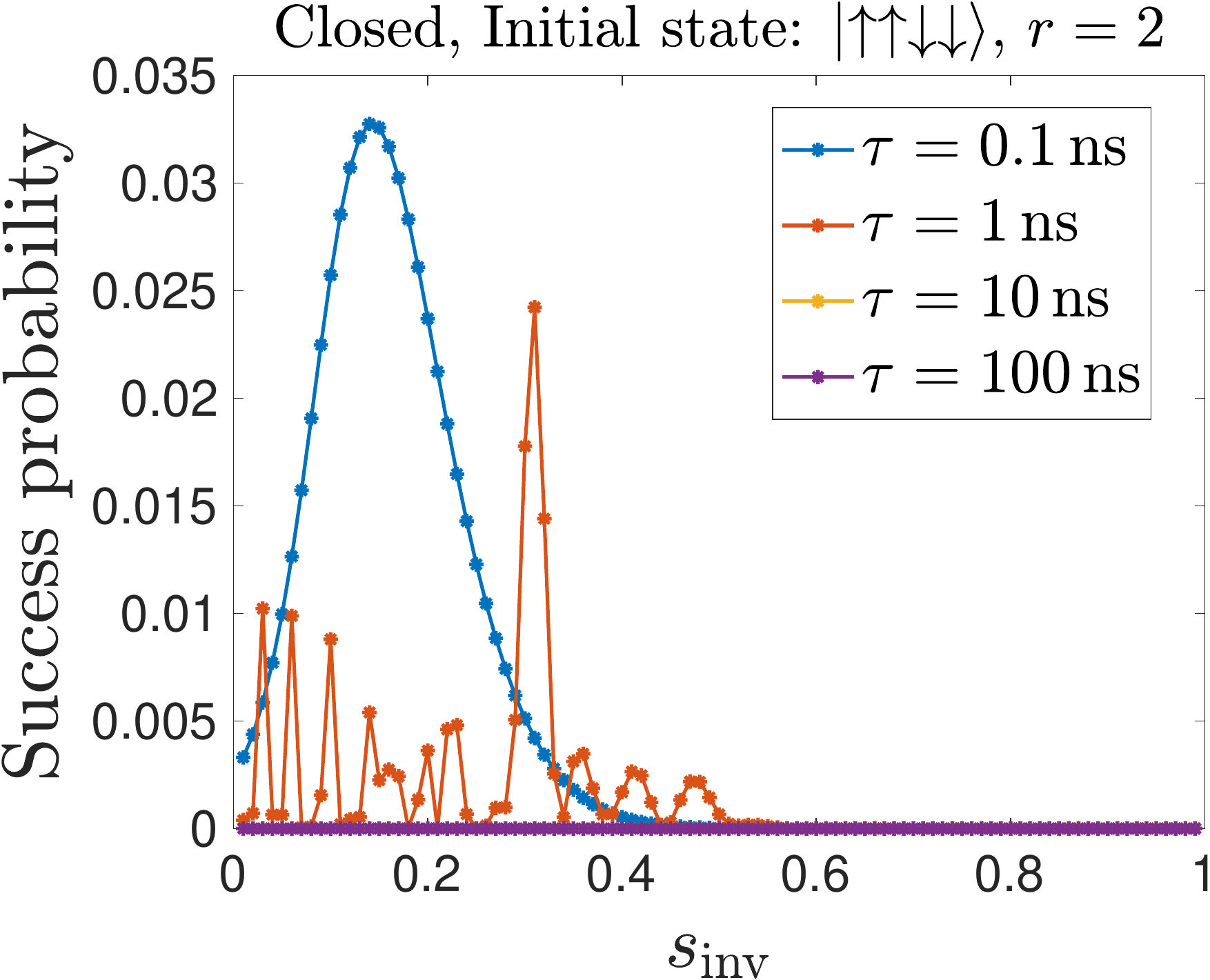}\label{fig:4qubitsunitary3}}
\subfigure[]{\includegraphics[width=0.4939\columnwidth]{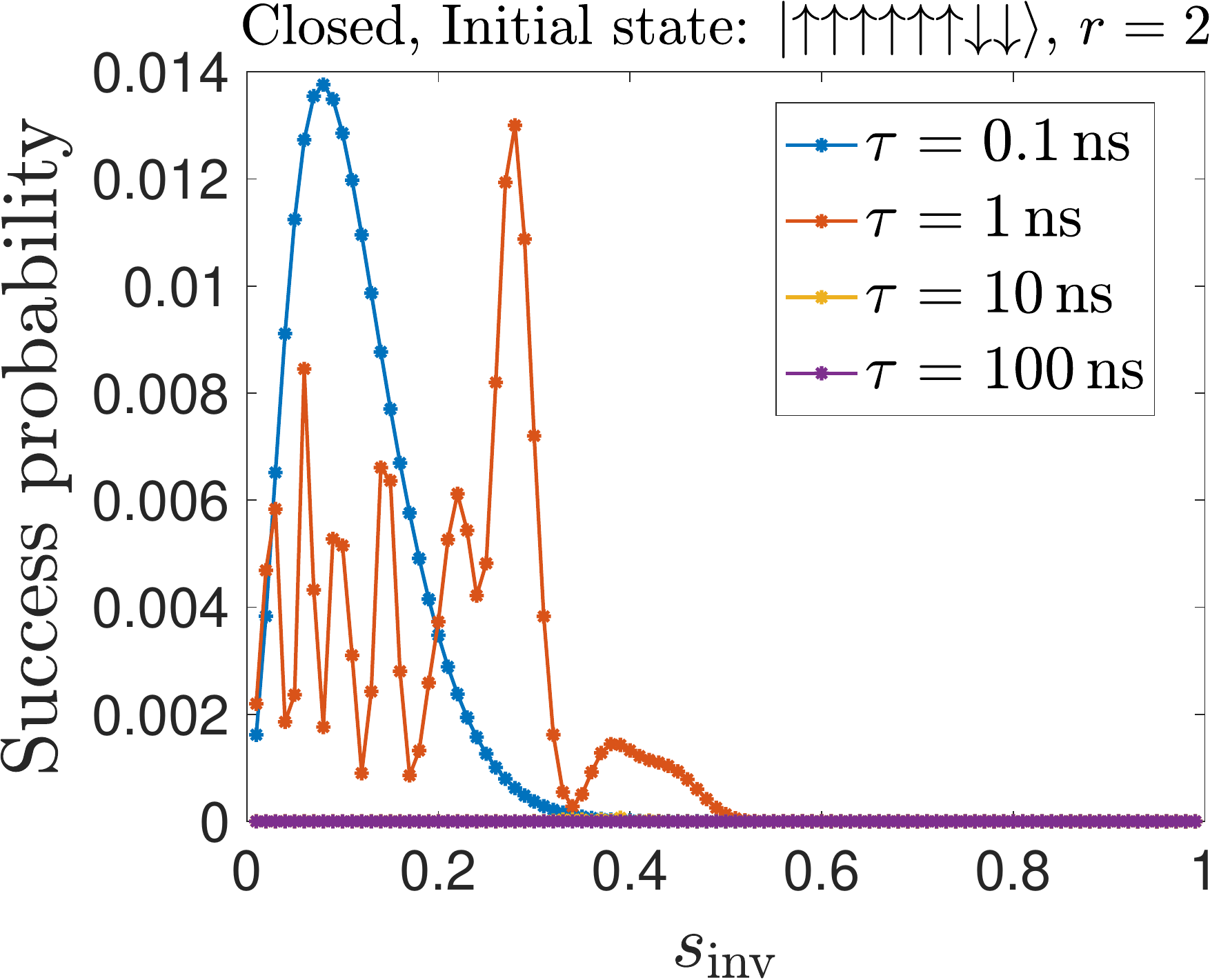}\label{fig:8qubitsunitary3}}
\caption{Total success probability with two spins down as the initial state. (a) $N=4$ and $r=1$, (b) $N=8$ and $r=1$, (c) $N=4$ and $r=2$, (d) $N=8$ and $r=2$.}
\label{fig:close0011_r1}
\end{figure}

Figure~\ref{fig:close0011_r1} (top row) shows that as expected, for $N=4$ and $r=1$, the success probability is overall lower than the case when the initial state is the first excited state, Fig.~\ref{fig:close0001_r1}. Only for the highest annealing rate (smallest $\tau$, most diabatic) is the maximum success probability similar to the case of the first excited state. For $N=8$, the success probability is much smaller, no matter how diabatic the process is. This indirectly confirms the dominant role played by thermal effects in our experimental results.

We plot in Fig.~\ref{fig:close0011_r1} (bottom row) the results for $r=2$ cycles, where we see that the success probability decreases compared to the $r=1$ case. This is consistent with the conclusions of Ref.~\cite{nishimori:reverse-pspin-2}. 

Recalling the experimental results reported in the previous section, we may therefore safely conclude that, as expected, the closed-system picture is far from the experimental reality of the D-Wave device.

\subsection{Open quantum system model}
\label{sec:AME}

For an open system, the state after the first cycle is
\begin{equation}
\rho(2t_{\text{inv}}) = V(2t_{\text{inv}},0)\rho(0) \,,
\end{equation}
where 
\begin{equation}
    V(t,0) = {\cal T}\exp\left[\int_0^{t}\mathcal{L}(t')dt'\right] \,.
\label{eq:vpropagator}    
\end{equation}
$\mathcal{L}(t)$ is the time-dependent Liouville superoperator. We use the adiabatic master equation (AME)~\cite{ABLZ:12-SI}, where $\mathcal{L}(t)$  at time $t$ satisfies
\begin{align}
	\frac{d\rho(t)}{dt} 
	&= \mathcal{L}(t)\rho(t)\\
	&= i \bigl[\rho(t), H(t) + H_\text{LS}(t)\bigr] + \mathcal{D}\bigl[\rho(t)\bigr] \,.
\end{align} 
Here, $ H_\text{LS}(t) $ is the Lamb shift term and $\mathcal{D} $ is the dissipator. Again, the time dependence of the integrand in Eq.~\eqref{eq:vpropagator} is incorporated into $s(t)$.

In the weak coupling limit, $\mathcal{D}$ can be expressed in a diagonal form with Lindblad operators $L_{i,\omega}(t)$:

\begin{eqnarray}
\label{eq:dissdiag}
{\mathcal{D}}_{}  [\rho(t)]  & = &\sum_{i}\sum_{\omega}\gamma_i(\omega)\left(
L_{i,\omega}(t) \rho(t) L^\dagger_{i,\omega}(t)\phantom{\frac{1}{2}} \right. \notag \\
&& \hspace{-0.5cm} \left. \qquad\qquad - \frac{1}{2} \left\{ L_{i,\omega}^{\dagger}(t) L_{i,\omega}(t) , \rho(t) \right\} \right) \,,
\end{eqnarray}
where the summation runs over the qubit index $i$ and the Bohr frequencies $\omega$ [all possible differences of the time-dependent eigenvalues of $H(t)$]. This dissipator expresses decoherence in the energy eigenbasis~\cite{Albash:2015nx}: quantum jumps occur only between the eigenstates of $H(t)$~\cite{yip:mcwf}.

At the end of $r$ cyles, the final state $\rho(2rt_{\text{inv}})$ is expressed as:
\begin{equation}
\rho(2rt_{\text{inv}}) = V(2rt_{\text{inv}}, 0)\rho(0)\,,
\end{equation}
with
\begin{equation}
    V(2rt_{\text{inv}}, 0) = \prod_{i=0}^{r-1}V(2(i+1)t_{\text{inv}}, 2it_{\text{inv}}) \,.
\end{equation}

We consider two different models of system-bath coupling: independent and collective dephasing~\cite{LidarWhaley:03}. 

In first case, we assume that the qubit system is coupled to independent, identical bosonic baths, with the bath and interaction Hamiltonians being 
\begin{subequations}
\begin{align}
\label{eq:SBm}
H_B &= \sum_{i=1}^N \sum_{k=1}^\infty \omega_k b_{k,i}^\dagger b_{k,i} \ ,  \\
H_{SB}^{\text{ind}} &= g \sum_{i=1}^N
\sigma_i^z \otimes \sum_k \left(b_{k,i}^\dagger + b_{k,i} \right) \ ,
\end{align}
\end{subequations}
where $b_{k,i}^\dagger$ and $b_{k,i}$ are, respectively, the raising and lowering operators for the $k$th oscillator mode with natural frequency $\omega_k$.  The rates appearing in Eq.~\eqref{eq:dissdiag} are given by
\begin{equation}
\gamma_{i}(\omega) = 2\pi \eta g^2\frac{\omega e^{-|\omega|/\omega_c}}{1-e^{-\beta\omega}}  \ ,
\end{equation}
arising from an Ohmic spectral density, and satisfying the Kubo-Martin Schwinger condition~\cite{kubo1957statistical, martin1959theory} ($\gamma_{i}(- \omega) = e^{-\beta\omega} \gamma_{i}(\omega)$, $\beta = 1/T$ being the inverse temperature).  We use $\eta g^2 = 10^{-3}$, the cutoff frequency $\omega_c = 1$ [THz], and the D-Wave device operating temperature $T= 12.1 \text{ [mK]} = 1.57$ [GHz]. With the independent dephasing assumption, the Lindblad operators are:
\begin{equation}
   L_{i,\omega}(t) = \sum_{\epsilon_b-\epsilon_a = \omega} | \varepsilon_a(t) \rangle \langle \varepsilon_a(t) | \sigma^z_{i} | \varepsilon_b(t) \rangle \langle \varepsilon_b(t) | \,,
\end{equation}
corresponding to dephasing in the instantaneous eigenbasis $\{|\varepsilon_a(t) \rangle\}$ of $H(t)$. Similarly to the closed system case, we rotate the density matrix and Lindblad operators into the instantaneous energy eigenbasis at each time step in our numerical simulations. This keeps the matrices sparse, without loss of accuracy. For $N = 8$ we truncate the system size to the lowest $n=18 = 1+1+8+8$ (i.e. total number of degenerate the ground states and first excited states at $s=1$) levels out of $256$.

We also consider the collective dephasing model, where all the qubits are coupled to a collective bath with the same coupling strength $g$, which preserves the spin symmetry. In this case, the interaction Hamiltonian becomes
\begin{align}
    H_{SB}^{\text{col}} = g S^z \otimes B,
\end{align}
where
\begin{align}
    S^z = \sum_{i} \sigma_i^z, \quad B = \sum_k \left(b_{k}^\dagger + b_{k} \right).
\end{align}
With this assumption, we can group together the Lindblad operators corresponding to different qubit $i$ into a single one:
\begin{equation}
   L_{\omega}(t) = \sum_{\epsilon_b-\epsilon_a = \omega} | \varepsilon_a(t) \rangle \langle \varepsilon_a(t) | S^z | \varepsilon_b(t) \rangle \langle \varepsilon_b(t) | \,.
\end{equation}
The resulting number of Lindblad operators is a factor of $N$ smaller than that of the independent system-bath coupling model.

The total success probability at the end of the $r$ cycles is
\begin{equation}
p(r) = \bra{\text{up}}\rho(2rt_{\text{inv}})\ket{\text{up}} + \bra{\text{down}}\rho(2rt_{\text{inv}})\ket{\text{down}} \,.
\end{equation}

Any relaxation during the reverse annealing dynamics to the global instantaneous ground state or the instantaneous first excited state of $H(t)$ is beneficial, the latter since it becomes degenerate with the ground states $\{\ket{\text{up}},\ket{\text{down}}\}$ of $H_0$ at the end of the anneal (see App.~\ref{append:spectrum}).

\subsubsection{Dependence on annealing time}
\label{sec:AME-tau}

\begin{figure}[h!]
\subfigure[]{\includegraphics[width=0.4939\columnwidth]{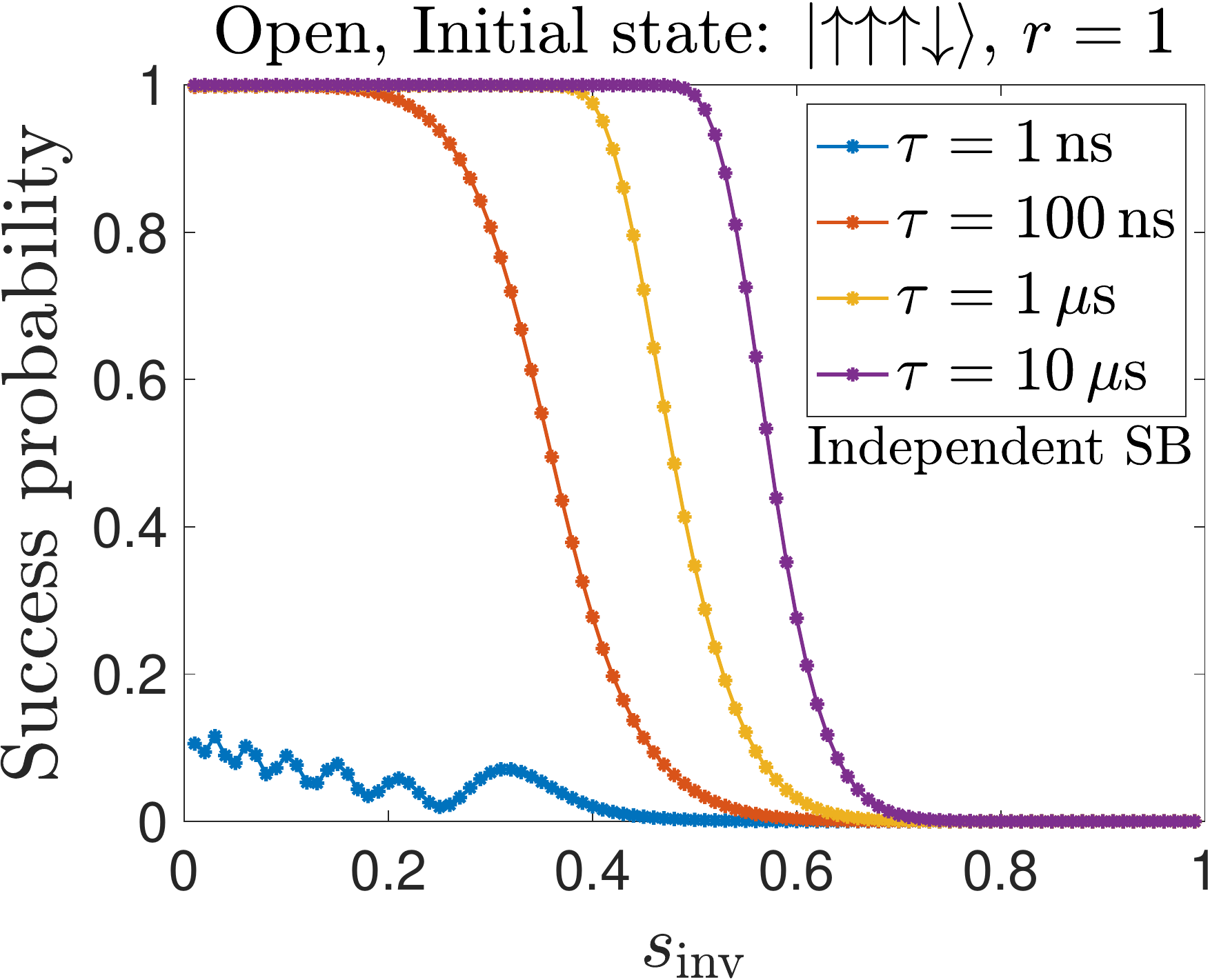}\label{fig:4qubitopena}}
\subfigure[]{\includegraphics[width=0.4939\columnwidth]{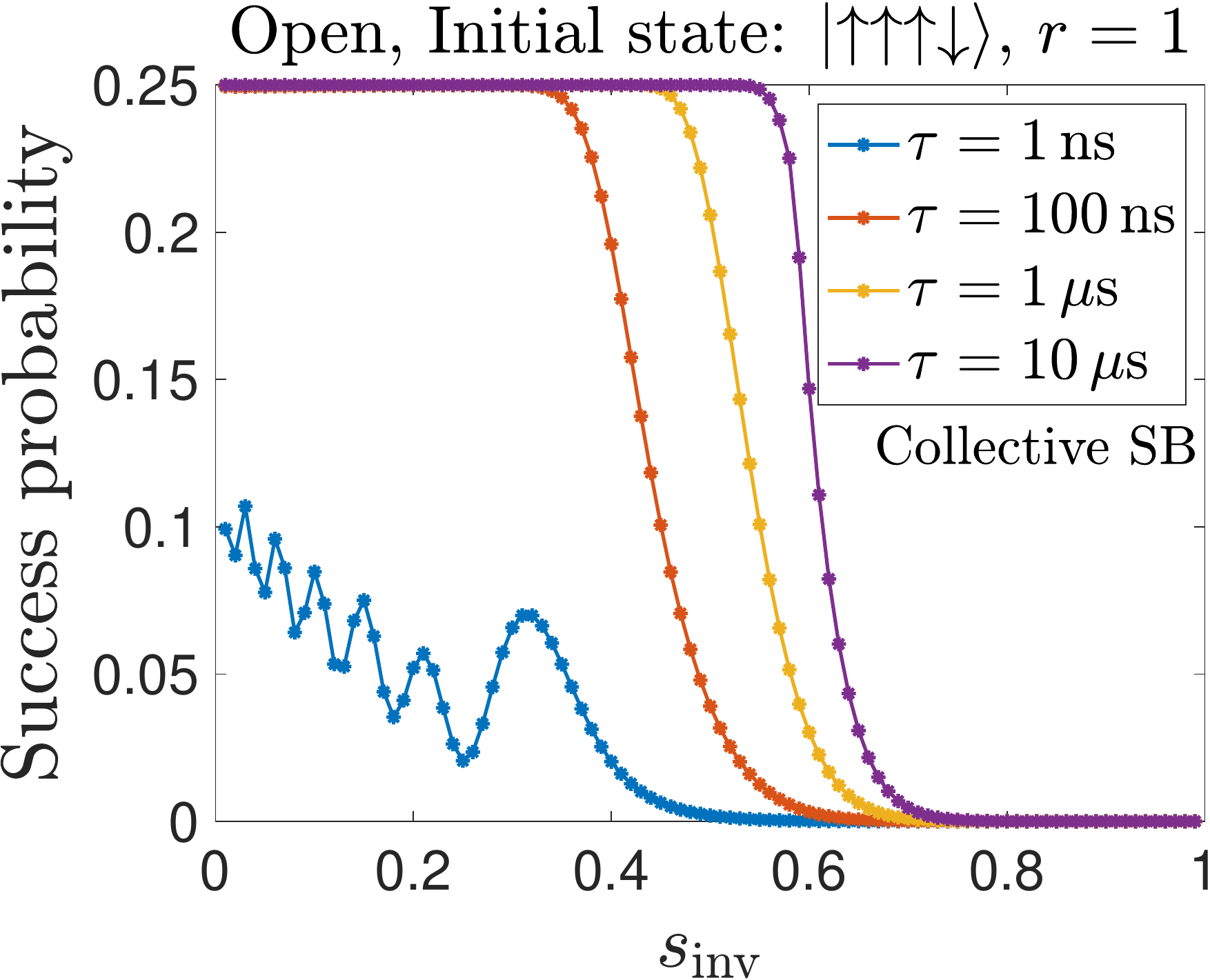}\label{fig:4qubitopenb}}
\subfigure[]{\includegraphics[width=0.4939\columnwidth]{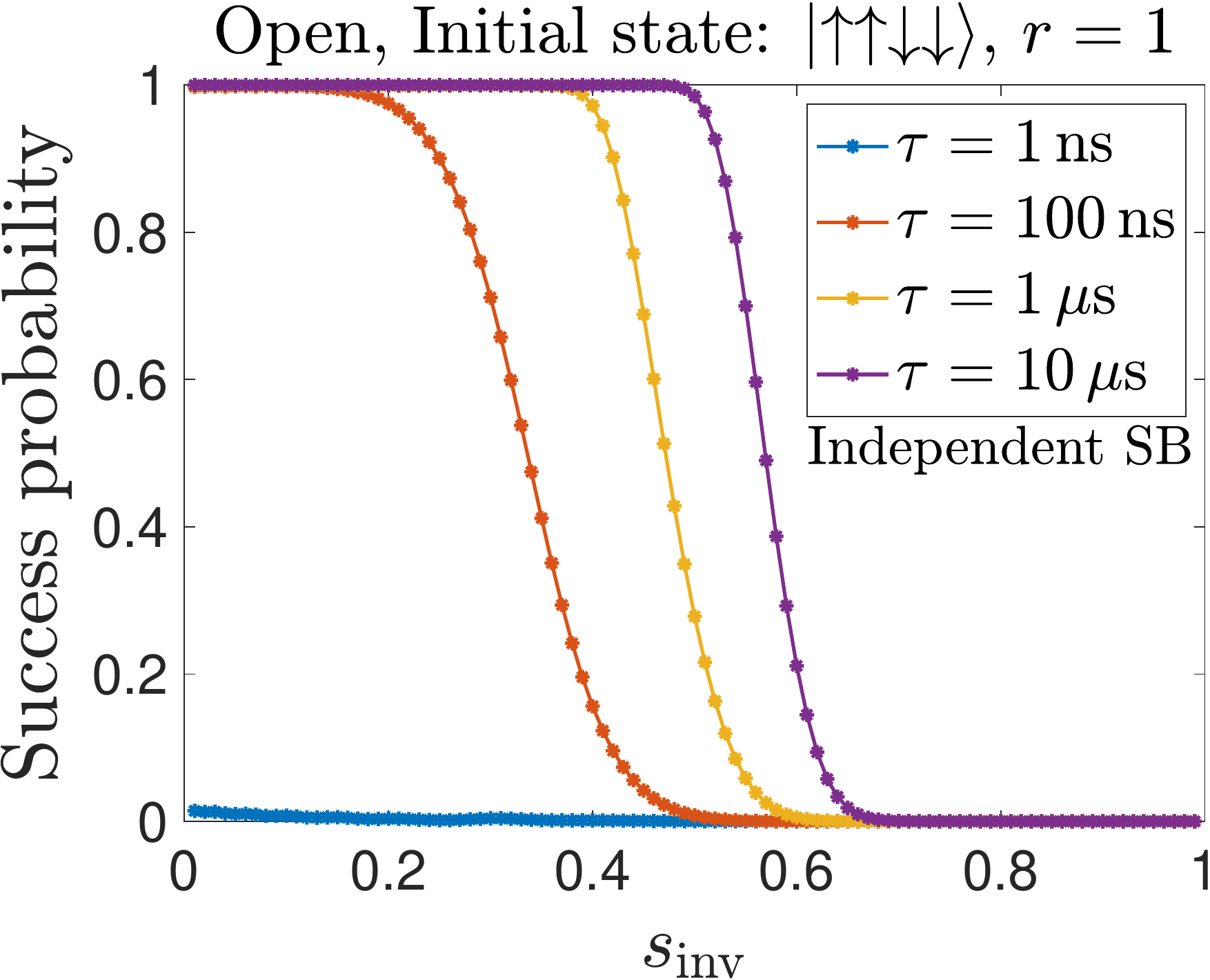}}
\subfigure[]{\includegraphics[width=0.4939\columnwidth]{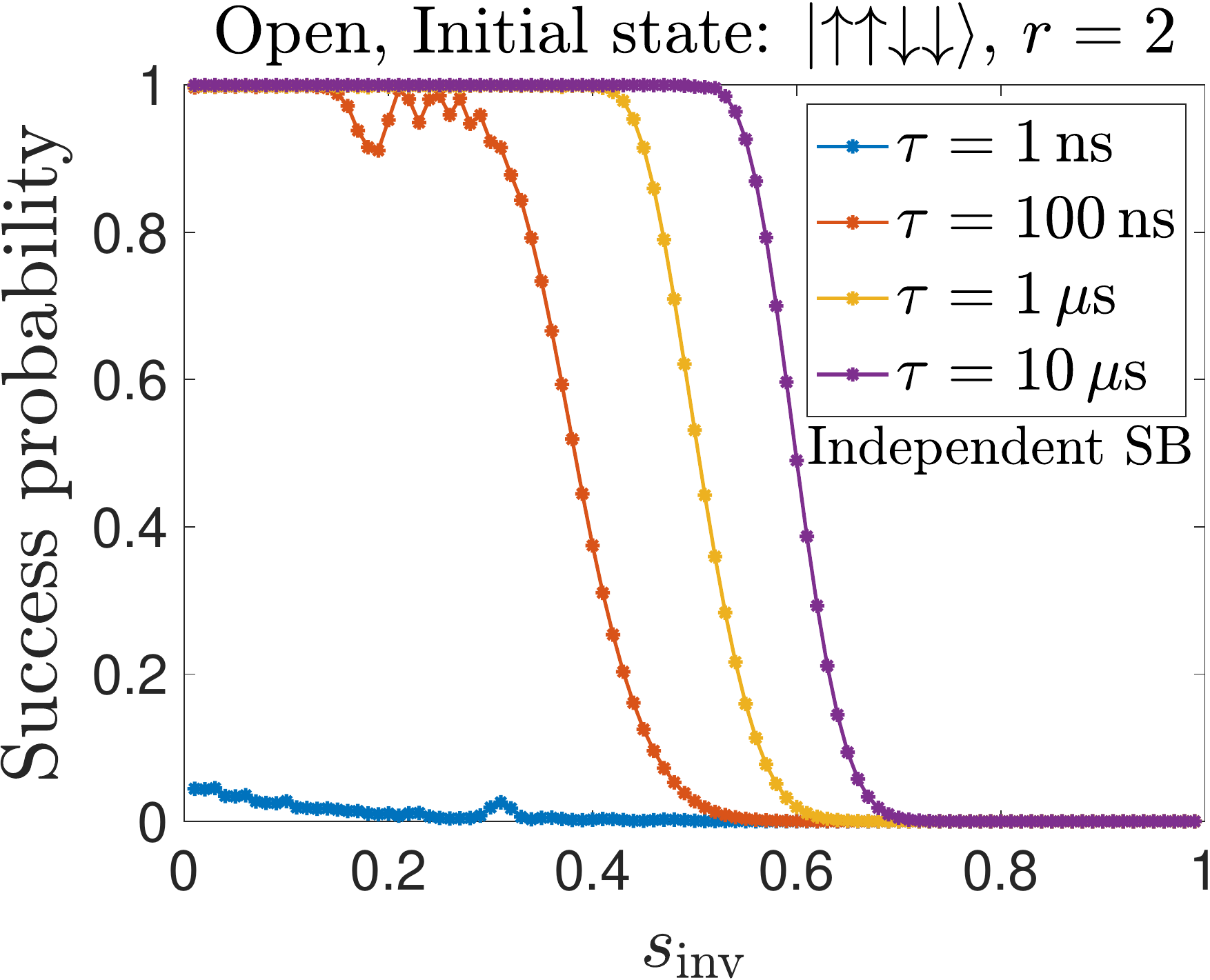}}
\caption{Top row: Success probability as a function of $s_{\mathrm{inv}}$ with (a) independent and (b) collective dephasing (SB denotes system-bath coupling). Initial state: $\ket{0001}~(m_0=0.5)$ and $r=1$, a single cycle. Bottom row: Success probability with different $\tau$. Initial state: $\ket{0011}~(m_0=0)$. (a) $r=1$, (b) $r=2$.}
\label{fig:4open_0001_r1}
\end{figure}

Figure~\ref{fig:4open_0001_r1} (top row) shows the simulation results for success probability as a function of $s_{\mathrm{inv}}$ with $N=4$, various $\tau$, and the initial state $\ket{0001}~(m_0=0.5)$, using the independent and collective dephasing models.
The independent dephasing model leads to a maximum success probability of $1$ for large $\tau$ and small $s_{\mathrm{inv}}$. However, the collective dephasing model leads to a maximum success probability of $1/4$. The reason is the same as in closed system simulations. For the initial state $\ket{0001}$, only $1/4$ of its population is in the subspace of maximum quantum spin number. 
However, the global (instantaneous) ground state and the first excited state of the annealing Hamiltonian $H(s)$ both belong to the maximum-spin subspace. Since the collective dephasing model preserves the spin symmetry and the dynamics is restricted to each subspace, at most $1/4$ of the initial population can be relaxed to the instantaneous ground state and the first excited state, and reach the correct solutions at the end of the anneal (see App.~\ref{append:successbound} for more details). It is also noteworthy that coherent oscillations are visible for $\tau=1$ [ns] (compare with Fig.~\ref{fig:close0001_r1}(a)), but not for the larger values of $\tau$ we have simulated. Recall that the experimental timescale in Sec.~\ref{sec:exp} is on the order of a [$\mu$s].

Comparing the simulation results in Fig.~\ref{fig:4open_0001_r1} and the experimental results in Fig.~\ref{fig:RA_20_diff_tau}(b), we observe that they share the same main features. Namely, the success probability increases with $\tau$; and, as $\tau$ increases the maximum success probability can be achieved with a larger inversion point $s_{\text{inv}}$. Most notably, the total success probability also drops to zero for sufficiently large $s_{\text{inv}}$ in our simulations. This is the freeze-out effect that is well captured by the adiabatic master equation (as first reported in Ref.~\cite{ABLZ:12-SI}), since  thermal relaxation is suppressed when the transverse field magnitude becomes so small that the system and system-bath Hamiltonians effectively commute.

Note that the experimental results in Sec.~\ref{sec:exp} have a maximum success probability as high as $1$, while our closed system simulations and open-system simulations with the collective dephasing model have success probabilities upper bounded by some constants $<1$, as already discussed. The high total success probability observed in our experiments is evidence that, as expected, the dynamics in the D-Wave device do not preserve spin symmetry. Collective system-bath coupling cannot explain the experimental results, leaving independent system-bath coupling as the only candidate consistent with the experiments according to our simulations.

\subsubsection{Dependence on initial condition and the number of cycles}
\label{sec:simulations-init}

For the initial state $\ket{0011}~(m_0=0)$ and with $r=1$, the independent dephasing model still gives a maximum success probability of $1$ as seen in Fig.~\ref{fig:4open_0001_r1}(c). The collective dephasing model (not shown) has a maximum success probability of $1/6$, since the initial state has $1/{{4}\choose{2}} = 1/6$ of its population in the maximum-spin subspace.

Comparing Fig.~\ref{fig:4open_0001_r1}(a) and (c), the dependence of the success probability on $s_{\text{inv}}$ is  similar to that of the initial state $\ket{0001}$. The $\tau = 1$ [ns] coherent oscillations visible for the latter are more attenuated for $\ket{0011}$, but this was also the case for the closed system simulations (contrast Fig.~\ref{fig:close0001_r1}(a) and Fig.~\ref{fig:close0011_r1}(a) at $\tau = 1$ [ns]). 
The main feature distinguishing Fig.~\ref{fig:4open_0001_r1}(a) and (c) is the shift to the left of the $\tau \geq 100$ [ns] curves for the initial state $\ket{0011}$, i.e., as $m_0$ is reduced from $0.5$ to $0$. This is consistent with our experimental results, as can be seen in Fig.~\ref{fig:RA_20_initial}(b).

In Fig.~\ref{fig:4open_0001_r1}(d), we consider the dependence on the number of cycles $r$, using the independent system-bath model. For $r=2$, we see that the results are similar to those of a single cycle. However, we do see a small improvement in the sense of a slight shift to the right of the curves with $\tau \geq 100$ [ns] compared with $r=1$, which is consistent with the experimental result shown in Fig.~\ref{fig:IRA}(d). We note that this improvement was not observed in the closed system case, as can be seen by contrasting Figs.~\ref{fig:close0011_r1}(a) and (c). Interestingly, there is also a small signature of coherent oscillations for $\tau \leq 100$ [ns].

Finally, we note that compared to the closed systems case, the results depend much less on the initial condition. 

\subsubsection{Size dependence and partial success probability}

\begin{figure}[h!]
\centering
\includegraphics[width=0.45\columnwidth]{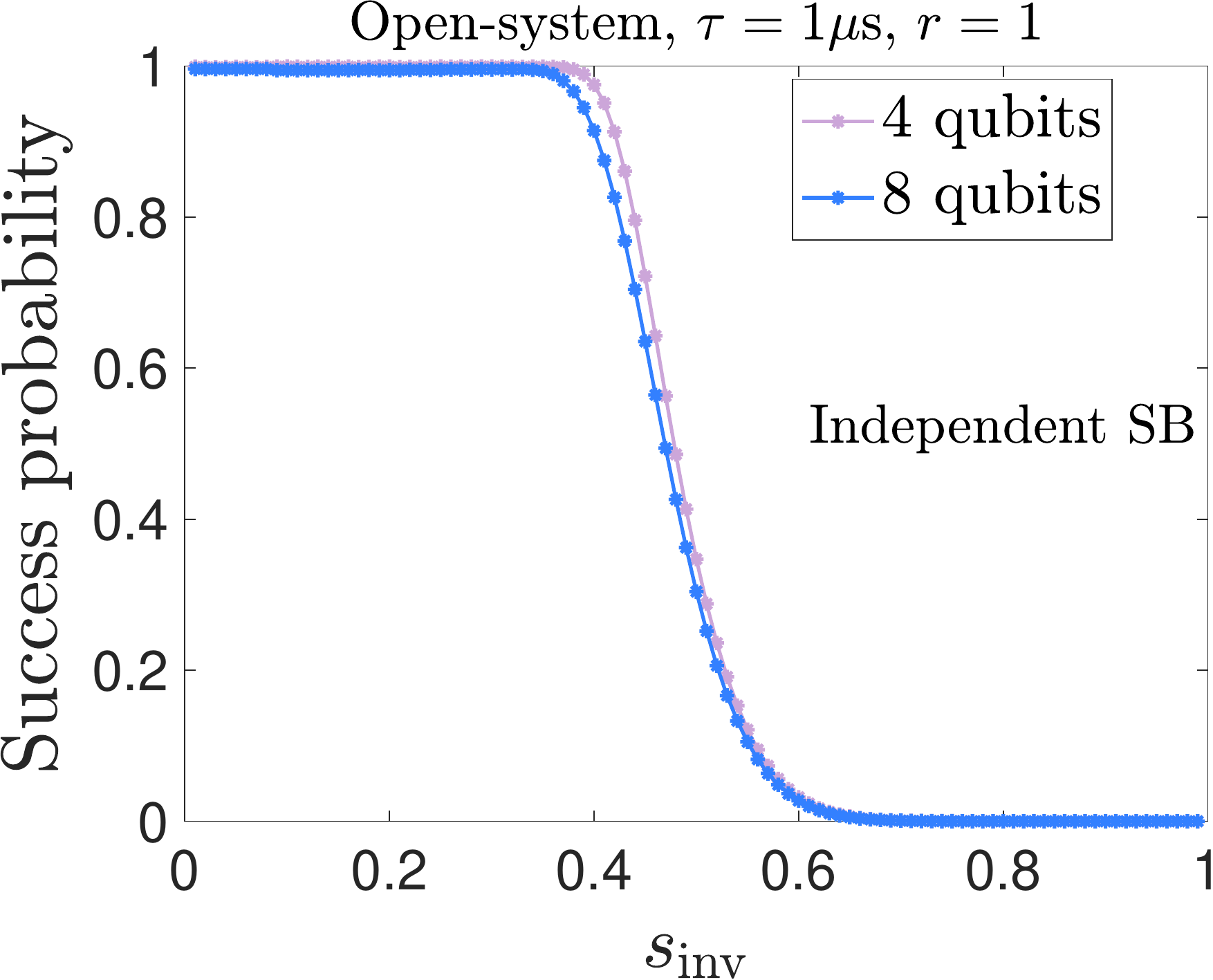}
\caption{Success probability as a function of $s_{\mathrm{inv}}$ for independent dephasing. The initial state has one spin down and $\tau$ is $1 [\mu$s]. The main plot shows the results for $N=4$ and $8$. The inset shows the partial success probabilities for $N=8$.}
\label{fig:4vs8open}
\end{figure}

In Fig.~\ref{fig:4vs8open}, we show the success probability for two different system sizes $N=4$ and $8$. The initial state has a single spin down. We observe that the results do not depend much on the system size, with the total success probabilities slightly larger (shifted to the right) for $N=4$. This is consistent with the experimental results for $N=4$ and $8$ shown in Fig.~\ref{fig:RA_20_size}(b).

While the total success probabilities of our open-system simulations with independent dephasing are generally consistent with the experimental total success probabilities, our open-system simulations always produce symmetric partial success probabilities as shown in Fig.~\ref{fig:4vs8open}, in stark contrast with the experiments, where the all-up state is strongly favored over the all-down state in the region around the minimum gap (see all the top panels in the figures in Sec.~\ref{sec:exp}). Clearly, this reflects a significant failure of the AME model.

One reason for this discrepancy is that our simulations do not include a mechanism such as the polarized spin-bath which we believe explains the experimentally observed asymmetry for small $s_{\mathrm{inv}}$ (see Sec.~\ref{sec:spinbathpol}). Moreover, our adiabatic master equation simulations result in a symmetric final all-up and all-down population 
since any relaxation event during the anneal (at any $s$) to either the global instantaneous ground state $\ket{\epsilon_0(s)}$ or the instantaneous first excited state $\ket{\epsilon_1(s)}$ of $H(s)$, which become degenerate at $s=1$, 
eventually populates $\ket{\epsilon_0(s=1)} = ({\ket{\uparrow}^{\otimes N} + \ket{\downarrow}^{\otimes N}})/{\sqrt{2}}$ or $\ket{\epsilon_1(s=1)} = ({\ket{\uparrow}^{\otimes N} - 
\ket{\downarrow}^{\otimes N}})/{\sqrt{2}}$. These two states 
have equal all-up and all-down populations. 
Similar observations are also found in Ref.~\cite{albash2015consistency}. 

\section{Semi-classical simulations}
\label{sec:classical}

In an effort to understand the asymmetric partial success probability observed in our experiments, we performed semi-classical simulations of the same problem using the spin-vector Monte Carlo (SVMC)~\cite{shin2014quantum} algorithm and a new variant with transverse-field-dependent
updates (SVMC-TF)~\cite{Albash2020ComparingRM}. For the $p$-spin problem, we replace the Hamiltonian of Eq.~\eqref{eq:H} by a classical Hamiltonian:
\begin{equation}
\mathcal{H}(s) = -\frac{A(s)}{2} \left(\sum_{i}^{N} \sin \theta_i\right) - \frac{B(s) N}{2}\left(\frac{1}{N}\sum_{i}^{N} \cos \theta_i\right)^{p}\,.
\label{eq:Hising}
\end{equation}
Each qubit $i$ is replaced by a classical $O(2)$ spin $\vec{M}_i=(\sin\theta_i,0,\cos\theta_i)$, $\theta_i \in [0, \pi]$. For the purpose of reverse annealing, we again need to specify the $t$ dependence of $s(t)$. The concept of time $t$ is here replaced by the number of Monte Carlo sweeps. we replace $\tau$ by a specified number of total sweeps. The total number of sweeps is then $2\tau(1-s_{\text{inv}})$, in analogy to the total annealing time in Eq.~\eqref{eq:t_a}. 

To simulate the effect of thermal hopping through this semi-classical landscape with inverse temperature $\beta$, we perform at each time step a spin update according to the Metropolis rule. In SVMC, a random angle $\theta_i^{'} \in [0, \pi]$ is picked for each spin $i$. Updates of the spin angles $\theta_i$ to $\theta_i^{'}$ are accepted according to the standard Metropolis rule associated with the change in energy ($\Delta E$) of the classical Hamiltonian. For the $p$-spin problem, $\Delta E$ cannot be expressed in a simple form as in case of the Ising problem Hamiltonian~\cite{shin2014quantum}. 

In SVMC-TF, the random angle $\theta_i^{'} =\theta_i + \epsilon_i(s)$ is picked in a restricted range
\begin{align}
    \epsilon_i(s) \in \left[-\min \left(1,\frac{A(s)}{B(s)}\right) \pi, \min \left(1,\frac{A(s)}{B(s)}\right) \pi\right]. 
\end{align}
The goal of SVMC-TF is to restrict the angle update for $A(s) < B(s)$, and imitate the freeze-out effect discussed in Sec.~\ref{sec:freezeout}. The full SVMC and SVMC-TF algorithms for reverse annealing are summarized in App.~\ref{append:svmc}, including the expression for $\Delta E$.



\subsection{Total success probability}
We report on SVMC and SVMC-TF simulations for $N=4$ and $8$. We choose the initial condition with a single spin down. In terms of angles, for $N=4$ the initial angles are $\{0,0,0,\pi\}$. We again use the temperature $T=12.1$ [mK]. The classical analogue of the annealing time is chosen to be $\tau = 10^3$ and $10^4$ sweeps. 

\begin{figure}[h!]
\subfigure[]{\includegraphics[width = 0.4939\columnwidth]{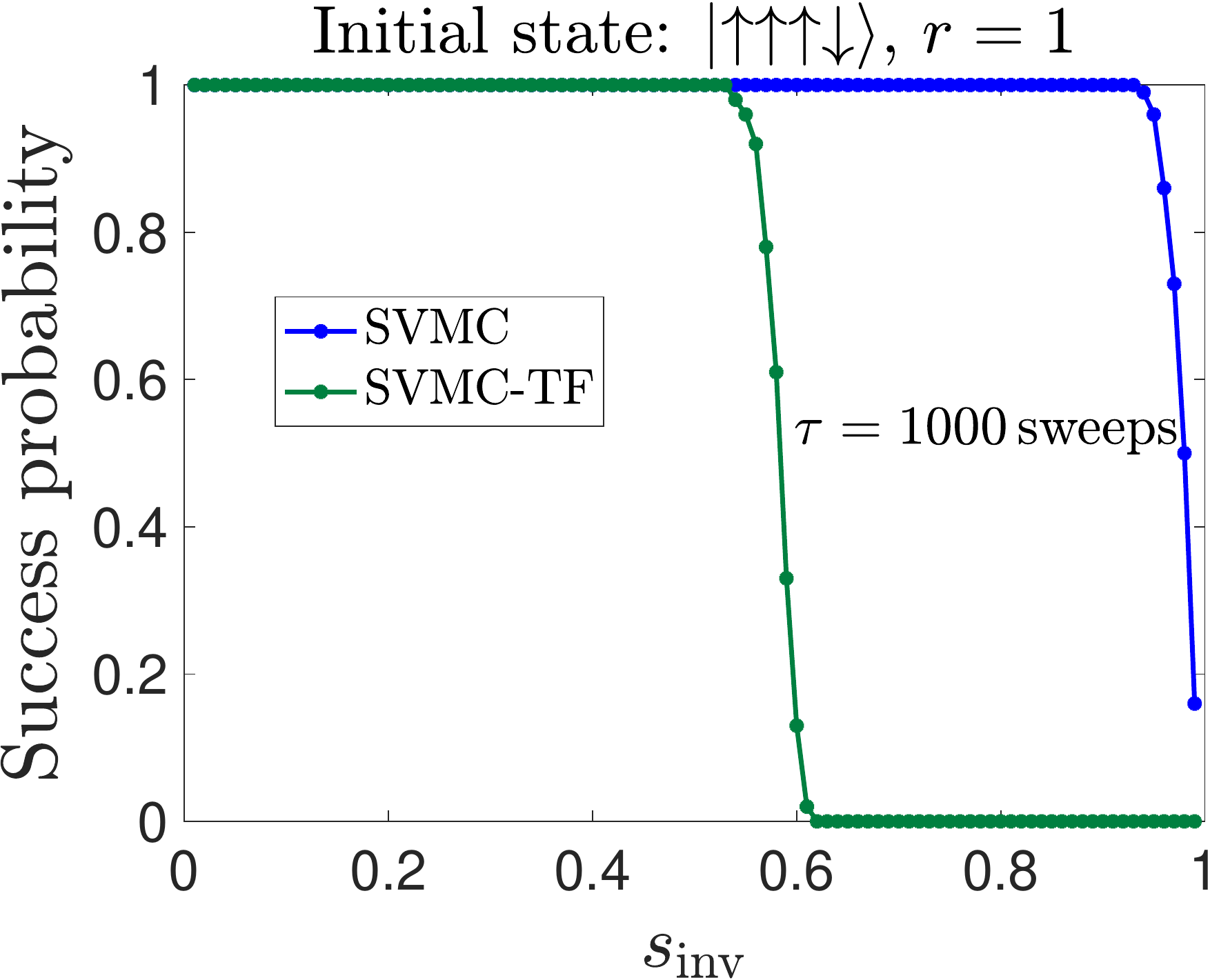}}
\subfigure[]{\includegraphics[width = 0.4939\columnwidth]{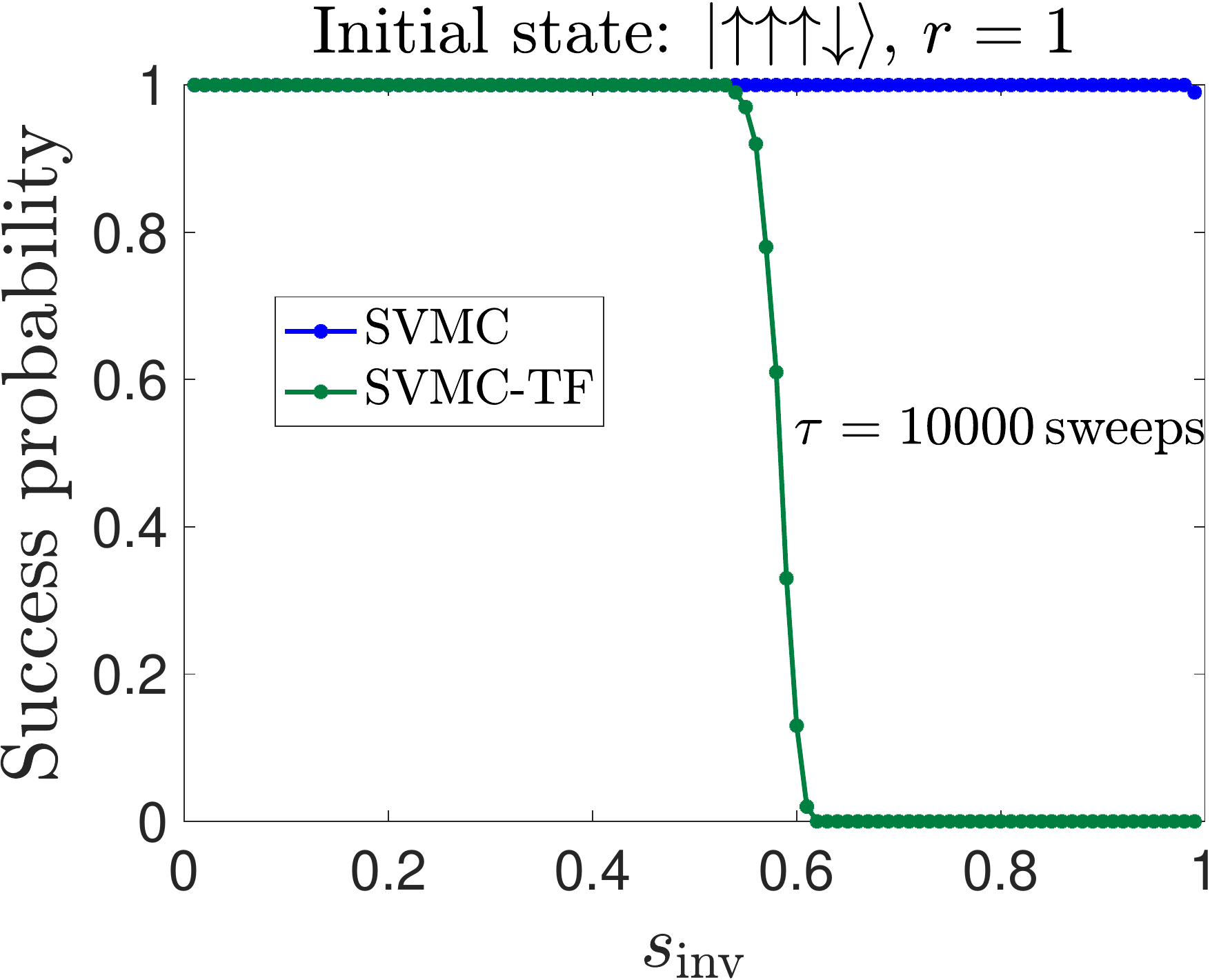}}
\caption{Total success probability for SVMC and SVMC-TF. (a) $\tau = 10^3$ sweeps, (b) $\tau = 10^4$ sweeps. Error bars are 2$\sigma$ over $10^4$ samples.}
\label{fig:comparealltotal}
\end{figure}

In Fig.~\ref{fig:comparealltotal} we display the simulation results for the total success probability using SVMC and SVMC-TF. The number of samples is $10^4$.
SVMC gives high total success probabilities even with large inversion points $s_{\text{inv}}$. A large $s_{\text{inv}}$ value means that during the whole reverse annealing process the ratio $A(s)/B(s)$ is small. In the D-Wave device, it is expected that for small $A(s)/B(s)$ the dynamics freezes, which makes it difficult to reach the ground state when the initial state is excited. With the number of sweeps increased from $\tau=10^3$ to $10^4$, we observe that even for very large inversion points $s_{\text{inv}}$ the total success probability of SVMC can be as high as $1$. This is because the angle updates in SVMC are completely random and thus with a sufficient number of sweeps, it is possible for the state to flip to the correct solution.

However, in SVMC-TF, the range of angle update is restricted for $A(s)/B(s) < 1$. The restricted angle updates (freeze-out effect) prevent the state from flipping to the correct solutions. Therefore the total success probability for large inversion points $s_{\text{inv}}$ is basically zero in SVMC-TF, regardless of the number of sweeps. This is also what we observe in the experimental data and in the adiabatic master equation simulations, as discussed in Sec.~\ref{sec:AME-tau}.

\subsection{Partial success probability}
\label{sec:PSP-SVMC}

\begin{figure}[h!]
\subfigure[]{\includegraphics[width = 0.4939\columnwidth]{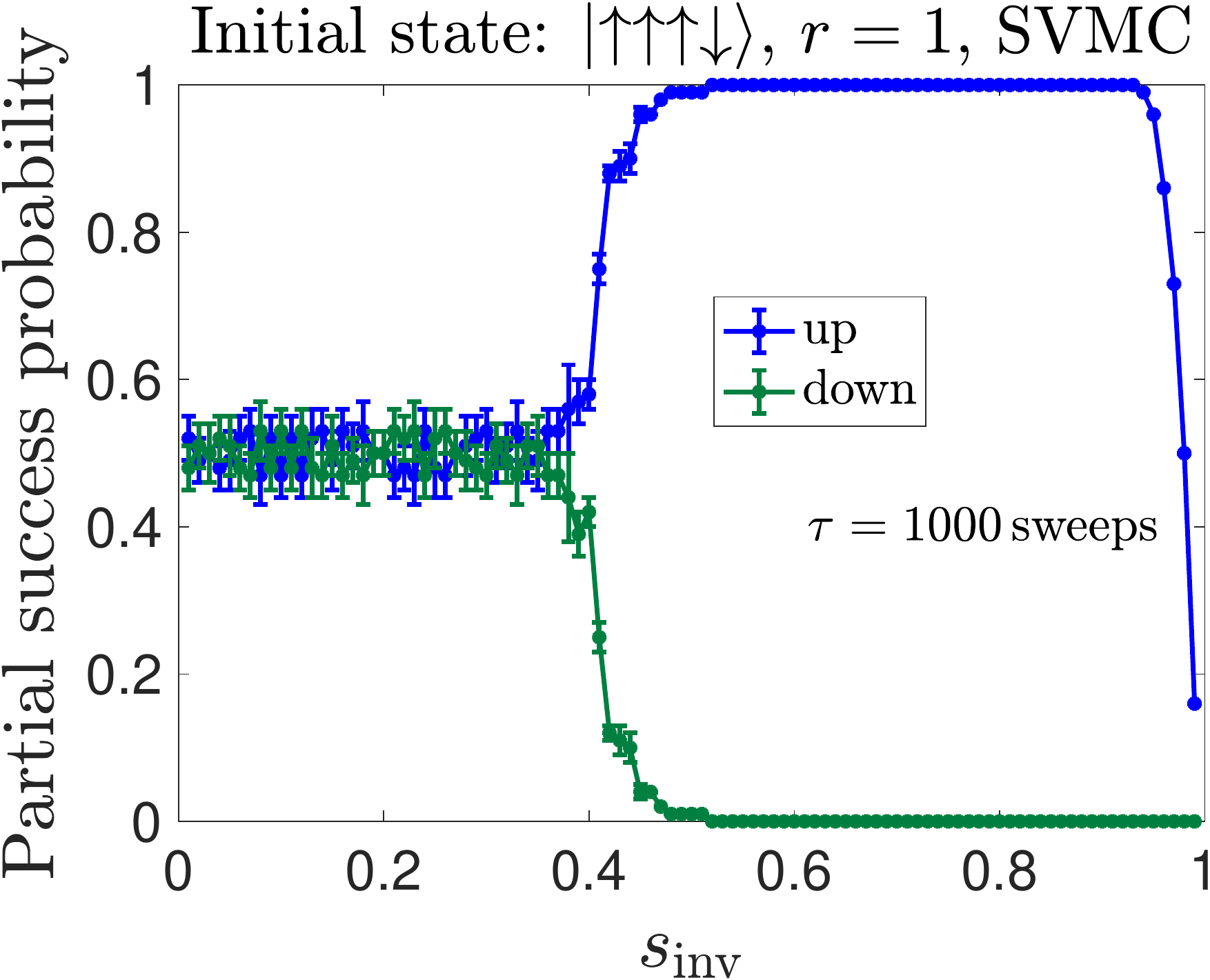}}
\subfigure[]{\includegraphics[width = 0.4939\columnwidth]{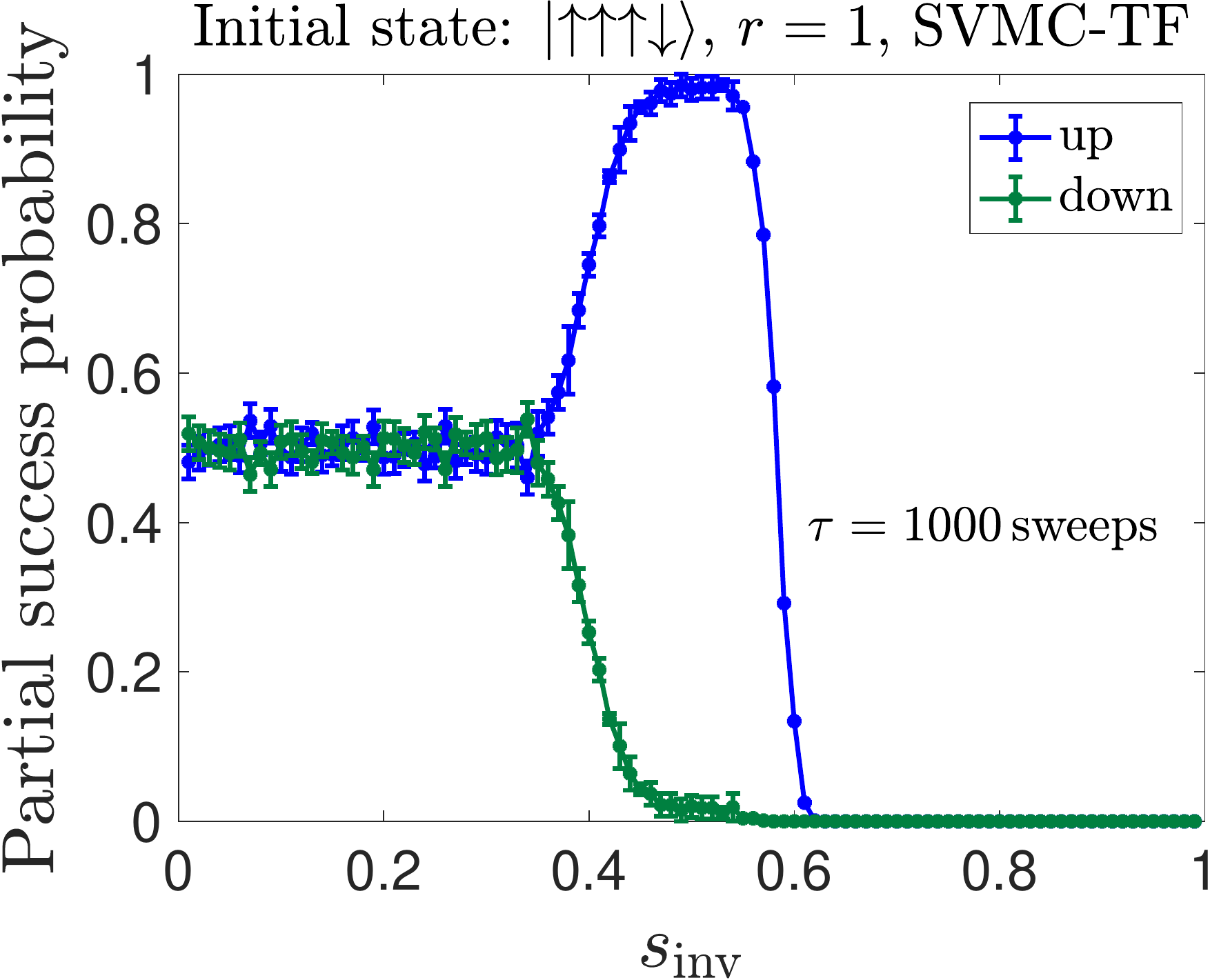}}
\subfigure[]{\includegraphics[width=0.4939\columnwidth]{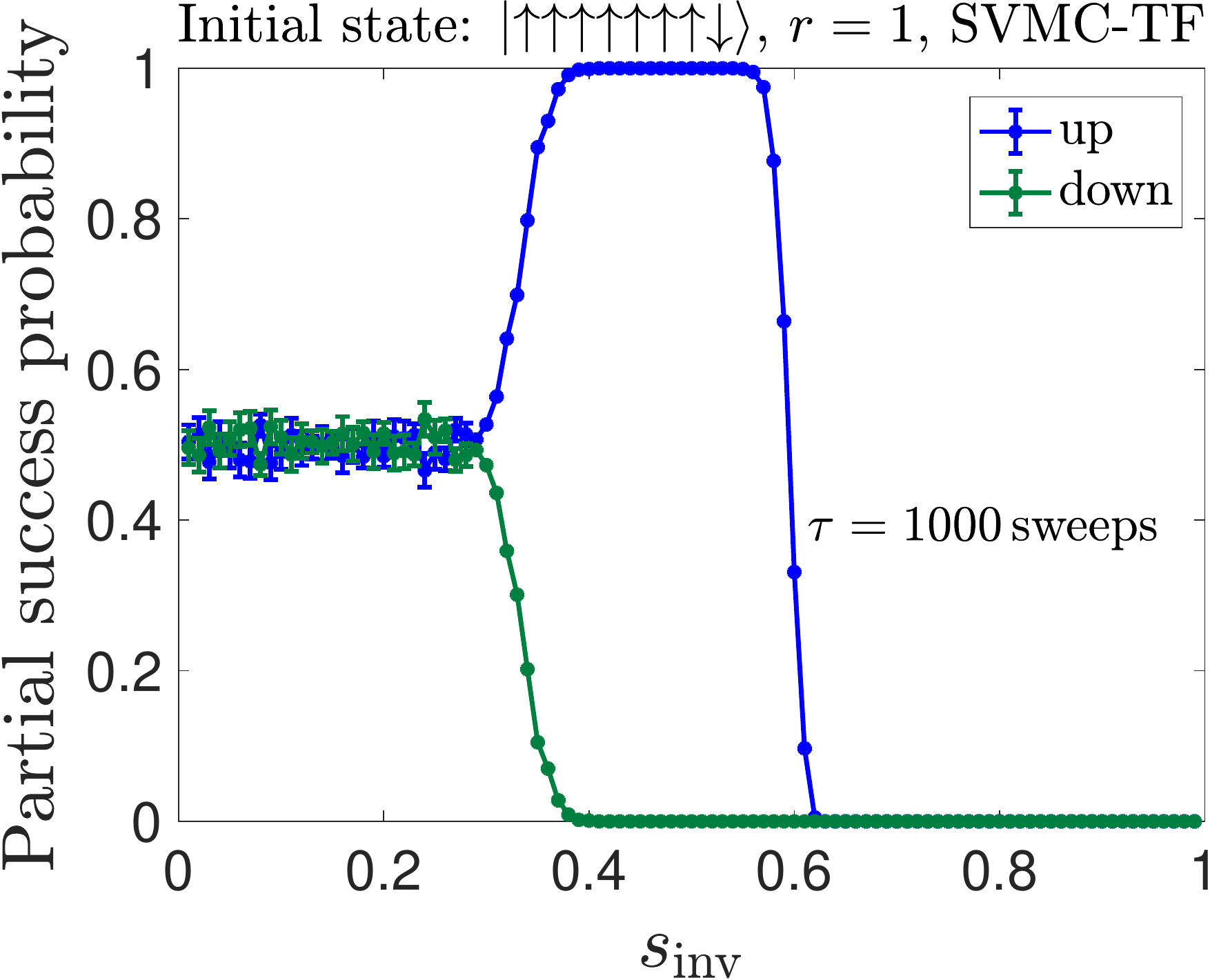}}
\caption{Partial success probability by (a) SVMC, $N=4 $and (b) SVMC-TF, $N=4$. $\tau = 10^3$ sweeps. (c) SVMC-TF, $N=8$. $\tau = 10^3$ sweeps. Error bars are 2$\sigma$ over $10^4$ samples.}
\label{fig:compareallpartial}
\end{figure}

We compare the partial success probability obtained from SVMC and SVMC-TF in Figs.~\ref{fig:compareallpartial}(a) and (b), respectively, for $N=4$ and $\tau = 10^3$. The SVMC-TF result for $N=8$ is plotted in Fig.~\ref{fig:compareallpartial}(c). Comparing the experimental data in Fig.~\ref{fig:RA_20_size}(a) and Fig.~\ref{fig:compareallpartial}, we observe a satisfactory agreement with SVMC-TF, in particular the correct trend for the unequal up and down partial success probabilities. Also, for a limited region of $s_{\text{inv}}$, the partial success probabilities of the all up states can be as high as $1$.
Classical SVMC-TF simulations also explain the D-Wave results of random systems in Ref.~\cite{Albash2020ComparingRM}.

For $N=8$, we observe a deviation from the experimental data shown in Fig.~\ref{fig:RA_20_size}(a) for $s_{\text{inv}} \lesssim 0.3$, where there exists a small but clear difference in the probabilities of all-up and all-down states, whereas the SVMC-TF data do not show such a trend. As discussed in Sec.~\ref{sec:spinbathpol}, we attribute the difference for $s_{\text{inv}} \lesssim 0.3$ to the spin-bath polarization effect, which is not modeled in our SVMC-TF simulations.

\begin{figure}[h!]
\subfigure[]{\includegraphics[width = 0.4939\columnwidth]{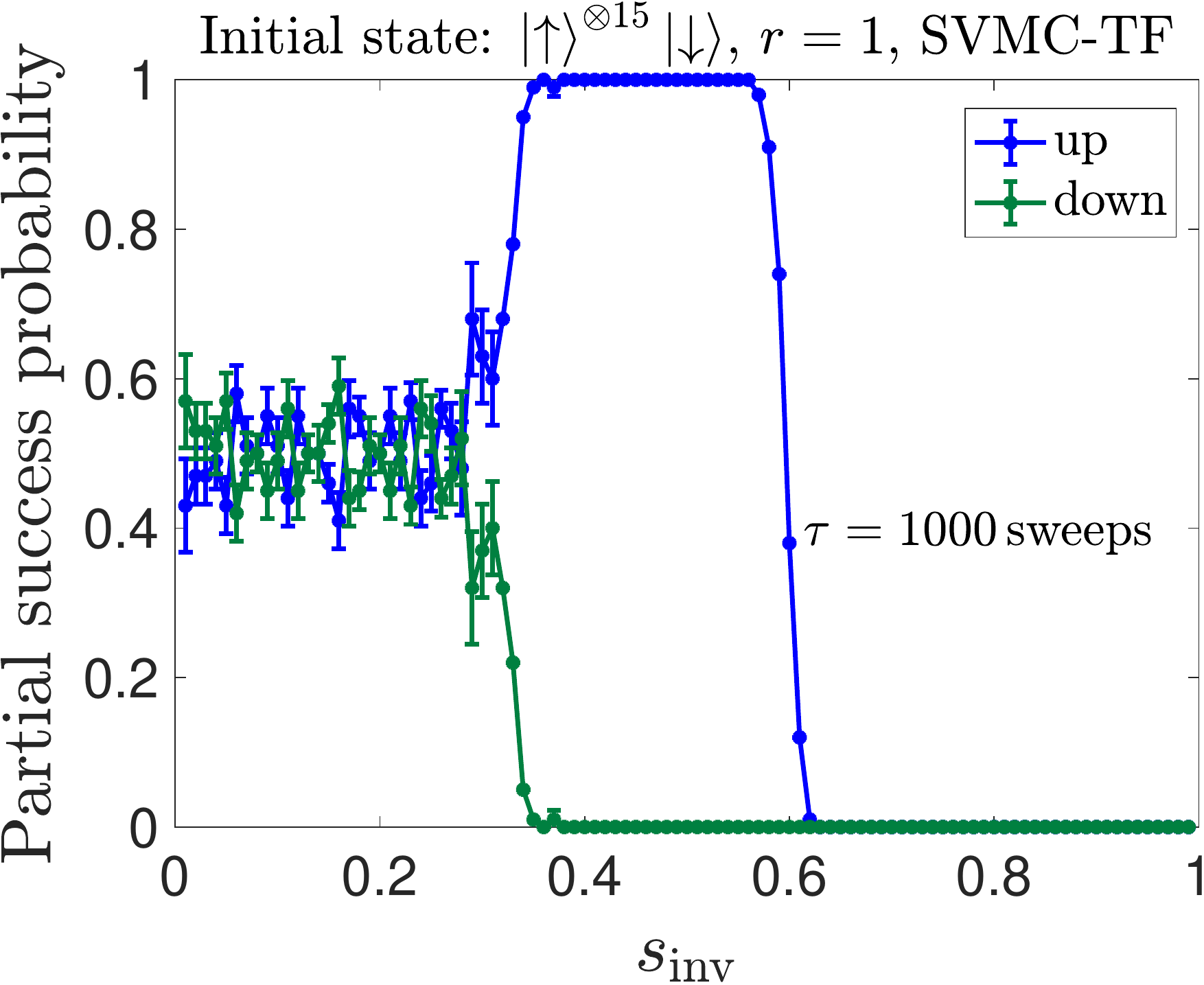}}
\subfigure[]{\includegraphics[width = 0.4939\columnwidth]{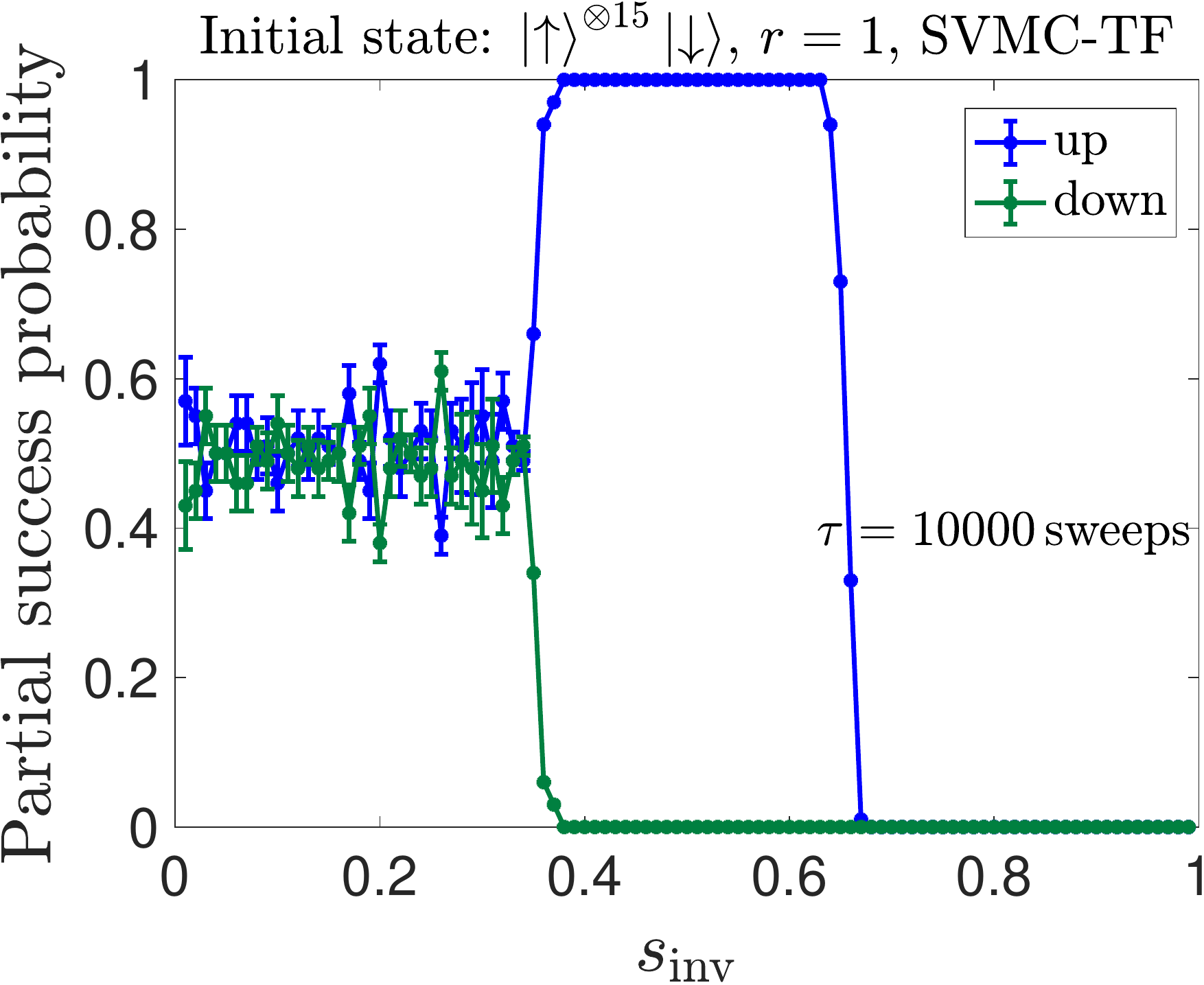}}
\subfigure[]{\includegraphics[width = 0.4939\columnwidth]{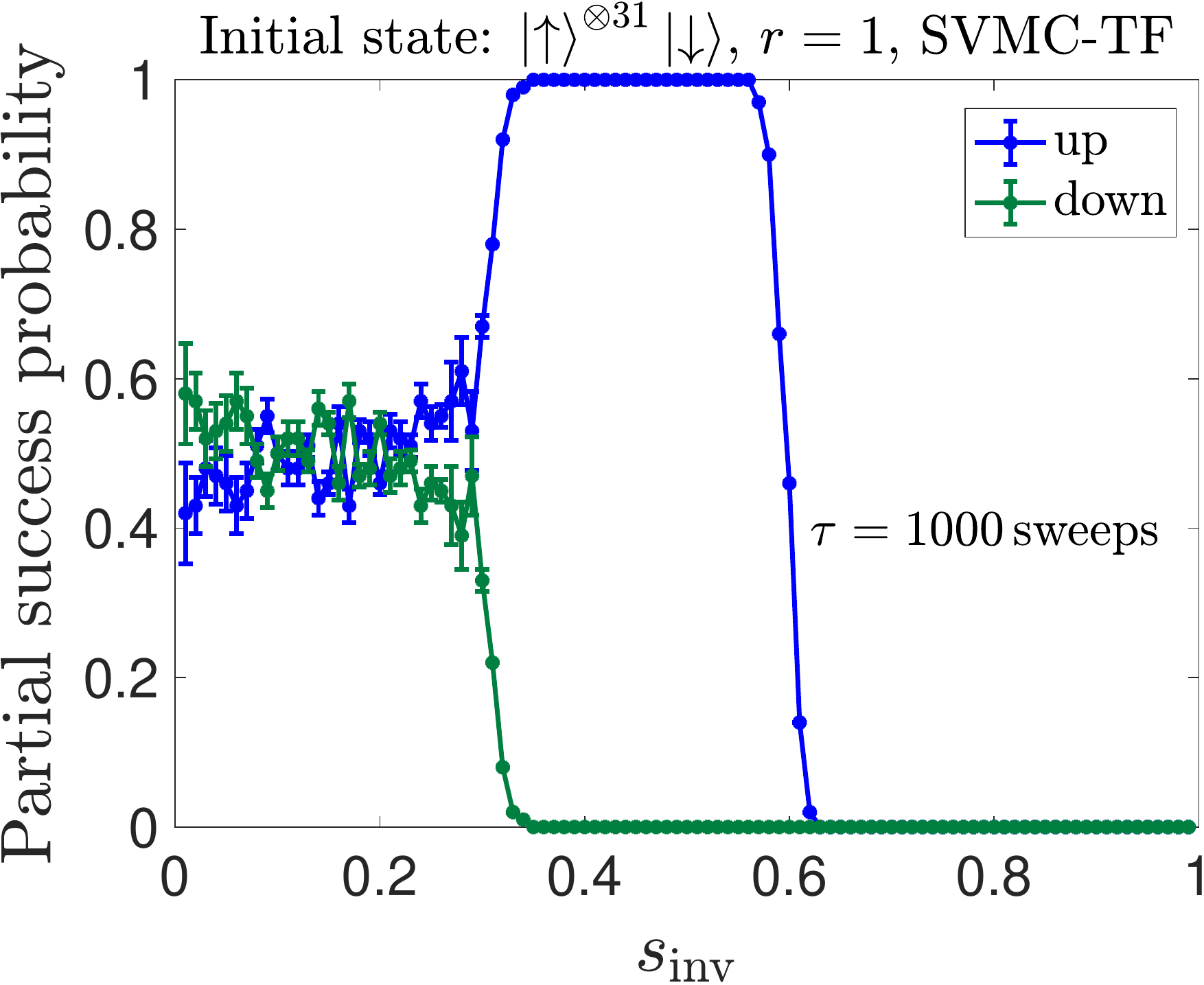}}
\subfigure[]{\includegraphics[width = 0.4939\columnwidth]{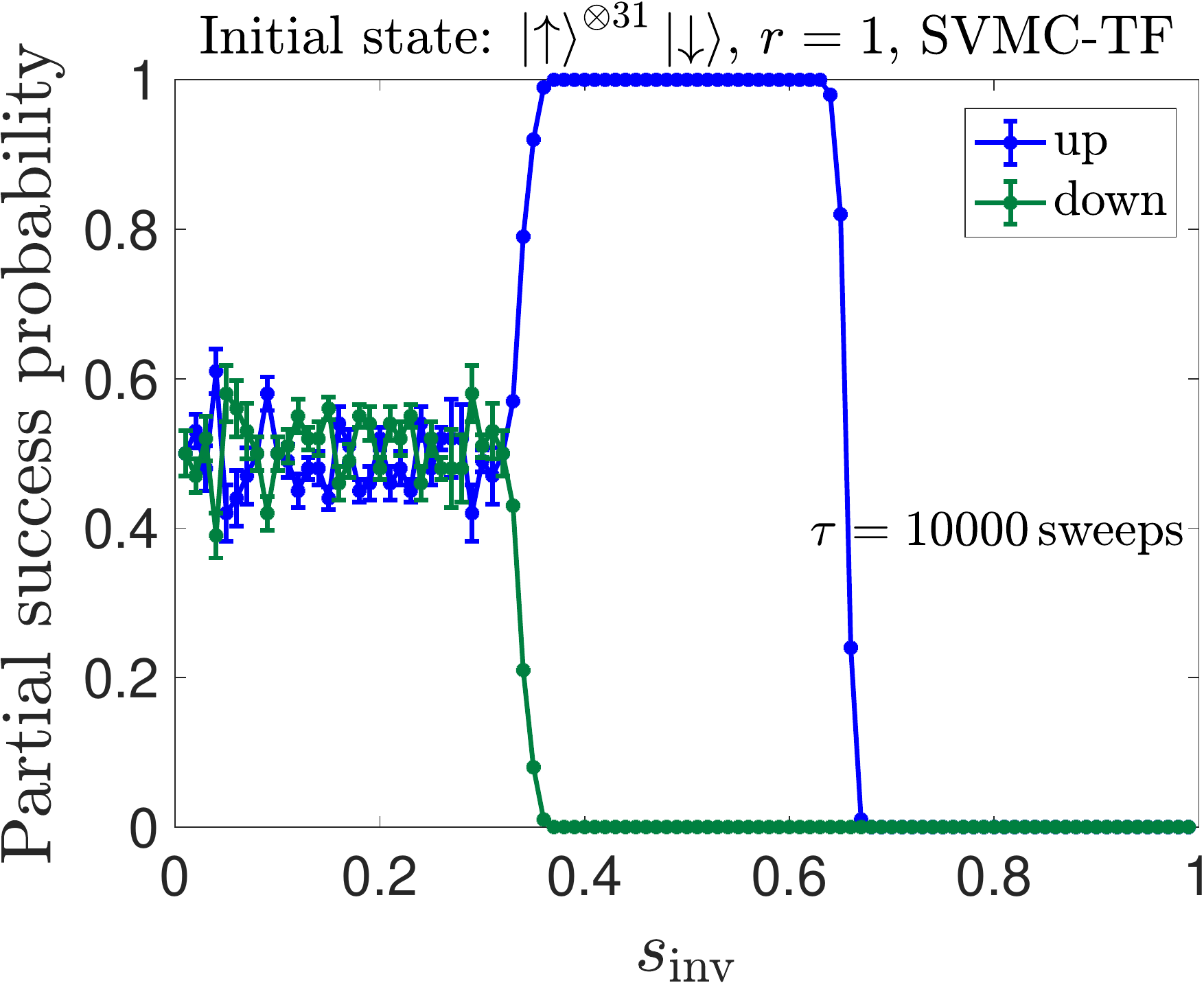}}
\caption{Partial success probability for SVMC-TF. (a) $N=16$, $\tau = 10^3$ sweeps, (b) $N=16$, $\tau = 10^4$ sweeps, (c) $N=32$, $\tau = 10^3$ sweeps, (b) $N=32$, $\tau = 10^4$ sweeps. Error bars are 2$\sigma$ over $10^4$ samples.}
\label{fig:partial16}
\end{figure}

In Fig.~\ref{fig:partial16} we display SVMC-TF reverse annealing simulation results of partial success probabilities for $N=16,32$ with $\tau = 10^3,10^4$ sweeps. For both sizes shown, the regime of high partial success probability for the all-up state is shifted slightly to higher $s_{\textrm{inv}}$ for $\tau = 10^4$ sweeps then for $\tau = 10^3$. This is consistent with the trend in the experimental results observed in Fig.~\ref{fig:RA_20_diff_tau}(a). However, the trend with system size is inconsistent with the experimental results shown in Fig.~\ref{fig:RA_20_size}(a): the results for $N=16$ and $N=32$ are virtually indistinguishable (apart from statistical fluctuations), which the experimental data shows that $N=16$ is not yet large enough for convergence. This (small) failure of the SVMC-TF model may hint at an interesting way in which to identify a ``quantum signature" in experimental quantum annealing~\cite{Boixo:2014yu,albash2015consistency}.

\section{\label{section:Conc}Conclusion of this chapter}
The previous chapter (Ref.~\cite{Passarelli2019}) has reported that the coupling with the environment improves the performance of reverse annealing for the $p$-spin model with $p=3$. In the present study, we have performed reverse annealing on a real device D-Wave 2000Q for the $p$-spin model with $p=2$ and have confirmed that the reverse annealing process yields high success probabilities as in the previous chapter (Ref.~\cite{Passarelli2019}).

We perform closed system simulations, and open-system simulations of quantum adiabatic master equation, which take into account the effect of decoherence in the instantaneous energy eigenbasis. The numerical simulations reveal that thermal relaxation indeed leads to high total success probabilities, which also match well with our experimental results. Experimental data of reverse annealing also in turn helps us understand the noise mechanism in D-Wave machines. Based on the experimental data, we make careful analysis on the simulation methods and the correct choice of system-bath coupling model.  We argue that reverse annealing would fail in a closed system setting or open system simulation with the assumption of collective dephasing. 

To further support the experimental results of asymmetrical partial success probabilities, we perform semi-classical simulations of spin-vector Monte Carlo. Our results show that the semi-classical thermal effect also benefits the performance of reverse annealing in real-world quantum annealing devices. We hope that this realization would be a step forward in the understanding and development of devices where thermal effects lead to quantum acceleration.

\section{Acknowledgments of this chapter}
All experimental data is provided by Yuki Bando. The experimental section is written by Yuki Bando. We thank Masayuki Ohzeki for useful comments and Sigma-i Co., Ltd. for providing machine time on D-Wave 2000Q.
The research is based partially upon work supported by the Office of the Director of National Intelligence (ODNI), Intelligence Advanced Research Projects Activity (IARPA) and the Defense Advanced Research Projects Agency (DARPA), via the U.S. Army Research Office contract W911NF-17-C-0050. The views and conclusions contained herein are those of the authors and should not be interpreted as necessarily representing the official policies or endorsements, either expressed or implied, of the ODNI, IARPA, DARPA, or the U.S. Government. The U.S. Government is authorized to reproduce and distribute reprints for Governmental purposes notwithstanding any copyright annotation thereon. Computation for some of the work described in this chapter was supported by the University of Southern California Center for High-Performance Computing and Communications (hpcc.usc.edu).

%% file: chapter6.tex
\chapter{Adiabatic reverse annealing}
\label{chap: ara}
We now move on to the study of another variant of reverse annealing called adiabatic reverse annealing, originally introduced in~\cite{perdomo:sombrero, nishimori:reverse-pspin, nishimori:reverse-pspin-2}, in an open system setting. We focus on the $p$-spin Hamiltonian with $p=3$.

\section{Introduction and terminology}
The annealing Hamiltonian for adiabatic reverse annealing (ARA) is~\cite{nishimori:reverse-pspin-2, Crosson2020}: 
\begin{equation}
\label{eq:H}
  H(s, \lambda) = sH_{0} + (1-s)(1- \lambda)H_{\text{init}} + \Gamma (1-s)\lambda V_{\text{TF}} \,.
\end{equation}

The time-dependent parameters $s(t)$ and $\lambda(t)$ both
change from $0$ to $1$ as time $t$ proceeds from $0$ to $\tau$, where
$\tau$ is the total anneal time. We focus on the case of a linear function $s(t) = t/\tau$.

The notation of Eq.~\eqref{eq:H} is as follows. 
\begin{equation}
H_{0} = -N\left(\frac{1}{N}\sum_{i=1}^{N} \sigma^{z}_i\right)^{p}
\end{equation}
is the target problem Hamiltonian, with $p=3$ and
$N$ the number of qubits. 
\begin{equation}
H_{\text{init}} = -\sum_{i=1}^{N}\epsilon_i \sigma^{z}_i
\end{equation}
is the initial Hamiltonian, $\epsilon_i = \pm 1$. 
\begin{equation}
V_{\text{TF}} = -\sum_{i=1}^{N} \sigma^{x}_i
\end{equation}
is the transverse field Hamiltonian.
We want to perform (adiabatic) reverse annealing with various initial states $\ket{\psi(s=0)} = \ket{\epsilon_1, \epsilon_2, \dots, \epsilon_N}$ (the ground state of $H_{\text{init}}$). 

We further assume that $\lambda = s$. With this assumption, Eq.~\eqref{eq:H} is simplified as:
\begin{equation}
\label{eq:H_simp}
  H(s) = sH_{0} + (1-s)^2 H_{\text{init}} + \Gamma (1-s)s V_{\text{TF}}  \,.
\end{equation}

In the special case that $\lambda(t) = 1$, 
Eq.~\eqref{eq:H} is the quantum annealing (QA)  Hamiltonian.

\section{
Preliminary I: Adiabatic reverse annealing (ARA) Hamiltonian spectrum}
\begin{figure}[h!]
\centering
\includegraphics[width=0.6\linewidth]{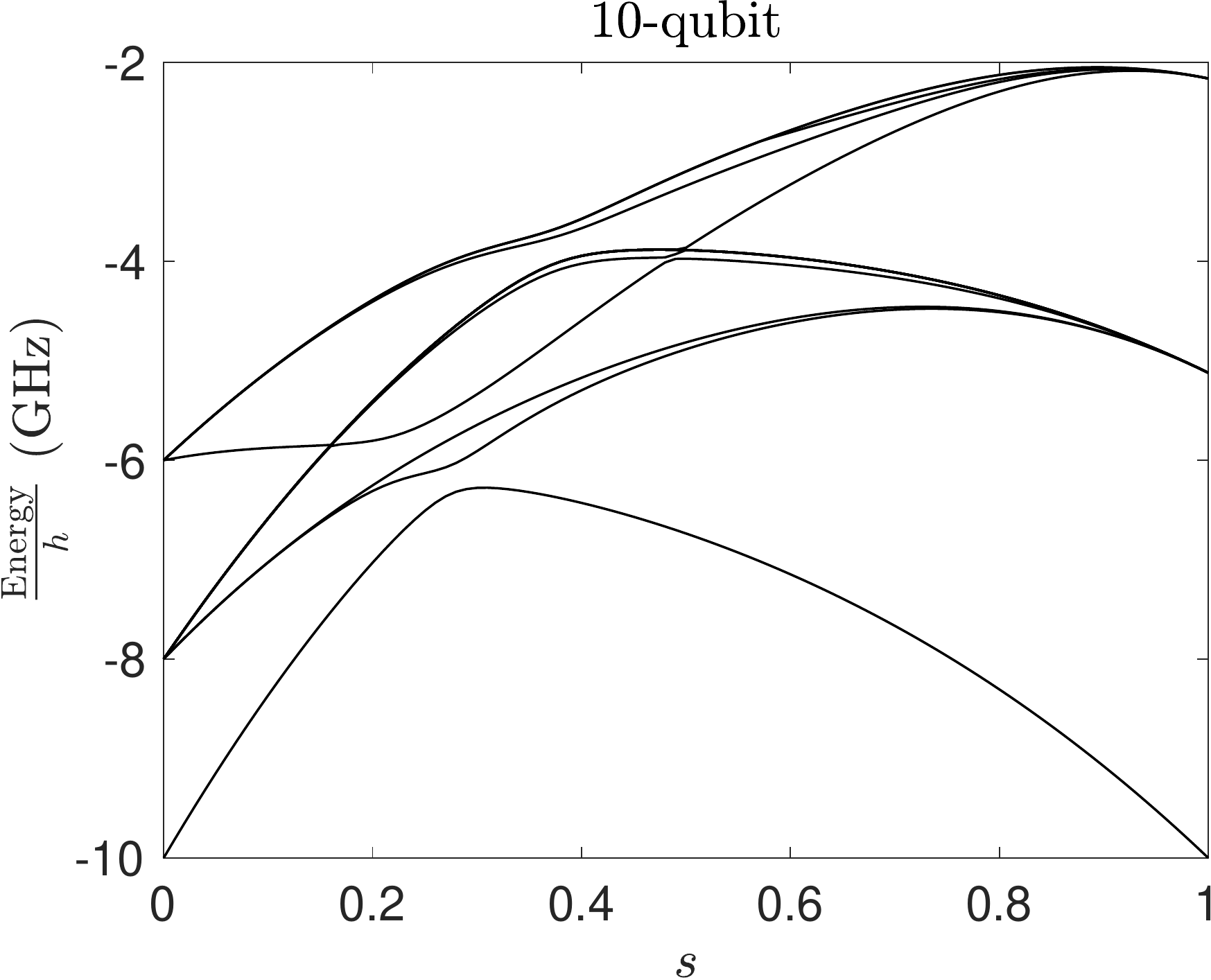}
\caption{The $16$ lowest lying energies of $10$ qubits, with $H_{\text{init}}$ having $N_\downarrow = 2$.
}
\label{fig: ARAspectrum}
\end{figure}

In this section, we want to explore how the spectrum changes by tuning certain parameters of Eq.~\eqref{eq:H_simp}. This is important in the calculations of diabatic transition rates, excitation and relaxation rates, and thus the understanding of open-system behavior of adiabatic reverse annealing (ARA).
The solution state of the $p$-spin problem with $N$ spins is $\ket{\uparrow}^{\otimes N}$. Consider the initial state with the number of spin-downs $N_\downarrow = 2$.
We plot the spectrum for $N=10$ in Fig.~\ref{fig: ARAspectrum}.

\subsection{Controlling the parameters: $\Gamma$}
The energy gap between instantaneous eigenenergies $\epsilon_{i}(s)$ and $\epsilon_{j}(s)$ is $\Delta_{ij}(s) = \epsilon_{i}(s) - \epsilon_{j}(s)$. The corresponding minimum energy gap is $\Delta_{ij} = \min_{s} \Delta_{ij}(s)$.

In adiabatic reverse annealing, we are interested in the minimum energy gap between the ground state $\epsilon_{0}(s)$ and the $1$st excited state $\epsilon_{1}(s)$. It is denoted by $\Delta$ ($=\Delta_{\min})$.  
\begin{equation}
    \Delta = \Delta_{10} = \min_{s} \Delta_{10}(s) = \min_{s} \epsilon_{1}(s) - \epsilon_{0}(s) \,.
\end{equation}

Consider the initial state with $N_{\uparrow}=1$. We want to explore how the gap properties change with $\Gamma$ in Eq.~\eqref{eq:H_simp}.
We calculate and plot in Fig.~\ref{fig: gammaplot} the value of $\Delta_{10}(s)$, for $N=8$ and $\Gamma = \{1, \cdots, 5\}$. In the inset we plot the dependence of $\Delta$ on $\Gamma$. In general $\Delta$ decreases as $\Gamma$ decreases. Also, the shape of the gap is sharper with smaller $\Gamma$. This allows diabatic transitions.

\begin{figure}[h!]
\centering
\includegraphics[width=0.6\linewidth]{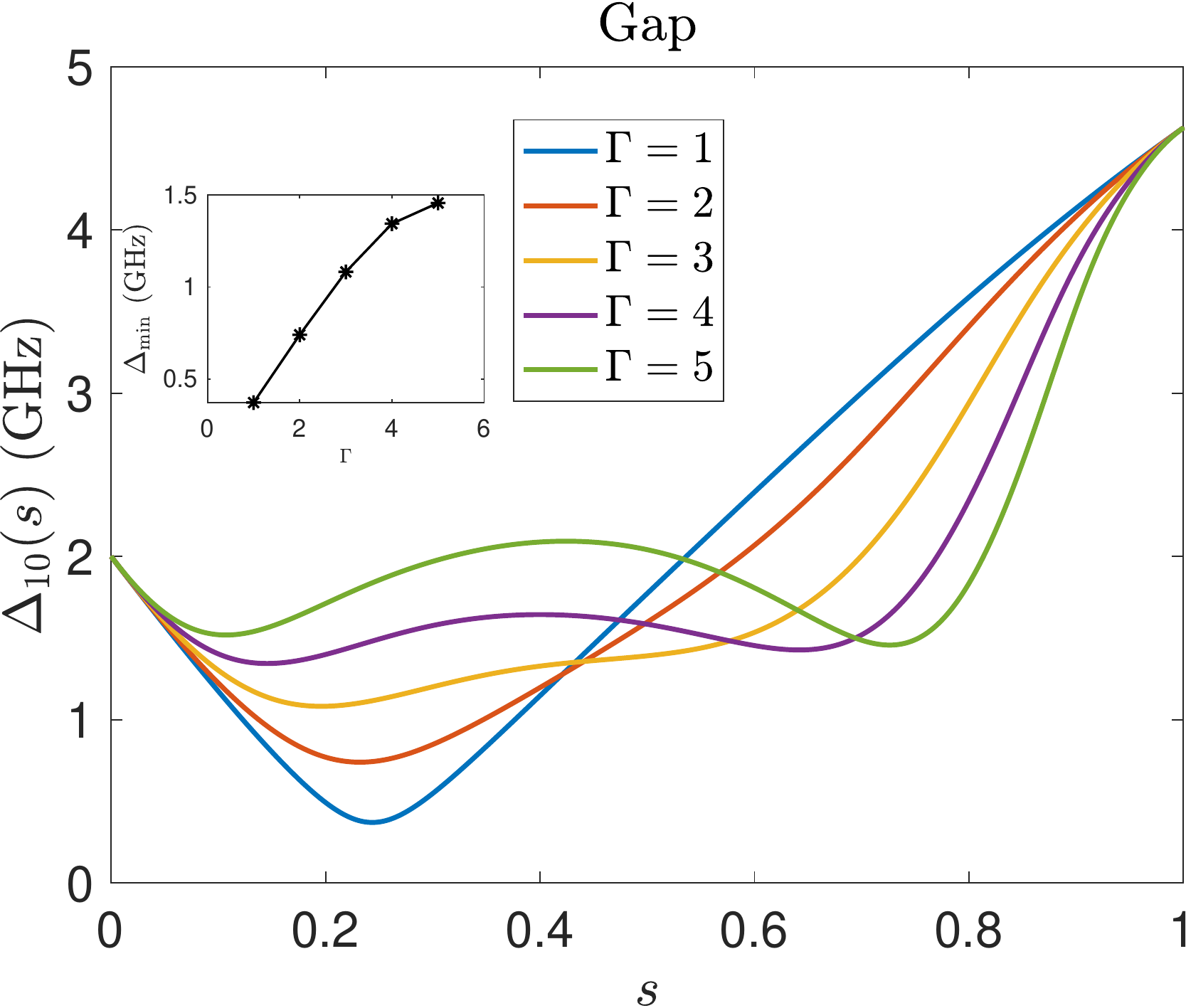}
\caption{$\Delta(s)$ for $\Gamma = \{1, \cdots, 5\}$ in units where $h=1$.}
\label{fig: gammaplot}
\end{figure}

\subsection{Controlling the parameters: Initial states $\epsilon_i$ or ($H_{\text{init}}$)}

We want to investigate how the gap properties change with different initial states. In this case we focus on $\Gamma = 1$.
\begin{figure}[h!]
\centering
\includegraphics[width=0.6\linewidth]{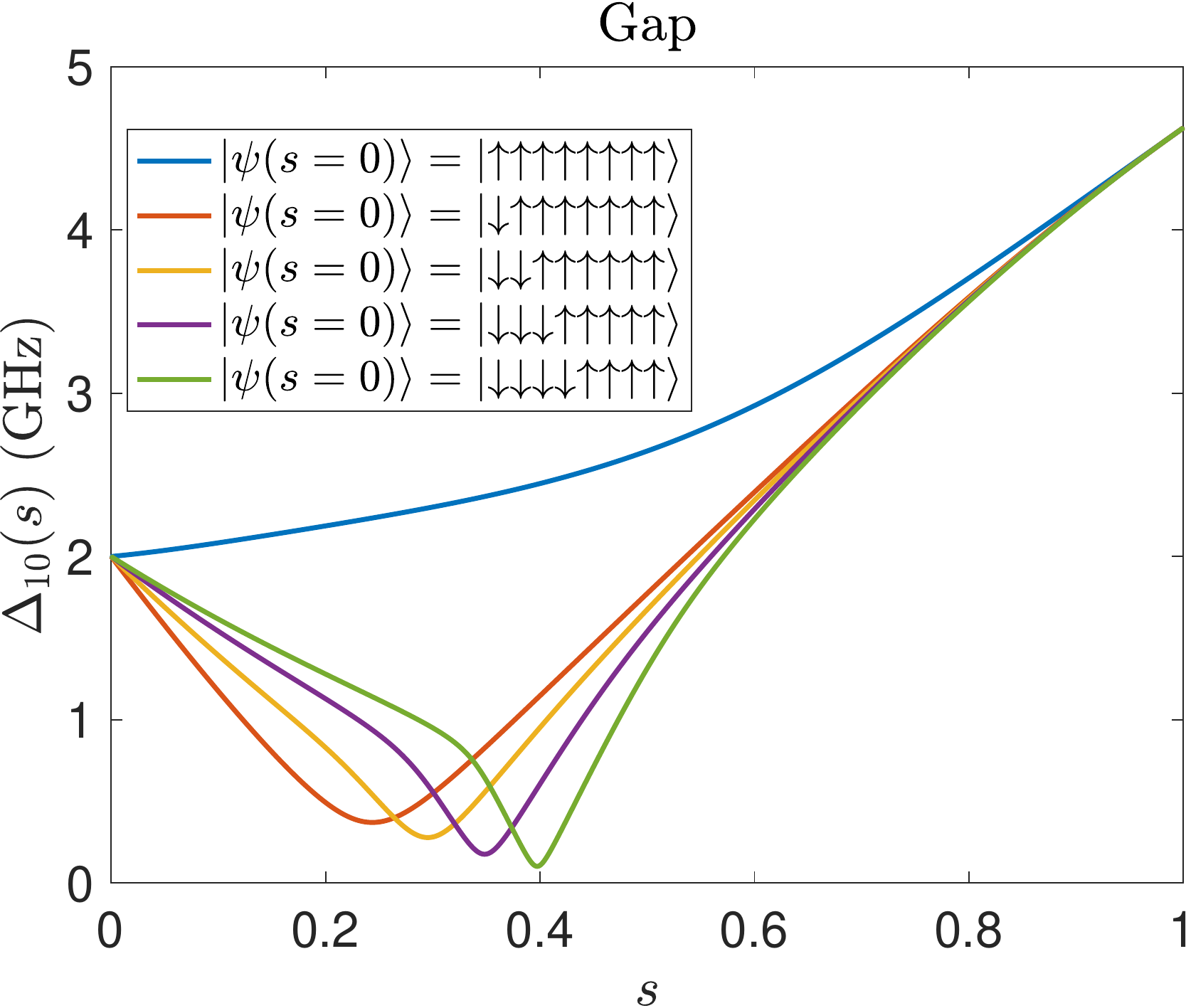}
\caption{$\Delta(s)$ for different $\epsilon_i$. $\Gamma = 1$ in units where $h=1$.}
\label{fig: gammaplot}
\end{figure}

From Fig.~\ref{fig: gammaplot}, we see that in general $\Delta$ decreases as the Hamming distance between the initial state and the target state increases.

\subsection{Controlling the parameters: $N$}
\begin{figure}[h!]
\centering
\includegraphics[width=0.6\linewidth]{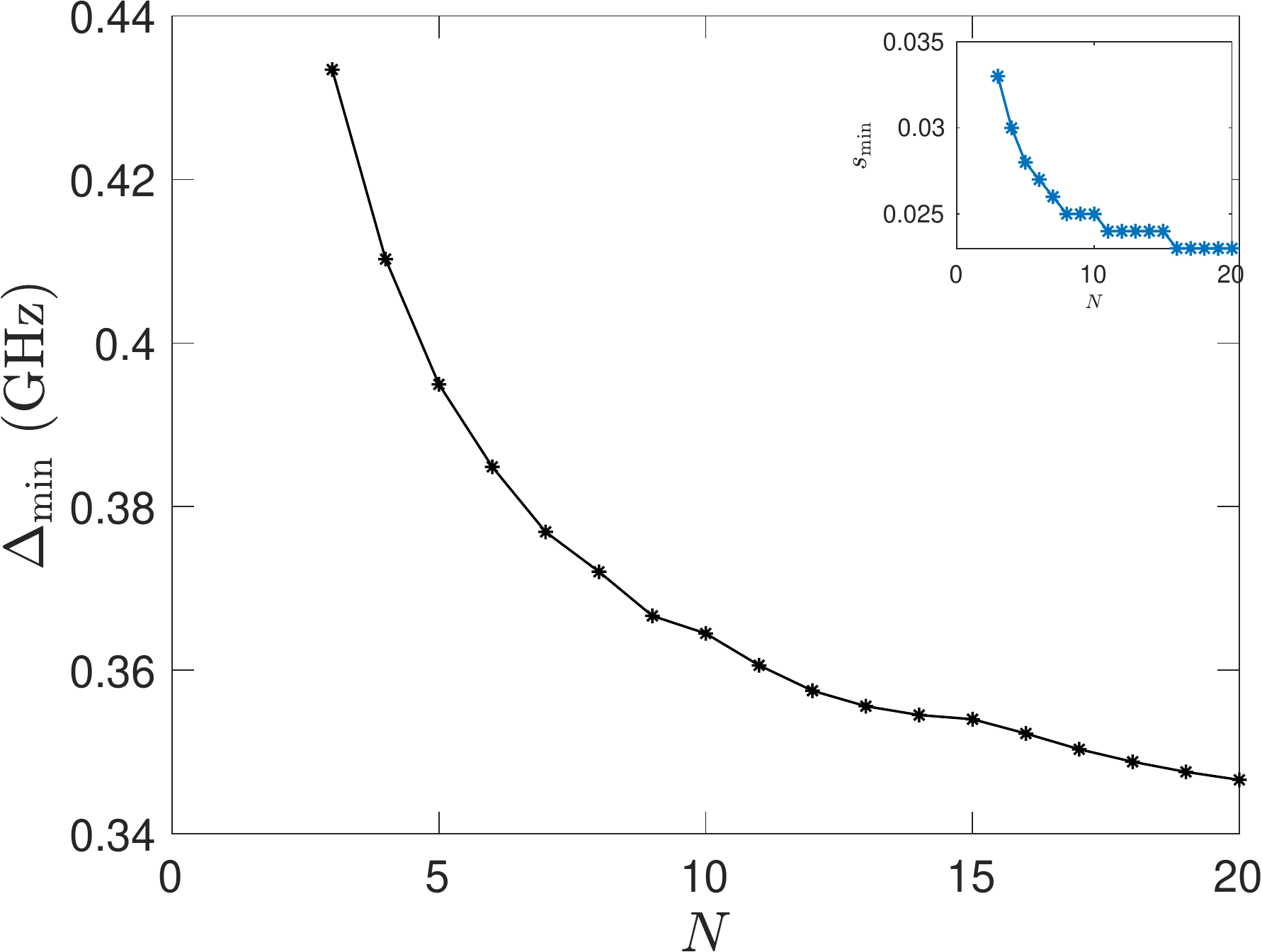}
\caption{$\Delta$ for $N = \{3, \cdots, 20\}$ in units where $h=1$. Inset: the value of $s_{\mathrm{min}}$ for each $N$.}
\label{fig: nplot}
\end{figure}

We want to investigate how the minimum gap changes with system size $N$. In this case we focus on the initial state of 
$N_{\downarrow}=1$ and $\Gamma = 1$. The value of $\Delta$ for each $N$ is plotted in Fig.~\ref{fig: nplot}.
We also plot the $s$ value where the minimum gap is found, i.e.,
\begin{equation}
    s_{\mathrm{min}} = \mathrm{argmin} \Delta_{10}(s) \,.
\end{equation}
 This is important for the calculation of relaxation rates for reverse annealing.

\section{Preliminary II: Time evolution}
Let $\ket{\psi(s=0)}$ ($\ket{\psi(t=0)}$) denote the initial state. The anneal time is $\tau$. The final state at the end of the adiabatic reverse annealing (ARA) and the success probability can be expressed as follows.

\subsection{Closed system}
The state after the ARA in a closed system is
\begin{equation}
\ket{\psi(\tau)} = U(\tau,0)\ket{\psi(0)}\,,
\end{equation}
where 
\begin{equation}
    U(\tau, 0) = {\cal T}\exp\left[-i\int_0^{\tau}H(t')dt'\right]
\end{equation}
is the unitary operator and ${\cal T}$ denotes the forward time-ordering. The time-dependence of the Hamiltonian (in the integrand) is given by the $s(t)$ function.


\subsection{Open system: Lindbladian dynamics}
For an open system, the state after ARA is
\begin{equation}
\rho(\tau) = V(\tau,0)\rho(0) \,,
\end{equation}
where 
\begin{equation}
    V(\tau,0) = {\cal T}\exp\left[\int_0^{\tau}\mathcal{L}(t')dt'\right] \,.
\label{eq:vpropagator}    
\end{equation}
$\mathcal{L}(t)$ is the time-dependent Liouville superoperator. For example, in adiabatic master equation, $\mathcal{L}(t)$  at time $t$ takes the form of
\begin{align}
\label{eq:ame}
	\frac{d\rho(t)}{dt} 
	&= \mathcal{L}(t)\rho(t)\\
	&= \iu \bigl[\rho(t), \ham(t) + \ham\ped{LS}(t)\bigr] + \diss\bigl[\rho(t)\bigr] \,.
\end{align} 
$ \ham\ped{LS}(t) $ is a Lamb shift term and $ \diss $ is the dissipator superoperator. Again, the time-dependence of the integrand in Eq.~\eqref{eq:vpropagator} depends on the $s(t)$ function.


\section{Preliminary III: Success metrics and classes of problems}
We use two success metrics for ARA: success probability and time to solution (TTS).
\subsection{Success probability}
The success probability refers to the probability of the final state being the target solution state. For the example of $p$-spin ($p=3$) model, the success probability at the end of ARA can be expressed as:

\begin{equation}
    \mathsf{Success\,probability\,\,} p_g(\tau) =
    \begin{cases}
|^{\otimes N}\langle \,\uparrow|\psi(\tau)\rangle|^2, \text{\quad\quad in a closed system.}\\
^{\otimes N}\langle\,\uparrow|\rho(\tau)|\uparrow\,\rangle ^{\otimes N} , \text{\quad in an open system.}
\end{cases}
\end{equation}

\subsection{Time to solution (TTS)}
The time to solution (TTS) is defined as~\cite{ronnow:speedup}:

\begin{equation}
    \mathsf{TTS}(\tau, p_d) = \tau \frac{\log (1-p_d)}{\log p_e(\tau)} \,.
\end{equation}

$p_d$ is the desired threshold probability and $p_e(\tau) = 1 - p_g(\tau)$.

\subsection{Classes of problems}
We want to explore four general classes of problems: 1. Adiabatic reverse annealing in an open system 
($\mathsf{ARA_\text{Open}}$), 2. Adiabatic reverse annealing in a closed system ($\mathsf{ARA_\text{Closed}}$), 3. Standard quantum annealing in an open system ($\mathsf{QA_\text{Open}}$), and 4. Standard quantum annealing in a closed system ($\mathsf{QA_\text{Closed}}$). The comparison between $\mathsf{ARA_\text{Closed}}$ and $\mathsf{QA_\text{Closed}}$ has been made in~\cite{nishimori:reverse-pspin-2} and it was shown that $\mathsf{ARA_\text{Closed}}$ can have an advantage over $\mathsf{QA_\text{Closed}}$. Since there are many studies (e.g.~\cite{passarelli:pspin} and~\cite{Passarelli2019}) showing that dissipation can give quantum enhancement, we want to investigate further if and under what conditions $\mathsf{ARA_\text{Open}}$ performs better than  $\mathsf{ARA_\text{Closed}}$. On an equal setting, we want also to compare if $\mathsf{ARA_\text{Open}}$ has any computational advantage over $\mathsf{QA_\text{Open}}$, in order to support the use of the formal protocol in a realistic experimental setting. Any advantage/enhancement is defined according to the two success metrics mentioned before.

We have already studied how the spectrum depends on $\Gamma$, $H_\text{init}$ and $N$. We want to evaluate how the performance of $\mathsf{ARA}$ and $\mathsf{QA}$ depends on these parameters. For each set of such parameters, we performed simulations in a range of anneal times $\tau$. 

As noted in~\cite{Crosson2020}, diabatic transition to higher excited states may be a shortcut of the route to the final solution. Therefore, we include in our studies anneal times $\tau$ shorter than the one set by adiabatic condition~\cite{albash:review-aqc}. We also explore the cases with small $\Gamma$ and large Hamming distance between the initial state and the target state, which result in a very small and sharp gap and thus diabatic transitions to higher excited states.

\section{Illustrative examples}
We perform numerical experiments for both closed and open system. We use the full Hamiltonian without sector decomposition. The coupling is $\eta = 1e^{-3}$ and the cutoff frequency is  $\omega_c = 1$THz. We use an independent dephasing bath $\hamsysbath\api{ind} = g \sum_{i} \sigma_i^z \otimes B_i$ and a collective dephasing bath  $\hamsysbath\api{col} = g \sum_{i} \sigma_i^z \otimes B$, and perform simulations of the adiabatic master equation Eq.~\eqref{eq:ame}. We begin by studying three examples in the following subsections (Sec.~\ref{subsectiona}-~\ref{subsectionc}) and illustrate the dynamics of ground state population during the anneal.

\subsection{Results of $8$-qubit simulation: $\tau = 40$ns. $\Gamma = 1$.}
\label{subsectiona}
\begin{figure}[h!]
\centering
\includegraphics[width = 12cm]{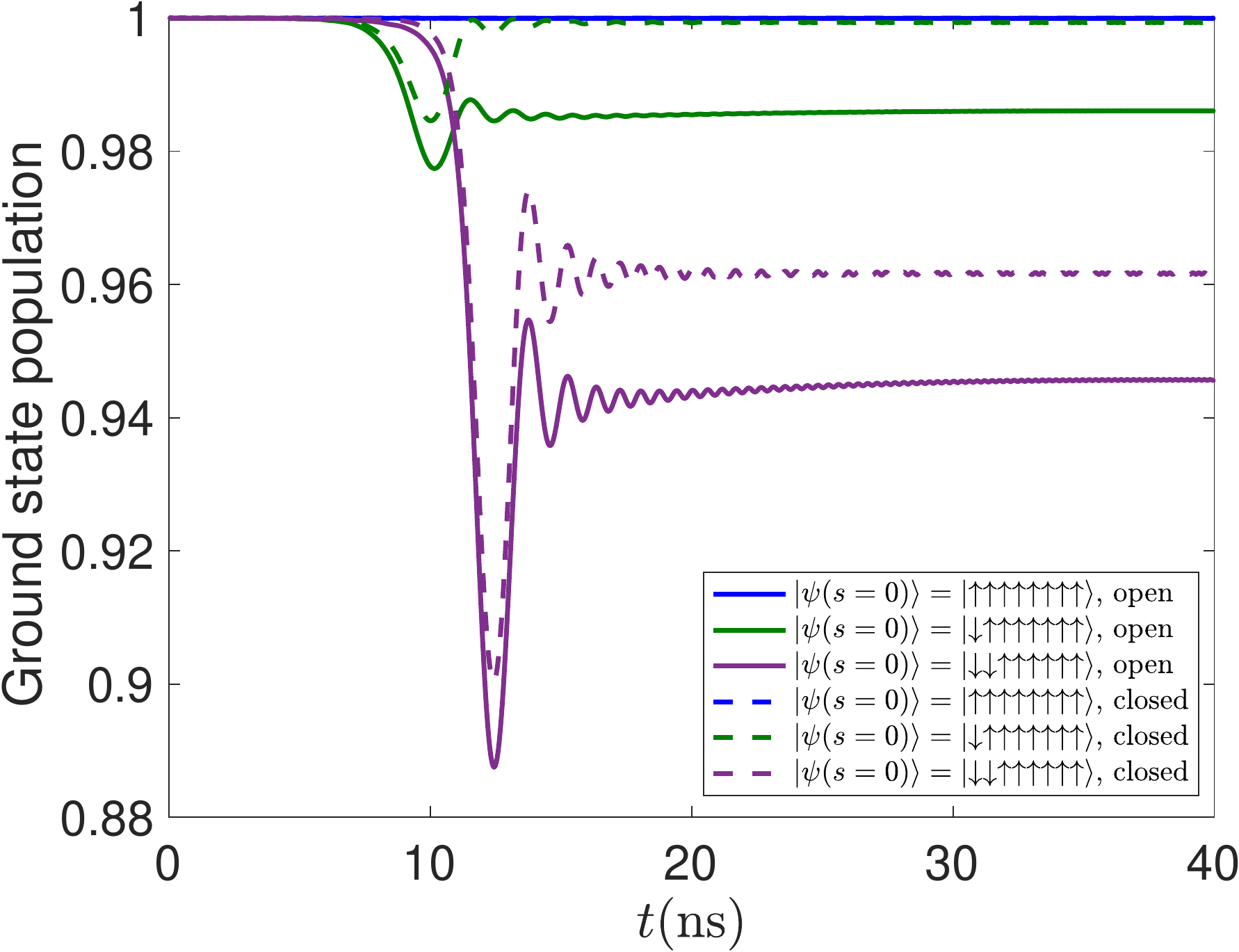}\label{fig:8qubitsame}
\caption{Open-system and close-system adiabatic reverse annealing (ARA) simulation results. $\tau = 40$ns. $\Gamma = 1$.}
\label{fig:8ame_727}
\end{figure}

 We simulate the closed and open-system evolution with different initial states. In this example we have $N=8$, $\tau = 40$ns. $\Gamma = 1$. The results are plotted in Fig.~\ref{fig:8ame_727}. The solid lines are open-system simulation results while the dotted lines are closed-system simulation results.

\textit{Conclusion: }
From Fig.~\ref{fig:8ame_727}, we see that open-system effect (thermal excitation) is detrimental in this case, i.e., for all initial states specified, the solid lines are below the dotted lines along the evolution and at the end of the evolution.

\subsection{Results of $4$-qubit simulation: $\tau = 4$ns. $\Gamma = 1$.}
\begin{figure}[h!]
\centering
\includegraphics[width = 12cm]{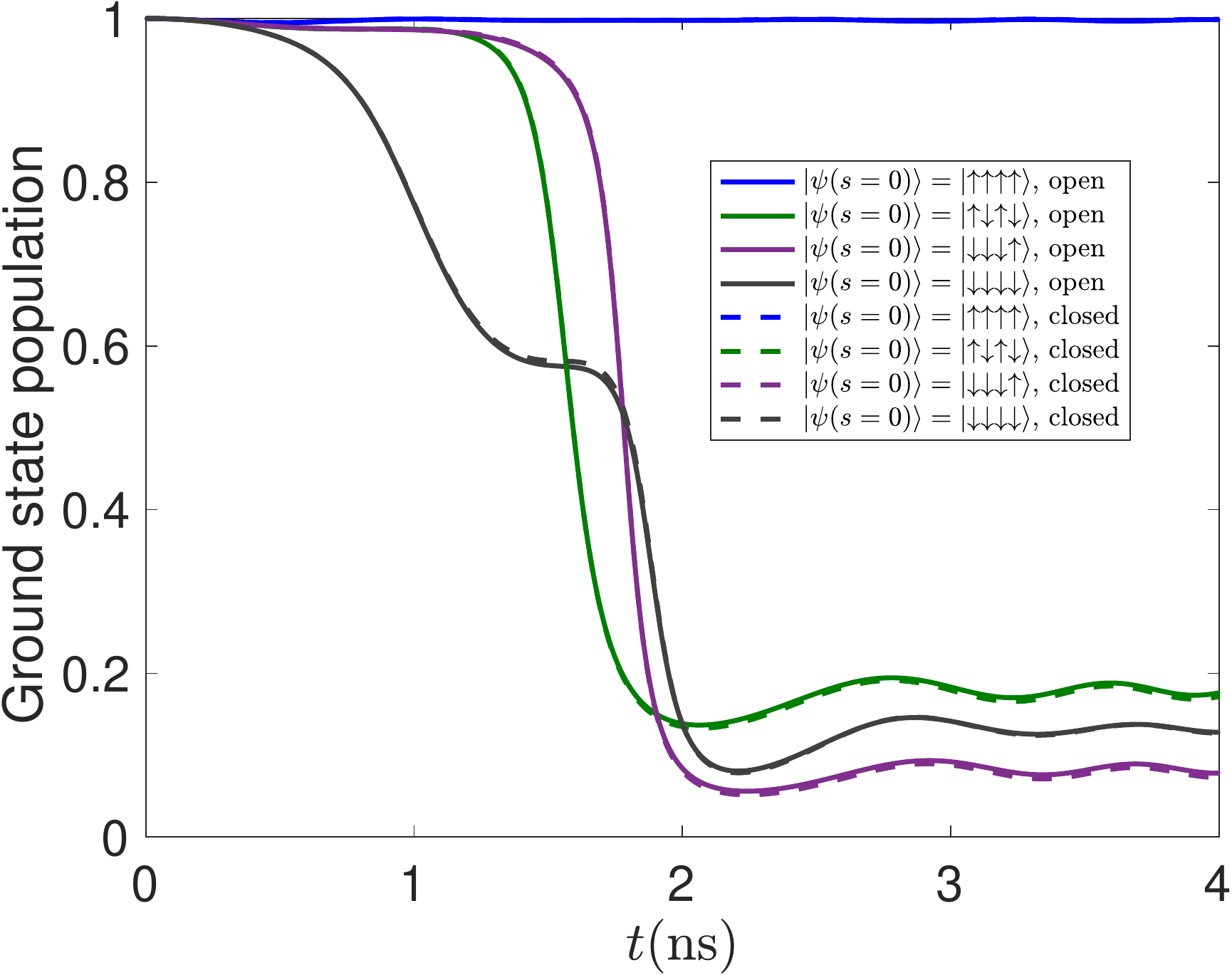}\label{fig:8qubitsame}
\caption{Open-system and close-system adiabatic reverse annealing (ARA) simulation results. $\tau = 4$ns. $\Gamma = 1$.}
\label{fig:4ame_723}
\end{figure}

In this example we have $N=4$, $\tau = 4$ns, and $\Gamma = 1$. The results are plotted in Fig.~\ref{fig:4ame_723}. The solid lines are open-system simulation results while the dotted lines are closed-system simulation results.

\textit{Conclusion: }
From Fig.~\ref{fig:4ame_723}, we see that open-system effect makes no difference, i.e. for all initial states specified, the solid lines almost overlap with the dotted lines along the evolution and at the end of the evolution. The total anneal time $\tau = 4$ns is too short for the environment (bath) to come into effect. The environment causes little difference to the success probabilities. This short anneal time also leads to observable diabatic transitions to higher excited states.

\subsection{Results of $4$-qubit simulation: $\tau = 400$ns. $\Gamma = 0.3$.}
\label{subsectionc}
\begin{figure}[h!]
\centering
\includegraphics[width = 12cm]{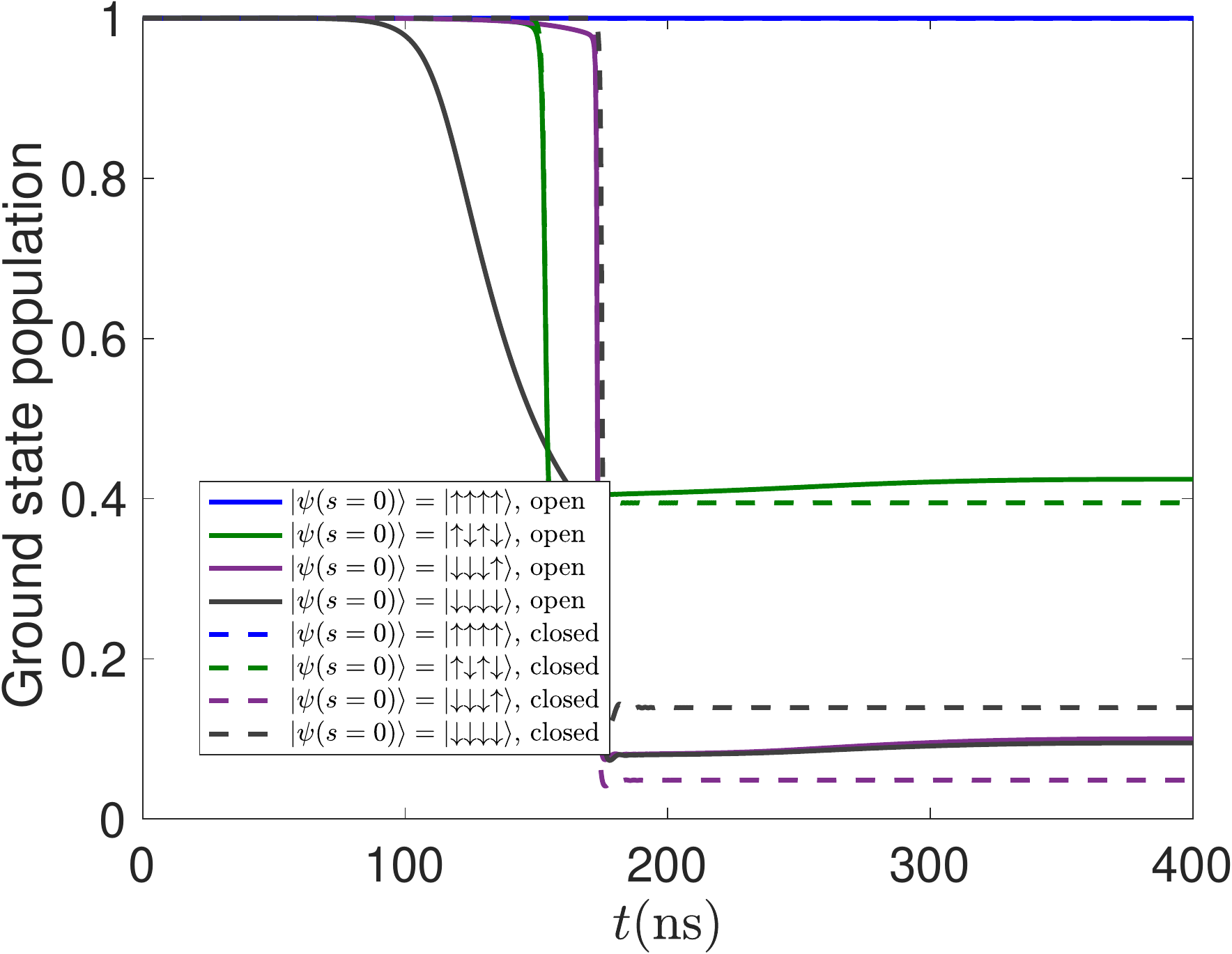}\label{fig:8qubitsame}
\caption{Open-system and close-system adiabatic reverse annealing (ARA) simulation results. $\tau = 400$ns. $\Gamma = 0.3$.}
\label{fig:4ame_724_2}
\end{figure}

In this example we have $N=4$, $\tau = 400$ns. $\Gamma = 0.3$. The results are plotted in Fig.~\ref{fig:4ame_724_2}. 

\textit{Conclusion: }
From Fig.~\ref{fig:4ame_724_2}, we observe sudden diabatic transitions, since $\Gamma = 0.3$ results in very small gaps for most of the $H_{\text{init}}$. Meanwhile, a relatively long anneal time of $\tau = 400$ns allows open-system relaxation mechanism to increase the (instantaneous and final success probability) ground state population for some of the initial states. 

\section{ARA: Random initial state $c=0.5$.}
\begin{figure}[h!]
\subfigure[]{\includegraphics[width = 8cm]{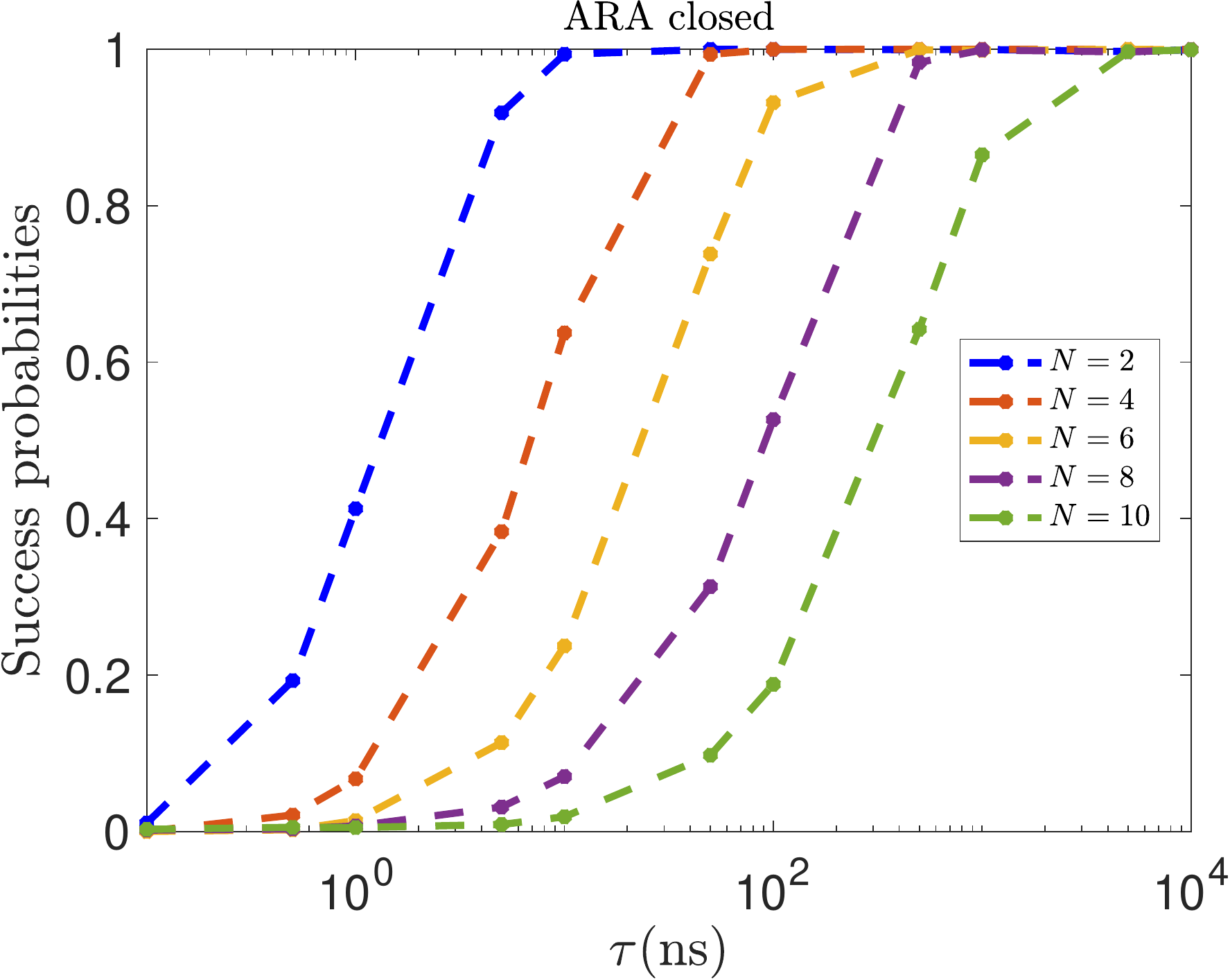}\label{fig:4qubitsunitarya}}
\subfigure[]{\includegraphics[width = 8cm]{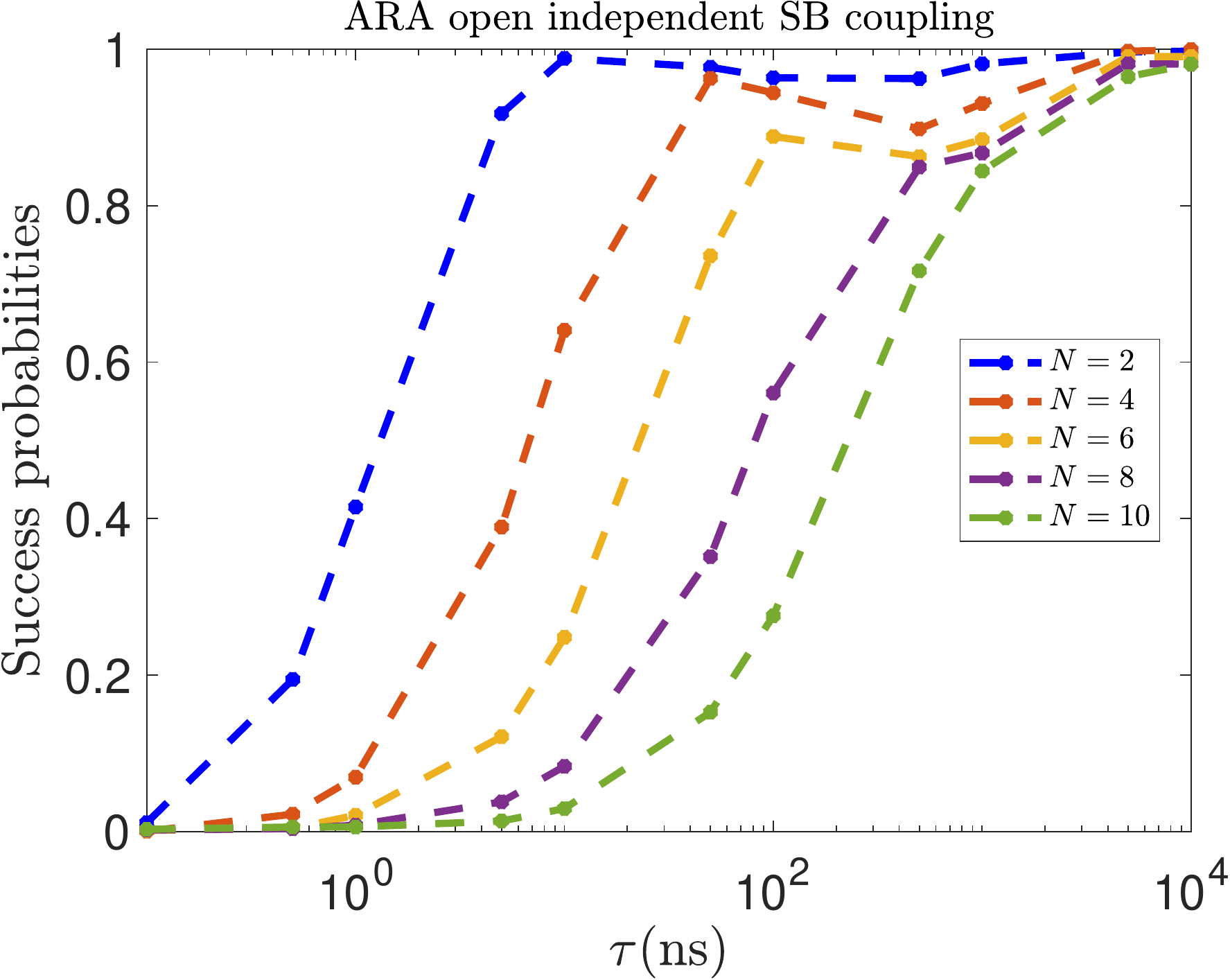}\label{fig:8qubitsunitary}}
\subfigure[]{\includegraphics[width = 8cm]{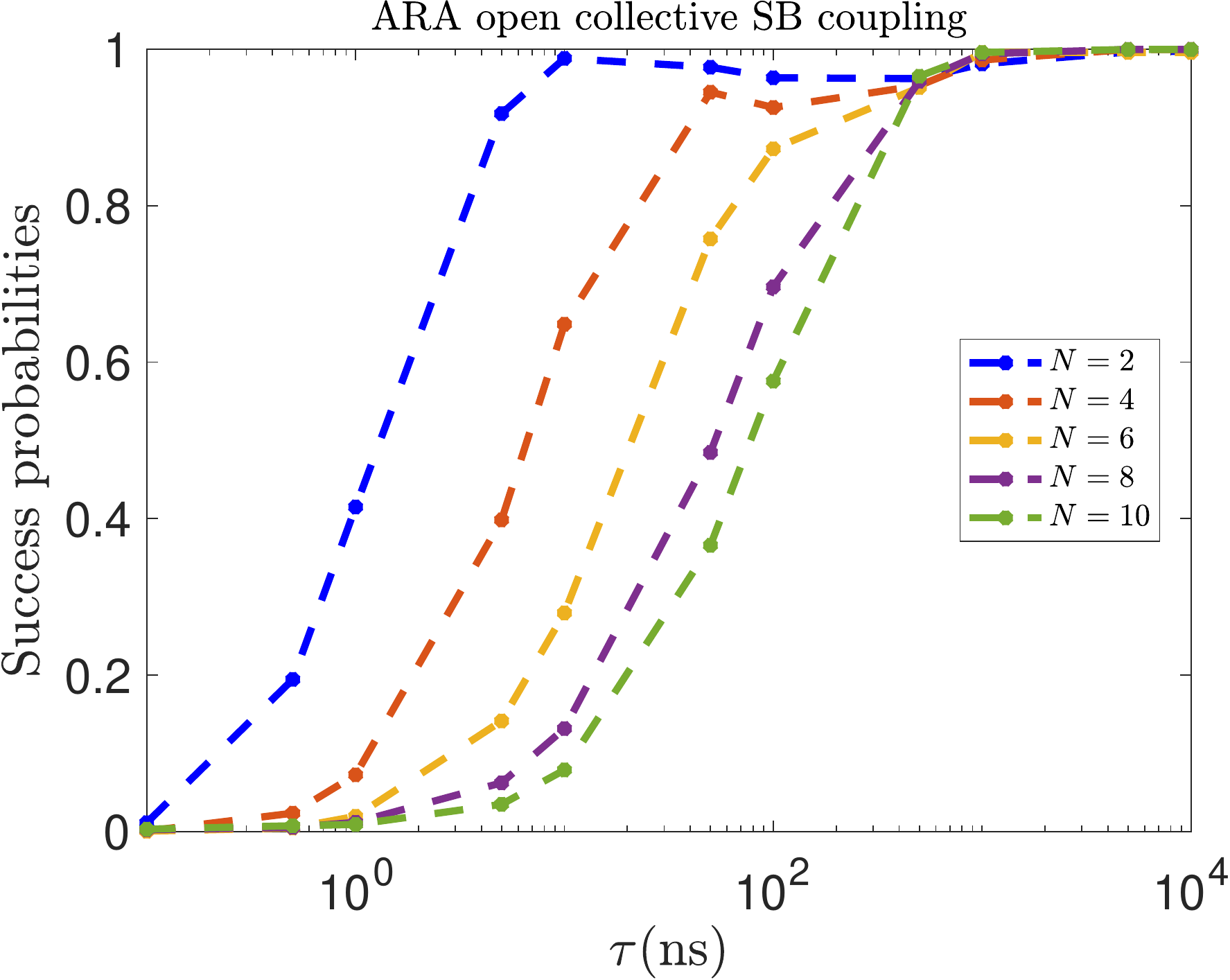}\label{fig:8qubitsunitary}}
\subfigure[]{\includegraphics[width = 8cm]{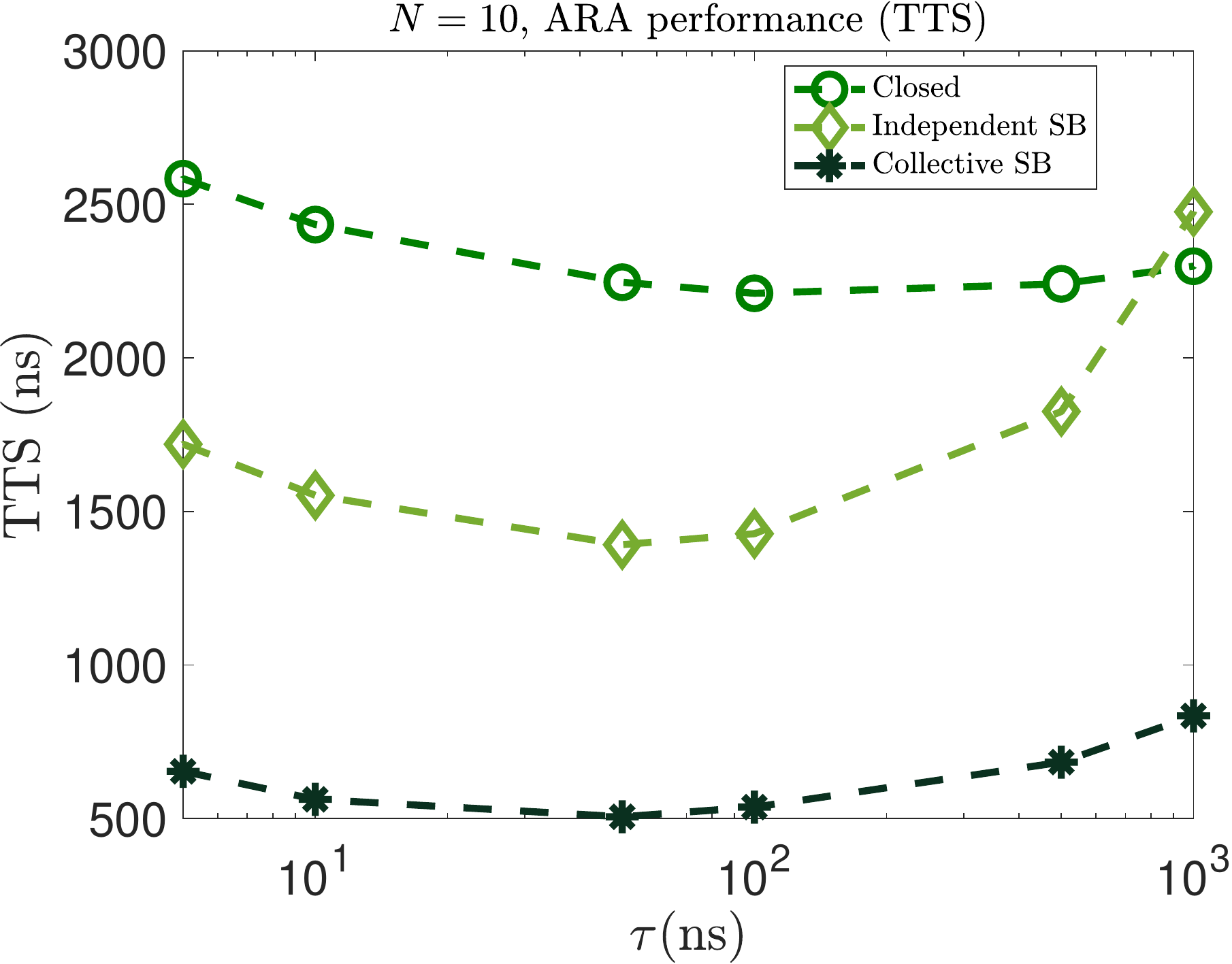}\label{fig:compareall}}
\caption{ARA performance for (a) closed system, (b) open system with independent system-bath coupling, (c) open system with collective system-bath coupling, (d) comparison of the 3 situations in terms of TTS ($N=10$).}
\end{figure}

As in previous studies, the success of ARA depends on whether our initial state is close to the correct solution or not. In this section, we study the most general case where the initial state is a random guess ($c=0.5$, where $c=N_{\uparrow}/N$) and want to see if $\mathsf{ARA_\text{Open}}$ can give an enhancement over $\mathsf{ARA_\text{Closed}}$. Instead of looking at ground state population along the anneal, we now focus only on the results of ARA at the end of anneal. A summary of the results as a function of $N$ and $\tau$ is shown in Fig.~\ref{fig:compareall}.

First of all, we observe that success probabilities go up monotonically with anneal time $\tau$ for the case of $\mathsf{ARA_\text{Closed}}$. As expected, the longer the anneal time $\tau$, the more adiabatic the evolution and the higher the success probability.
For $\mathsf{ARA_\text{Open}}$, however, we observe that there can exist an intermediate anneal time $\tau$ that gives a locally maximal success probability, and this is true for both SB coupling models. Such non-monotonic dependence of success probabilities in terms of anneal times $\tau$ is due to the environmental induced relaxation after diabatic transitions to higher excited states.

Secondly, at long anneal times, for the case of $\mathsf{ARA_\text{Closed}}$ the ground state is reached with with certainty (success probability $= 1$); while for the case of $\mathsf{ARA_\text{Open}}$, the Gibbs state is reached. Therefore, in all situations $\mathsf{ARA_\text{Closed}}$ always outperforms $\mathsf{ARA_\text{Open}}$ for long enough anneal times $\tau$. However, as shown in later section~\ref{sec:modelcompare}, the steady state reached by the two different system bath coupling model is different. 

To see if relaxation can provide any quantum enhancement, we do observe that for some $N$ $\mathsf{ARA_\text{Open}}$ performs better than $\mathsf{ARA_\text{Closed}}$ in the cases of intermediate anneal times $\tau$. We demonstrate in Fig.~\ref{fig:compareall} for the case of $N=10$ that $\mathsf{ARA}$ performs better in an open system settings in terms of shorter $\mathsf{TTS}$. For both  $\mathsf{ARA_\text{Closed}}$ and $\mathsf{ARA_\text{Open}}$ we observe optimal TTS. To avoid the misinterpretation of TTS, we avoid the calculation of $\mathsf{TTS}(\tau)$ where the success probability $p_g(\tau)$ is very close to $0$ and $1$. Lastly, we can see that in general the performance of both $\mathsf{ARA_\text{Closed}}$ and  $\mathsf{ARA_\text{Open}}$ worsens as $N$ increases, as expected due to the closure of the gap with increasing system size. 


\subsection{QA}
\begin{figure}[h!]
\subfigure[]{\includegraphics[width = 8cm]{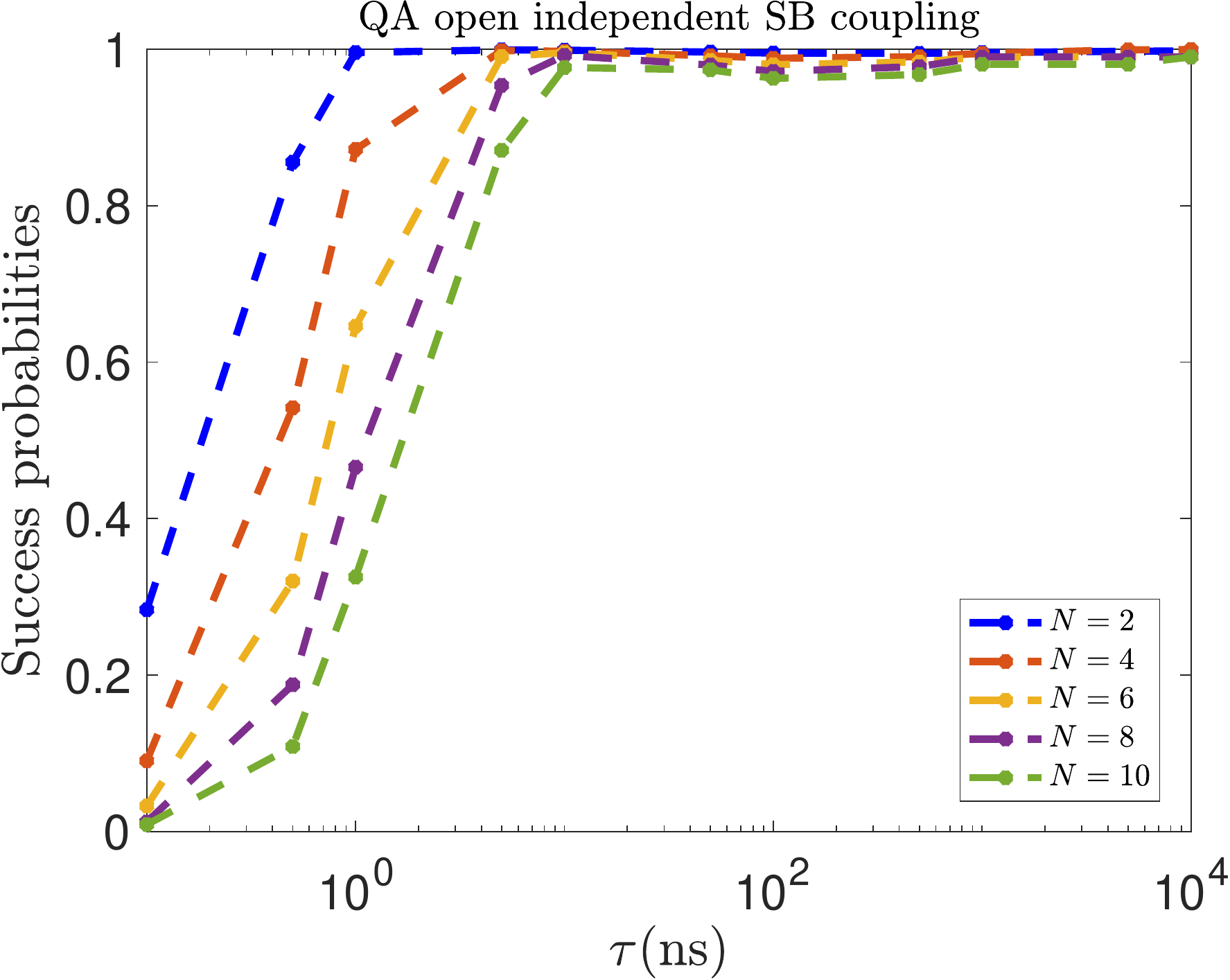}\label{fig:8qubitsunitary}}
\subfigure[]{\includegraphics[width = 8cm]{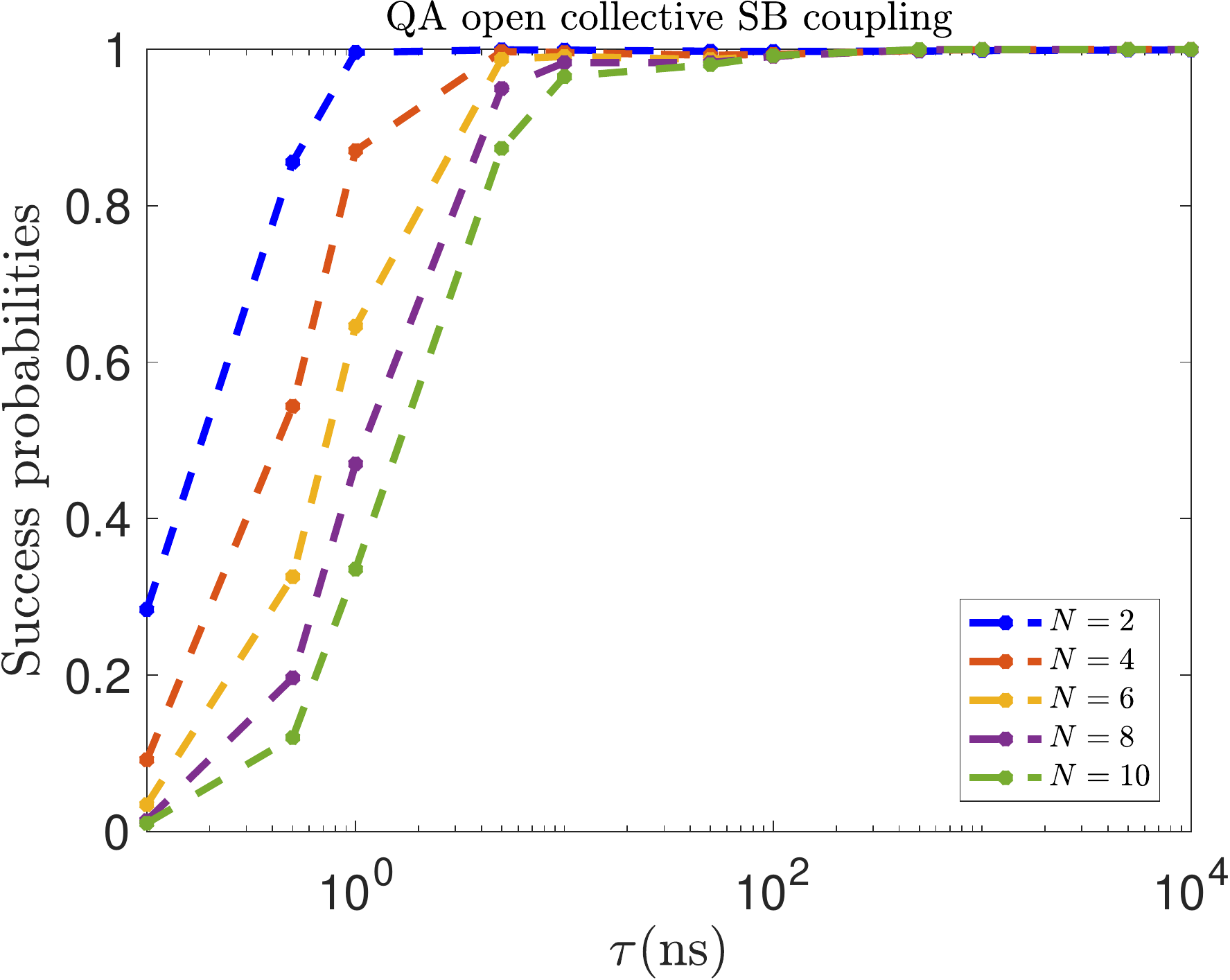}\label{fig:8qubitsunitary}}
\caption{Standard QA in an open system with two different system-bath coupling model.}
\label{fig:qaopen}
\end{figure}

 We show in Fig.~\ref{fig:qaopen} the case of quantum annealing in open system $\mathsf{ARA_\text{Open}}$. Note that the initial state of QA is the equal superposition  $\ket{+}^{\otimes N}$. Comparing Fig.~\ref{fig:qaopen} with Fig.~\ref{fig:compareall}, we can see easily that $\mathsf{QA_\text{Open}}$ achieves higher success probabilities for all $N$ and $\tau$. Therefore, for random initial states $c=0.5$,  $\mathsf{ARA_\text{Open}}$ cannot outperform $\mathsf{QA_\text{Open}}$.

\section{ARA performance in terms of $\Gamma$ and $H_{\mathrm{init}}$}
We now want to study the dependence of ARA performance in terms of $H_{\mathrm{init}}$ and $\Gamma$. For this purpose, we fix the number of qubits $N$ in our studies in this section.
\subsection{Min. Gap vs $H_{\mathrm{init}}$ vs $\Gamma$. $N=4$.}
\begin{figure}[h!]
\centering
\includegraphics[width=0.8\textwidth]{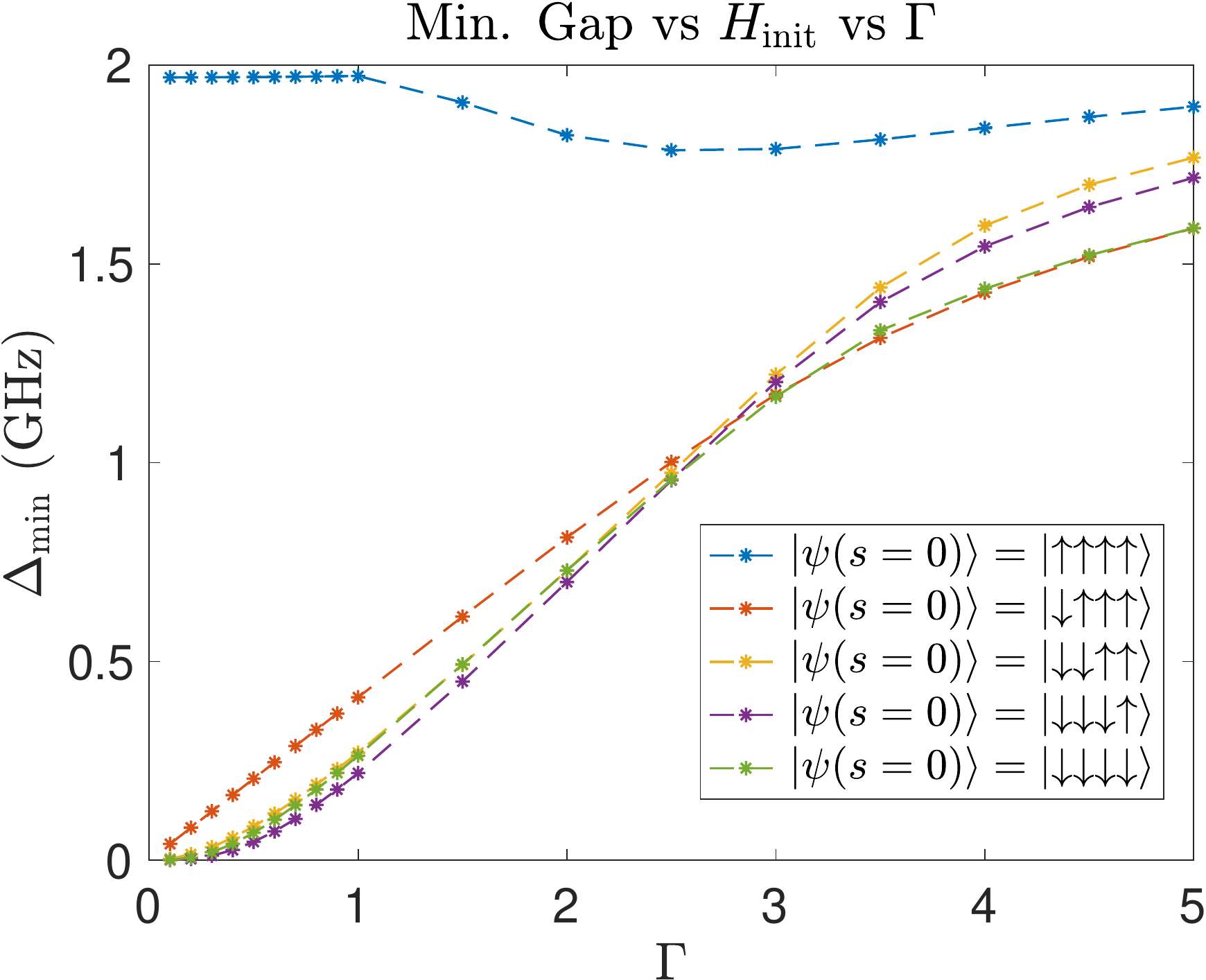}
\caption{The value of minimum gap as a function of $H_{\mathrm{init}}$ and $\Gamma$. $N=4$.}
\label{fig:spectrum_3d}
\end{figure}

We first plot in Fig.~\ref{fig:spectrum_3d} the value of minimum gap of ARA ($N=4$) as a function of initial state $H_{\mathrm{init}}$ and $\Gamma$. Except for initial states that are close to the correct solutions, in general the larger the transverse field $\Gamma$ the larger the minimum gap $\Delta_{\text{min}}$.

\subsection{Different Anneal times + Different $\Gamma = \{1,2,5\}$. $N=4$. $\mathsf{ARA_\text{Open}}$}
\begin{figure}[h!]
\subfigure[]{\includegraphics[width = 8cm]{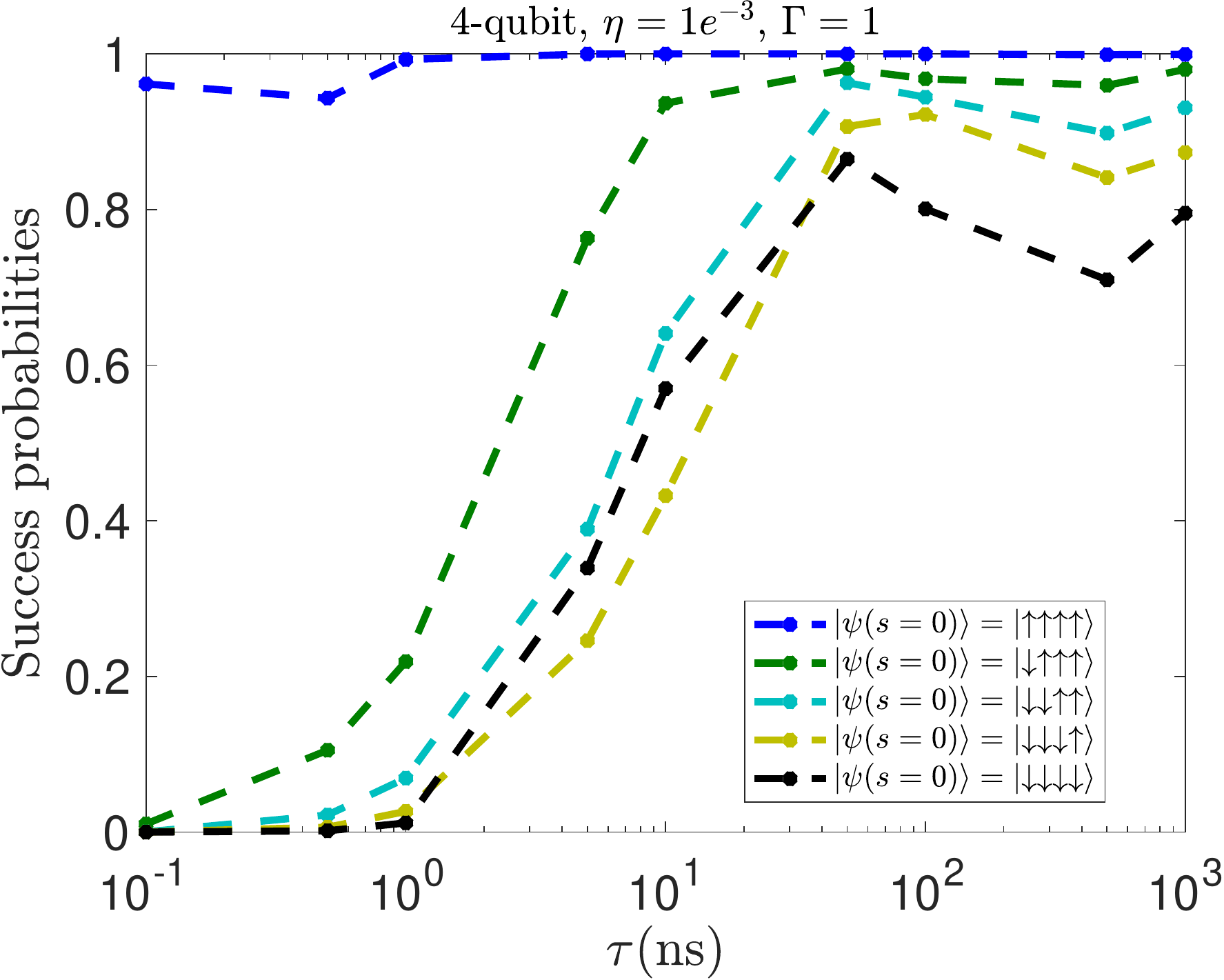}\label{fig:4qubitsunitarya}}
\subfigure[]{\includegraphics[width = 8cm]{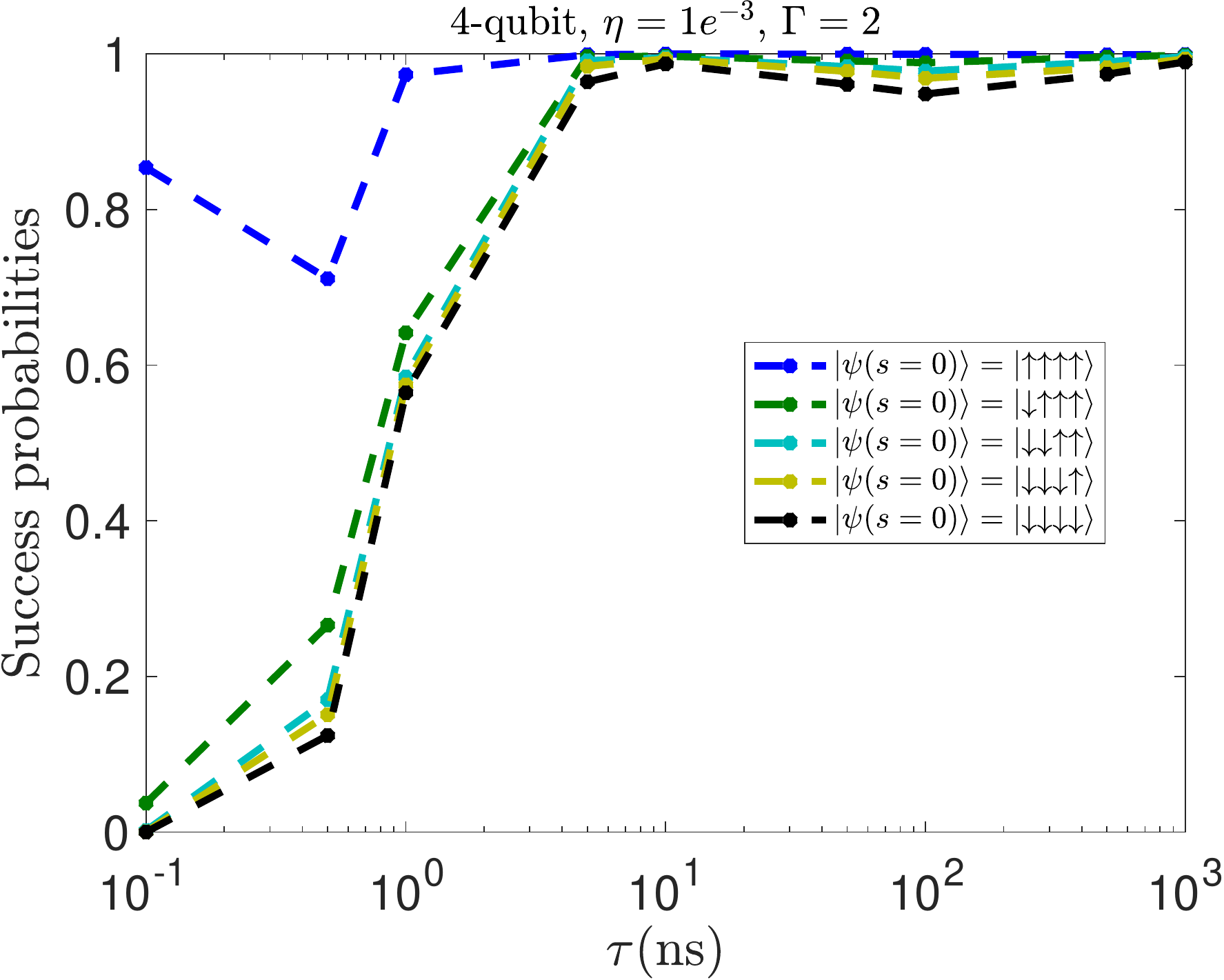}\label{fig:8qubitsunitary}}
\subfigure[]{\includegraphics[width = 8cm]{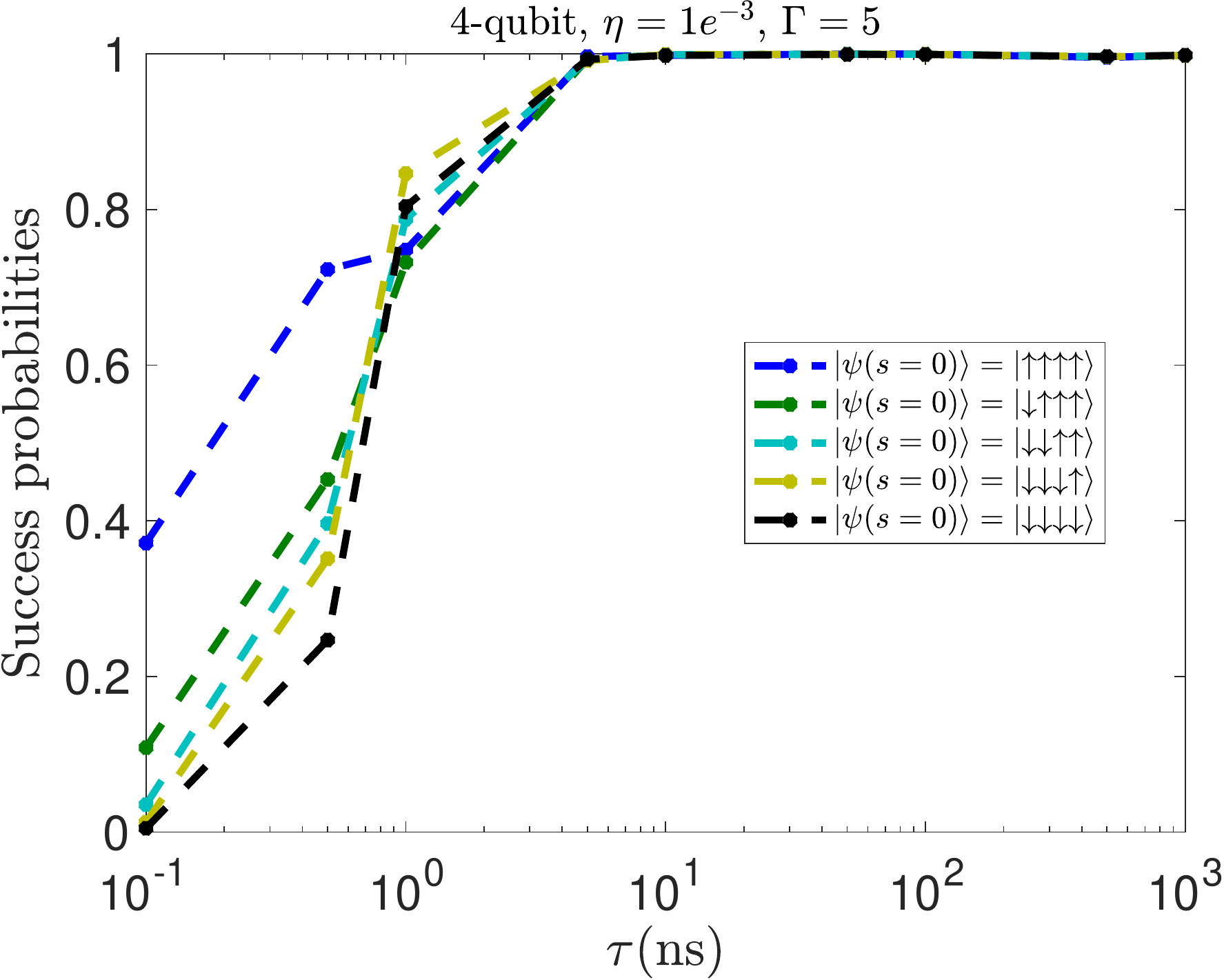}\label{fig:8qubitsunitary}}
\caption{$\mathsf{ARA_{\text{open}}}$ results (a) $\Gamma=1$, (b) $\Gamma=2$, (c) $\Gamma=5$. Independent system-bath coupling.}
\label{fig:ARAopen4qubits}
\end{figure}

\begin{figure}[h!]
\subfigure[]{\includegraphics[width = 8cm]{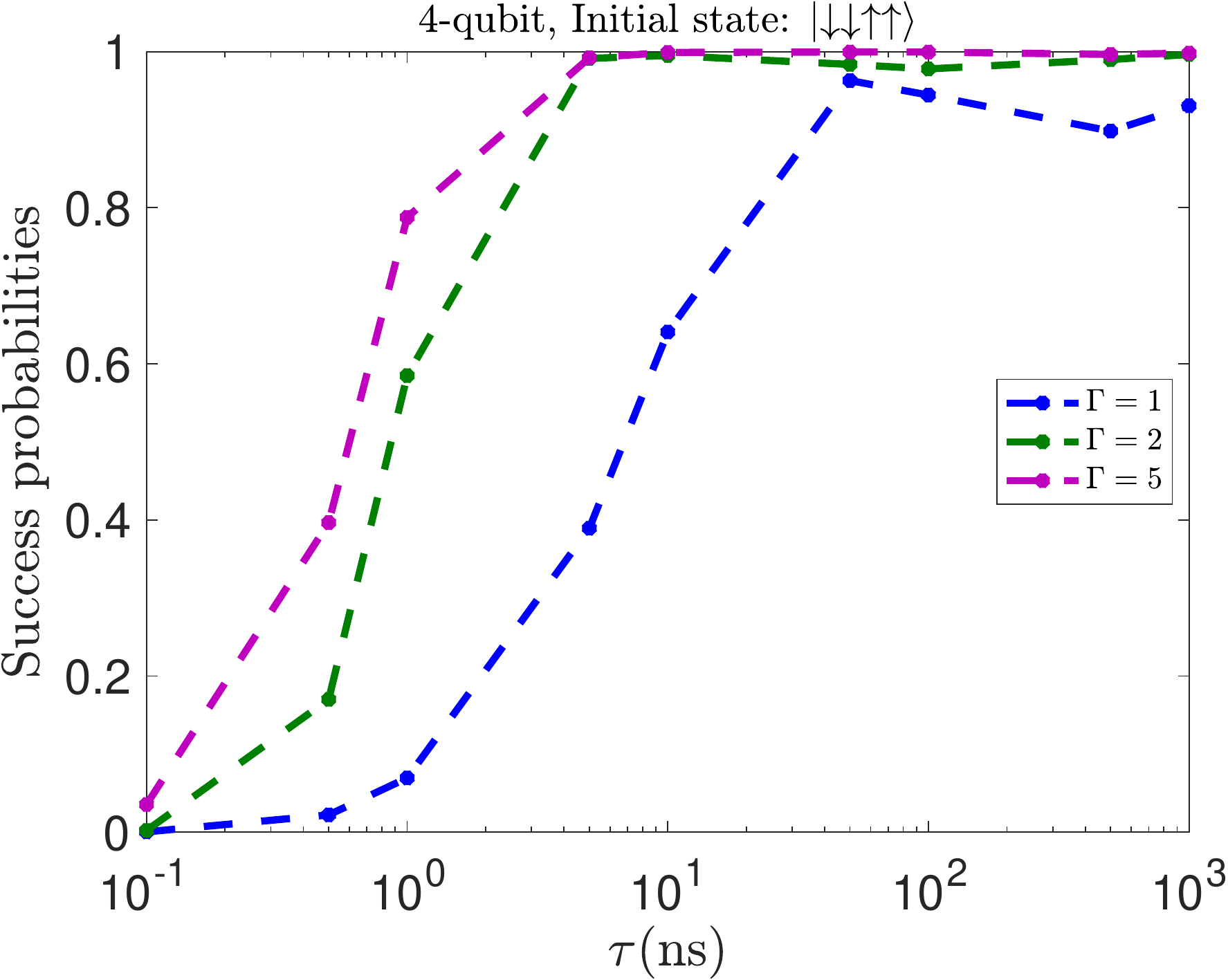}\label{fig:4qubitsunitarya}}
\subfigure[]{\includegraphics[width = 8cm]{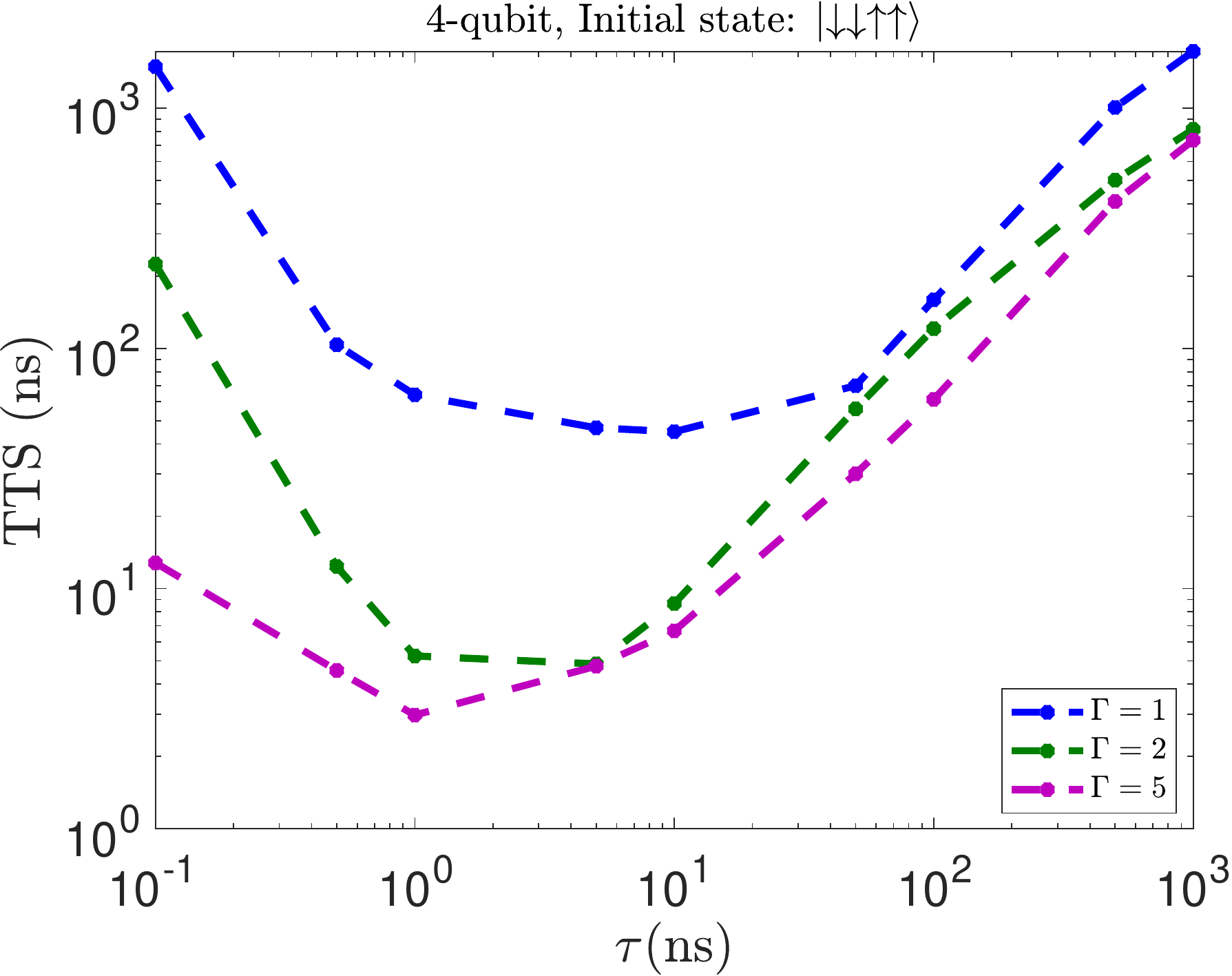}\label{fig:8qubitsunitary}}
\caption{$\mathsf{ARA_{\text{open}}}$ results, with initial state being $\ket{\downarrow\downarrow\uparrow\uparrow}$. (a) success probability, (b) $\mathsf{TTS}$. Independent system-bath coupling.}
\label{fig:ARAopen4qubitsuudd}
\end{figure}

We simulate the performance of $\mathsf{ARA_\text{Open}}$ as a function of initial state $H_{\mathrm{init}}$ and $\Gamma$, and plot the results in Fig.~\ref{fig:ARAopen4qubits} . We focus on the assumption of Independent system-bath coupling. In general the closer the initial state to the correct solution, the higher the success probabilities of $\mathsf{ARA_\text{Open}}$. Except for initial states that are close to the correct solutions, in general we want to increase the transverse field $\Gamma$ to enhance ARA performance because it results in a larger gap, in agreement with the observation in Fig.~\ref{fig:spectrum_3d}. We also plot this dependence on $\Gamma$ in Fig.~\ref{fig:ARAopen4qubitsuudd} in terms of success probability and $\mathsf{TTS}$. $\mathsf{TTS}$ is also shorter for larger $\Gamma$ and we observe an optimal $\mathsf{TTS}$ for all three values of $\Gamma$. 


\subsection{Different Anneal times + Different  $\Gamma<1$. $N=4$. $\mathsf{ARA_\text{Closed}}$ and $\mathsf{ARA_\text{Open}}$}
\begin{figure}[h!]
\subfigure[]{\includegraphics[width = 8cm]{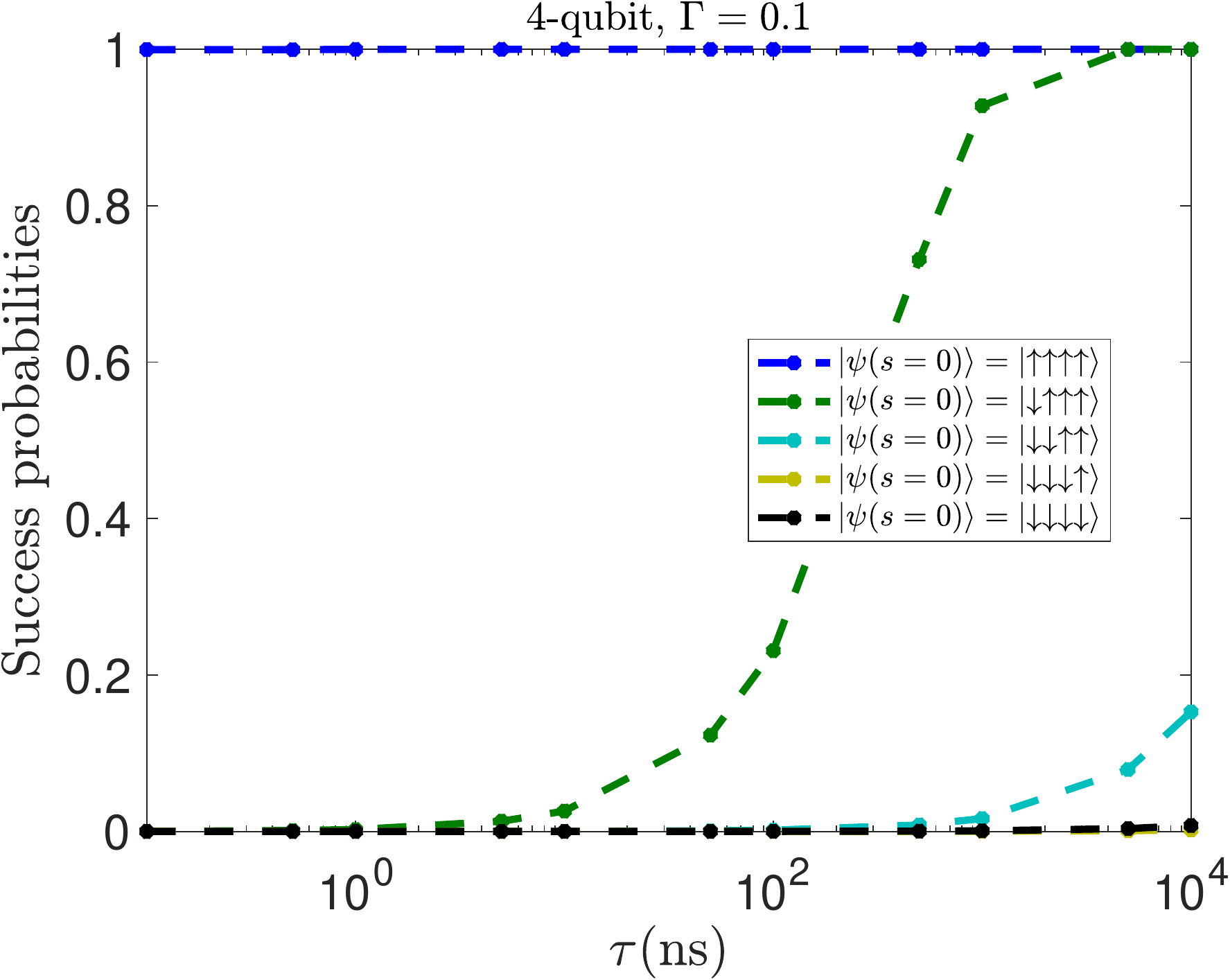}\label{fig:4qubitsunitarya}}
\subfigure[]{\includegraphics[width = 8cm]{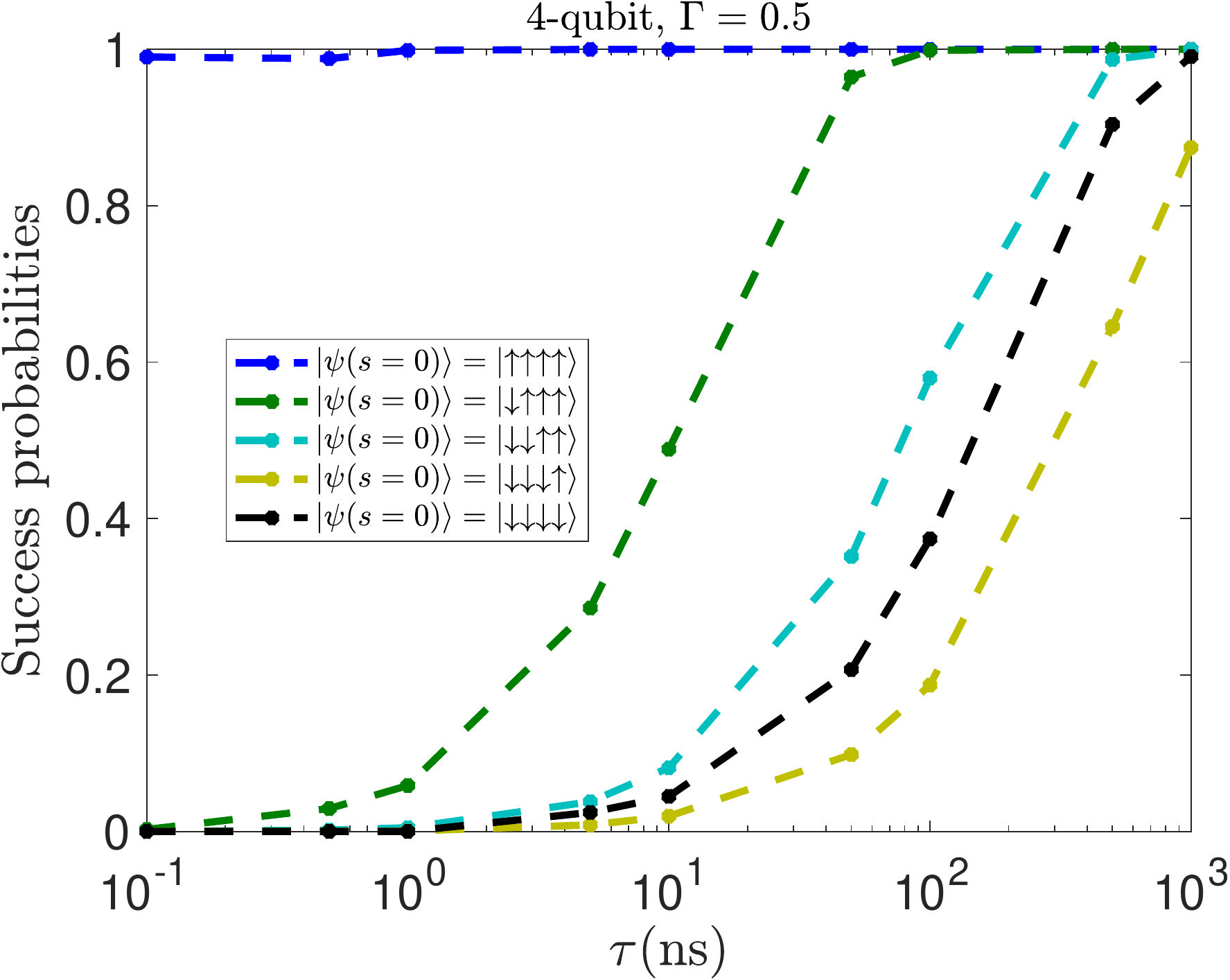}\label{fig:8qubitsunitary}}
\caption{$\mathsf{ARA_\text{Closed}}$  (a) $\Gamma = 0.1$, (b) $\Gamma = 0.5$. $N=4$.}
\label{fig:gammasmaller1_closed}
\end{figure}

\begin{figure}[h!]
\subfigure[]{\includegraphics[width = 8cm]{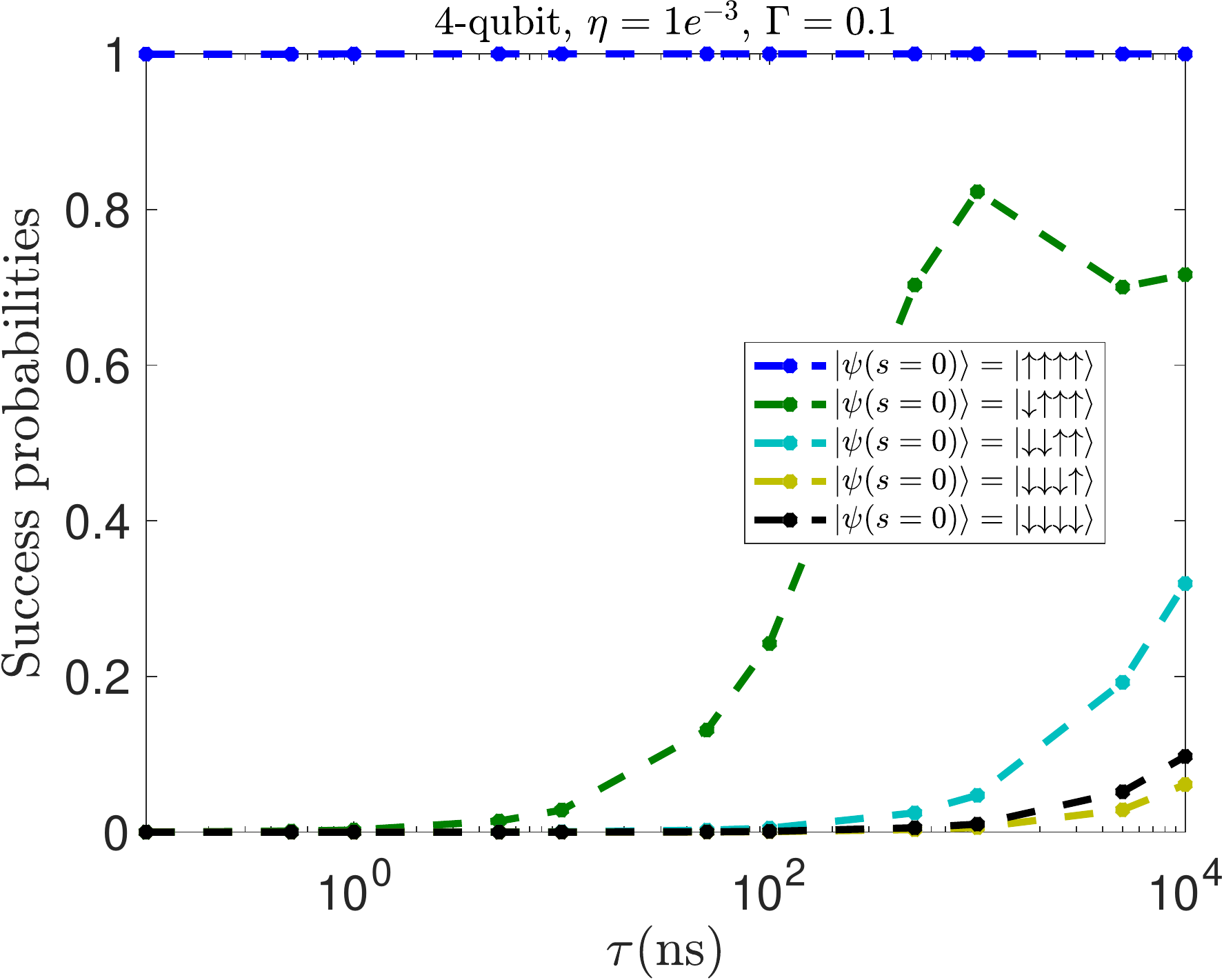}\label{fig:4qubitsunitarya}}
\subfigure[]{\includegraphics[width = 8cm]{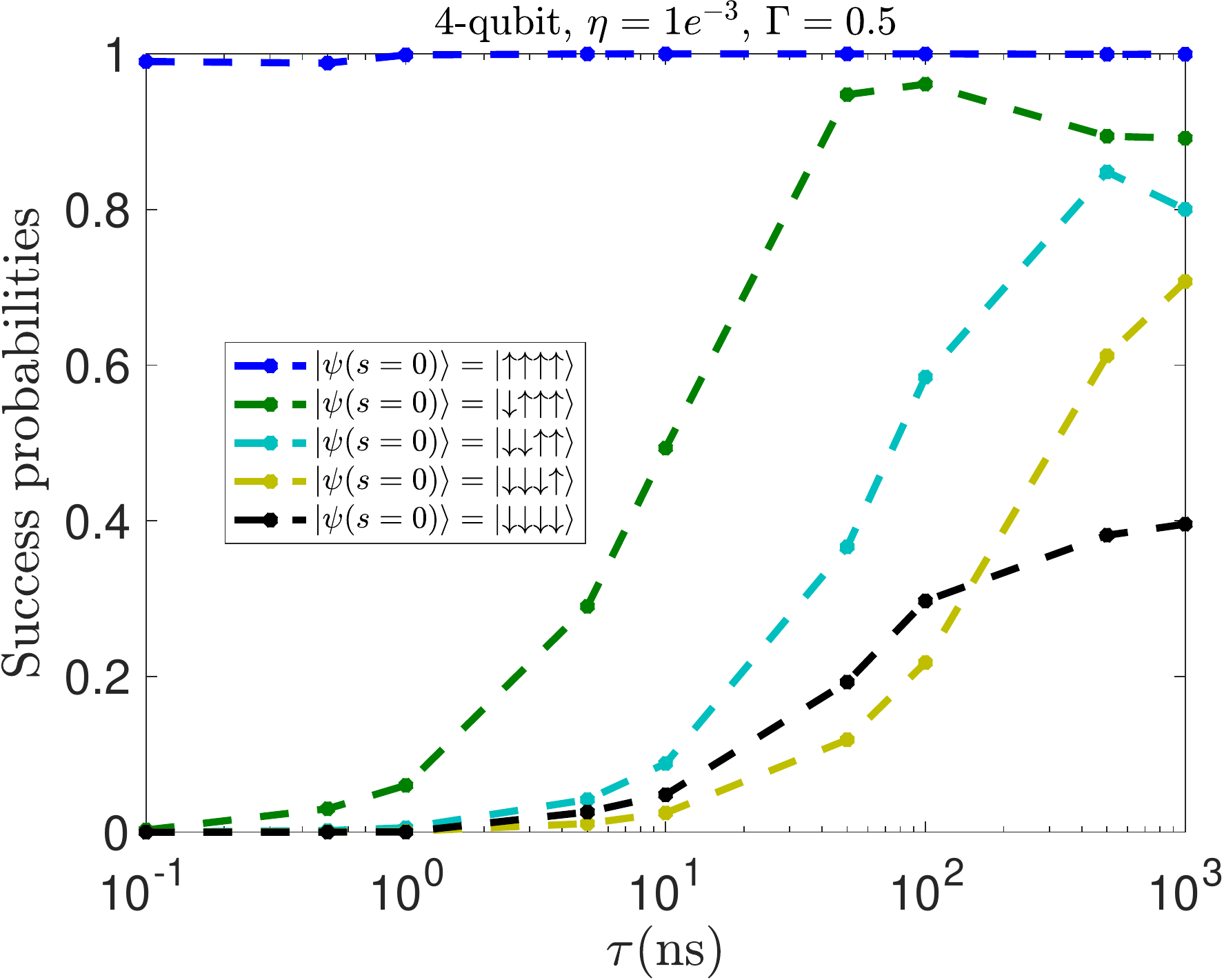}\label{fig:8qubitsunitary}}
\caption{$\mathsf{ARA_\text{Open}}$. (a) $\Gamma = 0.1$, (b) $\Gamma = 0.5$. $N=4$. Independent system-bath coupling.}
\label{fig:gammasmaller1_open}
\end{figure}

We study the case of $\Gamma<1$, which would results in a smaller gap (Fig.~\ref{fig:spectrum_3d}) except for the case where the initial state is equal to the correct solution. We plot in Fig.~\ref{fig:gammasmaller1_closed} and Fig.~\ref{fig:gammasmaller1_open} the simulated performance of $\mathsf{ARA_\text{Closed}}$ and $\mathsf{ARA_\text{Open}}$ respectively as a function of initial state $H_{\mathrm{init}}$ and $\Gamma$. Again we focus on the assumption of independent system-bath coupling. 

Comparing the results of $\mathsf{ARA_\text{Closed}}$ (Fig.~\ref{fig:gammasmaller1_closed}) and $\mathsf{ARA_\text{Open}}$ (Fig.~\ref{fig:gammasmaller1_open}), we again observe the non-monotonic dependence of success probabilities in terms of anneal times $\tau$ exclusively for $\mathsf{ARA_\text{Open}}$. We also see noticeable improvements of success probabilities due to open-system effect at intermediate anneal times $\tau$, especially if the initial states are far away from the correct solution. For both $\mathsf{ARA_\text{Closed}}$ and $\mathsf{ARA_\text{Open}}$, the larger $\Gamma$ the higher the success probabilities, except for the case where the initial state is equal to the correct solution.

\subsection{Different Anneal times. $\Gamma = 0.1$, $N_{\downarrow}=2$. $N=\{8,10\}$. $\mathsf{ARA_\text{Open}}$.}
We study the extreme case of $\Gamma = 0.1$ for larger $N$. Note that for such small $\Gamma = 0.1$, the transverse field is very weak and the Hamiltonian is dominated by the problem Hamiltonian $H_0$ and $H_{\text{init}}$. The resulting gap is very small and sharp and there is a large degree of diabatic transition to higher excited states. 

\begin{figure}[h!]
\subfigure[]{\includegraphics[width = 8cm]{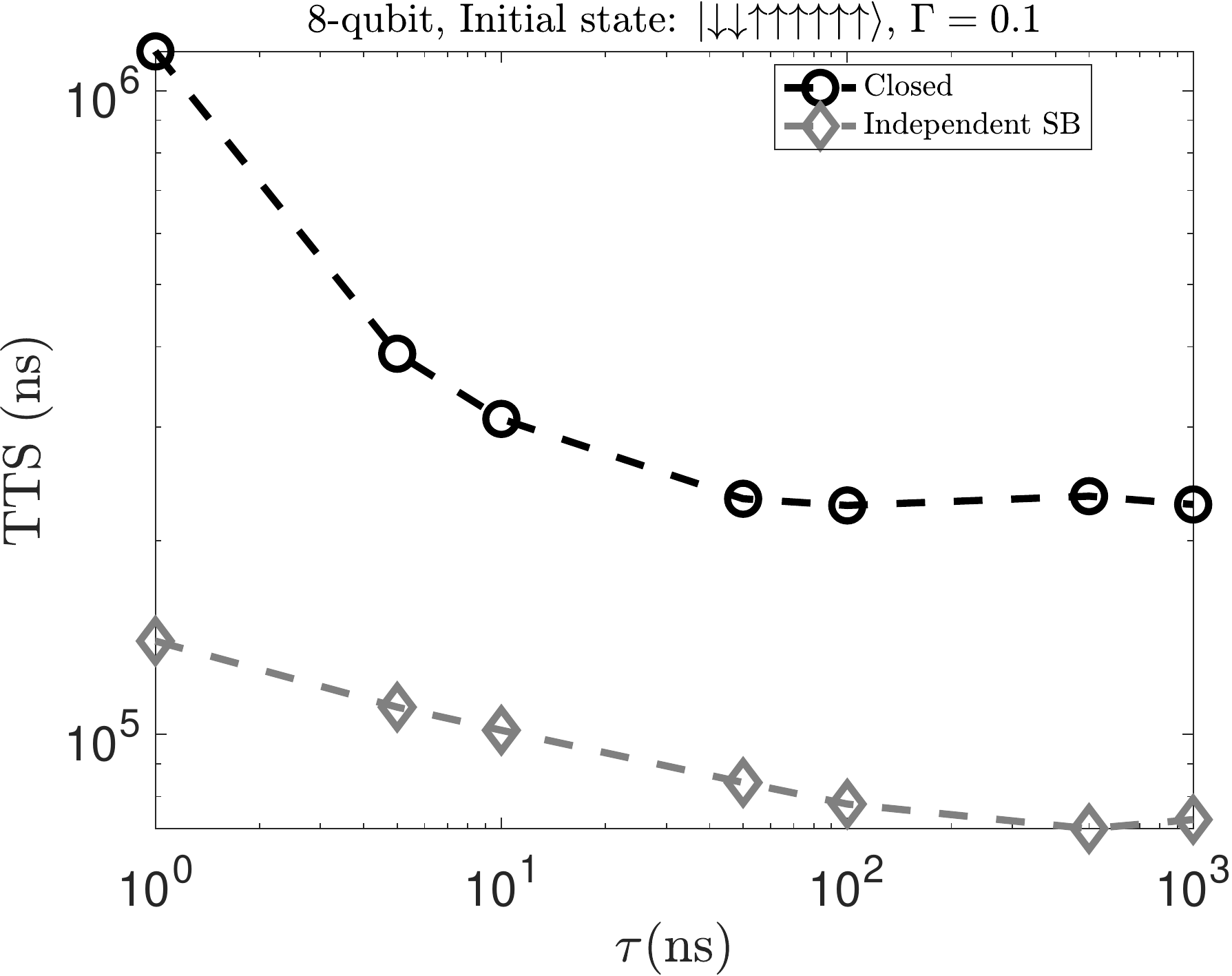}\label{fig:4qubitsunitarya}}
\subfigure[]{\includegraphics[width = 8cm]{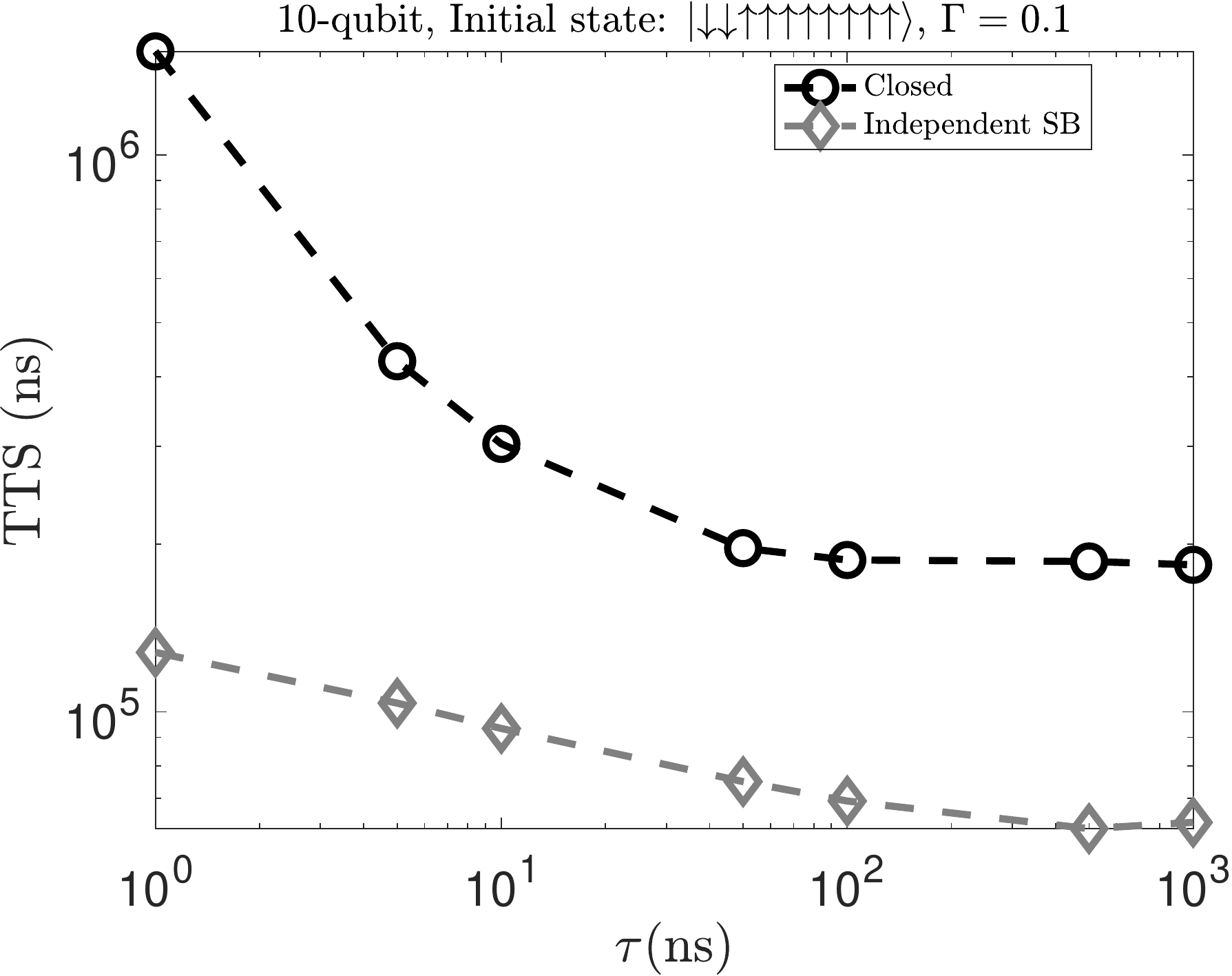}\label{fig:8qubitsunitary}}
\caption{(a) $N=8$, (b) $N=10$. $\Gamma = 0.1$ and $N_{\downarrow}=2$. Independent system-bath coupling.}
\label{fig:8_10_gamma01}
\end{figure}

We plot in Fig.~\ref{fig:8_10_gamma01} the $\mathsf{TTS}$ dependence of anneal times $\tau$, for $N=\{8,10\}$ both with $N_{\downarrow}=2$. Due to the significant diabatic transition to higher excited states, $\mathsf{ARA_\text{Open}}$ tends to perform better than $\mathsf{ARA_\text{Closed}}$ because of the relaxation to the ground state.

\section{Independent vs collective SB coupling}
\label{sec:modelcompare}
We investigate the difference between the  independent and collective system-bath coupling assumption.

\subsection{Long anneal times $\tau$}
As observed in Fig.~\ref{fig:compareall}, for long anneal times $\tau$ the success probabilities given by simulation with the collective SB coupling assumption are higher than those given by independent SB coupling.
In fact, the steady state solution $\rho(s = 1,\tau \rightarrow \infty)$ of the Lindbladian master equation in the weak coupling limit is the Gibbs state set by the problem Hamiltonian. 
\begin{equation}
\rho(s = 1,\tau \rightarrow \infty) = \frac{e^{- \beta H_S(1)}}{\mathcal{Z}} = \frac{e^{- \beta H_0}}{\mathcal{Z}}\ .
\label{eq:rhoS0}
\end{equation}
where $\mathcal{Z} = \textrm{Tr}\left(\exp\left(- \beta H_0 \right)\right)$.

Recall that the permutation symmetry allows us to write the Hamiltonian in terms of only a set of total spin operators of two subsystems,
\begin{equation}
 \hat S^{x,z}_1 \equiv \frac{1}{2}\sum_{i=1}^{|N_{\uparrow}|}\hat\sigma^{x,z}_i,\quad
 \hat S^{x,z}_2 \equiv \frac{1}{2}\sum_{i=|N_{\uparrow}|+1}^{N}\hat\sigma^{x,z}_i \,.
\label{ra:eq:s1s2}
\end{equation}


For collective SB coupling model, the dynamics occurs entirely in the two subspaces with maximum eigenvalues
of $S_1^2$ and $S^2_2$. The Lindbladian is defined according to the projection of $H(s)$ to such  subspaces.
For collective SB coupling model, the steady state is given by:
\begin{equation}
\rho'(s = 1,\tau \rightarrow \infty) = \frac{e^{- \beta H_0'}}{\mathcal{Z'}} 
\label{eq:rhoS0}
\end{equation}
where $\mathcal{Z'} = \textrm{Tr}\left(\exp\left(- \beta H_0' \right)\right)$ and $H_0' = PH_0P$. $P = \sum_{\alpha} \ket{\psi_{\alpha}}\bra{\psi_{\alpha}}$ and $\ket{\psi_{\alpha}} = \left|1;w_1\right> \otimes \left|2;w_2\right>$ is the basis that has the maximum eigenvalues of $S_1^2$ and $S_2^2$.

A simple proof below shows that for long anneal times the success probability given by collective SB coupling model is always higher than the one given by independent SB coupling model.
\begin{proof}
Since $H_0'$ is the projection of $H_0$ in the subspace with the largest values of \((\hat{\bf{S}}_1)^2\) and \((\hat{\bf{S}}_2)^2\), $\mathcal{Z'}<\mathcal{Z}$. The success (ground state) probability $\langle g | \rho'(s = 1, \tau \rightarrow \infty) | g \rangle = \frac{e^{- \beta e_g}}{\mathcal{Z'}} > \frac{e^{- \beta e_g}}{\mathcal{Z}} = \langle g | \rho(s = 1, \tau \rightarrow \infty) | g \rangle$, where $e_g$ is the ground state energy. 
\end{proof}

\subsection{Intermediate anneal times $\tau$}
\begin{figure}[h!]
\centering
\includegraphics[width=0.8\textwidth]{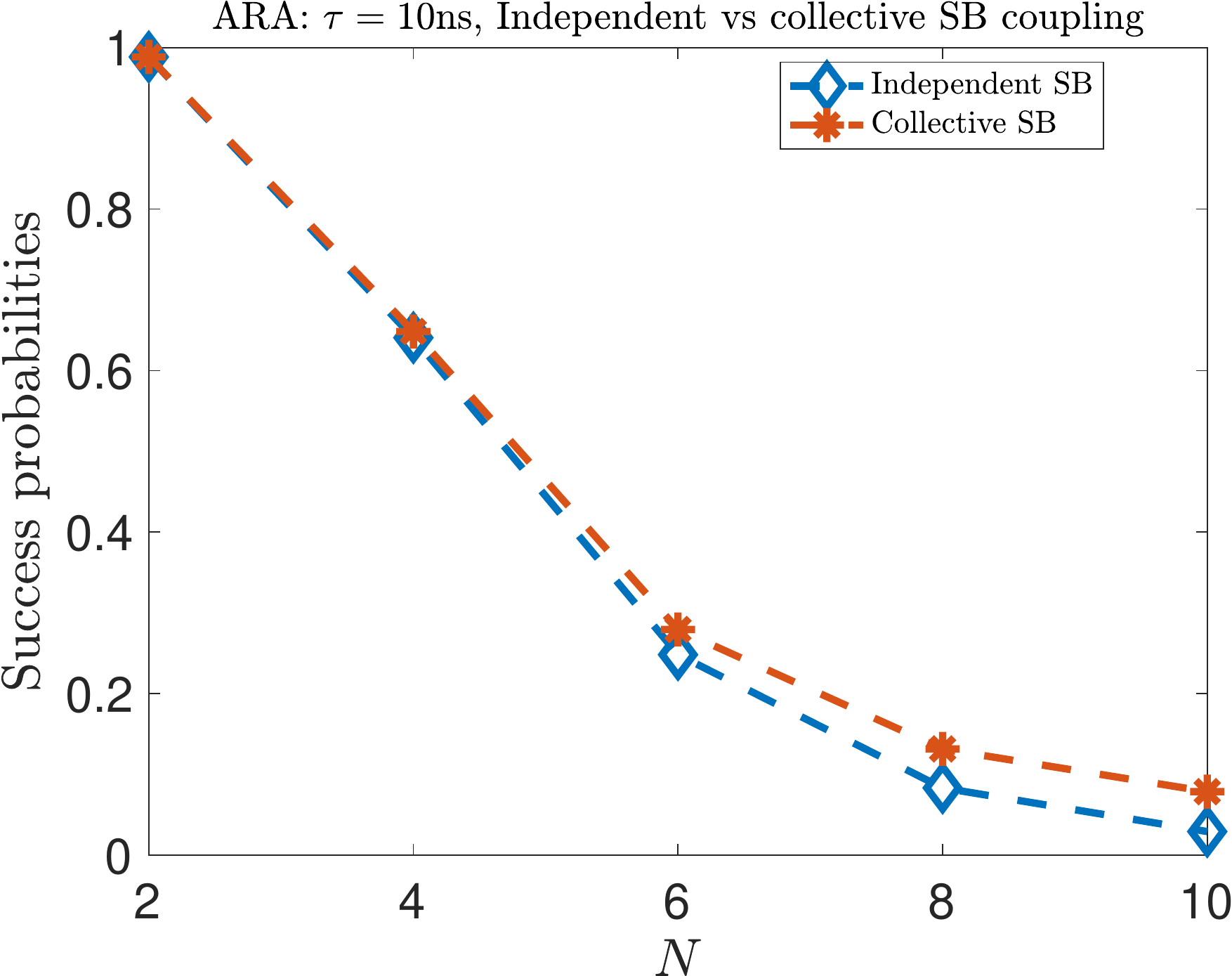}
\caption{Dependence of success probability as a function of $N$ due to the two different SB coupling model. $c=0.5$, $\Gamma=1$ and $\tau = 10$ns.}
\label{fig:indvscol_n}
\end{figure}

In general, since the assumption of independent SB coupling allows decoherence to many more eigenstates of $H(s)$ outside the maximal spin subspaces, it gives a lower success probability than the independent SB model for almost all anneal times $\tau$. For the example of $c=0.5$ and $\tau = 10$ns, we see in Fig.~\ref{fig:indvscol_n} that the difference of the success probability set by the two models diverges as system size $N$ increases. This is expected since the dimension reduction due to the restriction to maximal spin subspaces also scales exponentially in $N$.




\section{Conclusion of this chapter}
We studied the performance of ARA as a function of transverse field $\Gamma$, initial state $H_{\text{init}}$ and system size $N$. In general, except for the cases where the initial state is equal to the correct solution, ARA performs better with larger $\Gamma$. 

We observe non-monotonic dependence of success probabilities in terms of the anneal times $\tau$ for $\mathsf{ARA_\text{Open}}$. For long anneal times, $\mathsf{ARA_\text{Closed}}$ reaches the solution with certainty, while $\mathsf{ARA_\text{Open}}$ reaches the Gibbs state. For $\mathsf{ARA_\text{Open}}$, the difference due to independent system-bath coupling assumption and collective system-bath coupling assumption is studied for both long anneal times $\tau$ and intermediate anneal times $\tau$. In general, the former case always results in a lower success probability (or longer TTS).

$\mathsf{ARA_\text{Open}}$ tends for outperform $\mathsf{ARA_\text{Closed}}$ in the following conditions:
\begin{enumerate}
  \item Small $\Gamma$.
  \item Initial state far away from the correct solution.
  \item Large $N$.
  \item Intermediate anneal times $\tau$.
\end{enumerate}
The first three conditions result in a small and sharp gap for diabatic transitions for intermediate anneal times $\tau$. Because of the environmentally-induced relaxation back to the ground state after diabatic transitions to higher excited states, $\mathsf{ARA_\text{Open}}$ performs better than $\mathsf{ARA_\text{Closed}}$ with the above conditions, in agreement with the prospects of quantum enhancement for diabatic annealing~\cite{Crosson2020}.

%% file: Conclusion.tex
\chapter{Conclusion}
\label{chap: conclusions}
In the first part of this dissertation, we studied time-dependent quantum trajectory techniques which can be applied to three different areas: 1) simulation of the adiabatic master equation, 2) simulation of slow noise with $1/f$ power spectrum and a time-dependent system Hamiltonian, 3) simulation of weak measurements and feedback error correction protocols in quantum annealing. 

For the trajectories simulation of the adiabatic master equation, we  derived the time-dependent quantum trajectory theory and a corresponding numerical method, and performed simulations of several examples of problem gadgets with D-Wave annealers settings and relatively large problem sizes $N$. We also showed that this technique can provide up to a factor $N$ advantage over directly solving the master equation. The cost of running many trajectories, which is required to recover the master equation evolution, can be further minimized by running the trajectories in parallel. The quantum trajectories method provides insight into individual quantum jump trajectories and their statistics, thus shedding light on open system quantum adiabatic evolution beyond the master equation. For the simulation of $1/f$ noise in QA, based on the previously studied Bloch vector-magnetic field approach, we formulated a stochastic fluctuator Hamiltonian simulation method that includes a series of fluctuator terms with different flipping frequencies and Gaussian couplings that altogether constitute the $1/f$ power spectrum. We, for the first time, showed how to incorporate them into the inherently time-dependent annealing Hamiltonians, and construct an overall time-dependent stochastic Hamiltonian for parallelizable simulations. We studied in depth how noise with $1/f$ power spectrum affects quantum annealing, and we also studied  temperature effects, loss of coherence ($T_1$ and $T_2$) in QA under different fluctuator parameters. We allowed the noise fluctuators to take any form of multi-level operators and any noise-axis direction depending on the superconducting qubits and circuit decoherence process. We showed how to append these operators to two different superconducting circuit Hamiltonians. We fit our simulation results with experimental data, and showed how dynamical decoupling can extend the coherence of qubits under $1/f$ noise. For the simulation of weak measurement and feedback error correction protocols in quantum annealing, we derived the feedback master equation provided that the feedback is Markovian (feedback delay $\tau \rightarrow 0$) and further gave the timescale condition for feedback Markovianity. For realistic feedback which is subjected to non-negligible feedback delay and restrictions of the form of the feedback Hamiltonian, we studied the effectiveness of feedback error correction under such limitations and showed how to optimize the feedback delay time depending on the annealing schedule. 

In the second part of this dissertation we studied the open-system behavior of reverse annealing and showed how it can benefit from environment-induced relaxation. We presented a numerical study of the iterative reverse annealing (IRA) protocol applied to the $p$-spin model ($p=3$), including pausing, in an open system setting accounting for dephasing in the energy eigenbasis  which results in thermal relaxation. We considered both independent and collective dephasing and demonstrated that in both cases the open system dynamics can substantially enhance the performance of reverse annealing. We also studied the $p$-spin problem with $p=2$ with D-Wave annealer settings and examined the dependence of the performance on parameters such as problem size, anneal time, inversion points, etc. We performed open-system simulations that match the experimental data well. We found that only the independent dephasing model gives a correct description of reverse annealing in noisy D-Wave annealers, due to the spin symmetry breaking of classical inputs. Semi-classical simulations of spin-vector Monte Carlo also supplement our understanding of the experimental data well. For adiabatic reverse annealing (ARA), we performed simulations of the $p$-spin model ($p=3$) with an extra initialization Hamiltonian term. We found the conditions that result in a small and sharp gap for diabatic transitions with intermediate anneal times. Under such conditions, ARA performs better in an open-system setting, due to the environmentally-induced relaxation back to the ground state after diabatic transitions to higher excited states.

There is still much to understand about the topics we have studied, for example, the simulation of slow noise in superconducting qubits. First, we do not know how to simulate slow noise with $1/f^{\alpha}$ spectrum, with $\alpha \neq 1$, in particular using the stochastic Schr\"{o}odinger equation approach of time-dependent annealing and superconducting qubit Hamiltonian. The steady state of a qubit after Landau-Zener transition under $1/f$ telegraph noise is non-trivial~\cite{vestgaarden2008nonlinearly} and depends on the coupling strength and switching rate of the fluctuators. It would be interesting to come up with a similar observation in the case of quantum annealing and many qubits. As for weak measurements and feedback error correction, there are also questions worth further research, such as: how to perform the protocol for more than $1$ qubit, whether we should apply feedback based on the global excited states or the local excited states of each qubit, and how to adapt the protocol beyond the weak-coupling limit.

%% file: appendix_proof.tex


\section{Quantum trajectories for time-dependent adiabatic master equations}
\subsection{Error estimates}
\label{app:A}
Let us assume that the system is in a pure state at time $t$, i.e., $\rho_S(t) = \ketbra{\psi(t)}$, and let us consider a single time-step. In a single trajectory, the evolution of $\ket{\psi(t)}$ involves two possibilities: no jump or a jump. The ensemble average of trajectories after one finite time-step mainly involves two kinds of errors: the error associated with the norm of the state vector (Sec.~\ref{sssec:en}), and the error associated with the probability elements in a finite time-step (Sec.~\ref{sssec:ep}).

\subsubsection{Error associated with the norm squared of a no-jump trajectory}
\label{sssec:en}
The Schr\"{o}dinger equation of the effective Hamiltonian in the case of a no-jump trajectory is given by
\begin{equation}
\label{eqt:Schrapp2}
\frac{d\ket{\psi(t)}}{dt} = -i H_{\text{eff}}(t) \ket{\psi(t)} \ .
\end{equation}
The resulting state after one time-step $\Delta t$ is:
\begin{equation}
\label{eqt:euler} 
\ket{\tilde{\psi}(t+\Delta t)} =V_{\text{eff}}(t+\Delta t, t)\ket{\psi(t)}  \ ,
\end{equation}
where 
\beq 
V_{\text{eff}}(t+\Delta t, t) = {\cal T}\exp\left[-i\int_t^{t+\Delta t}H_{\text{eff}}(t')dt'\right]
\eeq 
is a non-unitary contractive evolution operator ($\vertiii{ V_{\text{eff}}(t+\Delta t, t)} \leq 1$ for every induced norm) and $\ket{\tilde{\psi}(t+\Delta t)}$ is unnormalized since $H_{\text{eff}}(t)$ is not hermitian.
We can expand the time-ordered exponential for $V_{\text{eff}}(t+\Delta t, t)$ as
\begin{align}
\label{Eqt: TDtaylor}
&{}{\cal T}\exp\left[-i\int_t^{t+\Delta t}H_{\text{eff}}(t')dt'\right]    \\
&=\mathds{1} - i \int_{t}^{t+\Delta t} dt' H_{\text{eff}}(t')   + \left(-i\right)^2 \frac{1}{2!} \mathcal{T}\left(\int_{t}^{t+\Delta t} dt' H_{\text{eff}}(t')\right)^2 +
\left(-i\right)^3 \frac{1}{3!}\mathcal{T}\left(\left(\int_{t}^{t+\Delta t} dt' H_{\text{eff}}(t')\right)^3\right) \cdots \nonumber
\end{align}
For any $H_{\text{eff}}(t)$ that is $C^{K}$ on an open interval containing $[0, t_f]$ ($H^{(k)}_{\text{eff}}(t)$ is continuous and bounded for $1\leq k \leq K$):
\begin{equation}
\int_{t}^{t+\Delta t} dt' H_{\text{eff}}(t') = 
\sum\limits_{k = 0}^{K-1}\frac{H^{(k)}_{\text{eff}}(t)}{(k+1)!}(\Delta t)^{k+1} + O((\Delta t)^{K+1})\, ,
\end{equation}
as $\Delta t \rightarrow 0$. Assume $H_{\text{eff}}(t)$ is $C^{K}$ and $K \geq 2$. Let $\dot{H}_{\eff}(t)\equiv H^{(1)}_{\text{eff}}(t)$ denote the time derivative of $H_\eff(t)$.
We have:
\bes
\begin{align}
\label{Eqt: integral}
\int_{t}^{t+\Delta t} dt' H_{\text{eff}}(t')&= H_{\text{eff}}(t)\Delta t + \frac{1}{2!}\dot{H}_{\text{eff}}(t) (\Delta t)^2 + O((\Delta t)^3) \,,\\
\mathcal{T}\left(\int_{t}^{t+\Delta t} dt' H_{\text{eff}}(t')\right)^2 &= 2\int_{t}^{t+\Delta t} dt' H_{\text{eff}}(t')  \int_{t}^{t'} dt'' H_{\text{eff}}(t'') \nonumber\\
&= H_{\text{eff}}(t)^2(\Delta t)^2 + \frac{1}{2}H_{\text{eff}}(t)\dot{H}_{\text{eff}}(t)(\Delta t)^3 + \frac{1}{2}\dot{H}_{\text{eff}}(t)H_{\text{eff}}(t)(\Delta t)^3 + O((\Delta t)^4)\ \,, \\
\mathcal{T}\left(\int_{t}^{t+\Delta t} dt' H_{\text{eff}}(t')\right)^3 &= H_{\text{eff}}(t)^3(\Delta t)^3 + O((\Delta t)^4) \,.
\end{align}
\ees
Therefore, we have, to second order in $\Delta t$:
\beq
\label{eq:A7}
V_{\text{eff}}(t+\Delta t, t) = \mathds{1} -i \left( H_{\text{eff}}(t)\Delta t + \frac{1}{2!}\dot{H}_{\text{eff}}(t) (\Delta t)^2 \right) + \left(-i\right)^2 \frac{1}{2!} H_{\text{eff}}(t)^2(\Delta t)^2  + O((\Delta t)^3)\ ,
\eeq
so that:
\begin{align}
&{} V^{\dagger}_{\text{eff}}(t+\Delta t, t) V_{\text{eff}}(t+\Delta t, t) \nonumber \\
=&{}  \mathds{1} + i\Delta t \left(H^{\dagger}_{\text{eff}}(t) - H_{\text{eff}}(t)\right) 
+ \frac{1}{2!}(\Delta t)^2\left[ i\dot{H}^{\dagger}_{\text{eff}}(t) - i\dot{H}_{\text{eff}}(t) 
- \left(H^{\dagger 2}_{\text{eff}}(t) + H^2_{\text{eff}}(t)\right)\right] + O((\Delta t)^3) \ . 
\end{align}
We approximate $N(t,\Delta t)\equiv \|\ket{\tilde{\psi}(t+\Delta t)} \|^2$ by:
\beq
\label{Eqt: Step2}
N(t,\Delta t) = \braket{\tilde{\psi}(t+\Delta t)|\tilde{\psi}(t+\Delta t)} 
\approx \bra{\psi(t)}\left(\mathds{1} + i \Delta t\left(H^{\dagger}_{\text{eff}}(t) - H_{\text{eff}}(t) \right)\right)\ket{\psi(t)} \equiv E(t,\Delta t)\, .
\eeq
The approximation error is given by:
\bes
\begin{align}
\delta (t,\Delta t) &\equiv N(t,\Delta t)-E(t,\Delta t)\\
 &= \frac{1}{2}(\Delta t)^2\bra{\psi(t)}\left[ i\dot{H}^{\dagger}_{\text{eff}}(t) - i\dot{H}_{\text{eff}}(t) 
- \left(H^{\dagger 2}_{\text{eff}}(t) + H^2_{\text{eff}}(t)\right)\right] \ket{\psi(t)}+ O((\Delta t)^3) \\
&\leq \frac{(\Delta t)^2}{2}\vertiii{i(\dot{H}^{\dagger}_{\text{eff}}(t) - \dot{H}_{\text{eff}}(t)) - ( H^{\dagger 2}_{\text{eff}}(t) + H^2_{\text{eff}}(t))} + \vertiii{O((\Delta t)^3)}\ ,
\label{Eqt: errornorm}
\end{align}
\ees
where $\|\cdot\|$ is the operator norm (largest singular value).

Eq.~\eqref{Eqt: errornorm} gives the relation between the error of the norm square approximation and the time-step. It is also helpful to consider the sources of error here as they will be used later. This error mainly arises from two sources during the truncations of Taylor expansion: 
\begin{enumerate}
\item The truncation of the Taylor expansion of the integral [Eq.~\eqref{Eqt: integral}] to keep only $H_{\text{eff}}(t)\Delta t$. This turns Eq.~\eqref{Eqt: TDtaylor} into $\exp\left(-iH_{\text{eff}}(t)\Delta t\right)$.  For this to hold, we require 
\begin{align}
\label{Eqt: case1}
\left\lVert \frac{1}{2}\dot{H}_{\text{eff}}(t) (\Delta t)^2 
+ O((\Delta t)^3) \right\rVert \ll  \lVert H_{\text{eff}}(t)\Delta t \rVert \, ,
\end{align}
implying:
\begin{align}
\Delta t \ll 2\frac{\vertiii{H_{\text{eff}}(t)}}{\vertiii{\dot{H}_{\text{eff}}(t)}}  \, ,
\label{Eqt: 1stcon}
\end{align}
assuming $\dot{H}_{\text{eff}}(t)\neq 0$; if it is then the condition becomes $\Delta t \ll \left(\frac{K!\vertiii{H_{\text{eff}}(t)}}{\vertiii{{H^{(K)}_{\text{eff}}(t)}}}\right)^{1/K}$ for the lowest value of $K$ such that $H^{(K)}_{\text{eff}}(t)\neq 0$, where the superscript denotes the $K$-th derivative.

\item Keeping only the first order term in Eq.~\eqref{eq:A7} afterwards, i.e., $\mathds{1} - iH_{\text{eff}}(t)$.  
This requires, in addition to Eq.~\eqref{Eqt: 1stcon}:
\begin{align}
\label{Eqt: case2}
\left\lVert\frac{1}{2}H^2_{\text{eff}}(t)(\Delta t)^2 + O((\Delta t)^3)
\right\rVert \ll  \vertiii{H_{\text{eff}}(t)\Delta t} \, ,
\end{align}
\end{enumerate}
implying, for all $t$ such that the denominators do not vanish:
 \beq 
 \label{eqt:Deltat2}
 \Delta t \ll  \min \left\{ \frac{\vertiii{H_{\text{eff}}(t)}}{\vertiii{\dot{H}_{\text{eff}}(t)}}, \frac{1}{\vertiii{H_{\text{eff}}(t)}} \right\} \ ,
 \eeq
where we ignored the factor of $2$.
In conclusion, the norm after one time-step is related to the jump probabilities as:
\bes
\begin{align}
N(t,\Delta t) &= 
 \bra{\psi(t)}\left(\mathds{1} + i (\Delta t)\left(H^{\dagger}_{\text{eff}}(t) - H_{\text{eff}} \right)\right)\ket{\psi(t)} \notag \\
 &\qquad + \delta(t,\Delta t) \\
&= 1 - \Delta t \sum\limits_{i}\braket{{A}_{i}^{\dagger}(t) {A}_{i} (t)} + \delta(t,\Delta t) \\
&= 1 - \Delta p(t) + \delta(t,\Delta t) \,,\label{Eqt: squareofnorm}
\end{align}
\ees
where we used Eq.~\eqref{eq:Heff} and defined the (approximate) jump probabilities as:
 \bes
 \label{Eqt: typeofprob}
\begin{align}
&{}\Delta p(t)  = \sum_i \Delta p_i(t) \\
&{}\Delta p_i(t) = \Delta t \braket{ {A}_{i}^{\dagger}(t) {A}_{i} (t)}  \, .
\label{eqt:typeofprob}
\end{align}
\ees
This explains Eqs.~\eqref{eqt:squareofnorm} and~\eqref{eq:jumprate}.

\subsubsection{Error associated with probability elements}
\label{sssec:ep}
We have defined the $\Delta p_i(t)$ and $\Delta p(t)$ in a fixed time-step in Eq.~\eqref{Eqt: typeofprob}.
Note that even in the time-independent case, where both the Hamiltonian and Lindblad operators are time-independent, $1- \Delta p(t)$ is never exactly equal to the norm squared of the state vector after one time-step, and $\Delta p_i(t)$ is not exactly the jump probability inside the time-step. They are only approximations. 

For a finite time-step $\Delta t$, with $p_0$ the probability of having no jump inside the interval $[t, t+\Delta t]$ and $p_1$ the probability of having one jump inside the interval $[t, t+\Delta t]$, we have:
\beq
p_{0} = e^{-\int_t^{t+\Delta t} \lambda(t') dt'} = 1-p_{1}  \,.
\label{eq:p_0}
\eeq
Note that as $\Delta t \rightarrow 0$, $p_{0} = 1 - \lambda(t)\Delta t + o(\Delta t)$, $p_{1} \simeq \lambda(t)\Delta t$.
We shall focus on the case where the time-step is sufficiently small such that the probability of two or more jumps occurring within a single time-step is negligible.\footnote{E.g., the quantum jump can be described by a Poisson process with a state-dependent inhomogeneous jump rate, with two or more jumps as successive one and no-jump processes, and $\sum_{n\geq 2}p_{n} = o(\Delta t)$ as $\Delta t\rightarrow 0$.} 

First, we can expand the exponential as
\begin{align}
p_0 
&=1 - \int_t^{t+\Delta t} 
\lambda(t')dt' + \frac{1}{2}\left(\int_t^{t+\Delta t} \lambda(t') dt'
\right)^2 \cdots \nonumber
\end{align}
During the time window from $t$ to $t+\Delta t$, $\lvert\psi(t')\rangle$ [required to calculate $\lambda(t')$] is obtained by solving the Schr\"{o}dinger equation with the effective Hamiltonians [Eq.~\eqref{eqt:Schrapp2}] and renormalizing the solution during the integration. As we show in Sec.~\ref{sec:C}, in any finite interval the norm squared of the unnormalized state vector [Eq.~\eqref{Eqt: Step2}] is equal to $p_0$. This is the reason why we can use the waiting time distribution (or Gillespie algorithm~\cite{gillespie1977exact}) as our simulation method in the main text.

Second, since
\begin{align}
\int_{t}^{t+\Delta t} dt' \lambda(t') = \sum\limits_{k = 0}^{\infty}\frac{\lambda^{(k)}(t)}{(k+1)!}(\Delta t)^{k+1} \ ,
\end{align}
we have
\begin{align}
\label{eqt:p0allterms}
= 1 - \lambda(t)\Delta t + e_p \ ,
\end{align} 
where the error $e_p$ associated with the probability elements in a fixed time-step is:
\begin{equation}
e_p = - \frac{1}{2}\frac{d}{dt}
\lambda(t)(\Delta t)^2 + \frac{1}{2}
\lambda^2(t)(\Delta t)^2 + O((\Delta t)^3) \,.
\end{equation}
This should be much smaller than the first order term $\lambda(t) \Delta t$. Therefore, we need 
\begin{align}
\Delta t \ll \left|\frac{ \lambda(t) }{  \lambda^2(t)-\frac{d}{dt} \lambda(t)}\right|  \, .
\label{Eqt: Poissoncon}
\end{align}
In the time-independent case, this reduces to 
\beq
\Delta t \ll \frac{1}{ \lambda} \ .
\eeq

\subsection{Proof of equivalence between the master equation and trajectories formulations}
\label{app:B}

Our goal in this section is to show how the master equation, Eq.~\eqref{eqt:effdiagonalform}, can be recovered from the quantum trajectories formulation, and to find a bound on the time-step $\Delta t$. This generalizes the proof for the time-independent case found in~\cite{molmer1993monte}.

\subsubsection{To jump or not to jump}
\label{sssec:2}
The probability elements $\Delta p(t)$ and $\Delta p_i (t)$ are important for determining whether a jump occurs and if a jump does occur, which jump type occurs. 
In order to determine if a jump occurs or not, we draw a random number $\epsilon$, uniformly distributed between 0 and 1. If $\Delta p(t) < \epsilon$, which is almost always the case since $\Delta p(t)$ is very small, no jump occurs. In the case of no jump, $\ket{\psi(t)}$  evolves according to the effective Schr\"{o}dinger equation, Eq.~\eqref{eqt:Schrapp2}. At time $t+\Delta t$ we simply renormalize the solution of Eq.~\eqref{eqt:euler}:
\bes
\begin{align}
\left|\psi(t + \Delta t)\right> &= \frac{1}{\sqrt{\braket{\tilde{\psi}(t+\Delta t)| \tilde{\psi}(t+\Delta t)}}}\ket{\tilde{\psi}(t+\Delta t)}\\&\stackrel{\eqref{Eqt: squareofnorm}}{=} \frac{1}{\sqrt{1 - \Delta p(t) + \delta}}\ket{\tilde{\psi}(t+\Delta t)} \ .
\end{align}
\ees

If $\Delta p(t) > \epsilon$, the state undergoes an abrupt jump and we choose the new wavefunction among the different states $A_{i}\ket{\psi(t)}$ and renormalize:
\bes
\begin{align}
\left|\psi(t + \Delta t)\right> &= \frac{{A}_{i}(t)\ket{\psi(t)}}{\sqrt{\bra{\psi(t)} {A}_{i}^{\dagger}(t) {A}_{i} (t)\ket{\psi(t)}}} \\
&\stackrel{\eqref{eqt:typeofprob}}{=} 
\sqrt{\frac{\Delta t}{\Delta p_i(t)}} {A}_{i}(t)\ket{\psi(t)}\ .
\end{align}
\ees
Which type of jumps occurs is determined according to the probability 
\bes
\begin{align}
\Pi_i(t) &= \frac{\Delta p_i(t)}{\Delta p(t)} = \frac{\braket{\psi(t)|{A}_{i}^{\dagger}(t){A}_{i}(t)|\psi(t)}\Delta t}{\sum\limits_{i}\braket{\psi(t)|{A}_{i}^{\dagger}(t){A}_{i}(t)|\psi(t)}\Delta t} \\
&= \frac{\braket{{A}_{i}^{\dagger}(t){A}_{i}(t)}}{\lambda(t)}\ ,
\label{eq:Pi(t)}
\end{align}
\ees
where 
\beq
\lambda(t) = \sum_{i}\langle {A}_{i}^{\dagger}(t){A}_{i}(t)\rangle 
\label{eq:lambda}
\eeq
is the time-dependent jump rate.

\subsubsection{Averaging over trajectories}
\label{sssec:3}

Let $H_{\text{eff}}(t)$ be $C^{K}$ with $K \geq 2$. 
We first express the mean value $\bar{\sigma}_{S}(t)$ as a sum over the non-Hermitian evolution [with probability $1 - \Delta p(t)$] and the jump trajectories [with probability $\Delta p(t)$], so that as $\Delta t \rightarrow 0$ we have: 
\bes
\begin{align}
\label{Eqt: averaging}
\bar{\sigma}_{S}(t + \Delta t) 
&= \left[1 - \Delta p(t)\right]\frac{\ket{\tilde{\psi}(t + \Delta t)}}{\sqrt{1 - \Delta p(t) + \delta}}\frac{\bra{\tilde{\psi}(t + \Delta t)}}{\sqrt{1 - \Delta p(t) + \delta}}  \nonumber \\
& + \Delta p(t) \sum\limits_{i} \Pi_{i}(t) \sqrt{\frac{\Delta t}{\Delta p_i(t)}}{A}_{i}(t)  \ket{\psi(t)}\sqrt{\frac{\Delta t}{\Delta p_i(t)}}\bra{\psi(t)}{A}_{i}^{\dagger}(t) \\
&\stackrel{\eqref{eq:Pi(t)}}{=}   \frac{1 - \Delta p(t)}{1 - \Delta p(t) + \delta}\ket{\tilde{\psi}(t + \Delta t)}\bra{\tilde{\psi}(t + \Delta t)}
+ \Delta t \sum\limits_{i}{A}_{i}(t)\sigma_{S}(t){A}_{i}^{\dagger}(t) 
\label{Eqt: noninnerproductform}
\end{align}
\ees
Combining Eq.~\eqref{eqt:euler} with Eq.~\eqref{eq:A7} we have:
\beq
\ket{\tilde{\psi}(t+\Delta t)} =
\left[ \mathds{1} -i \left( H_{\text{eff}}(t)\Delta t + \frac{1}{2!}\dot{H}_{\text{eff}}(t) (\Delta t)^2 \right) + \left(-i\right)^2 \frac{1}{2!} H_{\text{eff}}(t)^2(\Delta t)^2  + O((\Delta t)^3) \right] \ket{\psi(t)}  \ .
\eeq
Recall that in Eq.~\eqref{eqt:Deltat2} we gave conditions allowing us to neglect the $O((\Delta t)^2)$ terms.
Thus: 
\begin{align}
\label{eq:B3}
\bar{\sigma}_{S}(t + \Delta t) 
&=  \frac{1 - \Delta p(t)}{1 - \Delta p(t) + \delta}\left( \mathds{1} - i H_{\text{eff}}(t) \Delta t 
+ O((\Delta t)^2)\right)\bar{\sigma}_{S}(t) \left(\mathds{1} + i H^{\dagger}_{\text{eff}}(t) \Delta t
+O((\Delta t)^2)\right)\\
&\phantom{==}+ \Delta t \sum\limits_{i}{A}_{i}(t)\bar{\sigma}_{S}(t){A}_{i}^{\dagger}(t)\nonumber  \, .
\end{align} 
\begingroup
%
where we have replaced $\sigma_{S} (t)$ by $\bar{\sigma}_{S} (t)$ after averaging over many trajectories. 
Rearranging this expression into a form that exposes the terms that will become the master equation,
the expression for the averaged state at $t + \Delta t$ becomes:
\bes
\begin{align}
\bar{\sigma}_{S}(t + \Delta t) &= \bar{\sigma}_{S}(t) + i\Delta t\left(\bar{\sigma}_{S}(t)H_{\text{eff}}^{\dagger}(t) - H_{\text{eff}}(t)\bar{\sigma}_{S}(t) \right)+ \Delta t \sum_{i} {A}_{i}(t) \bar{\sigma}_{S}(t){A}_{i}^{\dagger}(t) \\
&\phantom{==}
- \frac{\delta}{1 - \Delta p(t) + \delta} \left[ \bar{\sigma}_{S}(t) +  i\bar{\sigma}_{S}(t)\left(H_{\text{eff}}^{\dagger}(t) -H_{\text{eff}}(t)\right) \bar{\sigma}_{S}(t) \Delta t \right] + O((\Delta t)^2)\ .
\label{eq:B5b}
\end{align}
\ees
Note that $\delta$ is $O((\Delta t)^2)$ as $\Delta t \rightarrow 0$ [Eq.~\eqref{Eqt: errornorm}], and $\Delta p(t) = \Delta t \sum_{i}\braket{\psi(t)|{A}_{i}^{\dagger}(t){A}_{i}(t)|\psi(t)}$ is $O(\Delta t)$ as $\Delta t \rightarrow 0$, so that:
\bes
\begin{align}
\frac{\delta}{1 - \Delta p(t) + \delta} &= 
\frac{O((\Delta t)^2)}{1 - O(\Delta t) + O((\Delta t)^2)} \\
&=O((\Delta t)^2)\, . 
\end{align} 
\ees
Therefore line~\eqref{eq:B5b} can be absorbed into $O((\Delta t)^2)$, and we are left with:
\begin{align}
\frac{\bar{\sigma}_{S}(t + \Delta t) - \bar{\sigma}_{S}(t)}{\Delta t} =&
-i\left(H_{\text{eff}}(t)\bar{\sigma}_{S}(t) - \bar{\sigma}_{S}(t)H_{\text{eff}}^{\dagger}(t) \right)\notag \\
&+  \sum_{i} {A}_{i}(t) \bar{\sigma}_{S}(t){A}_{i}^{\dagger}(t) 
+ O(\Delta t) \ ,
\end{align}
which becomes the master equation, Eq.~\eqref{eqt:effdiagonalform}, in the $\Delta t \to 0$ limit.

\subsubsection{Upper bound on $\Delta t$}
\label{ssec:upperbound}
The above proof takes $\Delta t \rightarrow 0$. We would like to know how small the time-step $\Delta t$ should be in order for the approximations made to be valid. In Eq.~\eqref{eq:B3}, we expanded the time-ordered exponential, and kept only the first order terms. This is equivalent to the criteria in Eqs.~\eqref{Eqt: case1} and ~\eqref{Eqt: case2}, summarized as a single condition in Eq.~\eqref{eqt:Deltat2}. 
As shown in Sec.~\ref{sssec:en}, this also automatically makes the error in the norm squared approximation $\delta$ small. We also need to satisfy Eq.~\eqref{Eqt: Poissoncon}, in order to accurately approximate the probability elements. Taken together, therefore:
 \beq 
 \Delta t \ll  \min \left\{ \frac{\vertiii{H_{\text{eff}}(t)}}{\vertiii{\dot{H}_{\text{eff}}(t)}}, \frac{1}{\vertiii{H_{\text{eff}}(t)}}, \left|\frac{ \lambda(t) }{  \lambda^2(t)-\dot \lambda(t)}\right| \right\} \ .
 \label{eq:B7}
 \eeq
In practice, choosing a constant time-step that satisfies Eq.~\eqref{eq:B7} in the whole timespan $[0, t_f]$ is sufficient, though one might prefer to implement an adaptive time-step tailored to the instantaneous value of the R.H.S. 
\subsection{On the validity of waiting times (quantum time-dependent operators)}
\label{sec:C}
Here we show the validity of using the waiting time distribution in the case of time-dependent operators. The argument presented here is based on Ref.~\cite{Breuer:2002} and we extend it to the time-dependent case.

Let us denote by $\ket{\psi(t)}$ and $\ket{\tilde{\psi}(t)}$ the normalized and unnormalized state vectors respectively, and let us assume they are equal at time $t$. This can happen when $t = 0$ or any time immediately after each jump. Let $t^+ \equiv t+\tau$, where $\tau$ can be as large as is possible until the next jump occurs,  and
\begin{equation}
V_{\text{eff}}(t^+, t) = {\cal T}\exp\left[-i\int_t^{t^+}H_{\text{eff}}(t')dt'\right] \ ,
\end{equation}
Then:
\bes
\begin{align}
\ket{\tilde{\psi}(t + \tau)} &=  
 V_{\text{eff}}(t^+, t)\ket{\psi(t)}  \ ,\\
\ket{\psi(t^+)} &=  
 \frac{V_{\text{eff}}(t^+, t)\ket{\psi(t)}}{\left\lVert V_{\text{eff}}(t^+, t)\ket{\psi(t)} \right\rVert} \,.
 \label{eq:C2b}
\end{align}
\ees
Then, starting from $t$, for any future $t^{+} > t$, we have
\bes
\begin{align}
\frac{d\phantom{,}}{dt^{+}}\left\lVert \ket{\tilde{\psi}(t^{+})} \right\rVert^2 &= \phantom{}\frac{d\phantom{,}}{dt^{+}}\left\lVert{\cal T}\exp\left[-i\int_t^{t^+}H_{\text{eff}}(t')dt'\right]\ket{\psi(t)}\right\rVert^2 \\
&= \frac{d\phantom{,}}{dt^{+}}\bra{\psi(t)}V_\eff^{\dagger}(t^+,t) V_\eff(t^+,t)\ket{\psi(t)} \\
&=\bra{\psi(t)}V_\eff^{\dagger}(t^+,t)(+i)H^{\dagger}_{\text{eff}}(t^{+})V_\eff^{}(t^+,t)\ket{\psi(t)} +
\bra{\psi(t)}V_\eff^{\dagger}(t^+,t)(-i)H_{\text{eff}}(t^{+})V_\eff^{}(t^+,t)\ket{\psi(t)} \\
&=-\bra{\psi(t)}V_\eff^{\dagger}(t^+,t)\left(\sum\limits_{i}{A}^{\dagger}_{i}(t^{+}){A}_{i}(t^{+})\right)V_\eff^{}(t^+,t)\ket{\psi(t)} \\
&=  -\left\lVert V_{\text{eff}}(t^{+}, t)\ket{\psi(t)} \right\rVert^2 \sum\limits_{i} \bra{\psi(t^{+})} {A}^{\dagger}_{i}(t^{+}){A}_{i}(t^{+}) \ket{\psi(t^{+})}  \, ,
\end{align}
\ees
where in the last equality we used Eq.~\eqref{eq:C2b}.

Let $N(t^{+}) \equiv \| \ket{\tilde{\psi}(t^{+})} \|^2$, as in Eq.~\eqref{Eqt: Step2}.  We have
\begin{equation}
\frac{d\phantom{,}}{dt^{+}} N(t^{+}) = -N(t^{+})  \lambda(t^+)   \, ,
\end{equation}
where $\lambda$ is the time-dependent jump rate [Eq.~\eqref{eq:lambda}].
The solution to this differential equation with the initial condition $N(t) = 1$ is
\begin{equation}
N(t^{+}) = \exp\left(-\int_t^{t^{+}} \lambda(t') dt'\right) = p_0(t^{+})  \, ,
\end{equation}
where $p_0$ is the probability of not having any jump inside the interval [Eq.~\eqref{eq:p_0}], which we have now shown to be equal to the the norm squared of the unnormalized state vector for any finite interval $[t, t+\tau]$. 

No commutators of operators at different times appears in the derivation. The use of the waiting time distribution is therefore valid for time-dependent operators as long as the correlation matrix is positive.

\subsection{Effect of the annealing schedule on the $8$-qubit chain example} 
\label{app:Chain}

We provide the functional form for the (D-Wave 1) annealing schedule functions $A(t)$ and $B(t)$ used in Sec.~\ref{sec:8qubitchain} in Fig.~\ref{fig:5a}.  We compare this non-linear schedule to a linear annealing schedule with an overall energy scale chosen to closely match the energy scale at which the first annealing schedule curves cross.  The dynamics associated with the linear annealing schedule [shown in Fig.~\ref{fig:5b}] is somewhat different than the non-linear one. This can be attributed to the differences between the two schedules. In order to ensure that the two sets of schedules intersect at the same energy scale, this requires the linear schedule to start from a significantly smaller energy scale.  This results in some important effects on the dynamics.  First, this energy scale is not large enough to ensure that the thermal state at $s = 0$ has negligible weight on excited states.  Since we use the ground state as the initial state of the simulation, which is now sufficiently different from the thermal state at $s=0$, the dissipative dynamics causes visible changes in the state immediately, which can be see as both the dip in Fig~\ref{fig:5a} near $s=0$ and the large number of excitations at the beginning of the anneal in Fig.~\ref{fig:5c}.  Second, even after this initial dip, the lower energy scale associated with the linear schedule means that thermal depopulation of the ground state occurs generally sooner relative to the non-linear schedule studied in the main text, although at a lower rate because of the linear form of the schedule.  Furthermore, because the transverse field remains significant compared to the Ising term for longer in the anneal than in the linear case, repopulation of the ground state due to thermal relaxation occurs for a longer period of time.  However, ultimately, the ground state probability at the end of the anneal is not significantly different than that shown in Fig.~\ref{fig:3qubits8qubits}. The jump statistics are shown in Fig.~\ref{fig:5c}, and closely resemble the non-linear case shown in the inset of Fig.~\ref{fig:8qubitssingle}.

\begin{figure*}[htb!] 
\centering
\subfigure[\ ]{\includegraphics[width=0.32\textwidth]{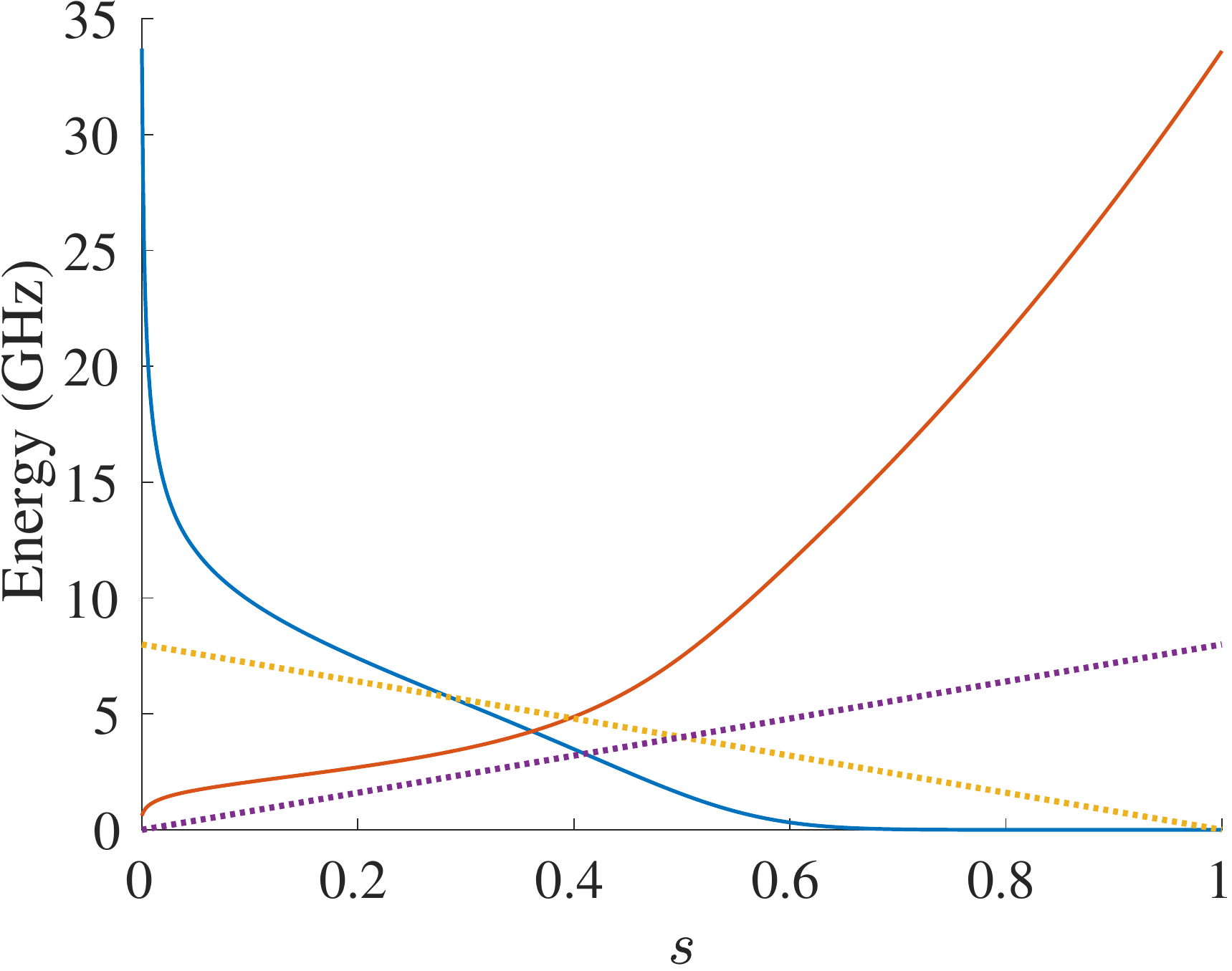}\label{fig:5a}} 
\subfigure[\ ]{\includegraphics[width=0.32\textwidth]{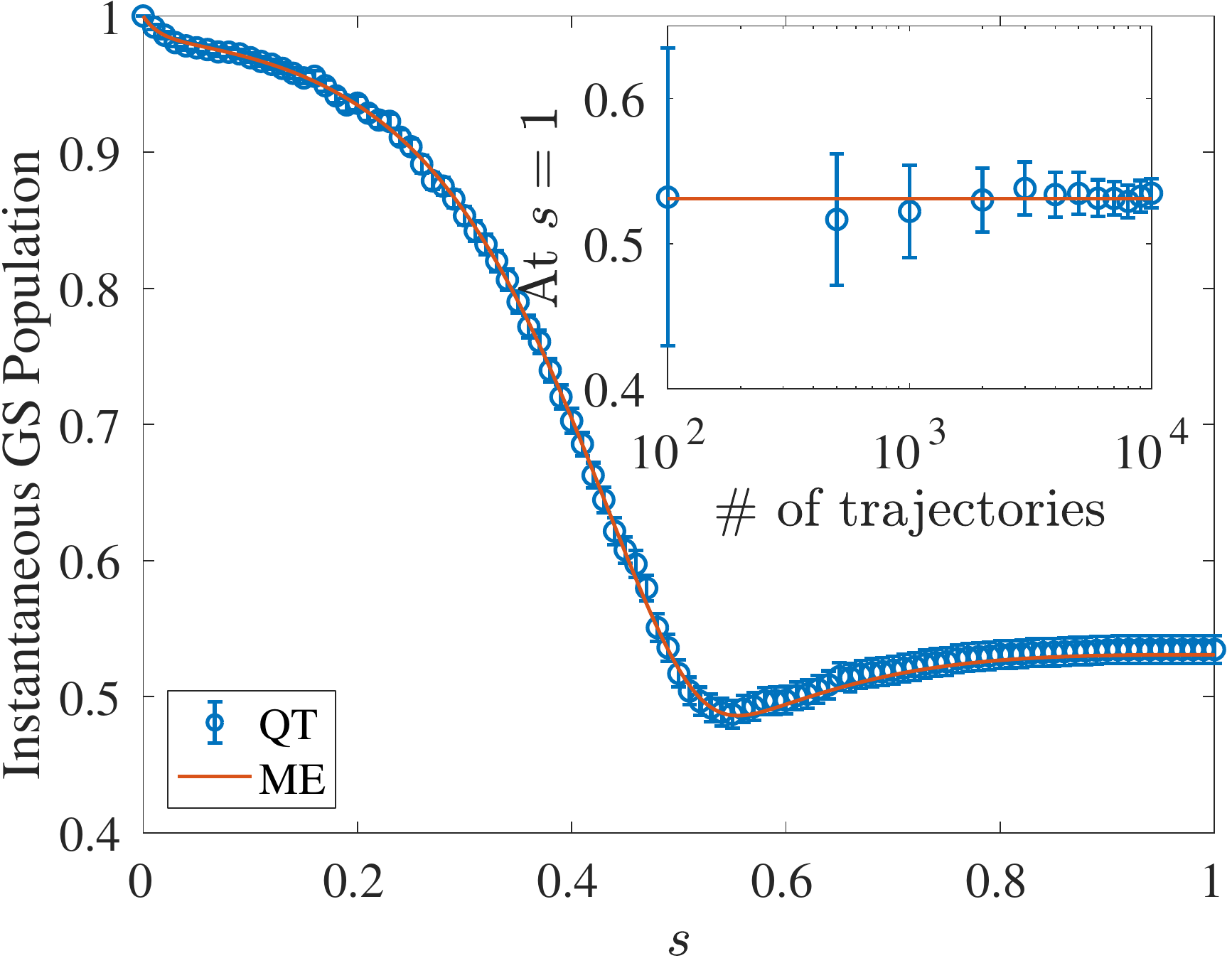}\label{fig:5b}} 
\subfigure[\ ]{\includegraphics[width=0.32\textwidth]{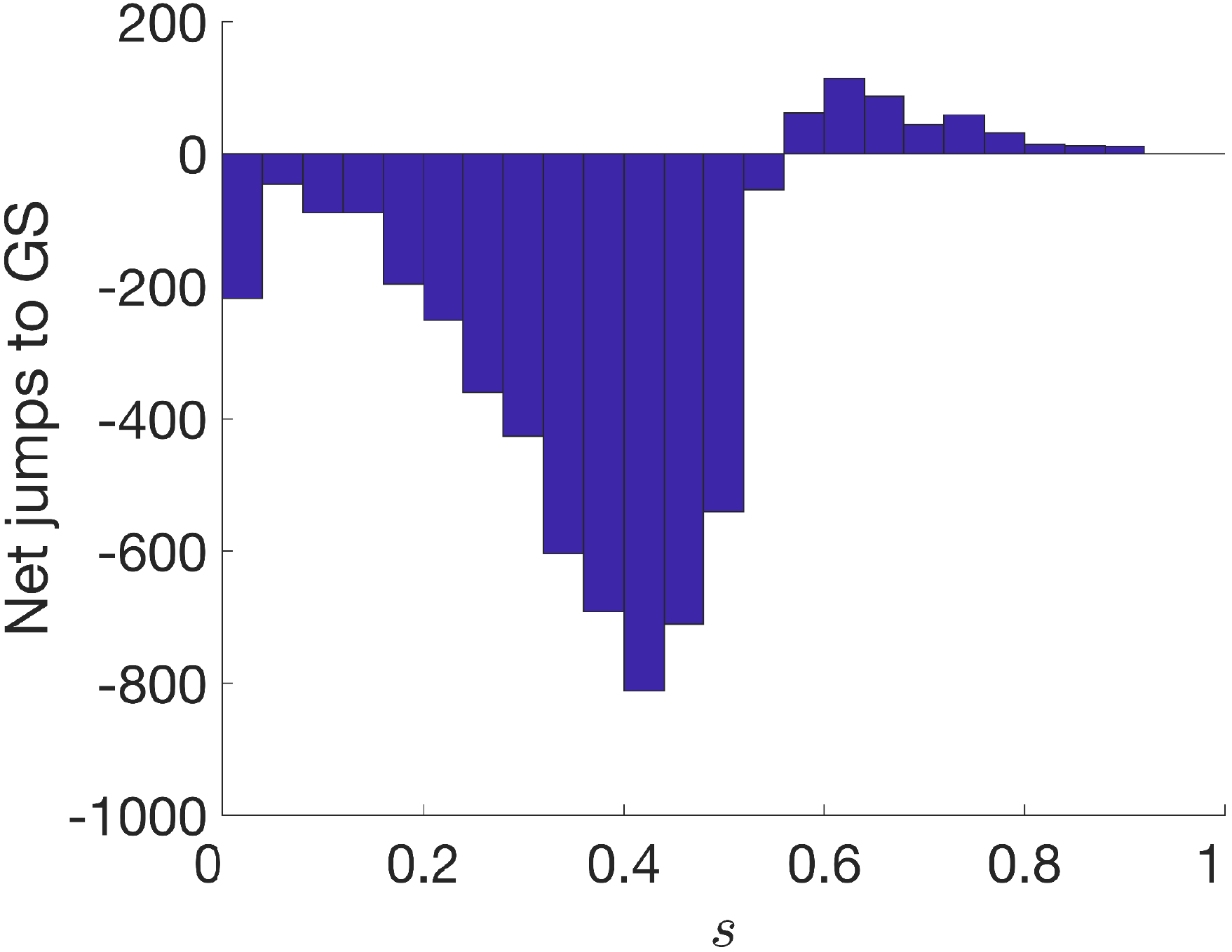}\label{fig:5c}} 
\caption{(a): Dotted lines: Linear schedule; Solid lines: D-Wave schedule used in Fig.~\ref{fig:3qubits8qubits}; (b): Same as in Fig.~\ref{fig:3qubits8qubits} (the evolution of the population in the instantaneous ground state for the $8$-qubit problem), but with a linear schedule. Inset: the convergence of the ground state population towards the AME results. (c): A histogram of the net number of jumps to the instantaneous ground state (GS). It shares the same pattern as in the non-linear schedule case (the inset of Fig.~\ref{fig:8qubitssingle}), but the net number of jumps out of the ground state is smaller due to the smaller energy scale in the linear schedule.}
\label{fig:8qubitslinear}
\end{figure*}

\subsection{Proof that eigenstates of $H_S$ are modified only under jumps}
\label{app:eigenstate-proof}

Recall that the Lindblad operators are defined in Eq.~\eqref{eq:Lindblad2} of the main text as:
\begin{align}
L_{\alpha, \omega}(t) &= \sum_{a,b} \delta_{\omega, \eps_b(t) - \eps_a(t)} \bra{\eps_a(t)} A_\alpha \ket{\eps_b(t)} 
| \eps_a(t) \rangle  \langle \eps_b(t)| \,.
\end{align}
After inverting Eq.~\eqref{eq:L} we have 
\begin{equation}
A_{i,\omega}(t) = \sum_{\alpha}u^{*}_{i,\alpha}(\omega)L_{\alpha, \omega}(t) 
\end{equation}
for the new Lindblad operators corresponding to a diagonalized $\gamma$ matrix of decay rates.

Assume for simplicity that the $\gamma$ matrix is already diagonal, so that $u$ is just an identity transformation and $i$ is a relabeling of $\alpha$; then
\begin{align}
A_{i,\omega}(t) &= \sum_{a,b} \delta_{\omega, \eps_b(t) - \eps_a(t)} \bra{\eps_a(t)} A_\alpha \ket{\eps_b(t)} 
| \eps_a(t) \rangle  \langle \eps_b(t)| \nonumber \\
A^{\dagger}_{i,\omega}(t) &= \sum_{a,b} \delta_{\omega, \eps_b(t) - \eps_a(t)} \bra{\eps_b(t)} A_\alpha \ket{\eps_a(t)} 
| \eps_b(t) \rangle  \langle \eps_a(t)| \,.
\end{align}

Consider the drift term 
\[
-\frac{1}{2}\sum_{i} \left[ A_{i}^{\dagger}(t)A_{i}(t) - \braket{A_{i}^{\dagger}(t)A_{i}(t)}\right]\ket{\psi(t)}dt\ .
\] 
Since $A_i(t)$ comes from the redefinition where the index $i$ includes the Bohr frequencies [Eq.~\eqref{eqt:Aired}]: $\sqrt{\gamma'_i(\omega)}A_{i,\omega}(t)\rightarrow A_i(t)$, the drift term becomes the following after we reintroduce the Bohr frequencies:
\begin{equation}
- \frac{1}{2}\sum_{i}\sum_{\omega} \gamma'_i(\omega)\left[ A_{i,\omega}^{\dagger}(t)A_{i,\omega}(t) - \bra{\psi(t)} A_{i,\omega}^{\dagger}(t)A_{i,\omega}(t) \ket{\psi(t)} \right]\ket{\psi(t)}dt \,. 
\label{eq:explicit}
\end{equation}

\subsubsection{When $\ket{\psi(t)}$ is an eigenstate of $H_{S}(t)$}
If the $\ket{\psi(t)}$ is an eigenstate of $H_{S}(t)$ (denoted as $\ket{\eps_b(t)}$ here),
\begin{align}
A_{i,\omega}(t)\ket{\psi(t)} = A_{i,\omega}(t)\ket{\eps_b(t)} &=  \sum_{a} \delta_{\omega, \eps_b(t) - \eps_a(t)} \bra{\eps_a(t)} A_\alpha \ket{\eps_b(t)} 
| \eps_a(t) \rangle \nonumber \\
\bra{\psi(t)}A^{\dagger}_{i,\omega}(t) = \bra{\eps_b(t)}A^{\dagger}_{i,\omega}(t) &= \sum_{a} \delta_{\omega, \eps_b(t) - \eps_a(t)} \bra{\eps_b(t)} A_\alpha \ket{\eps_a(t)} \langle \eps_a(t)| \,.
\end{align}

Concentrate on the term of the parenthesis inside each summation of Eq.~\eqref{eq:explicit}, i.e.
\begin{equation}
\left[ A_{i,\omega}^{\dagger}(t)A_{i,\omega}(t) - \bra{\eps_b(t)} A_{i,\omega}^{\dagger}(t)A_{i,\omega}(t) \ket{\eps_b(t)}  \right]\ket{\eps_b(t)} \,.
\label{eq:paraterm}
\end{equation}
The first term is:
\begin{align}
&\phantom{{}={}} A_{i,\omega}^{\dagger}(t) A_{i,\omega}(t)\ket{\eps_b(t)}  \nonumber\\
&= \left(\sum_{a^{\prime},b^{\prime}} \delta_{\omega, \eps_{b^{\prime}}(t) - \eps_{a^{\prime}}(t)} \bra{\eps_{b^{\prime}}(t)} A_\alpha \ket{\eps_{a^{\prime}}(t)} 
| \eps_{b^{\prime}}(t) \rangle  \langle \eps_{a^{\prime}}(t)| \right) \sum_{a} \delta_{\omega, \eps_b(t) - \eps_a(t)} \bra{\eps_a(t)} A_\alpha \ket{\eps_b(t)} 
| \eps_a(t) \rangle \nonumber\\
&= \sum_{a,b^{\prime}} \delta_{\omega, \eps_{b^{\prime}}(t) - \eps_{a}(t)} \delta_{\omega, \eps_b(t) - \eps_a(t)}\bra{\eps_{b^{\prime}}(t)} A_\alpha \ket{\eps_{a}(t)}  \bra{\eps_a(t)} A_\alpha \ket{\eps_b(t)}
| \eps_{b^{\prime}}(t) \rangle  \nonumber\\
&= \sum_{a,b^{\prime}} \delta_{0, \eps_{b^{\prime}}(t) - \eps_{b}(t)} \delta_{\omega, \eps_b(t) - \eps_a(t)}\bra{\eps_{b^{\prime}}(t)} A_\alpha \ket{\eps_{a}(t)}  \bra{\eps_a(t)} A_\alpha \ket{\eps_b(t)}
| \eps_{b^{\prime}}(t) \rangle   \, ,
\label{eq:firstterm}
\end{align}
where the sum over $b^{\prime}$ denotes the sum over $\ket{\eps_{b^{\prime}}}$ sharing the same energy as $\ket{\eps_{b}}$.
The second term is:
\begin{align}
&\phantom{{}={}} \bra{\eps_b(t)} A_{i,\omega}^{\dagger}(t)A_{i,\omega}(t) \ket{\eps_b(t)} \ket{\eps_b(t)}  \nonumber\\
&= \left(\sum_{a^{\prime}} \delta_{\omega, \eps_b(t) - \eps_{a^{\prime}}(t)} \bra{\eps_b(t)} A_\alpha \ket{\eps_{a^{\prime}}(t)} \langle \eps_{a^{\prime}}(t)| \right) \left(\sum_{a} \delta_{\omega, \eps_b(t) - \eps_a(t)} \bra{\eps_a(t)} A_\alpha \ket{\eps_b(t)} 
| \eps_a(t) \rangle \right)\ket{\eps_b(t)} \nonumber\\
&= \left(\sum_{a} \delta_{\omega, \eps_b(t) - \eps_{a}(t)} \bra{\eps_b(t)} A_\alpha \ket{\eps_{a}(t)} \bra{\eps_a(t)} A_\alpha \ket{\eps_b(t)} \right)\ket{\eps_b(t)} \,.\nonumber\\
\label{eq:secondterm}
\end{align}

Subtracting Eq.~\eqref{eq:secondterm} from Eq.~\eqref{eq:firstterm} yields the drift term [Eq.~\eqref{eq:paraterm}], which is not zero, but a linear combination of degenerate eigenstates with the same energy $\eps_b(t)$. Before the jump happens the environment leads to the redistribution of $\ket{\eps_b(t)}$ to other states in the same energy manifold. (The Lamb shift $H_{\text{\text{LS}}}(t) = \sum_{i,\omega}S_{i}(\omega) A^\dagger_{i,\omega}(t) A_{i,\omega}(t) $ also yields the same effect.) Since they all share the same energy, this does not affect the overlap with the ground state. If the evolution by $H_{\text{S}}(t)$ is adiabatic, such a linear combination will stay in the same energy manifold and this explains the square-pulse like behavior in the overlapping with the ground state in Fig.~\ref{fig:8qubitssingle} of the main text.

\subsubsection{No degeneracy in $\eps_b(t)$}
If there are no degenerate states with energy $\eps_b(t)$, 
\begin{align}
&\phantom{{}={}} A_{i,\omega}^{\dagger}(t) A_{i,\omega}(t)\ket{\eps_b(t)}  = \sum_{a} \delta_{\omega, \eps_b(t) - \eps_a(t)}\bra{\eps_{b}(t)} A_\alpha \ket{\eps_{a}(t)}  \bra{\eps_a(t)} A_\alpha \ket{\eps_b(t)}
| \eps_{b}(t) \rangle 
\end{align}
This cancels with $\bra{\eps_b(t)} A_{i,\omega}^{\dagger}(t)A_{i,\omega}(t) \ket{\eps_b(t)} \ket{\eps_b(t)}$ [Eq.~\eqref{eq:secondterm}] and the drift term [Eq.~\eqref{eq:paraterm}] becomes zero.
\newline

\subsection{Derivation of Eq.~\eqref{eq:inctunnrate}}
\label{app:lindbladfermi}

When the state is $\ket{\psi_1(t)}$, its jump rate $\lambda_{1 \rightarrow 0}(t) $ to $\ket{\psi_0(t)}$ comprises the summation of Lindblad terms responsible for the $1 \rightarrow 0$ transition:
\begin{equation}
\lambda_{1 \rightarrow 0}(t) = \sum_{\alpha \in \{1\rightarrow 0\}}\braket{{A}_\alpha^{\dagger}(t){A}_{\alpha}(t)} \,.
\end{equation}
The summation is over the number of qubits $n$. Since each qubit is coupled to its own environment with an independent noise source, the $\gamma$ matrix in Eq.~\eqref{eq:unitarytransform} is already diagonal. From Eq.~\eqref{eq:Lindblad2}
we know that
\begin{equation}
L_{\alpha, \omega_{10}}(t) = \sum_{a,b} \delta_{\omega_{10}, \eps_b(t) - \eps_a(t)} \bra{\eps_a(t)} \sigma^{z}_{\alpha} \ket{\eps_b(t)} | \eps_a(t) \rangle  \langle \eps_b(t)|    \,.
\end{equation}

Assume that Bohr frequency $\omega_{10}(t)$ is due only to the $1\rightarrow 0$ transition (even if it is not, the other terms would be annihilated by the matrix element $\langle \psi_1(t) |\dots |\psi_1(t) \rangle$). The Lindblad operators ${A}_{\alpha}(t)$ have the form:
\begin{equation}
{A}_{\alpha}(t) = \sqrt{\gamma_{\alpha}(\omega_{10})}\langle\psi_0(t)|\sigma^{z}_{\alpha} |\psi_1(t)\rangle| \psi_0(t) \rangle  \langle \psi_1(t)|  \,.
\end{equation}
Therefore:
\begin{align}
\lambda_{1 \rightarrow 0}(t) &= \sum_{\alpha}  \langle \psi_1(t) | {A}^{\dagger}_{\alpha}(t)  {A}_{\alpha}(t)   |\psi_1(t) \rangle    \\
&=  \sum_{\alpha=1}^{n} \gamma_\alpha(\omega_{10})|\langle\psi_0(t)|\sigma^{z}_{\alpha} |\psi_1(t)\rangle|^2 \nonumber  \,.
\end{align}
Here $\gamma_\alpha(\omega_{10})$ is evaluated with respect to the Ohmic spectral density.

\section{Experimental iterative reverse annealing}
\subsection{\label{append:spectrum}Spectrum of the $p$-spin problem with $p=2$ and scaling of the minimum gap with $N$}
The spectrum of $p$-spin problem and the ordering of energies of different spin sectors can be found in~\cite{Bapst2012}. For the particular case of $n=4, p=2$ and the D-Wave 2000Q annealing schedule, we plot the spectrum in Fig.~\ref{fig:myspectrum}. From Fig.~\ref{fig:myspectrum} we can see that the instantaneous first excited state $\epsilon_{1}(s)$ and the instantaneous ground state $\epsilon_{0}(s)$ converge at $s=1$.

The energy gap between instantaneous eigenenergies $\epsilon_{i}(s)$ and $\epsilon_{j}(s)$ is $\Delta_{ij}(s) = \epsilon_{i}(s) - \epsilon_{j}(s)$. The corresponding minimum energy gap is $\Delta_{ij} = \min_{s} \Delta_{ij}(s)$. For $p=2$, we are interested instead in the minimum energy gap between the ground state $\epsilon_{0}(s)$ and the second excited state $\epsilon_{2}(s)$, since the $1$st excited state $\epsilon_{1}(s)$ and the ground state converge at $s=1$. It is denoted by $\Delta$.  
\begin{equation}
    \Delta = \Delta_{20} = \min_{s} \Delta_{20}(s) = \min_{s} \epsilon_{2}(s) - \epsilon_{0}(s) \,.
\end{equation}
We calculate and plot in Figure~\ref{fig:RA_20_size}(c) the value of $\Delta$, for $N = \{4, \cdots, 22\}$. We also show the $s$ where the minimum gap is found, i.e.
\begin{equation}
    s_{\mathrm{min}} = \mathrm{argmin} \Delta_{20}(s) \,.
\end{equation}

\begin{figure}[h!]
    \centering
    \includegraphics[width=0.6\linewidth]{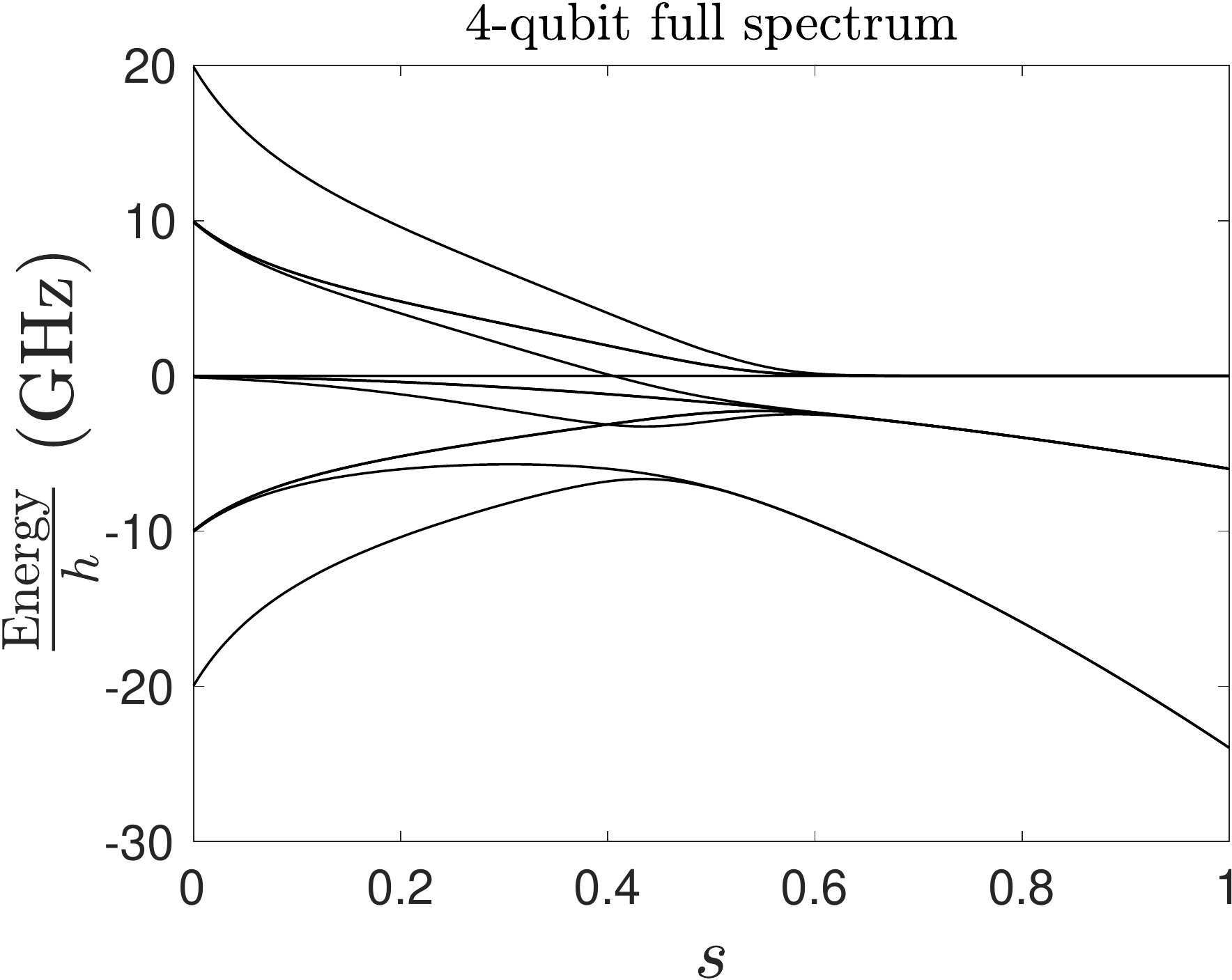}
    \caption{Full spectrum of 4 qubits.}
    \label{fig:myspectrum}
\end{figure}


\subsection{\label{append:successbound}Bound on success probabilities, for closed system and open-system dynamics with collective system-bath interaction}
If the dynamics preserve spin symmetry (for example in closed system Schr\"{o}dinger equation and open-system simulation with the collective system-bath coupling assumption), there exists a natural upper bound on the maximum success probabilities achievable in reverse annealing. The upper bound is the population of the initial state in the maximum spin sector.

For the example of $N=4$, there are two degenerate ground states ($\ket{0000}$ ($\ket{0} \equiv \ket{\uparrow}$, $\ket{1} \equiv \ket{\downarrow}$) and $\ket{1111}$) of the problem Hamiltonian ($H_0$) and they both belong to the maximum spin subspace of $S=S_{\text{max}}=\frac{N}{2}=2$. While the computational basis state with one spin down, for example, $\ket{0001}$ is a first excited state of $H_0$, it is not a basis of the maximum spin subspace $S=2$. A superposition of the computational basis states with one spin down, i.e. $\frac{\ket{0001}+\ket{0010}+\ket{0100}+\ket{1000}}{\sqrt{4}}$, however belong to the maximum spin subspace $S=2$, and this is the case for~\cite{Passarelli2019}. Unfortunately, the D-Wave machine does not allow such an initialization. 

In general, suppose that the initial computational basis state is $\ket{\psi(t=0)}_{\text{comp}}$, with a particular (normalized) magnetization $m_0$ (Eq.~\ref{eq:mo}). It is represented as a linear combination of states, each of which has a fixed value of total spin momentum $S$ and (normalized) magnetization $m_0$,
\begin{equation}
\label{eq:compinitialstate}
    \ket{\psi(t=0)}_{\text{comp}} = \sum_{S=S_{\text{min}}}^{S_{\text{max}}}a_S\ket{S, m_{S}=m_0}. 
\end{equation}
From the theory of addition of angular momentum, we know that $S_{\text{min}} = N/2-\lfloor N/2 \rfloor$ and $S_{\text{max}} = N/2$. The total spin momentum (integer or half-integer) $S \in \{S_{\text{min}}, S_{\text{min}} + 1, \dots, S_{\text{max}}\}$. $\ket{S, m_{S}=m_0}$ is a simultaneous eigenstate of $\bold{S}^2 = \sum_{\alpha\in\{x,y,z\}}S^{\alpha}$ and $S^z=\frac{1}{2}\sum_{i=1}^{N}\sigma_i^z$ (or $H_p$), with eigenvalues:
\begin{align}
    \bold{S}^2\ket{S, m_{S}=m_0} &= S(S+1)\ket{S, m_{S}=m_0} \,,\\
    S^z\ket{S, m_{S}=m_0} &= \Big(\frac{N}{2}m_0\Big)\ket{S, m_{S}=m_0}
\end{align}
In Eq.~(\ref{eq:compinitialstate}), $|a_S|^2$ is the initial state's population in that spin subspace. 


For the $p$-spin Hamiltonian, in a closed system the state in each spin subspace develops independently due to the preservation of spin symmetry of the Hamiltonian~\cite{nishimori:reverse-pspin-2}. At the end of a single cycle of reverse annealing ($r=1$), we have from Eq.~(\ref{eq:compinitialstate}) that 
\begin{equation}
    U(2t_\text{inv},0)\ket{S, m_{S}=m_0} = \sum_{M=-S}^{S} c_{M} \ket{S, m_{S}=2M/N},
\end{equation}
where $c_{M}$ is the relative phase developed in the other bases of the spin $S$ subspace at the end of the time evolution. The (unnormalized) magnetization $M = (\frac{N}{2}m_{S}) \in \{-S, -S+1, \dots, S\}$.

Therefore, the final state of $1$-cycle reverse annealing (denoted by $\ket{\psi}_{\text{fin}}$) can be expressed as:
\begin{align}
    \ket{\psi}_{\text{fin}} &= U(2t_\text{inv},0)\ket{\psi(t=0)}_{\text{comp}} \nonumber\\
    &= \sum_{S=S_{\text{min}}}^{S_{\text{max}}}a_S\left(\sum_{M=-S}^{S} c_M \ket{S, m_{S}=2M/N}\right)\,.
\end{align}
We know that the all-up state ($\ket{0}^{\otimes N} = \left|\uparrow\right>^{\otimes N} = \ket{\text{up}}$) and all-down state ($\ket{1}^{\otimes N} = \left|\downarrow\right>^{\otimes N} = \ket{\text{down}}$) are both solution states of $H_0$, and moreover they lie in the maximum spin subspace $S=S_{\text{max}}$. Particularly, $\ket{\text{up}} = \ket{S_{\text{max}}, 1}$ and  $\ket{\text{down}} = \ket{S_{\text{max}}, -1}$. Therefore, projecting $\ket{\psi}_{\text{fin}}$ onto the $\ket{\text{up}}$ gives:
\begin{align}
    &\phantom{==}\braket{\text{up}|\psi}_{\text{fin}}\nonumber\\
    &=  \sum_{S=S_{\text{min}}}^{S_{\text{max}}}a_S\left(\sum_{M=-S}^{S} c_{M} \braket{\text{up}|S, m_{S}=2M/N}\right)\\
    &= a_{S_\text{max}}c_{S_\text{max}}\,.
\end{align}
Similarly, $\braket{\text{down}|\psi}_{\text{fin}} = a_{S_\text{max}}c_{-S_\text{max}}$. 

The total success probability is thus bounded by: 
\begin{align}
&\phantom{==}|\braket{\text{up}|\psi}_{\text{fin}}|^2 + |\braket{\text{down}|\psi}_{\text{fin}}|^2 \nonumber\\
&= |a_{S_\text{max}}|^2(|c_{S_\text{max}}|^2+|c_{-S_\text{max}}|^2) \nonumber\\
&\leq |a_{S_\text{max}}|^2 \,,
\end{align}
where equality is met with 
$(|c_{S_\text{max}}|^2+|c_{-S_\text{max}}|^2)=1$.
\newline \rightline{$\square$}

The upper bound $|a_{S_\text{max}}|^2$ is the population of the initial state in the maximum spin subspace. For the initial state of $\ket{0001}$, we have $a_{S_\text{max}} = \frac{1}{\sqrt{4}}$ since $\ket{S=S_\text{max}=2, m_{S}=0.5} = \frac{\ket{0001}+\ket{0010}+\ket{0100}+\ket{1000}}{\sqrt{4}}$. Therefore, the total success probability is bounded by $|a_{S_\text{max}}|^2 = \frac{1}{4}$. For the other examples in the main text, the initial state of $\ket{0011}$ has $a_{S_\text{max}} = \frac{1}{\sqrt{{{4}\choose{2}}}} = \frac{1}{\sqrt{6}}$; while the initial state of $\ket{00000001}$ has $a_{S_\text{max}} = \frac{1}{\sqrt{8}}$.

By expressing the above equations in terms of density matrix, we can also generalize the same conclusion to open-system simulation of collective system-bath coupling, where the population of each spin-sector is preserved during dynamics.

\subsection{\label{append:svmc}SVMC and SVMC-TF for reverse annealing}
We outline the algorithm of SVMC and SVMC-TF with application to reverse annealing.
\begin{spacing}{0.8}
\begin{algorithm}[H]
\scriptsize
\label{alg:1}
\caption{SVMC and SVMC-TF ($p$-spin reverse annealing)}
\hspace*{\algorithmicindent}
\begin{algorithmic}[1]
\State $\mathcal{H}(s) = -\frac{A(s)}{2} (\sum_{i}^{N} \sin \theta_i) - \frac{B(s) N}{2}\left(\frac{1}{N}\sum_{i}^{N} \cos \theta_i\right)^{p}$, with a specified reverse annealing $s=t$ dependence $s(t)$.
\Procedure{SVMC}{}
\For{$k=1$ to $K$ (number of samples)}
    \State Initial computational basis state: $\ket{0}_i\rightarrow \theta^{k}_{i,t}: 0$, $\ket{1}_i\rightarrow \theta^{k}_{i,t}: \pi$.
    \For{$t=1$ to $T$ (total number of sweeps)}
        \For{$i=1$ to $n$ (number of qubits)}
            \State Randomly choose a new angle $\theta_{i,t}^{'k} \in [0, \pi]$.
            \State Calculate $\Delta E$.
                \If{$\Delta E \leq 0$}
                    \State $\theta_{i,t}^{k} \rightarrow \theta_{i,t}^{'k}$
                \ElsIf {$p < \exp (-\beta\Delta E)$ where $p\in [0,1]$ is drawn with uniform probability.}
                    \State $\theta_{i,t}^{k} \rightarrow \theta_{i,t}^{'k}$
                \EndIf
    	 \EndFor
	\EndFor
	\State Take the mean of $K$ samples: $\bar{\theta}_{i,t} = (\sum_{k=1}^{K}\theta_{i,t}^{'k})/K$.
\EndFor
\EndProcedure
\State \Return $\bar{\theta}_{i,t}$.
\Procedure{SVMC-TF}{}
\For{$k=1$ to $K$ (number of samples)}
    \State Initial computational basis state: $\ket{0}_i\rightarrow \theta^{k}_{i,t}: 0$, $\ket{1}_i\rightarrow \theta^{k}_{i,t}: \pi$.
    \For{$t=1$ to $T$ (total number of sweeps)}
        \For{$i=1$ to $n$ (number of qubits)}
            \State $\theta_{i,t}^{'k} = \theta_{i,t}^{k} + \epsilon_i(s(t/T))$, 
            \State a random $\epsilon_i(s(t/T) \in [-\min \left(1,\frac{A(s(t/T)}{B(s(t/T)}\right) \pi, \min \left(1,\frac{A(s(t/T)}{B(s(t/T)}\right) \pi]$
            \State Calculate $\Delta E$.
                \If{$\Delta E \leq 0$}
                    \State $\theta_{i,t}^{k} \rightarrow \theta_{i,t}^{'k}$
                \ElsIf {$p < \exp (-\beta\Delta E)$ where $p\in [0,1]$ is drawn with uniform probability.}
                    \State $\theta_{i,t}^{k} \rightarrow \theta_{i,t}^{'k}$
                \EndIf
    	 \EndFor
	\EndFor
	\State Take the mean of $K$ samples: $\bar{\theta}_{i,t} = (\sum_{k=1}^{K}\theta_{i,t}^{'k})/K$.
\EndFor
\EndProcedure
\State \Return $\bar{\theta}_{i,t}$.

\Procedure{Projection onto computational basis}{}
\If{$0\leq\theta_{i,t}^{k}\leq \pi/2$}
    \State $\ket{\psi^k(t)}_i = \ket{0}$
\ElsIf {$\pi/2\leq\theta_{i,t}^{k}\leq \pi$}
    \State $\ket{\psi^k(t)}_i = \ket{1}$
\EndIf
\EndProcedure
\State Remark: $\Delta E$, for example, due to the update of qubit $2$, can be expressed in the following form:
\begin{align*}
\phantom{=}&\Delta E_2 \\
= 
&\left(-\frac{A(s)}{2} \left(\sum_{\substack{i=1\\i \ne 2}}^{N}\sin \theta_i + \sin \theta^{'}_2\right) - \frac{B(s)N}{2}\left(\frac{1}{N}\left(\sum_{\substack{i=1\\i \ne 2}}^{N} \cos \theta_i + \cos \theta^{'}_2\right)\right)^{2}  \right) \\
- &\left(-\frac{A(s)}{2} \left(\sum_{i=1}^{N}\sin \theta_i\right) - \frac{B(s)N}{2}\left(\frac{1}{N}\left(\sum_{i=1}^{N} \cos \theta_i\right)\right)^{2} \right) \,.
\end{align*}
\end{algorithmic}
\end{algorithm}
\end{spacing}